\documentclass[12pt, a4paper]{article}
\usepackage{tikz-cd}
\usetikzlibrary{arrows,snakes,shapes.arrows,decorations.markings}
     \tikzstyle{bbc}=[draw,circle,fill=black,scale=.75]
     \tikzstyle{rc}=[circle,fill=red,scale=.6]
     \tikzstyle{wc}=[draw,circle,scale=.75]

\usepackage{color}

\usepackage{latexsym}
\usepackage{amsmath, mathdots, amscd,bbm}
\usepackage{multirow}
\usepackage{amssymb}
\usepackage{euscript}
\usepackage{amsfonts}
\usepackage{verbatim}
\usepackage{epsfig}
\usepackage{hyperref}
\usepackage[margin=2cm,bottom=2.0cm]{geometry}
\usepackage{float}
\usepackage{cite}
\usepackage{fancyhdr} % Required for custom headers
\usepackage{lastpage} % Required to determine the last page for the footer
\usepackage{extramarks} % Required for headers and footers
\usepackage{graphicx} % Required to insert images
\usepackage{lipsum} % Used for inserting dummy 'Lorem ipsum' text into the template
\usepackage{color}
\usepackage{tikz}
\usepackage{xypic}
\usepackage{cancel,slashed}
\numberwithin{equation}{section}

\def\be{\begin{eqnarray}}
\def\ee{\end{eqnarray}}
\def\0{\nonumber}

\providecommand{\Tr}{\textnormal{Tr}}

\providecommand{\mod}{\textnormal{mod}}

\newcommand{\al}[1]{\begin{align}#1\end{align}}

\def\bbZ{\mathbb{Z}}
\def\0{\nonumber}

\begin{document}
\begin{titlepage}

\begin{flushright}

\end{flushright}
 
\vskip 1cm
\begin{center}
 
{\LARGE\bf \boldmath $\mathcal{N}=2$\; Orbi-S-Folds} 
 
 \vskip 2cm
 
{\large Simone Giacomelli,$^{1,2}$ Raffaele Savelli,$^{3}$ and Gianluca Zoccarato$^{3}$}

 \vskip 0.9cm
 
 {\it  $^1$ Dipartimento di Fisica, Universit\`a degli Studi di Milano-Bicocca,\\ Piazza della Scienza 3, I-20126 Milano, Italy \\[2mm]
 
$^2$ INFN, sezione di Milano-Bicocca,\\ Piazza della Scienza 3,  I-20126 Milano, Italy \\[2mm]
 
 $^3$ Dipartimento di Fisica, Universit\`a di Roma ``Tor Vergata'' \\ \& INFN, sezione di Roma ``Tor Vergata'' \\ Via della Ricerca
Scientifica 1, I-00133 Roma, Italy
 }
 \vskip 2cm
 
\abstract{\noindent We introduce a new class of non-compact backgrounds of Type IIB string theory preserving eight supercharges by combining S-folds and non-perturbative 7-branes wrapping orbifolds, and study the four-dimensional superconformal field theories arising at low energy on $D3$-branes probing them. We draw a precise correspondence between this setup and the torus compactification of six-dimensional orbi-instanton theories with a Stiefel-Whitney twist, and use it to determine the main features of such strongly-coupled systems, like central charges, spectra of Coulomb-branch operators, networks of Higgs-branch flows. Finally, with the aim to improve our understanding of the landscape of $\mathcal{N}=2$ superconformal field theories, and possibly to extend their classification beyond rank two, we provide a detailed catalogue of all the rank-three theories that our framework gives access to.}

\end{center}

\end{titlepage}

\tableofcontents

\section{Introduction}\label{Sec:Intro}

The study of supersymmetric quantum field theories with eight supercharges continues to witness an elegant and fruitful marriage between physics and geometry. What makes this context so attractive is the relative ease to reach a remarkably detailed description of models whose dynamics is most of the times intrinsically strongly coupled. This is possible thanks to a vast arsenal of techniques made available essentially by the success of string theory in dealing with singular spaces, which permits handy and oftentimes multifaceted geometric realizations of such theories (see e.g.~\cite{Akhond:2021xio,Argyres:2022mnu} for recent reviews).

Among the many candidate setups for engineering field theories, $D3$-branes probing F-theory backgrounds has always played a pivotal role, because of its versatility in combining the standard ingredients of weakly-coupled braneworld scenarios with non-trivial behaviors in the space of coupling constants. Paradigmatic in this respect has been the discovery of S-folds, originally shown capable to produce four-dimensional (4d) field theories with the elusive $\mathcal{N}=3$ supersymmetry \cite{Garcia-Etxebarria:2015wns,Aharony:2016kai,Garcia-Etxebarria:2016erx}, but soon later recognized as a powerful construction that, when probed together with flat 7-branes, gives rise to a whole new class of $\mathcal{N}=2$ superconformal field theories (SCFTs) \cite{Apruzzi:2020pmv,Giacomelli:2020jel,Giacomelli:2020gee,Bourget:2020mez,Heckman:2020svr}.\footnote{See also \cite{Giacomelli:2023qyc}, which just began scratching the surface of the wild realm of $\mathcal{N}=1$ S-folds.}

There is only one more way to enrich such backgrounds without lowering the supersymmetry of the low-energy field theory on the probes: Adding an orbifold singularity $\mathbb{C}^2/\Gamma$, with $\Gamma\subset SU(2)$ discrete, along the four internal directions of the 7-brane worldvolume. The study of these F-theory configurations, limited to $\Gamma=\mathbb{Z}_k$, and of the resulting 4d SCFTs, has been the subject of \cite{Giacomelli:2022drw}, which however gave up the S-folds, focusing on product geometries like $\mathbb{C}^2/\Gamma\times K3$, with the second factor being the (local) elliptic fibration. This paper wants to extend the exploration of \cite{Giacomelli:2022drw} by incorporating the S-folds: Our point-like probes will generally be moving on singular spaces of the form
\be\label{TypeIIBOrbiSFolds}
\frac{\mathbb{C}^2\big/\,\mathbb{Z}_k\;\;\times\;\,\left(\mathbb{C}\times T^2\right)}{\mathbb{Z}_{h}}\,,
\ee
where $h={\rm lcm}(\ell,m)$, $\ell\leq6$ is the S-fold order, and $m=1,2,3,4,6$ encodes the 7-brane type.\footnote{$\mathbb{Z}_h$ acts on the four local coordinates as $(z_1,z_2,z_3,t)\to\left(e^{2\pi {\rm i}/\ell}\,z_1\,,\,e^{-2\pi {\rm i}/\ell}\,z_2\,,\,e^{2\pi {\rm i}/m}\,z_3\,,\,e^{-2\pi {\rm i}/m}\,t\right)$.}

Key to our analysis will be an alternative realization of S-fold theories in terms of six-dimensional (6d), minimally supersymmetric SCFTs (see e.g.~\cite{Apruzzi:2013yva,Heckman:2013pva,Gaiotto:2014lca, DelZotto:2014hpa,Heckman:2015bfa,Bhardwaj:2015xxa,Bhardwaj:2015oru}), compactified on a torus with almost-commuting holonomies, a.k.a.~Stiefel-Whitney (SW) twists \cite{Ohmori:2018ona,Heckman:2022suy}. The 6d theories, in turn, are engineered in M-theory on the worldvolume of $M5$-branes probing a $M9$-wall \cite{Horava:1996ma}. Such a correspondence follows directly from M/F-theory duality \cite{Ganor:1996mu,Seiberg:1996vs, Ganor:1996pc}, and the orbifold $\mathbb{C}^2/\Gamma$ can be harmlessly included into the picture along the $M9$-wall, transversely to the $M5$-branes, leading in 6d to the so-called orbi-instanton theories \cite{DelZotto:2014hpa, Mekareeya:2017jgc}. However, at the level of the low-energy field theory on the probes, the presence of the orbifold does have an impact on the correspondence, as already observed in the absence of S-folds/SW-twists and 7-branes/$M9$-wall \cite{Ohmori:2015pua,Ohmori:2015pia}: Membranes on vanishing cycles in the M-theory description yield an extra global symmetry ($SU(k)$ for $\Gamma=\mathbb{Z}_k$), which has no analog in the Douglas-Moore quiver description of Type IIB string theory \cite{Douglas:1996sw}. The reason is that the latter is based on a string-worldsheet analysis which is only reliable upon turning on a B-field along the vanishing cycles, thus breaking the extra flavor symmetry. In the M-theory picture, this operation corresponds to activating mass deformations for the symmetry naturally arising in that context.\footnote{In the limit of vanishing B-field, one must use a different duality frame to describe the field theory, and that is furnished by the 6d theory on the tensor branch.} The dual configurations in M-theory and type IIB string theory are summarized in Table \ref{tb:brane}.

The need of mass deformations to make contact with the SCFTs living on $D3$-branes has also been confirmed when 7-branes are present \cite{Giacomelli:2022drw}, and we will argue here that adding the S-fold makes no difference in this regard. The dictionary that we find is extremely easy: The M-theory setup corresponding to the Type IIB orbi-S-fold \eqref{TypeIIBOrbiSFolds} with maximal 7-brane (i.e.~$m=6$) involves an $M9$-wall wrapping $\mathbb{C}^2/\mathbb{Z}_{k\ell}$. For given $k$ and $\ell$ we find several infinite families of theories labelled by a choice of holonomy for the $E_8$ gauge field supported on the $M9$-wall in M-theory, or equivalently for the gauge field supported on the 7-brane in the Type IIB description. The total number of families grows with $k$ (for $\ell\leq 4$), with a rate of growth which depends on the value of $\ell$. The theories within each family are labelled by the number $N$ of probe branes ($M5$-branes in M-theory and $D3$-branes in Type IIB) which can be an arbitrary positive integer. As we increase $N$, the rank (i.e.~the dimension of the Coulomb branch) of the theory increases, and goes like $Nk$ at large $N$.

\begin{table}[t]
\begin{center}
\begin{tabular}{c|c c c}
\hline
Type IIB&$\mathbb R^{1,3}$&$\mathbb C^2/\mathbb Z_{k}$&$\mathbb C$\\
\hline
$N$ $D3$&$\times$&\\
S-fold$_\ell$&$\times$&\\
7-branes$_{6/\ell}$&$\times$&$\times$&\\
\hline
\end{tabular}
\begin{verbatim}


\end{verbatim}
\begin{tabular}{c|c c c c}
\hline
M-theory&$\mathbb R^{1,3}$& $T^2$ & $\mathbb C^2/\mathbb Z_{\ell k}$ & $\mathbb R$\\
\hline
$N$ $M5$&$\times$&$\times$&\\
$M9$&$\times$&$\times$&$\times$&\\
\hline
\end{tabular}
\end{center}
\caption{Diagrams summarizing the dual configurations in type IIB string theory and M-theory. The support of S-fold$_\ell$ (i.e.~the S-fold of type $\mathbb{Z}_\ell$) is in the directions where the action of $\mathbb{Z}_h$ in \eqref{TypeIIBOrbiSFolds} is trivial. The correspondence works when the parameter $m$ below \eqref{TypeIIBOrbiSFolds} is equal to $6$, and the $M9$ is associated to 7-branes of type $6/\ell$.}
\label{tb:brane}
\end{table}

For all these families of 4d SCFTs we introduce and study in detail two classes of mass deformations, which are sort of ``complementary'': One that indeed gets rid of the M-theory symmetry giving us access to the Type IIB picture, and another which instead decouples flavors arising from the 7-branes, thus partially resolving the singularity of the elliptic fibration. For both of them separately we will be able to predict how the key features of the SCFTs change, such as the spectrum of Coulomb-branch (CB) operators, the $a,c$ and flavor central charges and the pattern of Higgs-branch flows (encoded in the so-called Hasse diagram). Interestingly, we find that for $N=1$, namely the theory of lowest rank within each family, these two classes of mass deformations coincide and therefore the tree of mass deformations simplifies. For $N>1$ instead they differ, leading to a more intricate pattern of mass deformations. We will find explicit instances of this within the set of known rank-two theories.

It is a common feature of intrinsically strongly-coupled theories to admit a magnetic-quiver description \cite{Cabrera:2019izd,Benini:2010uu,Cabrera:2019dob,Cabrera:2018jxt, Cremonesi:2015lsa, Ferlito:2017xdq, Closset:2020scj, Closset:2020afy, Giacomelli:2020ryy}, and we will employ this powerful device to characterize RG flows between the 4d SCFTs: Higgsings are realized through quiver subtractions \cite{Cabrera:2018ann,Bourget:2019aer}, whereas mass deformations are implemented via suitable FI deformations \cite{vanBeest:2021xyt, Bourget:2023uhe}, as studied in \cite{Giacomelli:2022drw} for the simpler case of F-theory backgrounds without S-fold ($\ell=1$ in the notation of the present paper).

Finally, our framework is sufficiently general to allow us to identify a large amount of rank-three SCFTs. Motivated by the will to push beyond rank two the classification program of 4d $\mathcal{N}=2$ SCFTs initiated in \cite{Argyres:2015ffa, Argyres:2015gha, Argyres:2016xua, Argyres:2016yzz} and continued in \cite{Martone:2021ixp,Martone:2021drm,Bourget:2021csg}, we will tabulate the main properties of all the rank-three SCFTs arising from our construction and exhibit their trees of mass deformations. We would like to stress that our construction is algorithmic and the scan we perform at rank three could in principle be done at any rank.

The paper is organized as follows. In Section \ref{Sec:SWcompact} we review the SW compactification of 6d orbi-instanton theories which constitute the starting point of our analysis: For a bunch of examples at low orbifold order we derive magnetic quivers, CB spectrum, central charges, and Hasse diagrams. We also derive a precise rule to construct the magnetic quiver in general. Section \ref{Sec:IIBSfolds} deals with the first class of mass deformations, linking the M-theory picture to the Type IIB one: We propose a general rule to derive the magnetic quiver of the deformed theories and explain how to extract their main features from those before the deformation. In Section \ref{Sec:MassDef}, instead, we discuss the other class of mass deformations, lowering the type of 7-brane from the maximal one; we also touch upon theories obtained from removing the 7-brane completely and briefly comment on their 1-form symmetries. Section \ref{Sec:Rank3} contains our detailed compendium of rank-three SCFTs and in Section \ref{Sec:Disc} we draw our conclusions. Appendices \ref{App:Mth} and \ref{App:IIBth} host the specifics of all the theories that did not fit into Sections \ref{Sec:SWcompact} and \ref{Sec:IIBSfolds} respectively.

\section{Stiefel-Whitney compactifications of 6d SCFTs}\label{Sec:SWcompact}

In this section we review the background material needed for our analysis and we set the stage for the exploration of orbi-S-fold geometries. In Section \ref{Sec:OrbiInst} we summarize the properties of orbi-instanton theories, which can be thought of as the basic building blocks of 6d SCFTs (see \cite{Heckman:2018pqx}) and constitute the starting point of our construction. We will then review Stiefel-Whitney compactifications (which we also call twisted torus compactifications) of orbi-instanton theories down to 4d following \cite{Ohmori:2018ona, Heckman:2022suy} in Section \ref{Sec:Tred}, and summarize the properties of the resulting 4d theories derived in the above references in Section \ref{sec:4dprop}. In Section \ref{sec:magquivMth} we provide a precise rule to write down the magnetic quivers describing the Higgs branch of these theories, which will be used as a guidance for exploring orbi-S-fold theories throughout the paper. The content of this subsection is mostly new. Using these quivers we explore in Section \ref{sec:4dprop} the Hasse diagram of the 4d theories and in particular we show that they are connected via higgsing, in a way dictated by the Hasse diagram of the parent 6d theories. {Finally, we discuss the mapping between orbi-instanton twisted compactifications and S-fold geometries in Type IIB probed by $D3$-branes, which will be the focus of the rest of the paper.}

\subsection{Orbi-Instanton theories}\label{Sec:OrbiInst}

Our starting point will be a particular subclass of 6d SCFTs with $\mathcal N=(1,0)$, usually called orbi-instanton theories. These theories can be constructed in M-theory on the worldvolume of $N$ $M5$-branes probing an $E_8$ nine-brane wrapped on an ADE singularity. In the following we will always take the singularity to be of A type, which means the space wrapped by the nine-brane is the orbifold $\mathbb C^2/\mathbb Z_M$. A wide class of theories can be constructed by considering Higgs-branch deformations which can be characterized in terms of a nilpotent orbit $\sigma$ of $\mathfrak{su}_M$ and a homomorphism $\rho : \mathbb Z_M \rightarrow E_8$. In the following we will always take $\sigma$ to be trivial and construct theories by simply performing different choices of the homomorphism $\rho$.\footnote{In cases where $N$ is not sufficiently large there may be correlations between $\rho$ and $\sigma$, implying that in some instances $\sigma$ will end up being non-trivial.} In order to study these strongly coupled theories it is always convenient to go on the tensor branch, which in M-theory corresponds to moving the $M5$-branes off the nine-brane. The resulting theory is a quiver gauge theory, which is usually better represented in terms of the F-theory dual description. For instance the case of trivial $\rho$ corresponds to the following quiver in 6d
\begin{align}
[\mathfrak{e}_8] \, \, 1 \, \, 2 \, \, \overset{\mathfrak{su}_{2}}{2} \, \,{\overset{\mathfrak{su}_{3}}{2}} \, \,{\overset{\mathfrak{su}_{4}}{2}} \cdots \, \,\underset{[\mathfrak{su}(1)]}{{\overset{\mathfrak{su}_{M}}{2}}} \, \,{\overset{\mathfrak{su}_{M}}{2}} \cdots \, \,{\overset{\mathfrak{su}_{M}}{2}} [\mathfrak{su}_M]\,.
\end{align}
Let us briefly explain the notation: Enclosed in square parentheses are the factors of global symmetry, in the case above the global symmetry is $\mathfrak{e}_8 \oplus \mathfrak{su}_M \oplus \mathfrak{su}_1$.\footnote{In the upcoming discussion of compactification to 4d it will be extremely important to distinguish between the gauge group and its algebra, that is why we are writing the flavor symmetry in terms of its algebra for the time being. We refer the readers to \cite{Heckman:2022suy} for a guide on how to read the gauge group.} The $\mathfrak{su}_1$ factor simply denotes the presence of a single hypermultiplet. Then the quiver arranges itself as a series of nodes with potentially a gauge group algebra written on top of it: In the F-theory description a $n$-node corresponds to a $(-n)$-curve in the F-theory base, and in the M-theory description it corresponds to an interval between two $M5$-branes along the direction transverse to the M9. Between contiguous nodes there are always bi-fundamental hypermultiplets. The next step is to consider the effect of turning on a non-trivial homomorphism $\rho$: Luckily enough homomorphisms from $\mathbb Z_M$ to $E_8$ have been classified in terms of a partition of $M$ weighted by the Dynkin labels of the affine $E_8$ Dynkin diagram. That is, any such homomorphism $\rho$ can be specified by the tuple
\begin{align}\label{eq:rho6d}
(n_1,n_2,n_3, n_4, n_5,n_6, n_{4'}, n_{2'},n_{3'})\,,
\end{align}
subject to the constraint
\begin{align}
n_1 + 2(n_2+n_{2'}) + 3(n_3+n_{3'}) + 4 (n_4+n_{4'}) + 5 n_5 + 6 n_6 = M\,.
\end{align}
The algorithm that returns the 6d quiver given a homomorphism $\rho$ is described in \cite{Mekareeya:2017jgc} and we will not copy it here for the sake of simplicity.

\subsection{Twisted torus reduction and 4d SCFTs}\label{Sec:Tred}

In this section we will take the 6d theories discussed in the previous section and perform a $ T^2$ compactification in order to build 4d SCFTs. While already a trivial compactification leads to a wide and interesting class of 4d theories, we will be interested in turning on almost-commuting holonomies on the $ T^2$ in order to broaden the class of theories that we can build. Let us briefly describe what are almost-commuting holonomies: Consider a theory with symmetry group $\hat G/\Gamma$. Here $\hat G$ is simply connected and $\Gamma$ is a subgroup of the center of $\hat G$, that is $\Gamma \subset \mathcal Z(\hat G)$. Moreover, $\hat G$ comprises of the entire symmetry group of the theory, that is flavor and $R$-symmetry factors.\footnote{In the case we are interested in we will neglect any $R$-symmetry factor, as well as any Abelian factor. Turning on holonomies along $R$-symmetry would end up breaking supersymmetry which is not the case we would like to consider in this work.} In order to determine the properties of the resulting 4d theories, it proves convenient to consider the action of the discrete group $\Gamma$ at the level of the low-energy gauge theories on the tensor branch. In order to specify the action of $\Gamma$ on their elementary fields, we need to embed it in the (center of the) gauge group as well. Therefore, by slightly abusing the notation, $\hat{G}$ will be meant to also include the gauge-group factors.

We can consider a $ T^2$ compactification of the theory and turn on some discrete holonomies along the two legs of the torus and in order to make sure that the background is flat we need to require that the holonomies commute. The observation here is that the holonomies need to commute inside $\hat G / \Gamma$, \emph{not} inside $\hat G$, given that any element of $\Gamma$ acts trivially on our theory. Said differently, given two holonomies $P$ and $Q$ along the two 1-cycles of the torus we can require that
\begin{align}
P Q = \omega\, Q P\,, \qquad \omega \in \Gamma\,.
\end{align}
In order to build compactifications of 6d orbi-instanton theories with almost-commuting holonomies, it is necessary to have a non-trivial center-symmetry group $\Gamma$ in 6d. The answer highly depends on the choice of homomorphism $\rho$; however the result derived in \cite{Heckman:2022suy} is simply stated in terms of the tuple in equation \eqref{eq:rho6d}: If in this tuple the non-zero entries $n_i$ are such that $\gcd\{i\}=\ell$,\footnote{Note that by the notation $n_i$ we mean both the primed and the unprimed Kac labels.} then the 6d theory has center symmetry equal to $\mathbb Z_\ell$. In this case it happens that $M$ is a multiple of $\ell$, and we will write it in the following as $M \equiv \ell k$. This makes it clear that only the cases $\ell = 2, 3,4,5,6$ are possible. After the compactification with almost-commuting holonomies, the 4d theory can also be described by a linear quiver, however the effect of turning on the holonomies is to reduce each $\mathfrak{su}_L$ factor to $\mathfrak{su}_{L/\ell}$, and the same happens to the flavor factors (with some possible exceptions for flavor symmetry factors related to hypermultiplets in representations other than the fundamental). A small comment on notation: Oftentimes compactifications with almost-commuting holonomies have been also called twisted compactifications or Stiefel--Whitney (SW) compactifications in the literature, and we will use these terms interchangeably henceforth.

In the following we will discuss various properties of the resulting theories for different values of $\ell$ and low values of $k$.

\subsection{Magnetic quivers of twisted compactifications}\label{sec:magquivMth}

Here we start to systematically study the Higgs branch of the 4d theories obtained after compactification with almost-commuting holonomies. As anticipated before, we will focus on examples with various values of $\ell $ and low values of $k$. For each pair $\ell$ and $k$ there will be a few theories that can be constructed, however to streamline the presentation we will consider only the case with maximal Higgs branch in this section and list all the remaining cases in Appendix \ref{App:Mth}. Our tool will be the magnetic quiver of the 4d theory, and we will illustrate for each example how to derive the magnetic quiver of the theory after compactification with almost-commuting holonomies via Fayet--Ilioupoulos deformations. We will explain the reason why such FI deformations are necessary after going through the first example. By inspection of several examples we will be able to infer a general rule that will allow us to extrapolate the magnetic quiver for generic values of $k$ beyond the ones analyzed in this section and in Appendix \ref{App:Mth}. Further properties of the resulting 4d theories will be listed in Section \ref{sec:4dprop} and again in Appendix \ref{App:Mth} for the cases with non-maximal Higgs branch.

\subsubsection{Magnetic quivers for $k=2$ and $\ell = 2$}

The first case we consider has $k=2$ and $\ell =2 $. In order to have $\ell = 2$ it is only possible to turn on holonomies labelled by $n_2$, $n_{2'}$, $n_4$, $n_{4'}$, and $n_6$. We will label the theories obtained after twisted compactification following the conventions introduced in \cite{Heckman:2022suy}, albeit with a small change. Theories with $\ell =2$ will be called $\mathcal T_{E_6,2}^{(N)}(n_2,n_4,n_6,n_{4'},n_{2'})$ if $n_{4'}$ is even, and $\mathcal S_{E_6,2}^{(N)}(n_2,n_4,n_6,n_{4'},n_{2'})$ if $n_{4'}$ is odd. Compared to \cite{Heckman:2022suy} we added the $E_6$ subscript to indicate that there is a $E_6$-type 7-brane. In Section \ref{Sec:DecFlavors} we will discussed theories with a different kind of 7-brane, thus making it necessary to specify which kind of 7-brane  is present. Given that this is the very first case we consider we will be more detailed in the description of the derivation of the 4d magnetic quiver after twisted compactification. There exist five possible choices of homomorphism $\rho$ that give $k=2$, however the one we will consider in this section has $n_{4'} = 1$ with all the other affine $E_8$ Dynkin labels set to zero: Following our labeling convention this theory is $\mathcal S^{(N)}_{E_6,2} (0,0,0,1,0)$. This gives the 4d theory with maximal Higgs branch; all the remaining four theories can be obtained by this top theory via Higgs-branch flow and their properties will be discussed in Appendix \ref{App:Mth}. Our starting point in the derivation of the 4d magnetic quiver is the magnetic quiver of the 6d theory: The general rule to obtain the magnetic quiver of the 6d theory for any choice of $\rho$ was derived in \cite{Mekareeya:2017jgc} and for our case this yields the quiver
\begin{align}\label{eq:K2ell2max}
\begin{tikzpicture}
\filldraw[fill= white] (0,0) circle [radius=0.1] node[below] {\scriptsize 1};
\filldraw[fill= red] (1,0) circle [radius=0.1] node[below] {\scriptsize 2};
\filldraw[fill= white] (2,0) circle [radius=0.1] node[below] {\scriptsize 3};
\filldraw[fill= white] (3,0) circle [radius=0.1] node[below] {\scriptsize 4};
\filldraw[fill= white] (4,0) circle [radius=0.1] node[below] {\scriptsize N+4};
\filldraw[fill= white] (5,0) circle [radius=0.1] node[below] {\scriptsize 2N+4};
\filldraw[fill= red] (6,0) circle [radius=0.1] node[below] {\scriptsize 3N+4};
\filldraw[fill= white] (7,0) circle [radius=0.1] node[below] {\scriptsize 4N+4};
\filldraw[fill= white] (8,0) circle [radius=0.1] node[below] {\scriptsize 5N+4};
\filldraw[fill= white] (9,0) circle [radius=0.1] node[below] {\scriptsize 6N+4};
\filldraw[fill= white] (10,0) circle [radius=0.1] node[below] {\scriptsize 4N+2};
\filldraw[fill= white] (11,0) circle [radius=0.1] node[below] {\scriptsize 2N+1};
\filldraw[fill= red] (9,1) circle [radius=0.1] node[above] {\scriptsize 3N+2};
\draw [thick] (0.1, 0) -- (0.9,0) ;
\draw [thick] (1.1, 0) -- (1.9,0) ;
\draw [thick] (2.1, 0) -- (2.9,0) ;
\draw [thick] (3.1, 0) -- (3.9,0) ;
\draw [thick] (4.1, 0) -- (4.9,0) ;
\draw [thick] (5.1, 0) -- (5.9,0) ;
\draw [thick] (6.1, 0) -- (6.9,0) ;
\draw [thick] (7.1, 0) -- (7.9,0) ;
\draw [thick] (8.1, 0) -- (8.9,0) ;
\draw [thick] (9.1, 0) -- (9.9,0) ;
\draw [thick] (10.1, 0) -- (10.9,0) ;
\draw [thick] (9, 0.1) -- (9,0.9) ;
\end{tikzpicture}
\end{align}
The nodes colored in red will be important in the derivation of the 4d quiver, but before let us describe the general procedure following \cite{vanBeest:2021xyt}. We are interested in turning on some Fayet--Ilioupoulos terms on some nodes of the magnetic quiver and construct the resulting magnetic quiver after this deformation, however in general it is not possible to just turn on such FI terms on an arbitrary subset of nodes as it is necessary to ensure that the F-term and D-term equations have solutions lest we break supersymmetry. One way to ensure the existence of a solution for unitary quivers is to ensure that the ranks of the selected unitary nodes add up to zero (with some appropriate choice of signs).\footnote{Physically speaking the choice of the relative signs is related to the sign of the FI terms at the corresponding node.} In the case of the quiver written before we would like to turn on FI terms on the nodes colored in red, and indeed one can see that $3N+4-(3N+2)-2 = 0$. The quiver after the deformation can be obtained via a series of quiver subtractions \cite{Cabrera:2018ann,Bourget:2019aer}: Consider turning on a FI deformation on two $U(n)$ nodes, then the resulting magnetic quiver is a subtraction of a $U(n)$ quiver on the nodes connecting the two $U(n)$ nodes and a rebalancing via a single $U(n)$ node.\footnote{Recall that for a $U(n_c)$ node with $n_f$ flavor the balance is defined as $e = n_f-2 n_c$. When subtracting a quiver the balance of some nodes which are not affected by the subtraction can change, and therefore their original balance needs to be restored by connecting them to a rebalancing node. Nodes with $e=0$ are called balanced.} In the case FI terms are turned on on more than two nodes, follow a similar procedure on couple of nodes $U(n)$ and $U(n+k)$ subtracting a $U(n)$ quiver on the connecting nodes with $U(n)$ rebalancing, and continue following this procedure until no more nodes with FI terms turned on are present. This procedure may appear to be cumbersome, but it is easier to illustrate it by simply working it on the previous example. We start by subtracting a linear $U(2)$ quiver on nodes connecting the $U(2)$ and $U(3N+4)$ nodes. The result is
\begin{align}
\begin{tikzpicture}
\filldraw[fill= white] (2,0) circle [radius=0.1] node[below] {\scriptsize 1};
\filldraw[fill= white] (3,0) circle [radius=0.1] node[below] {\scriptsize 2};
\filldraw[fill= white] (4,0) circle [radius=0.1] node[below] {\scriptsize N+2};
\filldraw[fill= white] (5,0) circle [radius=0.1] node[below] {\scriptsize 2N+2};
\filldraw[fill= red] (6,0) circle [radius=0.1] node[below] {\scriptsize 3N+2};
\filldraw[fill= white] (7,0) circle [radius=0.1] node[below] {\scriptsize 4N+4};
\filldraw[fill= white] (8,0) circle [radius=0.1] node[below] {\scriptsize 5N+4};
\filldraw[fill= white] (9,0) circle [radius=0.1] node[below] {\scriptsize 6N+4};
\filldraw[fill= white] (10,0) circle [radius=0.1] node[below] {\scriptsize 4N+2};
\filldraw[fill= white] (11,0) circle [radius=0.1] node[below] {\scriptsize 2N+1};
\filldraw[fill= red] (9,1) circle [radius=0.1] node[above] {\scriptsize 3N+2};
\filldraw[fill= lime] (7,1) circle [radius=0.1] node[above] {\scriptsize 2};
\filldraw[fill= white] (6,1) circle [radius=0.1] node[above] {\scriptsize 1};
\draw [thick] (2.1, 0) -- (2.9,0) ;
\draw [thick] (3.1, 0) -- (3.9,0) ;
\draw [thick] (4.1, 0) -- (4.9,0) ;
\draw [thick] (5.1, 0) -- (5.9,0) ;
\draw [thick] (6.1, 0) -- (6.9,0) ;
\draw [thick] (7.1, 0) -- (7.9,0) ;
\draw [thick] (8.1, 0) -- (8.9,0) ;
\draw [thick] (9.1, 0) -- (9.9,0) ;
\draw [thick] (10.1, 0) -- (10.9,0) ;
\draw [thick] (9, 0.1) -- (9,0.9) ;
\draw [thick] (7, 0.1) -- (7,0.9);
\draw [thick] (6.1, 1) -- (6.9,1);
\end{tikzpicture}
\end{align}
The lime node is the $U(2)$ rebalancing node, which is connected to the $U(1)$ node and the $U(4N+4)$ nodes of the original diagram because these two needed to be rebalanced. We can now finish and turn on the deformation in the two $U(3N+2)$ nodes subtracting a $U(3N+2)$ quiver between the two. 
\begin{align}\label{K2ell2maxFI}
\begin{tikzpicture}
\filldraw[fill= white] (0,0) circle [radius=0.1] node[below] {\scriptsize 1};
\filldraw[fill= white] (1,0) circle [radius=0.1] node[below] {\scriptsize 2};
\filldraw[fill= white] (2,0) circle [radius=0.1] node[below] {\scriptsize N+2};
\filldraw[fill= white] (3,0) circle [radius=0.1] node[below] {\scriptsize 2N+2};
\filldraw[fill= lime] (4,0) circle [radius=0.1] node[below] {\scriptsize 3N+2};
\filldraw[fill= white] (5,0) circle [radius=0.1] node[below] {\scriptsize 4N+2};
\filldraw[fill= white] (6,0) circle [radius=0.1] node[below] {\scriptsize 3N+2};
\filldraw[fill= white] (7,0) circle [radius=0.1] node[below] {\scriptsize 2N+2};
\filldraw[fill= white] (8,0) circle [radius=0.1] node[below] {\scriptsize N+2};
\filldraw[fill= lime] (9,0) circle [radius=0.1] node[below] {\scriptsize 2};
\filldraw[fill= white] (10,0) circle [radius=0.1] node[below] {\scriptsize 1};
\filldraw[fill= white] (5,1) circle [radius=0.1] node[left] {\scriptsize 2N+1};
\draw [thick] (0.1, 0) -- (0.9,0) ;
\draw [thick] (1.1, 0) -- (1.9,0) ;
\draw [thick] (2.1, 0) -- (2.9,0) ;
\draw [thick] (3.1, 0) -- (3.9,0) ;
\draw [thick] (4.1, 0) -- (4.9,0) ;
\draw [thick] (5.1, 0) -- (5.9,0) ;
\draw [thick] (6.1, 0) -- (6.9,0) ;
\draw [thick] (7.1, 0) -- (7.9,0) ;
\draw [thick] (8.1, 0) -- (8.9,0) ;
\draw [thick] (9.1, 0) -- (9.9,0) ;
\draw [thick] (5, 0.1) -- (5,0.9) ;
\end{tikzpicture}
\end{align}
In lime we colored the two rebalancing nodes, the $U(2)$ from the first step, and the $U(3N+2)$ from the second step. Note that in doing the last step we did \emph{not} rebalance the rebalancing $U(2)$ node. Now we come to the reason why such FI deformations were necessary: We know that our 6d theory has a $\mathbb Z_2$ symmetry that allows for turning on some almost-commuting holonomies upon $T^2$ compactification. However the quiver \eqref{eq:K2ell2max} clearly does not have such symmetry, but the quiver \eqref{K2ell2maxFI} obtained after the FI deformations \emph{does} have a $\mathbb Z_2$ symmetry (the rebalancing nodes are absolutely identical to any other node and we highlighted them in order to make the presentation more clear). Given the manifest $\mathbb Z_2$ symmetry we can fold the quiver \eqref{K2ell2maxFI} obtaining the quiver
\begin{align}\label{eq:foldK2ell2}
\begin{tikzpicture}
\filldraw[fill= white] (0,0) circle [radius=0.1] node[below] {\scriptsize 1};
\filldraw[fill= white] (1,0) circle [radius=0.1] node[below] {\scriptsize 2};
\filldraw[fill= white] (2,0) circle [radius=0.1] node[below] {\scriptsize N+2};
\filldraw[fill= white] (3,0) circle [radius=0.1] node[below] {\scriptsize 2N+2};
\filldraw[fill= white] (4,0) circle [radius=0.1] node[below] {\scriptsize 3N+2};
\filldraw[fill= white] (5,0) circle [radius=0.1] node[below] {\scriptsize 4N+2};
\filldraw[fill= white] (6,0) circle [radius=0.1] node[below] {\scriptsize  2N+1};
\draw [thick] (0.1, 0) -- (0.9,0) ;
\draw [thick] (1.1, 0) -- (1.9,0) ;
\draw [thick] (2.1, 0) -- (2.9,0) ;
\draw [thick] (4.1, 0.05) -- (4.9,0.05) ;
\draw [thick] (4.1, -0.05) -- (4.9,-0.05) ;
\draw [thick] (3.1, 0) -- (3.9,0) ;
\draw [thick] (5.1, 0) -- (5.9,0) ;
\draw [thick] (4.4,0) -- (4.6,0.2);
\draw [thick] (4.4,0) -- (4.6,-0.2);
\end{tikzpicture}
\end{align}
We claim that \eqref{eq:foldK2ell2} is the quiver of the 4d theory after twisted compactification. From the magnetic quiver \eqref{eq:foldK2ell2} we can immediately extract two important properties: The flavor symmetry\footnote{To extract the non-Abelian part of the flavor symmetry from a magnetic quiver just eliminate all non-balanced nodes and the flavor symmetry has Dynkin diagram the remaining nodes. The Abelian part (which we neglect here) is $U(1)^r$ where $r+1$ is the number of unbalanced nodes.} is $\mathfrak{su}_2 \oplus \mathfrak{su}_4 \oplus \mathfrak{su}_2$  which matches the results of \cite{Heckman:2022suy}, and the quaternionic dimension of the Higgs branch\footnote{For a quiver with $U(m_i)$ nodes we have that $\text{dim } {\mathbb H} = (\sum_i m_i) -1$.} is $12N+11$ again matching with \cite{Heckman:2022suy}.\footnote{We will discuss how to extract the dimension of the Higgs branch from the results of \cite{Heckman:2022suy} in Section \ref{sec:4dprop}.} In the quiver \eqref{eq:foldK2ell2} we can see an oriented double connection between the $U(3N+2)$ and the $U(4N+2)$ nodes. This means that the node $U(4N+2)$ sees $2 (3N+2)$ fundamental hypers coming from the $U(3N+2)$ node, but the $U(3N+2)$ node only sees $4N+2$ fundamental hypers coming from the $U(4N+2)$ node. This implies for instance that the balance of the $U(3N+2)$ node is $2N+2+4N+2-2(3N+2) = 0$ while the balance of the $U(4N+2)$ node is $2N+1 + 2(3N+2)-2(4N+2) = 1$.

\subsubsection{Magnetic quivers for $k=3$ and $\ell = 2$}

We will briefly outline the construction of the magnetic quiver for the theories with $k=3$ and $\ell = 2$. In this case for $k=3$ we find nine choices of holonomies, and again we will focus on the theory with maximal Higgs branch, with the remaining eight theories listed in Appendix \ref{App:Mth}. In this case the theory with maximal Higgs branch has $n_{4'} = n_{2'} = 1$, that is the theory $\mathcal S_{E_6,2}^{(N)}(0,0,0,1,1)$, with all the other labels equal to zero. The magnetic quiver of the 6d theory is 
\begin{align}
\begin{tikzpicture}
\filldraw[fill= white] (-4,0) circle [radius=0.1] node[below] {\scriptsize 1};
\filldraw[fill= white] (-3,0) circle [radius=0.1] node[below] {\scriptsize 2};
\filldraw[fill= red] (-2,0) circle [radius=0.1] node[below] {\scriptsize 3};
\filldraw[fill= white] (-1,0) circle [radius=0.1] node[below] {\scriptsize 4};
\filldraw[fill= white] (0,0) circle [radius=0.1] node[below] {\scriptsize 5};
\filldraw[fill= white] (1,0) circle [radius=0.1] node[below] {\scriptsize 6};
\filldraw[fill= white] (2,0) circle [radius=0.1] node[below] {\scriptsize N+6};
\filldraw[fill= white] (3,0) circle [radius=0.1] node[below] {\scriptsize 2N+6};
\filldraw[fill= red] (4,0) circle [radius=0.1] node[below] {\scriptsize 3N+6};
\filldraw[fill= white] (5,0) circle [radius=0.1] node[below] {\scriptsize 4N+6};
\filldraw[fill= white] (6,0) circle [radius=0.1] node[below] {\scriptsize 5N+6};
\filldraw[fill= white] (7,0) circle [radius=0.1] node[below] {\scriptsize 6N+6};
\filldraw[fill= white] (8,0) circle [radius=0.1] node[below] {\scriptsize 4N+3};
\filldraw[fill= white] (9,0) circle [radius=0.1] node[below] {\scriptsize 2N+1};
\filldraw[fill= red] (7,1) circle [radius=0.1] node[above] {\scriptsize 3N+3};
\draw [thick] (-3.1, 0) -- (-3.9,0) ;
\draw [thick] (-2.1, 0) -- (-2.9,0) ;
\draw [thick] (-1.1, 0) -- (-1.9,0) ;
\draw [thick] (0.1, 0) -- (0.9,0) ;
\draw [thick] (-0.1, 0) -- (-0.9,0) ;
\draw [thick] (1.1, 0) -- (1.9,0) ;
\draw [thick] (2.1, 0) -- (2.9,0) ;
\draw [thick] (3.1, 0) -- (3.9,0) ;
\draw [thick] (4.1, 0) -- (4.9,0) ;
\draw [thick] (5.1, 0) -- (5.9,0) ;
\draw [thick] (6.1, 0) -- (6.9,0) ;
\draw [thick] (7.1, 0) -- (7.9,0) ;
\draw [thick] (8.1, 0) -- (8.9,0) ;
\draw [thick] (7, 0.1) -- (7,0.9) ;
\end{tikzpicture}
\end{align}
Again we can follow the same procedure outlined before and turn on FI terms on the nodes highlighted in red. The resulting quiver will have a $\mathbb Z_2$ symmetry, and after folding we find
\begin{align}
 \begin{tikzpicture}
\filldraw[fill= white] (0,0) circle [radius=0.1] node[below] {\scriptsize 1};
\filldraw[fill= white] (1,0) circle [radius=0.1] node[below] {\scriptsize 2};
\filldraw[fill= white] (2,0) circle [radius=0.1] node[below] {\scriptsize 3};
\filldraw[fill= white] (3,0) circle [radius=0.1] node[below] {\scriptsize N+3};
\filldraw[fill= white] (4,0) circle [radius=0.1] node[below] {\scriptsize 2N+3};
\filldraw[fill= white] (5,0) circle [radius=0.1] node[below] {\scriptsize 3N+3};
\filldraw[fill= white] (6,0) circle [radius=0.1] node[below] {\scriptsize 4N+3};
\filldraw[fill= white] (7,0) circle [radius=0.1] node[below] {\scriptsize  2N+1};
\draw [thick] (0.1, 0) -- (0.9,0) ;
\draw [thick] (1.1, 0) -- (1.9,0) ;
\draw [thick] (2.1, 0) -- (2.9,0) ;
\draw [thick] (5.1, 0.05) -- (5.9,0.05) ;
\draw [thick] (5.1, -0.05) -- (5.9,-0.05) ;
\draw [thick] (3.1, 0) -- (3.9,0) ;
\draw [thick] (4.1, 0) -- (4.9,0) ;
\draw [thick] (5.4,0) -- (5.6,0.2);
\draw [thick] (5.4,0) -- (5.6,-0.2);
\draw [thick] (6.1,0) -- (6.9,0);
\end{tikzpicture}
\end{align}
Notice that this has a skeleton similar to the quiver \eqref{eq:foldK2ell2} albeit with a longer tail on the left hand side. This is a reflection of the fact that any theory with any values of $\ell $ and $k$ will have a $\mathfrak{su}_k$ as a factor in the global symmetry, and this factor is always stored in the tail of the magnetic quiver. Therefore with higher values of $k$ we will always expect a tail with $k$ nodes attached to the quiver. In this case we can also match the symmetry and dimension of the Higgs branch already computed in \cite{Heckman:2022suy}.

\subsubsection{Magnetic quivers for $k=2$ and $\ell = 3$}

We now move away from $\ell = 2$ and start studying the cases with $\ell = 3$. In order to have $\ell =3$ the holonomies available are $n_3$, $n_{3'}$, and $n_6$. For $\ell =3$ we will call theories as $\mathcal T^{(N)}_{D_4,3} (n_3,n_6,n_{3'})$ if $n_{3'} = 0 \mod 3$, $\mathcal S^{(N)}_{D_4,3} (n_3,n_6,n_{3'})$ if $n_{3'} = 1 \mod 3$, and $\mathcal R^{(N)}_{D_4,3} (n_3,n_6,n_{3'})$ if $n_{3'} = 2 \mod 3$.
 For $k=2$ there are four possible choices of holonomies, and the one giving the maximal Higgs branch has $n_{3'}=2$, that is the theory $\mathcal R^{(N)}_{D_4,3}(0,0,2)$,with the others three theories again connected via Higgs-branch flow to this maximal theory. The properties of the remaining cases will be discussed in Appendix \ref{App:Mth}.
As before we will start from the magnetic quiver of the 6d theory before twisted toroidal compactification
\begin{align}
\begin{tikzpicture}
\filldraw[fill= white] (-4,0) circle [radius=0.1] node[below] {\scriptsize 1};
\filldraw[fill= blue] (-3,0) circle [radius=0.1] node[below] {\scriptsize 2};
\filldraw[fill= white] (-2,0) circle [radius=0.1] node[below] {\scriptsize 3};
\filldraw[fill= red] (-1,0) circle [radius=0.1] node[below] {\scriptsize 4};
\filldraw[fill= white] (0,0) circle [radius=0.1] node[below] {\scriptsize 5};
\filldraw[fill= white] (1,0) circle [radius=0.1] node[below] {\scriptsize 6};
\filldraw[fill= red] (2,0) circle [radius=0.1] node[below] {\scriptsize N+6};
\filldraw[fill= blue] (3,0) circle [radius=0.1] node[below] {\scriptsize 2N+6};
\filldraw[fill= white] (4,0) circle [radius=0.1] node[below] {\scriptsize 3N+6};
\filldraw[fill= blue] (5,0) circle [radius=0.1] node[below] {\scriptsize 4N+6};
\filldraw[fill= white] (6,0) circle [radius=0.1] node[below] {\scriptsize 5N+6};
\filldraw[fill= blue] (7,0) circle [radius=0.1] node[below] {\scriptsize 6N+6};
\filldraw[fill= red] (8,0) circle [radius=0.1] node[below] {\scriptsize 4N+4};
\filldraw[fill= white] (9,0) circle [radius=0.1] node[below] {\scriptsize 2N+2};
\filldraw[fill= red] (7,1) circle [radius=0.1] node[above] {\scriptsize 3N+2};
\draw [thick] (-3.1, 0) -- (-3.9,0) ;
\draw [thick] (-2.1, 0) -- (-2.9,0) ;
\draw [thick] (-1.1, 0) -- (-1.9,0) ;
\draw [thick] (-0.1, 0) -- (-0.9,0) ;
\draw [thick] (0.1, 0) -- (0.9,0) ;
\draw [thick] (1.1, 0) -- (1.9,0) ;
\draw [thick] (2.1, 0) -- (2.9,0) ;
\draw [thick] (3.1, 0) -- (3.9,0) ;
\draw [thick] (4.1, 0) -- (4.9,0) ;
\draw [thick] (5.1, 0) -- (5.9,0) ;
\draw [thick] (6.1, 0) -- (6.9,0) ;
\draw [thick] (7.1, 0) -- (7.9,0) ;
\draw [thick] (8.1, 0) -- (8.9,0) ;
\draw [thick] (7, 0.1) -- (7,0.9) ; 
\end{tikzpicture}
\end{align}
In this case it will be necessary to perform two FI deformations in order to reach a magnetic quiver with a $\mathbb Z_3$ symmetry that can be appropriately folded. We have highlighted in red the FI deformation that ought to be turned on first, and in blue the FI deformation that should be turned as a second step.  Since this is the first case that requires two steps, we will outline it in detail: After turning on the FI deformation on the red nodes we get the following quiver
\begin{align}
\begin{tikzpicture}
\filldraw[fill= white] (-4,0) circle [radius=0.1] node[below] {\scriptsize 1};
\filldraw[fill= blue] (-3,0) circle [radius=0.1] node[below] {\scriptsize 2};
\filldraw[fill= white] (-2,0) circle [radius=0.1] node[below] {\scriptsize 3};
\filldraw[fill= lime] (-1,0) circle [radius=0.1] node[below] {\scriptsize 4};
\filldraw[fill= blue] (0,0) circle [radius=0.1] node[below] {\scriptsize N+4};
\filldraw[fill= white] (1,0) circle [radius=0.1] node[below] {\scriptsize 2N+4};
\filldraw[fill= blue] (2,0) circle [radius=0.1] node[below] {\scriptsize 3N+4};
\filldraw[fill= white] (3,0) circle [radius=0.1] node[below] {\scriptsize 4N+4};
\filldraw[fill= lime] (4,0) circle [radius=0.1] node[below] {\scriptsize 3N+2};
\filldraw[fill= white] (5,0) circle [radius=0.1] node[below] {\scriptsize 2N+2};
\filldraw[fill= lime] (6,0) circle [radius=0.1] node[below] {\scriptsize N+2};
\filldraw[fill= white] (7,0) circle [radius=0.1] node[below] {\scriptsize 2};
\filldraw[fill= white] (8,0) circle [radius=0.1] node[below] {\scriptsize 1};
\filldraw[fill= blue] (3,1) circle [radius=0.1] node[above] {\scriptsize 2N+2};
\draw [thick] (-3.1, 0) -- (-3.9,0) ;
\draw [thick] (-2.1, 0) -- (-2.9,0) ;
\draw [thick] (-1.1, 0) -- (-1.9,0) ;
\draw [thick] (-0.1, 0) -- (-0.9,0) ;
\draw [thick] (0.1, 0) -- (0.9,0) ;
\draw [thick] (1.1, 0) -- (1.9,0) ;
\draw [thick] (2.1, 0) -- (2.9,0) ;
\draw [thick] (3.1, 0) -- (3.9,0) ;
\draw [thick] (4.1, 0) -- (4.9,0) ;
\draw [thick] (5.1, 0) -- (5.9,0) ;
\draw [thick] (6.1, 0) -- (6.9,0) ;
\draw [thick] (7.1, 0) -- (7.9,0) ;
\draw [thick] (3, 0.1) -- (3,0.9) ; 
\end{tikzpicture}
\end{align}
Here we highlighted in lime the rebalancing nodes introduced in the process of turning on the FI deformation, and in blue the nodes to be turned on at the next step. Turning on the final deformation we find a quiver that admits a $\mathbb Z_3$ symmetry, and the folded result is
\begin{align}
\begin{tikzpicture}
\filldraw[fill= white] (0,0) circle [radius=0.1] node[below] {\scriptsize 1};
\filldraw[fill= white] (1,0) circle [radius=0.1] node[below] {\scriptsize 2};
\filldraw[fill= white] (2,0) circle [radius=0.1] node[below] {\scriptsize N+2};
\filldraw[fill= white] (3,0) circle [radius=0.1] node[below] {\scriptsize 2N+2};
\filldraw[fill= white] (4,0) circle [radius=0.1] node[below] {\scriptsize 3N+2};
\draw [thick] (0.1, 0) -- (0.9,0) ;
\draw [thick] (1.1, 0) -- (1.9,0) ;
\draw [thick] (2.1, 0) -- (2.9,0) ;
\draw [thick] (3.1, 0.07) -- (3.9,0.07) ;
\draw [thick] (3.1, -0.07) -- (3.9,-0.07) ;
\draw [thick] (3.1, 0) -- (3.9,0) ;
\draw [thick] (3.4,0) -- (3.6,0.2);
\draw [thick] (3.4,0) -- (3.6,-0.2);
\end{tikzpicture}
\end{align}
As usual this quiver meets the expected properties of flavor symmetry and dimension of Higgs branch computed in \cite{Heckman:2022suy}.

\subsubsection{Magnetic quivers for $k=2$ and $\ell > 3$}

For the $\ell = 4$ the available holonomies in 6d are only $n_4$ and $n_{4'}$. We will call these theories $\mathcal T_{\mathcal H_2,4}^{(N)}(n_4,n_{4'})$ if $n_{4'}$ is even, and $\mathcal S_{\mathcal H_2,4}^{(N)}(n_4,n_{4'})$ if $n_{4'}$ is odd. At $k=2$ there are three distinct theories. As usual we will focus on the maximal theory, which in this context has holonomy $n_{4'} = 2$, that is theory $\mathcal S_{\mathcal H_2,4}^{(N)}(1,1)$. The 6d quiver is 
\begin{align}
\begin{tikzpicture}
\filldraw[fill= white] (-6,0) circle [radius=0.1] node[below] {\scriptsize 1};
\filldraw[fill= white] (-5,0) circle [radius=0.1] node[below] {\scriptsize 2};
\filldraw[fill= white] (-4,0) circle [radius=0.1] node[below] {\scriptsize 3};
\filldraw[fill= white] (-3,0) circle [radius=0.1] node[below] {\scriptsize 4};
\filldraw[fill= white] (-2,0) circle [radius=0.1] node[below] {\scriptsize 5};
\filldraw[fill= white] (-1,0) circle [radius=0.1] node[below] {\scriptsize 6};
\filldraw[fill= white] (0,0) circle [radius=0.1] node[below] {\scriptsize 7};
\filldraw[fill= white] (1,0) circle [radius=0.1] node[below] {\scriptsize 8};
\filldraw[fill= white] (2,0) circle [radius=0.1] node[below] {\scriptsize N+7};
\filldraw[fill= white] (3,0) circle [radius=0.1] node[below] {\scriptsize 2N+6};
\filldraw[fill= white] (4,0) circle [radius=0.1] node[below] {\scriptsize 3N+5};
\filldraw[fill= white] (5,0) circle [radius=0.1] node[below] {\scriptsize 4N+4};
\filldraw[fill= white] (6,0) circle [radius=0.1] node[below] {\scriptsize 5N+4};
\filldraw[fill= white] (7,0) circle [radius=0.1] node[below] {\scriptsize 6N+4};
\filldraw[fill= white] (8,0) circle [radius=0.1] node[below] {\scriptsize 4N+2};
\filldraw[fill= white] (9,0) circle [radius=0.1] node[below] {\scriptsize 2N+1};
\filldraw[fill= white] (7,1) circle [radius=0.1] node[above] {\scriptsize 3N+2};
\draw [thick] (-0.1-5, 0) -- (-0.9-5,0) ;
\draw [thick] (-0.1-4, 0) -- (-0.9-4,0) ;
\draw [thick] (-0.1-3, 0) -- (-0.9-3,0) ;
\draw [thick] (-0.1-2, 0) -- (-0.9-2,0) ;
\draw [thick] (-0.1-1, 0) -- (-0.9-1,0) ;
\draw [thick] (0.1, 0) -- (0.9,0) ;
\draw [thick] (-0.1, 0) -- (-0.9,0) ;
\draw [thick] (1.1, 0) -- (1.9,0) ;
\draw [thick] (2.1, 0) -- (2.9,0) ;
\draw [thick] (3.1, 0) -- (3.9,0) ;
\draw [thick] (4.1, 0) -- (4.9,0) ;
\draw [thick] (5.1, 0) -- (5.9,0) ;
\draw [thick] (6.1, 0) -- (6.9,0) ;
\draw [thick] (7.1, 0) -- (7.9,0) ;
\draw [thick] (8.1, 0) -- (8.9,0) ;
\draw [thick] (7, 0.1) -- (7,0.9) ;
\end{tikzpicture}
\end{align}
This will require multiple deformations to get a quiver with $\mathbb Z_4$ symmetry. For simplicity we just quote the end result
\begin{align}
\begin{tikzpicture}
\filldraw[fill= white] (0,0) circle [radius=0.1] node[below] {\scriptsize 1};
\filldraw[fill= white] (1,0) circle [radius=0.1] node[below] {\scriptsize 2};
\filldraw[fill= white] (2,0) circle [radius=0.1] node[below] {\scriptsize N+1};
\filldraw[fill= white] (3,0) circle [radius=0.1] node[below] {\scriptsize 2N+1};
\draw [thick] (0.1, 0) -- (0.9,0) ;
\draw [thick] (1.1, 0) -- (1.9,0) ;
\draw [thick] (2.1, 0.06) -- (2.9,0.06) ;
\draw [thick] (2.1, 0.02) -- (2.9,0.02) ;
\draw [thick] (2.1,-0.02) -- (2.9,-0.02) ;
\draw [thick] (2.1, -0.06) -- (2.9,-0.06) ;
\draw [thick] (2.4,0) -- (2.6,0.2);
\draw [thick] (2.4,0) -- (2.6,-0.2);
\end{tikzpicture}
\end{align}
As in the previous cases discussed this quiver meets the expected properties of flavor symmetry and dimension of Higgs branch computed in \cite{Heckman:2022suy}.

For $\ell = 5$ only the $n_5$ holonomy is available, so that we always have $k = n_5$. We will call these theories $\mathcal T^{(N)}_{\mathcal H_0,5}(n_5)$. For the sake of shortness we will quote only the magnetic quiver of the theory $\mathcal T^{(N)}_{\mathcal H_0,5}(2)$
\begin{align}
\begin{tikzpicture}
\draw [thick] (1.1, 0) -- (1.9,0) ;
\draw [thick] (2, .1) -- (3,.1) ;
\draw [thick] (2, -.1) -- (3,.-.1) ;
\draw [thick] (2, -.1) -- (3,.-.1) ;
\draw [thick] (2.1, 0) -- (2.9,0) ;
\draw [thick] (2+0.07, -0.05) -- (3-0.07,-0.05) ;
\draw [thick] (2+0.07, 0.05) -- (3-0.07,0.05) ;
\draw [thick] (2.4,0) -- (2.6,0.2);
\draw [thick] (2.4,0) -- (2.6,-0.2);
\draw [thick] (3.05, 0.05) to [out=45,in=315,looseness=15] (3.05,-0.05);
\filldraw[fill= white] (1,0) circle [radius=0.1] node[below] {\scriptsize 1};
\filldraw[fill= white] (2,0) circle [radius=0.1] node[below] {\scriptsize 2};
\filldraw[fill= white] (3,0) circle [radius=0.1] node[below] {\scriptsize N};
\end{tikzpicture}
\end{align}

For $\ell = 6$ only the $n_6$ holonomy is available. We will call these theories $\mathcal T^{(N)}_{\varnothing,6}(n_6)$. For the sake of shortness we will quote only the magnetic quiver of the theory $\mathcal T^{(N)}_{\varnothing,6}(2)$

\begin{align}
\begin{tikzpicture}
\draw [thick] (1.1, 0) -- (1.9,0) ;
\draw [thick] (2, .1) -- (3,.1) ;
\draw [thick] (2, -.1) -- (3,.-.1) ;
\draw [thick] (2, -.1) -- (3,.-.1) ;
\draw [thick] (2,0.06) -- (3,0.06);
\draw [thick] (2,0.02) -- (3,0.02);
\draw [thick] (2,-0.06) -- (3,-0.06);
\draw [thick] (2,-0.02) -- (3,-0.02);
\draw [thick] (2.4,0) -- (2.6,0.2);
\draw [thick] (2.4,0) -- (2.6,-0.2);
%\draw[thick] (3.11,0) circle [radius=0.1];
\draw [thick] (3.05, 0.05) to [out=45,in=315,looseness=15] (3.05,-0.05);
\filldraw[fill= white] (1,0) circle [radius=0.1] node[below] {\scriptsize 1};
\filldraw[fill= white] (2,0) circle [radius=0.1] node[below] {\scriptsize 2};
\filldraw[fill= white] (3,0) circle [radius=0.1] node[below] {\scriptsize N};
\end{tikzpicture}
\end{align}

\subsubsection{General magnetic quivers}

The procedure discussed thus far is successful in deriving the magnetic quiver for generic values of $k$ and $ \ell$, however the whole process of introducing FI deformations becomes increasingly cumbersome as we increase the value of $k$. We would like to present an algorithm that allows to construct the magnetic quiver in one fell swoop for any choice of 6d holonomies without any reference to the magnetic quiver of the parent 6d theory. To start discussing the algorithm let us look at the magnetic quiver obtained thus far, remembering that for generic values of $k$ we expect a $T(SU(k))$ tail attached to it. Let us begin for $\ell = 2$, where we obtain the following quiver for generic $k$
\begin{align}\label{eq:ell2quiv}
\begin{tikzpicture}
\filldraw[fill= gray] (1,0) circle [radius=0.1] node[below] {\scriptsize {$k$}};
\filldraw[fill= white] (2,0) circle [radius=0.1] node[below] {\scriptsize {N}$+\delta_1$ };
\filldraw[fill= white] (2,0) circle [radius=0.1] node[above] {\scriptsize $n_2$ };
\filldraw[fill= white] (3,0) circle [radius=0.1] node[below] {\scriptsize {2N}$+\delta_2$};
\filldraw[fill= white] (3,0) circle [radius=0.1] node[above] {\scriptsize $n_4$ };
\filldraw[fill= white] (4,0) circle [radius=0.1] node[below] {\scriptsize {3N}$+\delta_3$};
\filldraw[fill= white] (4,0) circle [radius=0.1] node[above] {\scriptsize $n_6$ };
\filldraw[fill= white] (5,0) circle [radius=0.1] node[below] {\scriptsize {4N}$+\delta_4$};
\filldraw[fill= white] (5,0) circle [radius=0.1] node[above] {\scriptsize $n_{4'}$};
\filldraw[fill= white] (6,0) circle [radius=0.1] node[below] {\scriptsize  {2N}$+\delta_5$};
\filldraw[fill= white] (6,0) circle [radius=0.1] node[above] {\scriptsize $n_{2'}$};
\draw [thick] (1.1, 0) -- (1.9,0) ;
\draw [thick] (2.1, 0) -- (2.9,0) ;
\draw [thick] (4.1, 0.05) -- (4.9,0.05) ;
\draw [thick] (4.1, -0.05) -- (4.9,-0.05) ;
\draw [thick] (3.1, 0) -- (3.9,0) ;
\draw [thick] (5.1, 0) -- (5.9,0) ;
\draw [thick] (4.4,0) -- (4.6,0.2);
\draw [thick] (4.4,0) -- (4.6,-0.2);
\end{tikzpicture}
\end{align}
We chose to denote with a gray circle the $T(SU(k))$ tail. Upon eliding the $T(SU(k))$ tail we just recover the twisted affine diagram of $E_6$, as expected from the fact that we have an $E_6$ 7-brane stack. The same will be true in all other cases we discuss in this section. Note that in writing the quiver we labeled each node above with one label of the $E_8$ holonomy in 6d compatible with the $\mathbb Z_2 $ symmetry. Then for a specific choice of 6d holonomy labeled by the tuple $[n_2,n_4,n_6,n_{4'},n_{2'}]$ the corresponding node must have a balance equal to the label of the 6d holonomy. To make the algorithm more clear, let us discuss the case of the holonomy $[0,0,0,1,0]$. Since $n_{4'}=1$ and all the others zero, the node attached to $n_{4'}$ needs to have balance equal to 1, and all the other nodes need to be balanced. Solving these equations fixes all parameters $\delta_i$ but one, and it is usually convenient to write in general $N+\delta_1 = L+ k$. In the example discussed this gives the quiver 
\begin{align}\label{eq:ell2quiv2}
\begin{tikzpicture}
\filldraw[fill= gray] (1,0) circle [radius=0.1] node[below] {\scriptsize {$2$}};
\filldraw[fill= white] (2,0) circle [radius=0.1] node[below] {\scriptsize { L+2} };
\filldraw[fill= white] (3,0) circle [radius=0.1] node[below] {\scriptsize {2L+2}};
\filldraw[fill= white] (4,0) circle [radius=0.1] node[below] {\scriptsize {3L+2}};
\filldraw[fill= white] (5,0) circle [radius=0.1] node[below] {\scriptsize {4L+2}};
\filldraw[fill= white] (6,0) circle [radius=0.1] node[below] {\scriptsize  {2L+1}};
\draw [thick] (1.1, 0) -- (1.9,0) ;
\draw [thick] (2.1, 0) -- (2.9,0) ;
\draw [thick] (4.1, 0.05) -- (4.9,0.05) ;
\draw [thick] (4.1, -0.05) -- (4.9,-0.05) ;
\draw [thick] (3.1, 0) -- (3.9,0) ;
\draw [thick] (5.1, 0) -- (5.9,0) ;
\draw [thick] (4.4,0) -- (4.6,0.2);
\draw [thick] (4.4,0) -- (4.6,-0.2);
\end{tikzpicture} 
\end{align}
This quiver has the right balancing rules, however there is one parameter that need to be fixed, which is equivalent to fixing $L$ in terms of $N$, that is writing $L = N + \alpha$ and fixing $\alpha$. We do this in the following fashion: We require that the fundamental quiver
\begin{align}
\begin{tikzpicture}
\filldraw[fill= white] (2,0) circle [radius=0.1] node[below] {\scriptsize 1};
\filldraw[fill= white] (3,0) circle [radius=0.1] node[below] {\scriptsize 2};
\filldraw[fill= white] (4,0) circle [radius=0.1] node[below] {\scriptsize 3};
\filldraw[fill= white] (5,0) circle [radius=0.1] node[below] {\scriptsize 4};
\filldraw[fill= white] (6,0) circle [radius=0.1] node[below] {\scriptsize 2};
\draw [thick] (2.1, 0) -- (2.9,0) ;
\draw [thick] (4.1, 0.05) -- (4.9,0.05) ;
\draw [thick] (4.1, -0.05) -- (4.9,-0.05) ;
\draw [thick] (3.1, 0) -- (3.9,0) ;
\draw [thick] (5.1, 0) -- (5.9,0) ;
\draw [thick] (4.4,0) -- (4.6,0.2);
\draw [thick] (4.4,0) -- (4.6,-0.2);
\end{tikzpicture} 
\end{align}
can be subtracted from \eqref{eq:ell2quiv2} at most $N$ times. In the case at hand this fixes $L = N$, indeed recovering the quiver \eqref{eq:foldK2ell2}.

A similar rule can be deduced for the other possible values of $\ell$. For $\ell = 3$ the starting point is
\begin{align}\label{eq:ell3quiv}
\begin{tikzpicture}
\filldraw[fill= gray] (1,0) circle [radius=0.1] node[below] {\scriptsize $k$};
\filldraw[fill= white] (2,0) circle [radius=0.1] node[below] {\scriptsize N+$\delta_1$};
\filldraw[fill= white] (2,0) circle [radius=0.1] node[above] {\scriptsize $n_3$};
\filldraw[fill= white] (3,0) circle [radius=0.1] node[below] {\scriptsize 2N+$\delta_2$};
\filldraw[fill= white] (3,0) circle [radius=0.1] node[above] {\scriptsize $n_6$};
\filldraw[fill= white] (4,0) circle [radius=0.1] node[below] {\scriptsize 3N+$\delta_3$};
\filldraw[fill= white] (4,0) circle [radius=0.1] node[above] {\scriptsize $n_{3'}$};
\draw [thick] (1.1, 0) -- (1.9,0) ;
\draw [thick] (2.1, 0) -- (2.9,0) ;
\draw [thick] (3.1, 0.07) -- (3.9,0.07) ;
\draw [thick] (3.1, -0.07) -- (3.9,-0.07) ;
\draw [thick] (3.1, 0) -- (3.9,0) ;
\draw [thick] (3.4,0) -- (3.6,0.2);
\draw [thick] (3.4,0) -- (3.6,-0.2);
\end{tikzpicture}
\end{align}
All parameters $\delta_i$ can be fixed by requiring the appropriate balancing in terms of parent $E_8$ holonomies. To fix the last parameter make sure that the following quiver
\begin{align}
\begin{tikzpicture}
\filldraw[fill= white] (2,0) circle [radius=0.1] node[below] {\scriptsize 1};
\filldraw[fill= white] (3,0) circle [radius=0.1] node[below] {\scriptsize 2};
\filldraw[fill= white] (4,0) circle [radius=0.1] node[below] {\scriptsize 3};
\draw [thick] (2.1, 0) -- (2.9,0) ;
\draw [thick] (3.1, 0.07) -- (3.9,0.07) ;
\draw [thick] (3.1, -0.07) -- (3.9,-0.07) ;
\draw [thick] (3.1, 0) -- (3.9,0) ;
\draw [thick] (3.4,0) -- (3.6,0.2);
\draw [thick] (3.4,0) -- (3.6,-0.2);
\end{tikzpicture}
\end{align}
can be subtracted at most $N$ times. Similarly, for $\ell = 4$ we can start from
\begin{align}
\begin{tikzpicture}
\filldraw[fill= gray] (1,0) circle [radius=0.1] node[below] {\scriptsize $k$};
\filldraw[fill= white] (2,0) circle [radius=0.1] node[below] {\scriptsize N+$\delta_1$};
\filldraw[fill= white] (2,0) circle [radius=0.1] node[above] {\scriptsize $n_4$};
\filldraw[fill= white] (3,0) circle [radius=0.1] node[below] {\scriptsize 2N+$\delta_2$};
\filldraw[fill= white] (3,0) circle [radius=0.1] node[above] {\scriptsize $2n_{4'}$};
\draw [thick] (1.1, 0) -- (1.9,0) ;
\draw [thick] (2.1, 0.06) -- (2.9,0.06) ;
\draw [thick] (2.1, 0.02) -- (2.9,0.02) ;
\draw [thick] (2.1,-0.02) -- (2.9,-0.02) ;
\draw [thick] (2.1, -0.06) -- (2.9,-0.06) ;
\draw [thick] (2.4,0) -- (2.6,0.2);
\draw [thick] (2.4,0) -- (2.6,-0.2);
\end{tikzpicture}
\end{align}
One parameter can be fixed by requiring the appropriate balancing rules in terms of $E_8$ holonomies, and the remaining parameter can be fixed by requiring that the quiver
\begin{align}
\begin{tikzpicture}
\filldraw[fill= white] (2,0) circle [radius=0.1] node[below] {\scriptsize 1};
\filldraw[fill= white] (3,0) circle [radius=0.1] node[below] {\scriptsize 2};
\draw [thick] (2.1, 0.06) -- (2.9,0.06) ;
\draw [thick] (2.1, 0.02) -- (2.9,0.02) ;
\draw [thick] (2.1,-0.02) -- (2.9,-0.02) ;
\draw [thick] (2.1, -0.06) -- (2.9,-0.06) ;
\draw [thick] (2.4,0) -- (2.6,0.2);
\draw [thick] (2.4,0) -- (2.6,-0.2);
\end{tikzpicture}
\end{align}
can be subtracted at most $N$ times. For $\ell =5 $ and $\ell = 6$ the resulting quivers are so simple that we can just present them for any value of $k$: For $\ell = 5$ we have
\begin{align}\label{eq:z5quiver}
\begin{tikzpicture}
\draw [thick] (2, .1) -- (3,.1) ;
\draw [thick] (2, -.1) -- (3,.-.1) ;
\draw [thick] (2, -.1) -- (3,.-.1) ;
\draw [thick] (2.1, 0) -- (2.9,0) ;
\draw [thick] (2+0.07, -0.05) -- (3-0.07,-0.05) ;
\draw [thick] (2+0.07, 0.05) -- (3-0.07,0.05) ;
\draw [thick] (2.4,0) -- (2.6,0.2);
\draw [thick] (2.4,0) -- (2.6,-0.2);
%\draw[thick] (3.11,0) circle [radius=0.1];
\draw [thick] (3.05, 0.05) to [out=45,in=315,looseness=15] (3.05,-0.05);
\filldraw[fill= gray] (2,0) circle [radius=0.1] node[below] {\scriptsize $k$};
\filldraw[fill= white] (3,0) circle [radius=0.1] node[below] {\scriptsize N};
\end{tikzpicture}
\end{align}
where $k = n_5$, 
and for $\ell = 6$ we have
\begin{align}\label{eq:z6quiver}
\begin{tikzpicture}
\draw [thick] (2, .1) -- (3,.1) ;
\draw [thick] (2, -.1) -- (3,.-.1) ;
\draw [thick] (2, -.1) -- (3,.-.1) ;
\draw [thick] (2,0.06) -- (3,0.06);
\draw [thick] (2,0.02) -- (3,0.02);
\draw [thick] (2,-0.06) -- (3,-0.06);
\draw [thick] (2,-0.02) -- (3,-0.02);
\draw [thick] (2.4,0) -- (2.6,0.2);
\draw [thick] (2.4,0) -- (2.6,-0.2);
%\draw[thick] (3.11,0) circle [radius=0.1];
\draw [thick] (3.05, 0.05) to [out=45,in=315,looseness=15] (3.05,-0.05);
\filldraw[fill= gray] (2,0) circle [radius=0.1] node[below] {\scriptsize $k$};
\filldraw[fill= white] (3,0) circle [radius=0.1] node[below] {\scriptsize N};
\end{tikzpicture}
\end{align}
where $k= n_6$.

For all cases we can actually provide an equivalent algorithm much akin to the one introduced in \cite{Mekareeya:2017jgc}. We start with $\ell = 2$: For any choice of 6d holonomies compatible with a $\mathbb Z_2$ quotient we find that the magnetic quiver after twisted compactification is
\begin{align}
\begin{tikzpicture}
\filldraw[fill= gray] (1,0) circle [radius=0.1] node[below] {\scriptsize {$k$}};
\filldraw[fill= white] (2,0) circle [radius=0.1] node[below] {\scriptsize {N}${}_1$ };
\filldraw[fill= white] (3,0) circle [radius=0.1] node[below] {\scriptsize {N}${}_{2}$};
\filldraw[fill= white] (4,0) circle [radius=0.1] node[below] {\scriptsize {N}${}_{3}$};
\filldraw[fill= white] (5,0) circle [radius=0.1] node[below] {\scriptsize {N}${}_4$};
\filldraw[fill= white] (6,0) circle [radius=0.1] node[below] {\scriptsize  {N}${}_5$};
\draw [thick] (1.1, 0) -- (1.9,0) ;
\draw [thick] (2.1, 0) -- (2.9,0) ;
\draw [thick] (4.1, 0.05) -- (4.9,0.05) ;
\draw [thick] (4.1, -0.05) -- (4.9,-0.05) ;
\draw [thick] (3.1, 0) -- (3.9,0) ;
\draw [thick] (5.1, 0) -- (5.9,0) ;
\draw [thick] (4.4,0) -- (4.6,0.2);
\draw [thick] (4.4,0) -- (4.6,-0.2);
\end{tikzpicture}
\end{align}
Here $k = n_2 + n_{2'} + 2n_4 + 2n_{4'} + 3n_6$. The tuple of integers $\underline N$ is given by
\begin{align}
\underline N = N_2 \underline{d}_2  + \sum_{i} n_i \underline{q}_i\,,
\end{align}
where
\begin{align}
N_2 &= N-  \left\lfloor \frac{n_{4'}}{2} \right\rfloor - n_2 -n_4 -n_6\,,\\
d_2 & = [1,2,3,4,2]\,, \\
q_2 & = [1,2,3,4,2]\,, \qquad q_4  = [2,2,3,4,2]\,, \qquad  q_6  = [3,3,3,4,2]\,,\\
q_{2'} &= [1,1,1,1,0]\,, \qquad q_{4'} = [2,2,2,2,0]\,.
\end{align}
We can produce similar rules for other values of $\ell$. For $\ell = 3$ the quiver has the form
\begin{align}
\begin{tikzpicture}
\filldraw[fill= gray] (1,0) circle [radius=0.1] node[below] {\scriptsize $k$};
\filldraw[fill= white] (2,0) circle [radius=0.1] node[below] {\scriptsize N${}_1$};
\filldraw[fill= white] (3,0) circle [radius=0.1] node[below] {\scriptsize N${}_2$};
\filldraw[fill= white] (4,0) circle [radius=0.1] node[below] {\scriptsize N${}_3$};
\draw [thick] (1.1, 0) -- (1.9,0) ;
\draw [thick] (2.1, 0) -- (2.9,0) ;
\draw [thick] (3.1, 0.07) -- (3.9,0.07) ;
\draw [thick] (3.1, -0.07) -- (3.9,-0.07) ;
\draw [thick] (3.1, 0) -- (3.9,0) ;
\draw [thick] (3.4,0) -- (3.6,0.2);
\draw [thick] (3.4,0) -- (3.6,-0.2);
\end{tikzpicture}
\end{align}
Here $k = n_3 + n_{3'} + 2 n_6$. The tuple of integers $\underline N$ is given by
\begin{align}
\underline N = N_3 \underline{d}_3  + \sum_{i} n_i \underline{q}_i\,,
\end{align}
where
\begin{align}
N_3 &= N-  \left\lfloor \frac{n_{3'}}{3} \right\rfloor - n_3  -n_6\,,\\
d_3 & = [1,2,3]\,, \\
q_3 & = [1,2,3]\,, \qquad q_6  = [1,0,0]\,, \qquad q_{3'} = [1,1,1]\,.
\end{align}
Finally for $\ell = 4$ the quiver is
\begin{align}
\begin{tikzpicture}
\filldraw[fill= gray] (1,0) circle [radius=0.1] node[below] {\scriptsize $k$};
\filldraw[fill= white] (2,0) circle [radius=0.1] node[below] {\scriptsize N${}_1$};
\filldraw[fill= white] (3,0) circle [radius=0.1] node[below] {\scriptsize N${}_2$};
\draw [thick] (1.1, 0) -- (1.9,0) ;
\draw [thick] (2.1, 0.06) -- (2.9,0.06) ;
\draw [thick] (2.1, 0.02) -- (2.9,0.02) ;
\draw [thick] (2.1,-0.02) -- (2.9,-0.02) ;
\draw [thick] (2.1, -0.06) -- (2.9,-0.06) ;
\draw [thick] (2.4,0) -- (2.6,0.2);
\draw [thick] (2.4,0) -- (2.6,-0.2);
\end{tikzpicture}
\end{align}
Here $k= n_4+n_{4'}$.  The tuple of integers $\underline N$ is given by
\begin{align}
\underline N = N_4 \underline{d}_4  + \sum_{i} n_i \underline{q}_i\,,
\end{align}
where
\begin{align}
N_4 &= N-  \left\lfloor \frac{n_{4'}}{2} \right\rfloor - n_2 \,,\\
d_4& = [1,2]\,, \\
q_4 & = [1,2]\,, \qquad q_{4'}  = [1,1]\,.
\end{align}
The cases of $\ell = 5$ and $\ell =6$ were already given in equations \eqref{eq:z5quiver} and \eqref{eq:z6quiver} respectively for generic $k$.

\subsection{From 6d orbi-instanton to S-folded 7-branes}

The shape of the magnetic quivers we derived in the previous subsection is justified by the connection between the SW compactification of the 6d orbi-instanton theories and the theories on $D3$-branes probing 7-branes in the presence of S-folds. In particular, the choices of $E_8$ holonomy in 6d that are compatible with the SW twist are in 1:1 correspondence with the choices of holonomy for the corresponding maximal S-folded 7-brane. The latter, in turn, are described by assigning Kac labels to the associated twisted affine Dynkin diagrams \cite{Giacomelli:2020gee}. Such Kac labels parametrize an automorphism of order $M$ of the gauge algebra supported on the 7-brane; this automorphism has order $\ell'$, a divisor of $\ell$, as an outer automorphism. According to Kac's theorem (see \cite{Kac}, Theorem 8.6), we have
\be
\sum_{\text{nodes}}d_in_i=\frac{M}{\ell'}\,,
\ee
where the $n_i$ are associated to admissible holonomy choices and $d_i$ are the Dynkin labels of the corresponding nodes in the twisted affine Dynkin diagram. In this subsection we will explicitly describe this map of holonomies.

\paragraph{$\ell=2$.}
In this case, the map of holonomies is between the affine $E_8$ Dynkin diagram and the twisted affine $E_6$ Dynkin diagram, because the maximal 7-brane compatible with the $\ell=2$ S-fold is the $E_6$-type one. The map in terms of Kac labels is the following
\begin{align}
\begin{tikzpicture}
\filldraw[fill= white] (4,0) circle [radius=0.1] node[below] {\scriptsize $n_1$};
\filldraw[fill= green] (5,0) circle [radius=0.1] node[below] {\scriptsize $n_2$};
\filldraw[fill= white] (6,0) circle [radius=0.1] node[below] {\scriptsize $n_3$};
\filldraw[fill= green] (7,0) circle [radius=0.1] node[below] {\scriptsize $n_4$};
\filldraw[fill= white] (8,0) circle [radius=0.1] node[below] {\scriptsize $n_5$};
\filldraw[fill= green] (9,0) circle [radius=0.1] node[below] {\scriptsize $n_6$};
\filldraw[fill= green] (10,0) circle [radius=0.1] node[below] {\scriptsize $n_{4'}$};
\filldraw[fill= green] (11,0) circle [radius=0.1] node[below] {\scriptsize $n_{2'}$};
\filldraw[fill= white] (9,1) circle [radius=0.1] node[above] {\scriptsize $n_{3'}$};
\draw [thick] (4.1, 0) -- (4.9,0) ;
\draw [thick] (5.1, 0) -- (5.9,0) ;
\draw [thick] (6.1, 0) -- (6.9,0) ;
\draw [thick] (7.1, 0) -- (7.9,0) ;
\draw [thick] (8.1, 0) -- (8.9,0) ;
\draw [thick] (9.1, 0) -- (9.9,0) ;
\draw [thick] (10.1, 0) -- (10.9,0) ;
\draw [thick] (9, 0.1) -- (9,0.9) ;
\draw[thick,<->] (12, 0) -- (13.5,0) ;
\filldraw[fill= white] (15-0.5,0) circle [radius=0.1] node[below] {\scriptsize $n_2$};
\filldraw[fill= white] (16-0.5,0) circle [radius=0.1] node[below] {\scriptsize $n_4$ };
\filldraw[fill= white] (17-0.5,0) circle [radius=0.1] node[below] {\scriptsize $n_6$};
\filldraw[fill= white] (18-0.5,0) circle [radius=0.1] node[below] {\scriptsize $n_{4'}$};
\filldraw[fill= white] (19-0.5,0) circle [radius=0.1] node[below] {\scriptsize $n_{2'}$ };
\draw [thick] (15.1-0.5, 0) -- (15.9-0.5,0) ;
\draw [thick] (16.1-0.5, 0) -- (16.9-0.5,0) ;
\draw [thick] (17.1-0.5, 0.05) -- (17.9-0.5,0.05) ;
\draw [thick] (17.1-0.5, -0.05) -- (17.9-0.5,-0.05) ;
\draw [thick] (18.1-0.5, 0) -- (18.9-0.5,0) ;
\draw [thick] (17.4-0.5,0) -- (17.6-0.5,0.2);
\draw [thick] (17.4-0.5,0) -- (17.6-0.5,-0.2);
\end{tikzpicture}
\end{align}
where the admissible $E_8$ Kac labels are highlighted by green nodes. The Kac labels of the diagram on the right parametrize an automorphism of the $E_6$ algebra of order $M$, which has order $\ell'=\ell=2$ as an outer automorphism.

\paragraph{$\ell=3$.}
In this case, the map of holonomies is between the affine $E_8$ Dynkin diagram and the twisted affine $D_4$ Dynkin diagram, because the maximal 7-brane compatible with the $\ell=3$ S-fold is the $D_4$-type one. The map in terms of Kac labels is the following
\begin{align}
\begin{tikzpicture}
\filldraw[fill= white] (4,0) circle [radius=0.1] node[below] {\scriptsize $n_1$};
\filldraw[fill= white] (5,0) circle [radius=0.1] node[below] {\scriptsize $n_2$};
\filldraw[fill= green] (6,0) circle [radius=0.1] node[below] {\scriptsize $n_3$};
\filldraw[fill= white] (7,0) circle [radius=0.1] node[below] {\scriptsize $n_4$};
\filldraw[fill= white] (8,0) circle [radius=0.1] node[below] {\scriptsize $n_5$};
\filldraw[fill= green] (9,0) circle [radius=0.1] node[below] {\scriptsize $n_6$};
\filldraw[fill= white] (10,0) circle [radius=0.1] node[below] {\scriptsize $n_{4'}$};
\filldraw[fill= white] (11,0) circle [radius=0.1] node[below] {\scriptsize $n_{2'}$};
\filldraw[fill= green] (9,1) circle [radius=0.1] node[above] {\scriptsize $n_{3'}$};
\draw [thick] (4.1, 0) -- (4.9,0) ;
\draw [thick] (5.1, 0) -- (5.9,0) ;
\draw [thick] (6.1, 0) -- (6.9,0) ;
\draw [thick] (7.1, 0) -- (7.9,0) ;
\draw [thick] (8.1, 0) -- (8.9,0) ;
\draw [thick] (9.1, 0) -- (9.9,0) ;
\draw [thick] (10.1, 0) -- (10.9,0) ;
\draw [thick] (9, 0.1) -- (9,0.9) ;
\draw[thick,<->] (12, 0) -- (13.5,0) ;
\filldraw[fill= white] (14.5,0) circle [radius=0.1] node[below] {\scriptsize $n_3$};
\filldraw[fill= white] (15.5,0) circle [radius=0.1] node[below] {\scriptsize $n_6$};
\filldraw[fill= white] (16.5,0) circle [radius=0.1] node[below] {\scriptsize $n_{3'}$};
\draw [thick] (14.6, 0) -- (15.4,0) ;
\draw [thick] (15.6, 0.07) -- (16.4,0.07) ;
\draw [thick] (15.6, -0.07) -- (16.4,-0.07) ;
\draw [thick] (15.6, 0) -- (16.4,0) ;
\draw [thick] (15.9,0) -- (16.1,0.2);
\draw [thick] (15.9,0) -- (16.1,-0.2);
\end{tikzpicture}
\end{align}
where the admissible $E_8$ Kac labels are highlighted by green nodes. The Kac labels of the diagram on the right parametrize an automorphism of the $D_4$ algebra of order $M$, which has order $\ell'=\ell=3$ as an outer automorphism.

\paragraph{$\ell=4$.}
In this case, the map of holonomies is between the affine $E_8$ Dynkin diagram and the twisted affine $A_2$ Dynkin diagram, because the maximal 7-brane compatible with the $\ell=4$ S-fold is the $\mathcal{H}_2$-type one. The map in terms of Kac labels is the following
\begin{align}
\begin{tikzpicture}
\filldraw[fill= white] (4,0) circle [radius=0.1] node[below] {\scriptsize $n_1$};
\filldraw[fill= white] (5,0) circle [radius=0.1] node[below] {\scriptsize $n_2$};
\filldraw[fill= white] (6,0) circle [radius=0.1] node[below] {\scriptsize $n_3$};
\filldraw[fill= green] (7,0) circle [radius=0.1] node[below] {\scriptsize $n_4$};
\filldraw[fill= white] (8,0) circle [radius=0.1] node[below] {\scriptsize $n_5$};
\filldraw[fill= white] (9,0) circle [radius=0.1] node[below] {\scriptsize $n_6$};
\filldraw[fill= green] (10,0) circle [radius=0.1] node[below] {\scriptsize $n_{4'}$};
\filldraw[fill= white] (11,0) circle [radius=0.1] node[below] {\scriptsize $n_{2'}$};
\filldraw[fill= white] (9,1) circle [radius=0.1] node[above] {\scriptsize $n_{3'}$};
\draw [thick] (4.1, 0) -- (4.9,0) ;
\draw [thick] (5.1, 0) -- (5.9,0) ;
\draw [thick] (6.1, 0) -- (6.9,0) ;
\draw [thick] (7.1, 0) -- (7.9,0) ;
\draw [thick] (8.1, 0) -- (8.9,0) ;
\draw [thick] (9.1, 0) -- (9.9,0) ;
\draw [thick] (10.1, 0) -- (10.9,0) ;
\draw [thick] (9, 0.1) -- (9,0.9) ;
\draw[thick,<->] (12, 0) -- (13.5,0) ;
\filldraw[fill= white] (14.5,0) circle [radius=0.1] node[below] {\scriptsize ${n}_4$};
\filldraw[fill= white] (15.5,0) circle [radius=0.1] node[below] {\scriptsize $2{n}_{4'}$};
\draw [thick] (14.6, 0.06) -- (15.4,0.06) ;
\draw [thick] (14.6, 0.02) -- (15.4,0.02) ;
\draw [thick] (14.6,-0.02) -- (15.4,-0.02) ;
\draw [thick] (14.6, -0.06) -- (15.4,-0.06) ;
\draw [thick] (14.9,0) -- (15.1,0.2);
\draw [thick] (14.9,0) -- (15.1,-0.2);
\end{tikzpicture}
\end{align}
where the admissible $E_8$ Kac labels are highlighted by green nodes. The Kac labels of the diagram on the right parametrize an automorphism of the $A_2$ algebra of order $M$, which has order $\ell'=\frac{\ell}{2}=2$ as an outer automorphism.\\

For non-maximal choices of 7-branes (which we will discuss in Section \ref{Sec:MassDef}), the $T(SU(k))$ tail is attached to the (twisted)-affine Dynkin diagram of the corresponding type. Here is the full list where we also indicate the Dynkin labels of the nodes in the diagrams:
\be\label{xxxx}
\begin{array}{|c|c|c|c|c|} 
\hline
\ell & G & \ell' & G^{(\ell')} & \text{Dynkin Diagram} \\ 
\hline 
2& E_6 & 2 & E_6^{(2)} & \begin{tikzpicture}\filldraw[fill= white] (15-0.5,0) circle [radius=0.1] node[below] {\scriptsize $1$};
\filldraw[fill= white] (16-0.5,0) circle [radius=0.1] node[below] {\scriptsize $2$ };
\filldraw[fill= white] (17-0.5,0) circle [radius=0.1] node[below] {\scriptsize $3$};
\filldraw[fill= white] (18-0.5,0) circle [radius=0.1] node[below] {\scriptsize $2$};
\filldraw[fill= white] (19-0.5,0) circle [radius=0.1] node[below] {\scriptsize $1$ };
\node[] at (14,0.5) {};
\draw [thick] (15.1-0.5, 0) -- (15.9-0.5,0) ;
\draw [thick] (16.1-0.5, 0) -- (16.9-0.5,0) ;
\draw [thick] (17.1-0.5, 0.05) -- (17.9-0.5,0.05) ;
\draw [thick] (17.1-0.5, -0.05) -- (17.9-0.5,-0.05) ;
\draw [thick] (18.1-0.5, 0) -- (18.9-0.5,0) ;
\draw [thick] (17.4-0.5,0) -- (17.6-0.5,0.2);
\draw [thick] (17.4-0.5,0) -- (17.6-0.5,-0.2);
\end{tikzpicture}\\
\hline 
2& D_4 & 2 & D_4^{(2)} & \begin{tikzpicture}
\filldraw[fill= white] (16-0.5,0) circle [radius=0.1] node[below] {\scriptsize $1$ };
\filldraw[fill= white] (17-0.5,0) circle [radius=0.1] node[below] {\scriptsize $1$};
\filldraw[fill= white] (18-0.5,0) circle [radius=0.1] node[below] {\scriptsize $1$};
\filldraw[fill= white] (19-0.5,0) circle [radius=0.1] node[below] {\scriptsize $1$ };
\node[] at (15,0.5) {};
\draw [thick] (16.1+0.5, 0) -- (16.9+0.5,0) ;
\draw [thick] (17.1-1.5, 0.05) -- (17.9-1.5,0.05) ;
\draw [thick] (17.1-1.5, -0.05) -- (17.9-1.5,-0.05) ;
\draw [thick] (17.4-1.5,0) -- (17.6-1.5,0.2);
\draw [thick] (17.4-1.5,0) -- (17.6-1.5,-0.2);
\draw [thick] (17.1+0.5, 0.05) -- (17.9+0.5,0.05) ;
\draw [thick] (17.1+0.5, -0.05) -- (17.9+0.5,-0.05) ;
\draw [thick] (17.6+0.5,0) -- (17.4+0.5,0.2);
\draw [thick] (17.6+0.5,0) -- (17.4+0.5,-0.2);
\end{tikzpicture}\\
\hline 
2& A_2& 1 & A_2^{(1)} & \begin{tikzpicture}\filldraw[fill= white] (14.5,0) circle [radius=0.1] node[below] {\scriptsize $1$};
\filldraw[fill= white] (15.5,0) circle [radius=0.1] node[below] {\scriptsize $1$};
\filldraw[fill= white] (15,1) circle [radius=0.1] node[left] {\scriptsize $1$};
\draw [thick] (14.6, 0) -- (15.4,0) ;
\draw [thick] (14.53,0.07) -- (14.97,0.93) ;
\draw [thick] (15.03,0.93) -- (15.47,0.07) ;
\end{tikzpicture}\\
\hline 
3& D_4 & 3 & D_4^{(3)} & \begin{tikzpicture}\filldraw[fill= white] (14.5,0) circle [radius=0.1] node[below] {\scriptsize $1$};
\filldraw[fill= white] (15.5,0) circle [radius=0.1] node[below] {\scriptsize $2$};
\filldraw[fill= white] (16.5,0) circle [radius=0.1] node[below] {\scriptsize $1$};
\node[] at (14,0.5) {};
\draw [thick] (14.6, 0) -- (15.4,0) ;
\draw [thick] (15.6, 0.07) -- (16.4,0.07) ;
\draw [thick] (15.6, -0.07) -- (16.4,-0.07) ;
\draw [thick] (15.6, 0) -- (16.4,0) ;
\draw [thick] (15.9,0) -- (16.1,0.2);
\draw [thick] (15.9,0) -- (16.1,-0.2);
\end{tikzpicture}\\
\hline 
3& A_1& 1 & A_1^{(1)} & \begin{tikzpicture}\filldraw[fill= white] (14.5,0) circle [radius=0.1] node[below] {\scriptsize $1$};
\filldraw[fill= white] (15.5,0) circle [radius=0.1] node[below] {\scriptsize $1$};
\node[] at (14,0.5) {};
\draw [thick] (14.6, 0.05) -- (15.4,0.05) ;
\draw [thick] (14.6,-0.05) -- (15.4,-0.05) ;
\end{tikzpicture}\\
\hline 
4& A_2 & 2 & A_2^{(2)} & \begin{tikzpicture}\filldraw[fill= white] (14.5,0) circle [radius=0.1] node[below] {\scriptsize $2$};
\filldraw[fill= white] (15.5,0) circle [radius=0.1] node[below] {\scriptsize $1$};
\node[] at (14,0.5) {};
\draw [thick] (14.6, 0.06) -- (15.4,0.06) ;
\draw [thick] (14.6, 0.02) -- (15.4,0.02) ;
\draw [thick] (14.6,-0.02) -- (15.4,-0.02) ;
\draw [thick] (14.6, -0.06) -- (15.4,-0.06) ;
\draw [thick] (14.9,0) -- (15.1,0.2);
\draw [thick] (14.9,0) -- (15.1,-0.2);
\end{tikzpicture}\\
\hline
\end{array}
\ee 
As we can see from Table \eqref{xxxx}, the order as an outer automorphism $\ell'$ depends on the choice of 7-brane, while its order as an automorphism is inherited from the parent 6d theory and is always equal to $M$. We will explain in Section \ref{Sec:MassDef} how to determine the Kac labels for non-maximal 7-branes starting from those of the corresponding maximal 7-brane.

\subsection{Further properties of the 4d SCFTs}\label{sec:4dprop} 

In this section we would like to present some further properties of the SCFTs discussed in the previous section for the various values of $\ell$ and low values of $k$. We will discuss how to extract from the 6d description several properties like central charges, Coulomb-branch spectrum, and dimension of the Higgs branch. These properties were derived in \cite{Heckman:2022suy}, however it is worth reviewing them with an eye towards discussing them in the case of Type IIB string theory which will be the focus of Section \ref{Sec:IIBSfolds}.\footnote{A word of caution on the notation: In our setup we always take $N$ to be the number of tensor fields in 6d, or equivalently the number of $M5$-branes in M-theory. In \cite{Heckman:2022suy} $N-1$ was taken to be the number of tensor fields after the ramp of unitary gauge groups saturated in the 6d quiver. Formulas in this paper have been adapted from \cite{Heckman:2022suy} to reflect our definition of $N$.} For any choice of $\ell $ and $k$ we will present only the examples with Higgs branch of maximal dimension, with all the remaining cases discussed in Appendix \ref{App:Mth}.

\subsubsection{Central charges}

The computation of central charges done in \cite{Heckman:2022suy} follows the initial computations done in \cite{Ohmori:2018ona} for twisted compactifications. The whole analysis is an adaptation of the computation of \cite{Shapere:2008zf} to the case where S-folds are present. The idea is to move onto a generic point of the Coulomb branch of the theory so that the theory just becomes some $n_H$ hypermultiplets coupled to $n_V$ Abelian vector multiplet. Then using this it is possible to write the full central charges $a$, $c$, and the flavor central charge $\kappa_a$ of a simple factor of the flavor symmetry as
\begin{align}
a - a_{\text{generic}} &= 32 \left(\frac{3}{2\ell} -\frac{3}{4}\right) A -\frac{12}{\ell } B\,,\\
c - c_{\text{generic}} &= 32 \left(\frac{3}{\ell} -1\right) A -\frac{12}{\ell } B\,,\\
\kappa_a - \kappa_{a, \text{generic}} &= \frac{192}{\ell} C_a I_a\,.\label{fccM1}
\end{align}
In all three formulas the terms labelled $\text{generic}$ include the contributions that come from fields at a generic point on the Coulomb branch, and these terms can be computed from a weakly coupled analysis
\begin{align}
a_{\text{generic}} &= \frac{5}{24} n_V +\frac{1}{24} n_H\,,\\
c_{\text{generic}} &= \frac{1}{6} n_V +\frac{1}{12} n_H\,,\\
\kappa_{a, \text{generic}} &= 2 \hat k_{a}\,.\label{fccM2}
\end{align}
We need to explain what the number $\hat k_a$ is: For the case of the flavor symmetry factor coming from a 6d $\mathfrak{su}_p$ flavor factor on a (-2)-curve then one has $\hat k_a = p/\ell$. The case of a (-1)-curve is more delicate, but all cases are listed in Table 3 of \cite{Heckman:2022suy} in the $k_0$ column in terms of the residual 4d flavor symmetry after the twisted compactification. The terms $A$, $B$, and $C_a$ can be computed from the 6d anomaly polynomial. We refer to \cite{Heckman:2022suy} for a derivation of these coefficients. Finally the coefficient $I_a$ in the formula for the flavor central charge is the embedding index of the 4d flavor symmetry as a subalgebra of the 6d flavor symmetry.

Formulas for central charges for generic values of $N$ and various choices of 6d holonomies are always rather involved. They are all listed in \cite{Heckman:2022suy} and all formulas tend to take several lines. We therefore focus on the cases discussed in the theories analyzed in Section \ref{sec:magquivMth} providing values for the various central charges. To further simplify all formulas we will always list the combinations $4(2a-c)$ and $24(c-a)$ rather than the central charges $a$ and $c$ themselves.

\subsubsection{Coulomb-branch spectrum and Higgs-branch dimension}\label{CBspectrumMaxM}

Another very important property of the 4d theories that can be computed starting from the 6d construction is the spectrum of Coulomb-branch operators. The rule was first discussed in \cite{Ohmori:2018ona} and we will discuss its specialization to the case of orbi-instanton theories in 6d discussed in \cite{Heckman:2022suy}. Looking at the 6d quiver on the tensor branch, start by assigning to each curve a label $r$ in the following fashion: The rightmost curve will have $r=1$, and moving to the right increase $r$ by one for each tensor. Clearly the label $r$ will range from $1$ to $N$, and the (-1)-curve will have $r= N$. Then each curve will contribute separately to the Coulomb-branch spectrum. A (-2)-curve with label $r$ and gauge algebra in 6d given by $\mathfrak{su}_{\ell p_r}$ will give Coulomb-branch operators in 4d with the following conformal dimensions
\begin{align}
\Delta_r =  \{ 6r \}\,  \cup\, \{ 6r + d \, | \, d= 2, \dots, p_r\}\,.
\end{align}
The remaining curve will be the (-1)-curve, and the resulting spectrum depends on the gauge algebra $\mathfrak g$ in 6d. We list the dimensions of the operators for the various cases relevant to our analysis
\begin{align}
\mathfrak g = \varnothing\,, \quad &\rightarrow \quad \Delta_1 = \left\{\frac{6N}{\ell}\right\}\,,\\
\mathfrak g = \mathfrak{su}_{\ell p_1} \,, \quad &\rightarrow \quad \Delta_1 =  \{ 6N \}\,  \cup\, \{ 6N + d \, | \, d= 2, \dots, p_1\}\,,\\
\mathfrak g = \mathfrak{usp}_{4m+2} \,, \quad &\rightarrow \quad \Delta_1 =  \{ 6N \}\,  \cup\, \{ 6N + 2d \, | \, d= 2, \dots, m\}\,,\\
\mathfrak g = \mathfrak{usp}_{4m} \,, \quad &\rightarrow \quad \Delta_1 =  \{ 6N \}\,  \cup\, \{ 6N + 2d \, | \, d= 2, \dots, m-1\} \, \cup \, \{3N+m\}\,.
\end{align}
Note that for the case of symplectic groups only $\ell =2$ is allowed. One property of the Coulomb-branch operators spectrum is that the dimensions need to satisfy the Shapere--Tachikawa relation \cite{Shapere:2008zf}
\begin{align}\label{STrelation}
4(2a-c) &= \sum_i (2 \Delta_i -1)\,,
\end{align}
where the sum runs over all the Coulomb-branch operators. This is a sanity check that computations of central charges and spectrum of Coulomb-branch operators are correct, and this was confirmed in \cite{Heckman:2022suy}.\\

The results of \cite{Heckman:2022suy} can also give the dimension of the Higgs branch for most of the theories we analyzed. Indeed, if a theory is completely higgsable, that is on a generic point of the Higgs branch the theory only has a bunch of hypermultiplets, the dimension of the Higgs branch is simply given by the combination $24(c-a)$. This does not work for $\ell > 4$, and for $\ell < 5$ it does not work when the 7-brane is absent. There can be other instances where this rule does not work, and an analysis of the Hasse diagram can help find the counterexamples to this rule: We will discuss this in detail in Section \ref{Sec:IIBSfolds}.

\subsubsection{Properties of theories with low values of $k$}

Here we list the properties discussed thus far for all the theories introduced in Section \ref{sec:magquivMth}.

\paragraph{ Theories with $\ell=2$.} In Section \ref{sec:magquivMth} we discussed one theory with $k=2$ and one with $k=3$. For the former, that is the theory $\mathcal S_{E_6,2}^{(N)}(0,0,0,1,0)$, one has 
\begin{align}
4 (2a-c) &= 12 N^2 +14 N\,,\\
24 (c-a) &= 12 N +11\,,\\
\mathfrak{g} &= (\mathfrak{su}_2)_{6N+3} \oplus (\mathfrak{su}_4)_{12N+4} \oplus (\mathfrak{su}_2)_{16}\,,\\
\Delta &= \{\underbrace{(6k,6k+2)}_{k=1,\dots,N} \}\,,\qquad \text{dim}({\mathbb H}) = 12N+11\,.
\end{align}
In all examples we will write the flavor symmetry algebra as $\mathfrak g$ and attach to every non-Abelian factor a subscript with the value of the flavor central charge.\footnote{In general the flavor algebra can have some enhancements for low values of $N$. All cases are discussed in \cite{Heckman:2022suy}. Here we always assume that $N$ is sufficiently large as to avoid such enhancements.} Moving to the $k=3$ case, that is the theory $\mathcal S_{E_6,2}^{(N)} (0,0,0,1,1)$, one finds
\begin{align}
4 (2a-c) &= 18 N^2 +25 N\,,\\
24 (c-a) &= 12 N +18\,,\\
\mathfrak{g} &= (\mathfrak{su}_4)_{12N+6} \oplus (\mathfrak{su}_3)_{18}\,,\\
\Delta &= \{\underbrace{(6k,6k+2,6k+3)}_{k=1,\dots,N} \}\,,\qquad \text{dim}({\mathbb H}) = 12N+18\,.
\end{align}

\paragraph{ Theories with $\ell=3$.} In Section \ref{sec:magquivMth} we only discussed one theory with $k=2$, that is the theory $\mathcal R^{(N)}_{D_4,3}(0,0,2)$. Its central charges are
\begin{align}
4 (2a-c) &= 12 N^2 +14 N\,,\\
24 (c-a) &= 6 N +8\,,\\
\mathfrak{g} &= (\mathfrak{su}_3)_{12N+4} \oplus  (\mathfrak{su}_2)_{16}\,,\\
\Delta &= \{\underbrace{(6k,6k+2)}_{k=1,\dots,N} \}\,,\qquad \text{dim}({\mathbb H}) = 6N+8\,.
\end{align}

\paragraph{ Theories with  $\ell=4$.} In Section \ref{sec:magquivMth} we only discussed one theory with $k=2$, that is the theory $\mathcal T^{(N)}_{H_2,4}(0,2)$. Its central charges are
\begin{align}
4 (2a-c) &= 12 N^2 -7N -3\,,\\
24 (c-a) &=3N+3\,,\\
\mathfrak{g} &= (\mathfrak{su}_2)_{12N-8} \oplus  (\mathfrak{su}_2)_{16}\,,\\
\Delta &= \{\underbrace{(6k,6k+2)}_{k=1,\dots,N-1} ,3N/2\}\,,\qquad \text{dim}({\mathbb H}) = 3N+3\,.
\end{align}

\paragraph{ Central charges for $\ell=5$.} In Section \ref{sec:magquivMth} we only discussed one theory with $k=2$, that is the theory $\mathcal T_{\varnothing,5}^{(N)}(2)$. Its central charges are
\begin{align}
4 (2a-c) &= 12 N^2 -\frac{98}{5}N +6\,,\\
24 (c-a) &=\frac{6}{5}N+2\,,\\
\mathfrak{g} &= (\mathfrak{su}_2)_{16}\,,\\
\Delta &= \{\underbrace{(6k,6k+2)}_{k=1,\dots,N-2},6(N-1), 6N/5 \}\,,\qquad \text{dim}({\mathbb H}) = N+2\,.
\end{align}

\paragraph{ Central charges for $\ell=6$.} In Section \ref{sec:magquivMth} we only discussed one theory with $k=2$, that is the theory $\mathcal T_{\varnothing,6}^{(N)}(2)$. Its central charges are
\begin{align}
4 (2a-c) &= 12 N^2 -20N +6\,,\\
24 (c-a) &=2\,,\\
\mathfrak{g} &= (\mathfrak{su}_2)_{16}\,,\\
\Delta &= \{\underbrace{(6k,6k+2)}_{k=1,\dots,N-2},6(N-1), N \}\,,\qquad \text{dim}({\mathbb H}) = N+2\,.
\end{align}

\subsubsection{Hasse diagrams}\label{HasseMtheory}

In this section we would like to understand the Higgs-branch flow for all cases of $\ell $ and $k$ discussed thus far. We will see how all theories for fixed values of $\ell $ and $k$ are connected via Higgs-branch flow, and also how it is possible to also have jumps between theories with different values of $N$ via Higgs-branch flows. The technique to study Higgs-branch flows is quiver subtraction on the magnetic quivers: If the magnetic a theory $\mathcal K_2$ can be subtracted from the magnetic quiver of a theory $\mathcal K_1$, then there exists a Higgs-branch flow from theory $\mathcal K_1$ that gives at low energies theory $\mathcal K_2$, and the result of the subtraction gives information about the transverse slice. We will organize all the flows for fixed values of $\ell $ and $k$ into a Hasse diagram. In talking about Hasse diagrams we will need to use all the theories for fixed values of $\ell $ and $k$, including the ones that we have not presented in the main text; their properties can be extracted similarly to the ones we already discussed and they are all listed in Appendix \ref{App:Mth}. Let us remark that all the Hasse diagrams we find can be obtained also from the ones appearing in \cite{Fazzi:2022hal, Fazzi:2022yca} upon pruning the theories that do not have the $\mathbb Z_\ell$ symmetry. We will now discuss the cases for the theories discussed above, with the exception of the theories with $\ell = 5$ and $\ell =6$: For those values of $\ell$ there is a single theory for any value of $k$ and therefore no Hasse diagram to speak of.

\paragraph{Hasse diagram for $\ell = 2$ and $k=2$.} The first case we discuss is the Hasse diagram for $\ell =2$ and $k=2$. There are five theories in total and their Hasse diagram is
\begin{align}\label{E6ell2k2Hasse}
\begin{tikzpicture}
\node[]  (a1) at (0,0) {\scriptsize $\mathcal S^{(N)}_{E_6,2} (0,0,0,1,0)$}; 
\node[]  (a2) at (0,-1) {\scriptsize $\mathcal T^{(N)}_{E_6,2} (0,0,0,0,2)$}; 
\node[]  (a3) at (0,-2) {\scriptsize $\mathcal T^{(N)}_{E_6,2} (1,0,0,0,1)$}; 
\node[]  (a4) at (0,-3) {\scriptsize $\mathcal T^{(N)}_{E_6,2} (0,1,0,0,0)$}; 
\node[]  (a5) at (0,-4) {\scriptsize $\mathcal T^{(N)}_{E_6,2} (2,0,0,0,0)$}; 
\draw[->, thick] (a1)--(a2); 
\draw[->, thick] (a2)--(a3); 
\draw[->, thick] (a3)--(a4); 
\draw[->, thick] (a4)--(a5); 
\node[]  (b1) at (4,-3.5) {\scriptsize $\mathcal S^{(N-1)}_{E_6,2} (0,0,0,1,0)$}; 
\node[]  (b2) at (4,-4.5) {\scriptsize $\mathcal T^{(N-1)}_{E_6,2} (0,0,0,0,2)$}; 
\node[]  (b3) at (4,-5.5) {\scriptsize $\mathcal T^{(N-1)}_{E_6,2} (1,0,0,0,1)$}; 
\node[]  (b4) at (4,-6.5) {\scriptsize $\mathcal T^{(N-1)}_{E_6,2} (0,1,0,0,0)$}; 
\node[]  (b5) at (4,-7.5) {\scriptsize $\mathcal T^{(N-1)}_{E_6,2} (2,0,0,0,0)$}; 
\draw[->, thick] (b1)--(b2); 
\draw[->, thick] (b2)--(b3); 
\draw[->, thick] (b3)--(b4); 
\draw[->, thick] (b4)--(b5); 
\draw[->, thick] (a4)--(b1); 
\draw[->, thick] (a5)--(b3); 
\end{tikzpicture}
\end{align}

\paragraph{Hasse diagram for $\ell = 2$ and $k=3$.} For $\ell =2$ and $k=3$ there are nine theories in total and their Hasse diagram is

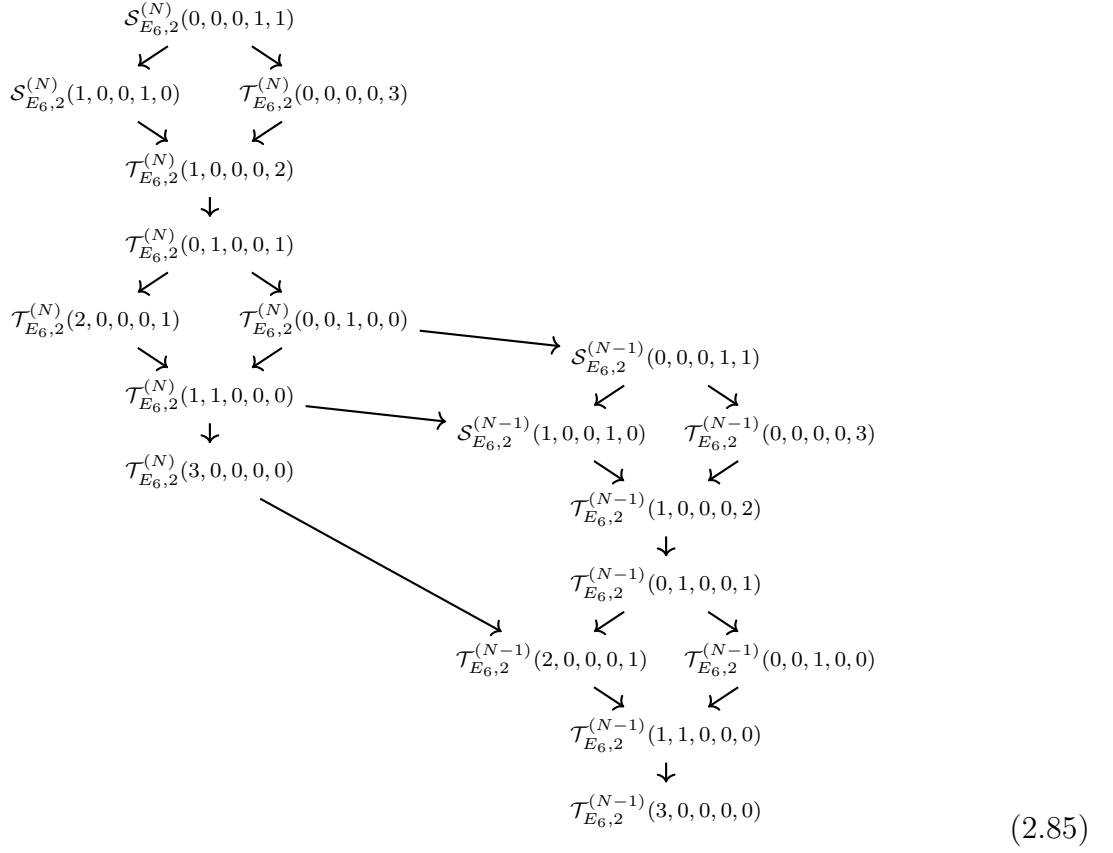
\begin{align}
\begin{tikzpicture}
\node[]  (a1) at (0,0) {\scriptsize $\mathcal S^{(N)}_{E_6,2} (0,0,0,1,1)$}; 
\node[]  (a2) at (-1.5,-1) {\scriptsize $\mathcal S^{(N)}_{E_6,2} (1,0,0,1,0)$}; 
\node[]  (a3) at (1.5,-1) {\scriptsize $\mathcal T^{(N)}_{E_6,2} (0,0,0,0,3)$}; 
\node[]  (a4) at (0,-2) {\scriptsize $\mathcal T^{(N)}_{E_6,2} (1,0,0,0,2)$}; 
\node[]  (a5) at (0,-3) {\scriptsize $\mathcal T^{(N)}_{E_6,2} (0,1,0,0,1)$}; 
\node[]  (a6) at (1.5,-4) {\scriptsize $\mathcal T^{(N)}_{E_6,2} (0,0,1,0,0)$}; 
\node[]  (a7) at (-1.5,-4) {\scriptsize $\mathcal T^{(N)}_{E_6,2} (2,0,0,0,1)$}; 
\node[]  (a8) at (0,-5) {\scriptsize $\mathcal T^{(N)}_{E_6,2} (1,1,0,0,0)$}; 
\node[]  (a9) at (0,-6) {\scriptsize $\mathcal T^{(N)}_{E_6,2} (3,0,0,0,0)$}; 
\draw[->, thick] (a1)--(a2); 
\draw[->, thick] (a1)--(a3); 
\draw[->, thick] (a2)--(a4); 
\draw[->, thick] (a3)--(a4); 
\draw[->, thick] (a4)--(a5); 
\draw[->, thick] (a5)--(a6); 
\draw[->, thick] (a5)--(a7); 
\draw[->, thick] (a7)--(a8);
\draw[->, thick] (a6)--(a8);  
\draw[->, thick] (a8)--(a9); 
\node[]  (b1) at (6,-4.5) {\scriptsize $\mathcal S^{(N-1)}_{E_6,2} (0,0,0,1,1)$}; 
\node[]  (b2) at (-1.5+6,-5.5) {\scriptsize $\mathcal S^{(N-1)}_{E_6,2} (1,0,0,1,0)$}; 
\node[]  (b3) at (1.5+6,-5.5) {\scriptsize $\mathcal T^{(N-1)}_{E_6,2} (0,0,0,0,3)$}; 
\node[]  (b4) at (6,-6.5) {\scriptsize $\mathcal T^{(N-1)}_{E_6,2} (1,0,0,0,2)$}; 
\node[]  (b5) at (6,-7.5) {\scriptsize $\mathcal T^{(N-1)}_{E_6,2} (0,1,0,0,1)$}; 
\node[]  (b6) at (1.5+6,-8.5) {\scriptsize $\mathcal T^{(N-1)}_{E_6,2} (0,0,1,0,0)$}; 
\node[]  (b7) at (-1.5+6,-8.5) {\scriptsize $\mathcal T^{(N-1)}_{E_6,2} (2,0,0,0,1)$}; 
\node[]  (b8) at (6,-9.5) {\scriptsize $\mathcal T^{(N-1)}_{E_6,2} (1,1,0,0,0)$}; 
\node[]  (b9) at (6,-10.5) {\scriptsize $\mathcal T^{(N-1)}_{E_6,2} (3,0,0,0,0)$}; 
\draw[->, thick] (b1)--(b2); 
\draw[->, thick] (b1)--(b3); 
\draw[->, thick] (b2)--(b4); 
\draw[->, thick] (b3)--(b4); 
\draw[->, thick] (b4)--(b5); 
\draw[->, thick] (b5)--(b6); 
\draw[->, thick] (b5)--(b7); 
\draw[->, thick] (b7)--(b8);
\draw[->, thick] (b6)--(b8);  
\draw[->, thick] (b8)--(b9); 
\draw[->, thick] (a6)--(b1); 
\draw[->, thick] (a8)--(b2); 
\draw[->, thick] (a9)--(b7); 
\end{tikzpicture}
\end{align}

\paragraph{Hasse diagram for $\ell = 3$ and $k=2$.} For $\ell =3$ and $k=2$ there are four theories in total and their Hasse diagram is

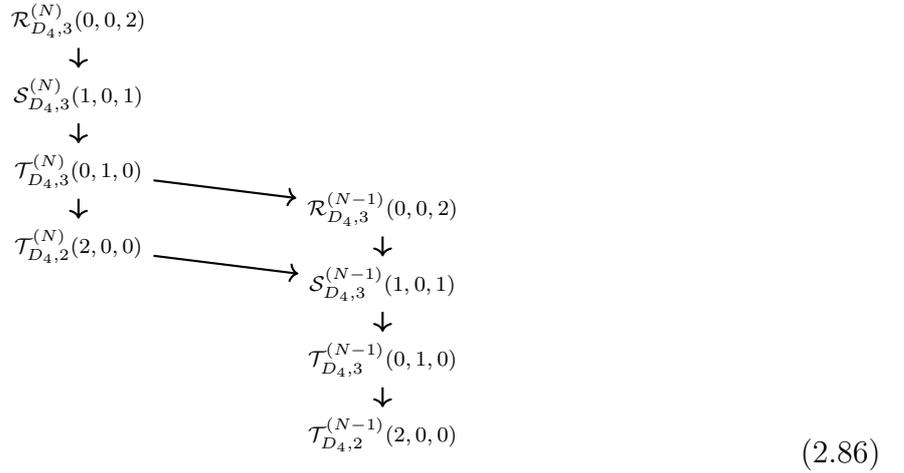
\begin{align}
\begin{tikzpicture}
\node[]  (a1) at (0,0) {\scriptsize $\mathcal R^{(N)}_{D_4,3} (0,0,2)$}; 
\node[]  (a2) at (0,-1) {\scriptsize $\mathcal S^{(N)}_{D_4,3} (1,0,1)$}; 
\node[]  (a3) at (0,-2) {\scriptsize $\mathcal T^{(N)}_{D_4,3} (0,1,0)$}; 
\node[]  (a4) at (0,-3) {\scriptsize $\mathcal T^{(N)}_{D_4,2} (2,0,0)$}; 
\draw[->, thick] (a1)--(a2); 
\draw[->, thick] (a2)--(a3); 
\draw[->, thick] (a3)--(a4); 
\node[]  (b1) at (4,-2.5) {\scriptsize $\mathcal R^{(N-1)}_{D_4,3} (0,0,2)$}; 
\node[]  (b2) at (4,-3.5) {\scriptsize $\mathcal S^{(N-1)}_{D_4,3} (1,0,1)$}; 
\node[]  (b3) at (4,-4.5) {\scriptsize $\mathcal T^{(N-1)}_{D_4,3} (0,1,0)$}; 
\node[]  (b4) at (4,-5.5) {\scriptsize $\mathcal T^{(N-1)}_{D_4,2} (2,0,0)$}; 
\draw[->, thick] (b1)--(b2); 
\draw[->, thick] (b2)--(b3); 
\draw[->, thick] (b3)--(b4); 
\draw[->, thick] (a3)--(b1); 
\draw[->, thick] (a4)--(b2); 
\end{tikzpicture}
\end{align}

\paragraph{Hasse diagram for $\ell = 4$ and $k=2$.} For $\ell =4$ and $k=2$ there are three theories in total and their Hasse diagram is

\begin{align}\label{HasseM8l4}
\begin{tikzpicture}
\node[]  (a1) at (0,0) {\scriptsize  $\mathcal S^{(N)}_{H_2,4} (1,1)$}; 
\node[]  (a2) at (0,-1) {\scriptsize $\mathcal T^{(N)}_{H_2,4} (0,2)$ }; 
\node[]  (a3) at (0,-2) {\scriptsize $\mathcal T^{(N)}_{H_2,4} (2,0)$}; 
\draw[->, thick] (a1)--(a2); 
\draw[->, thick] (a2)--(a3); 
\node[]  (b1) at (4,-2.5) {\scriptsize $\mathcal S^{(N-1)}_{H_2,4} (1,1)$}; 
\node[]  (b2) at (4,-1-2.5) {\scriptsize $\mathcal T^{(N-1)}_{H_2,4} (0,2)$ }; 
\node[]  (b3) at (4,-2-2.5) {\scriptsize $\mathcal T^{(N-1)}_{H_2,4} (2,0)$}; 
\draw[->, thick] (b1)--(b2); 
\draw[->, thick] (b2)--(b3); 
\draw[->, thick] (a3)--(b1); 
\end{tikzpicture}
\end{align}

\section{Type IIB Orbi-S-folds}\label{Sec:IIBSfolds}

The purpose of this section is to relate the family of four dimensional theories we have just discussed to theories living on the worldvolume of $D3$-branes probing a orbifold background in Type IIB string theory combined with a twist by the S-duality group, the so-called S-fold. The geometric setup  in F-theory involves a $\bbZ_{\ell}$ quotient acting on the $\mathbb{C}^3$ transverse to the $D3$ worldvolume times the $T^2$ fiber. Furthermore, we introduce a orbifold action $\bbZ_{k}$ on a $\mathbb{C}^2\subset\mathbb{C}^3$ and 7-branes wrapping the orbifold. All these ingredients indeed preserve eight supercharges on the $D3$-branes (see \cite{Apruzzi:2020pmv, Giacomelli:2023qyc}). 

As we have explained, all the models constructed previously have a $SU(k)$ global symmetry which we do not expect from the Type IIB perspective and indeed, the SW theories of the previous section are not directly identifiable with Type IIB theories. This feature was already discussed in \cite{Giacomelli:2022drw}, where it was found that we need to activate mass deformations for the $SU(k)$ symmetry to get the 4d theories on the worldvolume of the $D3$-brane. In the simpler setup in which we just have an orbifold singularity considered in \cite{Ohmori:2015pia} this mass deformation was interpreted as the activation of the B-field which enables us to carry out the worldsheet analysis which leads to the Douglas-Moore quiver \cite{Douglas:1996sw} on the worldvolume of the $D3$-brane. In this sense, we can regard the M-theory setup of the previous section as the ``strong-coupling limit'' of Type IIB, in which the B-field is switched off completely and the extra $SU(k)$ symmetry emerges. The activation of the mass deformation is therefore equivalent to turning on a non-trivial B-field along the vanishing cycles of the orbifold singularity, and in this regime the Type IIB description with $D3$ probes becomes more appropriate.
We will now explain how to implement the mass deformation and determine the properties of the resulting field theories.

A remark concerning notation: We will use the same notation of Section \ref{Sec:SWcompact} to indicate the various theories in each family, except that we will add a small circle on the letter to denote the deformed version of the theory, where the $SU(k)$ symmetry has been completely removed. The same notation will be adopted in Appendix \ref{App:IIBth}.

\subsection{From M-theory to Type IIB}\label{Sec:MtoIIB}

Let us start by explaining how to implement the $SU(k)$ mass deformation at the level of the magnetic quiver. As we have seen, in the quiver the $SU(k)$ symmetry is realized by a $T(SU(k))$ tail attached to a twisted affine Dynkin diagram. As we will now see, the effect of the mass deformation is to remove the tail and trade it for a single $U(1)$ node and our task is to understand how this connects to the other nodes in the Dynkin diagram. We will now discuss the various values of $\ell$ separately.

\subsubsection{The case $\ell=2$}

In this case the quiver involves the twisted $E_6$ Dynkin diagram, reflecting the fact that we are combining the $\bbZ_2$ S-fold with a 7-brane of $E_6$ type. 
The quiver has the form 
\begin{equation} 
\begin{tikzpicture}\label{E6Dynkin}
\filldraw[fill= white] (2,0) circle [radius=0.1] node[below] {\scriptsize A};
\filldraw[fill= white] (3,0) circle [radius=0.1] node[below] {\scriptsize B};
\filldraw[fill= white] (4,0) circle [radius=0.1] node[below] {\scriptsize C};
\filldraw[fill= white] (5,0) circle [radius=0.1] node[below] {\scriptsize D};
\filldraw[fill= white] (6,0) circle [radius=0.1] node[below] {\scriptsize  E};
%\filldraw[fill= white] (7,0) circle [radius=0.1] node[below] {\scriptsize  1};
\draw [thick] (2.1, 0) -- (2.9,0) ;
\draw [thick] (4.1, 0.05) -- (4.9,0.05) ;
\draw [thick] (4.1, -0.05) -- (4.9,-0.05) ;
\draw [thick] (3.1, 0) -- (3.9,0) ;
\draw [thick] (5.1, 0) -- (5.9,0) ;
\draw [thick] (4.4,0) -- (4.6,0.2);
\draw [thick] (4.4,0) -- (4.6,-0.2);
\end{tikzpicture} 
\end{equation} 
We start by considering a theory with $k=\ell=2$ and holonomy $(n_{2'}=2)$. Before folding, the quiver is star-shaped. The mass deformation which removes the tail is implemented by the following FI deformation: 
\begin{equation}\label{quiv1} 
\begin{tikzpicture} 
\filldraw[fill= red] (0,0) circle [radius=0.1] node[below] {\scriptsize 1};
\filldraw[fill= blue] (1,0) circle [radius=0.1] node[below] {\scriptsize 2};
\filldraw[fill= white] (2,0) circle [radius=0.1] node[below] {\scriptsize N+2};
\filldraw[fill= white] (3,0) circle [radius=0.1] node[below] {\scriptsize 2N+2};
\filldraw[fill= white] (4,0) circle [radius=0.1] node[below] {\scriptsize 3N+2};
\filldraw[fill= white] (5,0) circle [radius=0.1] node[below] {\scriptsize 4N+2};
\filldraw[fill= white] (6,0) circle [radius=0.1] node[below] {\scriptsize 3N+2};
\filldraw[fill= white] (7,0) circle [radius=0.1] node[below] {\scriptsize 2N+2};
\filldraw[fill= white] (8,0) circle [radius=0.1] node[below] {\scriptsize N+2};
\filldraw[fill= blue] (9,0) circle [radius=0.1] node[below] {\scriptsize 2};
\filldraw[fill= red] (10,0) circle [radius=0.1] node[below] {\scriptsize 1};
\filldraw[fill= white] (5,1) circle [radius=0.1] node[left] {\scriptsize 2N};
%\filldraw[fill= white] (5,2) circle [radius=0.1] node[left] {\scriptsize 1};
\draw [thick] (0.1, 0) -- (0.9,0) ;
\draw [thick] (1.1, 0) -- (1.9,0) ;
\draw [thick] (2.1, 0) -- (2.9,0) ;
\draw [thick] (3.1, 0) -- (3.9,0) ;
\draw [thick] (4.1, 0) -- (4.9,0) ;
\draw [thick] (5.1, 0) -- (5.9,0) ;
\draw [thick] (6.1, 0) -- (6.9,0) ;
\draw [thick] (7.1, 0) -- (7.9,0) ;
\draw [thick] (8.1, 0) -- (8.9,0) ;
\draw [thick] (9.1, 0) -- (9.9,0) ;
\draw [thick] (5, 0.1) -- (5,0.9) ;

\draw [->, thick] (5,-1)--(5,-2); 

\filldraw[fill= white] (2,-3) circle [radius=0.1] node[above] {\scriptsize N};
\filldraw[fill= white] (3,-3) circle [radius=0.1] node[above] {\scriptsize 2N};
\filldraw[fill= white] (4,-3) circle [radius=0.1] node[above] {\scriptsize 3N};
\filldraw[fill= white] (5,-3) circle [radius=0.1] node[above] {\scriptsize 4N};
\filldraw[fill= white] (6,-3) circle [radius=0.1] node[above] {\scriptsize 3N};
\filldraw[fill= white] (7,-3) circle [radius=0.1] node[above] {\scriptsize 2N};
\filldraw[fill= white] (8,-3) circle [radius=0.1] node[above] {\scriptsize N};
\filldraw[fill= white] (5,-4) circle [radius=0.1] node[below] {\scriptsize 2N};
\filldraw[fill= white] (4,-4) circle [radius=0.1] node[below] {\scriptsize 1};
\filldraw[fill= white] (6,-4) circle [radius=0.1] node[below] {\scriptsize 1};

\draw [thick] (2.1, -3) -- (2.9,-3) ;
\draw [thick] (3.1, -3) -- (3.9,-3) ;
\draw [thick] (4.1, -3) -- (4.9,-3) ;
\draw [thick] (5.1, -3) -- (5.9,-3) ;
\draw [thick] (6.1, -3) -- (6.9,-3) ;
\draw [thick] (7.1, -3) -- (7.9,-3) ;
\draw [thick] (4.1, -4) -- (4.9,-4) ;
\draw [thick] (5.1, -4) -- (5.9,-4) ;
\draw [thick] (5, -3.1) -- (5,-3.9) ;
\end{tikzpicture}
\end{equation} 
The Higgs branch of the resulting 4d SCFT, after the $SU(2)$ mass deformation, is described by folding the second quiver of Eq.~\eqref{quiv1}, i.e.~by the quiver
\begin{equation} 
\begin{tikzpicture}\label{k=4caseIV}
\filldraw[fill= white] (2,0) circle [radius=0.1] node[below] {\scriptsize N};
\filldraw[fill= white] (3,0) circle [radius=0.1] node[below] {\scriptsize 2N};
\filldraw[fill= white] (4,0) circle [radius=0.1] node[below] {\scriptsize 3N};
\filldraw[fill= white] (5,0) circle [radius=0.1] node[below] {\scriptsize 4N};
\filldraw[fill= white] (6,0) circle [radius=0.1] node[below] {\scriptsize  2N};
\filldraw[fill= white] (7,0) circle [radius=0.1] node[below] {\scriptsize 1};
%\filldraw[fill= white] (7,0) circle [radius=0.1] node[below] {\scriptsize  1};
\draw [thick] (2.1, 0) -- (2.9,0) ;
\draw [thick] (4.1, 0.05) -- (4.9,0.05) ;
\draw [thick] (4.1, -0.05) -- (4.9,-0.05) ;
\draw [thick] (3.1, 0) -- (3.9,0) ;
\draw [thick] (5.1, 0) -- (5.9,0) ;
\draw [thick] (6.1, 0.05) -- (6.9,0.05) ;
\draw [thick] (6.1, -0.05) -- (6.9,-0.05) ;
\draw [thick] (4.4,0) -- (4.6,0.2);
\draw [thick] (4.4,0) -- (4.6,-0.2);
\draw [thick] (6.6,0) -- (6.4,0.2);
\draw [thick] (6.6,0) -- (6.4,-0.2);
\end{tikzpicture} 
\end{equation} 
Notice that the deformation \eqref{quiv1} can manifestly be applied to any quiver which is foldable and contains a $T(SU(2))$ tail on each side. In particular it can be implemented for all $\ell=2$ SW models, for every choice of holonomy. For example, in the $k=2$ case with holonomy labelled by $n_{4'}=1$ the magnetic quiver has the form
\begin{equation} 
\begin{tikzpicture}\label{k=4caseVM}
\filldraw[fill= white] (0,0) circle [radius=0.1] node[below] {\scriptsize 1};
\filldraw[fill= white] (1,0) circle [radius=0.1] node[below] {\scriptsize 2};
\filldraw[fill= white] (2,0) circle [radius=0.1] node[below] {\scriptsize N+2};
\filldraw[fill= white] (3,0) circle [radius=0.1] node[below] {\scriptsize 2N+2};
\filldraw[fill= white] (4,0) circle [radius=0.1] node[below] {\scriptsize 3N+2};
\filldraw[fill= white] (5,0) circle [radius=0.1] node[below] {\scriptsize 4N+2};
\filldraw[fill= white] (6,0) circle [radius=0.1] node[below] {\scriptsize  2N+1};

%\filldraw[fill= white] (7,0) circle [radius=0.1] node[below] {\scriptsize  1};
\draw [thick] (0.1, 0) -- (0.9,0) ;
\draw [thick] (1.1, 0) -- (1.9,0) ;
\draw [thick] (2.1, 0) -- (2.9,0) ;
\draw [thick] (4.1, 0.05) -- (4.9,0.05) ;
\draw [thick] (4.1, -0.05) -- (4.9,-0.05) ;
\draw [thick] (3.1, 0) -- (3.9,0) ;
\draw [thick] (5.1, 0) -- (5.9,0) ;
\draw [thick] (4.4,0) -- (4.6,0.2);
\draw [thick] (4.4,0) -- (4.6,-0.2);
\end{tikzpicture} 
\end{equation}  
The procedure in \eqref{quiv1} then leads to the quiver 
\begin{equation} 
\begin{tikzpicture}\label{k=4caseV}
\filldraw[fill= white] (2,0) circle [radius=0.1] node[below] {\scriptsize N};
\filldraw[fill= white] (3,0) circle [radius=0.1] node[below] {\scriptsize 2N};
\filldraw[fill= white] (4,0) circle [radius=0.1] node[below] {\scriptsize 3N};
\filldraw[fill= white] (5,0) circle [radius=0.1] node[below] {\scriptsize 4N};
\filldraw[fill= white] (6,0) circle [radius=0.1] node[below] {\scriptsize  2N+1};
\filldraw[fill= white] (7,0) circle [radius=0.1] node[below] {\scriptsize 1};
%\filldraw[fill= white] (7,0) circle [radius=0.1] node[below] {\scriptsize  1};
\draw [thick] (2.1, 0) -- (2.9,0) ;
\draw [thick] (4.1, 0.05) -- (4.9,0.05) ;
\draw [thick] (4.1, -0.05) -- (4.9,-0.05) ;
\draw [thick] (3.1, 0) -- (3.9,0) ;
\draw [thick] (5.1, 0) -- (5.9,0) ;
\draw [thick] (6.1, 0.05) -- (6.9,0.05) ;
\draw [thick] (6.1, -0.05) -- (6.9,-0.05) ;
\draw [thick] (4.4,0) -- (4.6,0.2);
\draw [thick] (4.4,0) -- (4.6,-0.2);
\draw [thick] (6.6,0) -- (6.4,0.2);
\draw [thick] (6.6,0) -- (6.4,-0.2);
\end{tikzpicture} 
\end{equation} 
This operation is very easy to describe in general: We remove the $T(SU(2))$ tail and trade it for a single $U(1)$ node connected with an oriented edge to the node of the $E_6$ diagram labelled by $E$ in \eqref{E6Dynkin} and simultaneously we decrease by 2 units the rank of all the nodes in the Dynkin diagram except the node $E$. Indeed this can be done for every value of $N$ provided it is large enough, so that none of the nodes ends up having negative rank. In practice this restriction rules out only the case $N=1$ with $F_4$-preserving holonomy (namely $n_2=2$), which we know to correspond to a rank-1 theory we can study using the simpler $k=1$ $\ell=2$ S-fold geometry. 

Ler us now consider the $k=3$ example with holonomy labelled by $n_{2'}=3$, which leads to theories with $USp(8)$ global symmetry. The corresponding magnetic quiver is 
\begin{equation} 
\begin{tikzpicture}\label{k=6caseVIM}
\filldraw[fill= white] (-1,0) circle [radius=0.1] node[below] {\scriptsize 1};
\filldraw[fill= white] (0,0) circle [radius=0.1] node[below] {\scriptsize 2};
\filldraw[fill= white] (1,0) circle [radius=0.1] node[below] {\scriptsize 3};
\filldraw[fill= white] (2,0) circle [radius=0.1] node[below] {\scriptsize N+3};
\filldraw[fill= white] (3,0) circle [radius=0.1] node[below] {\scriptsize 2N+3};
\filldraw[fill= white] (4,0) circle [radius=0.1] node[below] {\scriptsize 3N+3};
\filldraw[fill= white] (5,0) circle [radius=0.1] node[below] {\scriptsize 4N+3};
\filldraw[fill= white] (6,0) circle [radius=0.1] node[below] {\scriptsize  2N};

%\filldraw[fill= white] (7,0) circle [radius=0.1] node[below] {\scriptsize  1};
\draw [thick] (-0.1, 0) -- (-0.9,0) ;
\draw [thick] (0.1, 0) -- (0.9,0) ;
\draw [thick] (1.1, 0) -- (1.9,0) ;
\draw [thick] (2.1, 0) -- (2.9,0) ;
\draw [thick] (4.1, 0.05) -- (4.9,0.05) ;
\draw [thick] (4.1, -0.05) -- (4.9,-0.05) ;
\draw [thick] (3.1, 0) -- (3.9,0) ;
\draw [thick] (5.1, 0) -- (5.9,0) ;
\draw [thick] (4.4,0) -- (4.6,0.2);
\draw [thick] (4.4,0) -- (4.6,-0.2);
\end{tikzpicture} 
\end{equation}  
and after the procedure \eqref{quiv1} we find 
\begin{equation} 
\begin{tikzpicture}\label{k=6caseVI}
\filldraw[fill= white] (1,0) circle [radius=0.1] node[below] {\scriptsize 1};
\filldraw[fill= white] (2,0) circle [radius=0.1] node[below] {\scriptsize N+1};
\filldraw[fill= white] (3,0) circle [radius=0.1] node[below] {\scriptsize 2N+1};
\filldraw[fill= white] (4,0) circle [radius=0.1] node[below] {\scriptsize 3N+1};
\filldraw[fill= white] (5,0) circle [radius=0.1] node[below] {\scriptsize 4N+1};
\filldraw[fill= white] (6,0) circle [radius=0.1] node[below] {\scriptsize  2N};
\filldraw[fill= white] (7,0) circle [radius=0.1] node[below] {\scriptsize 1};
%\filldraw[fill= white] (7,0) circle [radius=0.1] node[below] {\scriptsize  1};
\draw [thick] (1.1, 0) -- (1.9,0) ;
\draw [thick] (2.1, 0) -- (2.9,0) ;
\draw [thick] (4.1, 0.05) -- (4.9,0.05) ;
\draw [thick] (4.1, -0.05) -- (4.9,-0.05) ;
\draw [thick] (3.1, 0) -- (3.9,0) ;
\draw [thick] (5.1, 0) -- (5.9,0) ;
\draw [thick] (6.1, 0.05) -- (6.9,0.05) ;
\draw [thick] (6.1, -0.05) -- (6.9,-0.05) ;
\draw [thick] (4.4,0) -- (4.6,0.2);
\draw [thick] (4.4,0) -- (4.6,-0.2);
\draw [thick] (6.6,0) -- (6.4,0.2);
\draw [thick] (6.6,0) -- (6.4,-0.2);
\end{tikzpicture} 
\end{equation} 
As a consistency check, we can notice that \eqref{k=4caseIV}, \eqref{k=4caseV} and \eqref{k=6caseVI} all reduce for $N=1$ to the magnetic quiver of known rank-2 SCFTs which are mass deformations of the corresponding SW theories \cite{Bourget:2021csg}. 

At this stage we can proceed with the analysis of the deformation of the $T(SU(3))$ tail, which is always present for $k\geq3$. Considering again \eqref{k=6caseVIM}, the first step is to go to \eqref{k=6caseVI} and then we can proceed with another deformation to 
\begin{equation}\label{quiv2} 
\begin{tikzpicture} 
\filldraw[fill= red] (1,0) circle [radius=0.1] node[below] {\scriptsize 1};
\filldraw[fill= blue] (4,1) circle [radius=0.1] node[below] {\scriptsize 1};
\filldraw[fill= white] (2,0) circle [radius=0.1] node[below] {\scriptsize N+1};
\filldraw[fill= white] (3,0) circle [radius=0.1] node[below] {\scriptsize 2N+1};
\filldraw[fill= white] (4,0) circle [radius=0.1] node[below] {\scriptsize 3N+1};
\filldraw[fill= white] (5,0) circle [radius=0.1] node[below] {\scriptsize 4N+1};
\filldraw[fill= white] (6,0) circle [radius=0.1] node[below] {\scriptsize 3N+1};
\filldraw[fill= white] (7,0) circle [radius=0.1] node[below] {\scriptsize 2N+1};
\filldraw[fill= white] (8,0) circle [radius=0.1] node[below] {\scriptsize N+1};
\filldraw[fill= blue] (6,1) circle [radius=0.1] node[below] {\scriptsize 1};
\filldraw[fill= red] (9,0) circle [radius=0.1] node[below] {\scriptsize 1};
\filldraw[fill= white] (5,1) circle [radius=0.1] node[above] {\scriptsize 2N};
%\filldraw[fill= white] (5,2) circle [radius=0.1] node[left] {\scriptsize 1};
\draw [thick] (4.1, 1) -- (4.9,1) ;
\draw [thick] (1.1, 0) -- (1.9,0) ;
\draw [thick] (2.1, 0) -- (2.9,0) ;
\draw [thick] (3.1, 0) -- (3.9,0) ;
\draw [thick] (4.1, 0) -- (4.9,0) ;
\draw [thick] (5.1, 0) -- (5.9,0) ;
\draw [thick] (6.1, 0) -- (6.9,0) ;
\draw [thick] (7.1, 0) -- (7.9,0) ;
\draw [thick] (8.1, 0) -- (8.9,0) ;
\draw [thick] (5.1, 1) -- (5.9,1) ;
\draw [thick] (5, 0.1) -- (5,0.9) ;

\draw [->, thick] (5,-1)--(5,-2); 

\filldraw[fill= white] (2,-3) circle [radius=0.1] node[above] {\scriptsize N};
\filldraw[fill= white] (3,-3) circle [radius=0.1] node[above] {\scriptsize 2N};
\filldraw[fill= white] (4,-3) circle [radius=0.1] node[above] {\scriptsize 3N};
\filldraw[fill= white] (5,-3) circle [radius=0.1] node[above] {\scriptsize 4N};
\filldraw[fill= white] (6,-3) circle [radius=0.1] node[above] {\scriptsize 3N};
\filldraw[fill= white] (7,-3) circle [radius=0.1] node[above] {\scriptsize 2N};
\filldraw[fill= white] (8,-3) circle [radius=0.1] node[above] {\scriptsize N};
\filldraw[fill= white] (5,-4) circle [radius=0.1] node[below] {\scriptsize 2N-1};
\filldraw[fill= white] (6,-3.5) circle [radius=0.1] node[right] {\scriptsize 1};

\draw [thick] (2.1, -3) -- (2.9,-3) ;
\draw [thick] (3.1, -3) -- (3.9,-3) ;
\draw [thick] (4.1, -3) -- (4.9,-3) ;
\draw [thick] (5.1, -3) -- (5.9,-3) ;
\draw [thick] (6.1, -3) -- (6.9,-3) ;
\draw [thick] (7.1, -3) -- (7.9,-3) ;
\draw [thick] (5, -3.1) -- (5,-3.9) ;
\draw [thick] (5.03, -3.95) -- (5.97,-3.55) ;
\draw [thick] (5.97, -3.45) -- (5.03,-3.05) ;
\end{tikzpicture}
\end{equation} 
After folding the second quiver in Eq.~\eqref{quiv2}, we get 
\begin{equation}
\begin{tikzpicture}\label{k=6caseVI2}
\filldraw[fill= white] (2,0) circle [radius=0.1] node[below] {\scriptsize N};
\filldraw[fill= white] (3,0) circle [radius=0.1] node[below] {\scriptsize 2N};
\filldraw[fill= white] (4,0) circle [radius=0.1] node[below] {\scriptsize 3N};
\filldraw[fill= white] (5,0) circle [radius=0.1] node[below] {\scriptsize 4N};
\filldraw[fill= white] (6,0) circle [radius=0.1] node[below] {\scriptsize  2N-1};
\filldraw[fill= white] (5.5,1) circle [radius=0.1] node[above] {\scriptsize 1};
%\filldraw[fill= white] (7,0) circle [radius=0.1] node[below] {\scriptsize  1};
\draw [thick] (2.1, 0) -- (2.9,0) ;
\draw [thick] (4.1, 0.05) -- (4.9,0.05) ;
\draw [thick] (4.1, -0.05) -- (4.9,-0.05) ;
\draw [thick] (3.1, 0) -- (3.9,0) ;
\draw [thick] (5.1, 0) -- (5.9,0) ;
\draw [thick] (4.4,0) -- (4.6,0.2);
\draw [thick] (4.4,0) -- (4.6,-0.2);
\draw [thick] (5.02, 0.05) -- (5.48,0.95);
\draw [thick] (5.98, 0.05) -- (5.52,0.95) ;
\end{tikzpicture} 
\end{equation} 
and as a consistency check we can notice that for $N=1$ it reduces to the magnetic quiver of a known rank-2 SCFT \cite{Bourget:2021csg}. 
The rule in this case is to remove the $T(SU(3))$ tail and replace it with an Abelian node attached to the nodes $D$ and $E$ of the $E_6$ Dynkin diagram \eqref{E6Dynkin}. At the same time, we reduce by one the rank of the node $E$ and by three the rank of all other nodes.  

Finally, let us discuss the deformation of the $SU(4)$ tail, which exists only for $k\geq4$. In this case we propose the following rule: We remove the $SU(4)$ tail and replace it with a $U(1)$ node connected with a double link to the node $D$ of the $E_6$ Dynkin diagram \eqref{E6Dynkin}. We also decrease the rank of all the nodes by four, except node $D$ whose rank is decreased by two. This can be obtained via FI deformations. As a check of this proposal, we observe that for the $E_8$ holonomies labelled by $(n_2=1, n_{2'}=3)$ and $(n_2=n_{2'}=2)$ we recover for $N=1$ the magnetic quiver of known rank-2 SCFTs \cite{Bourget:2021csg}. 

Let us summarize our proposal for the tail deformations considered so far for a generic theory with $k=4$ and $\ell=2$: 
\be\label{defsell=2} 
\begin{tikzpicture} 
 \filldraw[fill= white] (-2,0) circle [radius=0.1] node[below] {\scriptsize 1};
 \filldraw[fill= white] (-1,0) circle [radius=0.1] node[below] {\scriptsize 2};
\filldraw[fill= white] (0,0) circle [radius=0.1] node[below] {\scriptsize 3};
\filldraw[fill= white] (1,0) circle [radius=0.1] node[below] {\scriptsize 4};
\filldraw[fill= white] (2,0) circle [radius=0.1] node[below] {\scriptsize A};
\filldraw[fill= white] (3,0) circle [radius=0.1] node[below] {\scriptsize B};
\filldraw[fill= white] (4,0) circle [radius=0.1] node[below] {\scriptsize C};
\filldraw[fill= white] (5,0) circle [radius=0.1] node[below] {\scriptsize D};
\filldraw[fill= white] (6,0) circle [radius=0.1] node[below] {\scriptsize E};

%\filldraw[fill= white] (7,0) circle [radius=0.1] node[below] {\scriptsize  1};
\draw [thick] (-1.1, 0) -- (-1.9,0) ;
\draw [thick] (-0.1, 0) -- (-0.9,0) ;
\draw [thick] (0.1, 0) -- (0.9,0) ;
\draw [thick] (1.1, 0) -- (1.9,0) ;
\draw [thick] (2.1, 0) -- (2.9,0) ;
\draw [thick] (4.1, 0.05) -- (4.9,0.05) ;
\draw [thick] (4.1, -0.05) -- (4.9,-0.05) ;
\draw [thick] (3.1, 0) -- (3.9,0) ;
\draw [thick] (5.1, 0) -- (5.9,0) ;
\draw [thick] (4.4,0) -- (4.6,0.2);
\draw [thick] (4.4,0) -- (4.6,-0.2); 

\draw[->, thick] (2,-0.5)--(2,-1.5); 

\filldraw[fill= white] (-1,-2) circle [radius=0.1] node[below] {\scriptsize  1};
\filldraw[fill= white] (0,-2) circle [radius=0.1] node[below] {\scriptsize 2};
\filldraw[fill= white] (1,-2) circle [radius=0.1] node[below] {\scriptsize A-2};
\filldraw[fill= white] (2,-2) circle [radius=0.1] node[below] {\scriptsize B-2};
\filldraw[fill= white] (3,-2) circle [radius=0.1] node[below] {\scriptsize C-2};
\filldraw[fill= white] (4,-2) circle [radius=0.1] node[below] {\scriptsize D-2};
\filldraw[fill= white] (5,-2) circle [radius=0.1] node[below] {\scriptsize E};
\filldraw[fill= white] (6,-2) circle [radius=0.1] node[below] {\scriptsize 1};
\draw [thick] (-0.1, -2) -- (-0.9,-2) ;
\draw [thick] (0.1, -2) -- (0.9,-2) ;
\draw [thick] (1.1, -2) -- (1.9,-2) ;
\draw [thick] (3.1, -1.95) -- (3.9,-1.95) ;
\draw [thick] (3.1, -2.05) -- (3.9,-2.05) ;
\draw [thick] (2.1, -2) -- (2.9,-2) ;
\draw [thick] (4.1, -2) -- (4.9,-2) ;
\draw [thick] (5.1, -1.95) -- (5.9,-1.95) ;
\draw [thick] (5.1, -2.05) -- (5.9,-2.05) ;
\draw [thick] (3.4,-2) -- (3.6,-1.8);
\draw [thick] (3.4,-2) -- (3.6,-2.2);
\draw [thick] (5.6,-2) -- (5.4,-1.8);
\draw [thick] (5.6,-2) -- (5.4,-2.2);

\draw[->, thick] (2,-2.5)--(2,-3.5); 

\filldraw[fill= white] (0,-4) circle [radius=0.1] node[below] {\scriptsize  1};
\filldraw[fill= white] (1,-4) circle [radius=0.1] node[below] {\scriptsize A-3};
\filldraw[fill= white] (2,-4) circle [radius=0.1] node[below] {\scriptsize B-3};
\filldraw[fill= white] (3,-4) circle [radius=0.1] node[below] {\scriptsize C-3};
\filldraw[fill= white] (4,-4) circle [radius=0.1] node[below] {\scriptsize D-3};
\filldraw[fill= white] (5,-4) circle [radius=0.1] node[below] {\scriptsize E-1};
\filldraw[fill= white] (4.5,-3) circle [radius=0.1] node[above] {\scriptsize 1};
\draw [thick] (0.1, -4) -- (0.9,-4) ;
\draw [thick] (1.1, -4) -- (1.9,-4) ;
\draw [thick] (3.1, -3.95) -- (3.9,-3.95) ;
\draw [thick] (3.1, -4.05) -- (3.9,-4.05) ;
\draw [thick] (2.1, -4) -- (2.9,-4) ;
\draw [thick] (4.1, -4) -- (4.9,-4) ;
\draw [thick] (3.4,-4) -- (3.6,-3.8);
\draw [thick] (3.4,-4) -- (3.6,-4.2);
\draw [thick] (4.02, -3.95) -- (4.48,-3.05);
\draw [thick] (4.98, -3.95) -- (4.52,-3.05) ;

\draw[->, thick] (2,-4.5)--(2,-5.5); 

\filldraw[fill= white] (1,-6) circle [radius=0.1] node[below] {\scriptsize A-4};
\filldraw[fill= white] (2,-6) circle [radius=0.1] node[below] {\scriptsize B-4};
\filldraw[fill= white] (3,-6) circle [radius=0.1] node[below] {\scriptsize C-4};
\filldraw[fill= white] (4,-6) circle [radius=0.1] node[below] {\scriptsize D-4};
\filldraw[fill= white] (5,-6) circle [radius=0.1] node[below] {\scriptsize E-2};
\filldraw[fill= white] (4,-5) circle [radius=0.1] node[above] {\scriptsize 1};
\draw [thick] (1.1, -6) -- (1.9,-6) ;
\draw [thick] (3.1, -5.95) -- (3.9,-5.95) ;
\draw [thick] (3.1, -6.05) -- (3.9,-6.05) ;
\draw [thick] (2.1, -6) -- (2.9,-6) ;
\draw [thick] (4.1, -6) -- (4.9,-6) ;
\draw [thick] (3.4,-6) -- (3.6,-5.8);
\draw [thick] (3.4,-6) -- (3.6,-6.2);
\draw [thick] (4.05, -5.95) -- (4.05,-5.05);
\draw [thick] (3.95, -5.95) -- (3.95,-5.05) ;
\end{tikzpicture} 
\ee  
As we have discussed before, string theoretically these relevant deformations implement the transition between the M-theory and Type IIB models. Alternatively, we can think of them as turning on the B-field along the vanishing cycles of the orbifold singularity in Type IIB string theory.

Indeed for $k>4$ the tail is bigger and therefore the effect of the above moves is to shorten the tail and reduce the rank of the nodes accordingly. In practice, in the first step we replace $T(SU(k))$ with $T(SU(k-2))$, in the second step we replace the tail with $T(SU(k-3))$ and finally with $T(SU(k-4))$. Notice that at each step the unbalancing of the nodes in the Dynkin diagram is inherited from that of the parent theory with the $T(SU(k))$ tail and therefore is dictated by the choice of holonomy, as we have explained in the previous section. 

Notice that if we choose a framing of the quivers in \eqref{defsell=2} by ungauging the Abelian node produced by the deformation, we find that nodes $D$ and $E$ are flavored and the sum (over the nodes of the Dynkin diagram) of the number of flavors times the Dynkin label of the node is equal to 2, 3 and 4 for the second, third and fourth quivers in \eqref{defsell=2} respectively. We therefore propose that in general the effect of the mass deformation is to trade the tail for an Abelian node connected to the Dynkin diagram in such a way that 
\be\label{abelianrule} \sum_{\text{nodes}}d_iF_i=k\,,\ee  
where $d_i$ and $F_i$ denote the Dynkin label and the number of flavors of the $i$-th node respectively.

\subsubsection{The case $\ell=3$} 

In this case the relevant Dynkin diagram is that of twisted affine $D_4$ 
\be\label{DynkinD4} 
\begin{tikzpicture}
\filldraw[fill= white] (2,-6) circle [radius=0.1] node[below] {\scriptsize A};
\filldraw[fill= white] (3,-6) circle [radius=0.1] node[below] {\scriptsize B};
\filldraw[fill= white] (4,-6) circle [radius=0.1] node[below] {\scriptsize C};
\draw [thick] (3.1, -5.95) -- (3.9,-5.95) ;
\draw [thick] (3.1, -6) -- (3.9,-6) ;
\draw [thick] (3.1, -6.05) -- (3.9,-6.05) ;
\draw [thick] (2.1, -6) -- (2.9,-6) ;
\draw [thick] (3.4,-6) -- (3.6,-5.8);
\draw [thick] (3.4,-6) -- (3.6,-6.2);
\end{tikzpicture}
\ee 
since we are combining the $\bbZ_3$ S-fold with a 7-brane of type $D_4$. Let us start from the case $k=2$, in which we have a $T(SU(2))$ tail in the magnetic quiver. We focus for concreteness on the holonomy $n_{3'}=2$. Starting from the corresponding magnetic quiver before folding it and turning on the deformation we find 
\be\label{defL=3}
\begin{tikzpicture} 
\filldraw[fill= red] (-1,0) circle [radius=0.1] node[above] {\scriptsize 1};
\filldraw[fill= blue] (0,0) circle [radius=0.1] node[above] {\scriptsize 2};
\filldraw[fill= white] (1,0) circle [radius=0.1] node[above] {\scriptsize N+2};
\filldraw[fill= white] (2,0) circle [radius=0.1] node[above] {\scriptsize 2N+2};
\filldraw[fill= white] (3,0) circle [radius=0.1] node[above] {\scriptsize 3N+2};
\filldraw[fill= white] (4,0) circle [radius=0.1] node[above] {\scriptsize 2N+2};
\filldraw[fill= white] (5,0) circle [radius=0.1] node[above] {\scriptsize N+2};
\filldraw[fill= blue] (6,0) circle [radius=0.1] node[above] {\scriptsize 2};
\filldraw[fill= red] (7,0) circle [radius=0.1] node[above] {\scriptsize 1};
\filldraw[fill= white] (3,-1) circle [radius=0.1] node[left] {\scriptsize 2N+2};
\filldraw[fill= white] (3,-2) circle [radius=0.1] node[left] {\scriptsize N+2};
\filldraw[fill= blue] (3,-3) circle [radius=0.1] node[left] {\scriptsize 2};
\filldraw[fill= red] (3,-4) circle [radius=0.1] node[left] {\scriptsize 1};
\draw [thick] (-0.1, 0) -- (-0.9,0) ;
\draw [thick] (0.1, 0) -- (0.9,0) ;
\draw [thick] (1.1, 0) -- (1.9,0) ;
\draw [thick] (2.1, 0) -- (2.9,0) ;
\draw [thick] (3.1, 0) -- (3.9,0) ;
\draw [thick] (4.1, 0) -- (4.9,0) ;
\draw [thick] (5.1, 0) -- (5.9,0) ;
\draw [thick] (6.1, 0) -- (6.9,0) ;
\draw [thick] (3, -0.1) -- (3,-0.9) ;
\draw [thick] (3, -1.1) -- (3,-1.9) ;
\draw [thick] (3, -2.1) -- (3,-2.9) ;
\draw [thick] (3, -3.1) -- (3,-3.9) ;  

\draw[->, thick] (7,-1)--(8,-1); 

\filldraw[fill= white] (8,0) circle [radius=0.1] node[below] {\scriptsize N};
\filldraw[fill= white] (9,0) circle [radius=0.1] node[below] {\scriptsize 2N};
\filldraw[fill= white] (10,0) circle [radius=0.1] node[below] {\scriptsize 3N};
\filldraw[fill= white] (11,0) circle [radius=0.1] node[below] {\scriptsize 2N};
\filldraw[fill= white] (12,0) circle [radius=0.1] node[below] {\scriptsize N};
\filldraw[fill= white] (10,-1) circle [radius=0.1] node[left] {\scriptsize 2N};
\filldraw[fill= white] (10,-2) circle [radius=0.1] node[left] {\scriptsize N};
\filldraw[fill= white] (9.5,1) circle [radius=0.1] node[above] {\scriptsize 1};
\filldraw[fill= white] (10.5,1) circle [radius=0.1] node[above] {\scriptsize 1};
\draw [thick] (8.1, 0) -- (8.9,0) ;
\draw [thick] (9.1, 0) -- (9.9,0) ;
\draw [thick] (10.1, 0) -- (10.9,0) ;
\draw [thick] (11.1, 0) -- (11.9,0) ;
\draw [thick] (10, -0.1) -- (10,-0.9) ;
\draw [thick] (10, -1.1) -- (10,-1.9) ;
\draw [thick] (9.6, 1) -- (10.4,1) ;
\draw [thick] (9.97, 0.05) -- (9.53,0.95) ;
\draw [thick] (10.03, 0.05) -- (10.47,0.95) ;  
\end{tikzpicture}
\ee 
Finally, if we turn on a deformation at the two Abelian nodes in the second quiver in \eqref{defL=3} and fold we get 
\be\label{quivk6L3} 
\begin{tikzpicture}
\filldraw[fill= white] (2,-6) circle [radius=0.1] node[below] {\scriptsize N};
\filldraw[fill= white] (3,-6) circle [radius=0.1] node[below] {\scriptsize 2N};
\filldraw[fill= white] (4,-6) circle [radius=0.1] node[below] {\scriptsize 3N};
\filldraw[fill= white] (5,-6) circle [radius=0.1] node[below] {\scriptsize 1};
\draw [thick] (3.1, -5.95) -- (3.9,-5.95) ;
\draw [thick] (3.1, -6) -- (3.9,-6) ;
\draw [thick] (3.1, -6.05) -- (3.9,-6.05) ;
\draw [thick] (2.1, -6) -- (2.9,-6) ;
\draw [thick] (3.4,-6) -- (3.6,-5.8);
\draw [thick] (3.4,-6) -- (3.6,-6.2);
\draw [thick] (4.1, -5.95) -- (4.9,-5.95);
\draw [thick] (4.1, -6.05) -- (4.9,-6.05) ;
\end{tikzpicture}
\ee 
Indeed the manipulation we have just considered can be implemented  for any SW theory with $\ell=3$, namely for every choice of holonomy and every value of $N$. The net effect is to remove the $T(SU(2))$ tail and replace it with a single $U(1)$ node connected with a double bond to the node labelled by $C$ in the twisted affine Dynkin diagram of $D_4$ \eqref{DynkinD4}. At the same time we have to decrease by two the rank of all the nodes in the dynkin diagram. As a consistency check of this procedure, we can notice that for $N=1$ we obtain the magnetic quiver of a known rank-2 theory. A similar consistency check is provided by the choice of holonomy $n_3=2$ and setting $N=2$ \cite{Bourget:2021csg}. 

Finally, let us discuss the $SU(3)$ mass deformation which involves removing the $T(SU(3))$ tail for the magnetic quiver. This step is possible only for $k\geq3$. Here we focus on the holonomy with $k=3$ $(n_3=1, n_{3'}=2)$, which reduces to the model discussed above for $N=1$. The magnetic quiver of the resulting 4d theory is 
\be\label{quivk9L3} 
\begin{tikzpicture}
\filldraw[fill= white] (-1,-6) circle [radius=0.1] node[below] {\scriptsize 1};
\filldraw[fill= white] (0,-6) circle [radius=0.1] node[below] {\scriptsize 2};
\filldraw[fill= white] (1,-6) circle [radius=0.1] node[below] {\scriptsize 3};
\filldraw[fill= white] (2,-6) circle [radius=0.1] node[below] {\scriptsize N+2};
\filldraw[fill= white] (3,-6) circle [radius=0.1] node[below] {\scriptsize 2N+2};
\filldraw[fill= white] (4,-6) circle [radius=0.1] node[below] {\scriptsize 3N+2};
\draw [thick] (3.1, -5.95) -- (3.9,-5.95) ;
\draw [thick] (3.1, -6) -- (3.9,-6) ;
\draw [thick] (3.1, -6.05) -- (3.9,-6.05) ;
\draw [thick] (2.1, -6) -- (2.9,-6) ;
\draw [thick] (3.4,-6) -- (3.6,-5.8);
\draw [thick] (3.4,-6) -- (3.6,-6.2);
\draw [thick] (-0.1, -6) -- (-0.9,-6);
\draw [thick] (0.1, -6) -- (0.9,-6);
\draw [thick] (1.1, -6) -- (1.9,-6) ;
\end{tikzpicture}
\ee 
If we look at the quiver before the folding the step we have described in \eqref{defL=3} becomes 
\be\label{defL=33}
\begin{tikzpicture} 
\filldraw[fill= red] (-2,0) circle [radius=0.1] node[above] {\scriptsize 1};
\filldraw[fill= blue] (-1,0) circle [radius=0.1] node[above] {\scriptsize 2};
\filldraw[fill= white] (0,0) circle [radius=0.1] node[above] {\scriptsize 3};
\filldraw[fill= white] (1,0) circle [radius=0.1] node[above] {\scriptsize N+2};
\filldraw[fill= white] (2,0) circle [radius=0.1] node[above] {\scriptsize 2N+2};
\filldraw[fill= white] (3,0) circle [radius=0.1] node[above] {\scriptsize 3N+2};
\filldraw[fill= white] (4,0) circle [radius=0.1] node[above] {\scriptsize 2N+2};
\filldraw[fill= white] (5,0) circle [radius=0.1] node[above] {\scriptsize N+2};
\filldraw[fill= white] (6,0) circle [radius=0.1] node[above] {\scriptsize 3};
\filldraw[fill= blue] (7,0) circle [radius=0.1] node[above] {\scriptsize 2};
\filldraw[fill= red] (8,0) circle [radius=0.1] node[above] {\scriptsize 1};
\filldraw[fill= white] (3,-1) circle [radius=0.1] node[left] {\scriptsize 2N+2};
\filldraw[fill= white] (3,-2) circle [radius=0.1] node[left] {\scriptsize N+2};
\filldraw[fill= white] (3,-3) circle [radius=0.1] node[left] {\scriptsize 3};
\filldraw[fill= blue] (3,-4) circle [radius=0.1] node[left] {\scriptsize 2};
\filldraw[fill= red] (3,-5) circle [radius=0.1] node[left] {\scriptsize 1};
\draw [thick] (-1.1, 0) -- (-1.9,0) ;
\draw [thick] (-0.1, 0) -- (-0.9,0) ;
\draw [thick] (0.1, 0) -- (0.9,0) ;
\draw [thick] (1.1, 0) -- (1.9,0) ;
\draw [thick] (2.1, 0) -- (2.9,0) ;
\draw [thick] (3.1, 0) -- (3.9,0) ;
\draw [thick] (4.1, 0) -- (4.9,0) ;
\draw [thick] (5.1, 0) -- (5.9,0) ;
\draw [thick] (6.1, 0) -- (6.9,0) ;
\draw [thick] (7.1, 0) -- (7.9,0) ;
\draw [thick] (3, -0.1) -- (3,-0.9) ;
\draw [thick] (3, -1.1) -- (3,-1.9) ;
\draw [thick] (3, -2.1) -- (3,-2.9) ;
\draw [thick] (3, -3.1) -- (3,-3.9) ; 
\draw [thick] (3, -4.1) -- (3,-4.9) ;  

\draw[->, thick] (8,-1)--(9,-1); 
\filldraw[fill= white] (9,0) circle [radius=0.1] node[below] {\scriptsize 1};
\filldraw[fill= white] (10,0) circle [radius=0.1] node[below] {\scriptsize N};
\filldraw[fill= white] (11,0) circle [radius=0.1] node[below] {\scriptsize 2N};
\filldraw[fill= white] (12,0) circle [radius=0.1] node[below] {\scriptsize 3N};
\filldraw[fill= white] (13,0) circle [radius=0.1] node[below] {\scriptsize 2N};
\filldraw[fill= white] (14,0) circle [radius=0.1] node[below] {\scriptsize N};
\filldraw[fill= white] (15,0) circle [radius=0.1] node[below] {\scriptsize 1};
\filldraw[fill= white] (12,-1) circle [radius=0.1] node[left] {\scriptsize 2N};
\filldraw[fill= white] (12,-2) circle [radius=0.1] node[left] {\scriptsize N};
\filldraw[fill= white] (12,-3) circle [radius=0.1] node[left] {\scriptsize 1};
\filldraw[fill= white] (11.5,1) circle [radius=0.1] node[above] {\scriptsize 1};
\filldraw[fill= white] (12.5,1) circle [radius=0.1] node[above] {\scriptsize 1};
\draw [thick] (14.1, 0) -- (14.9,0) ;
\draw [thick] (13.1, 0) -- (13.9,0) ;
\draw [thick] (12.1, 0) -- (12.9,0) ;
\draw [thick] (9.1, 0) -- (9.9,0) ;
\draw [thick] (10.1, 0) -- (10.9,0) ;
\draw [thick] (11.1, 0) -- (11.9,0) ;
\draw [thick] (12, -0.1) -- (12,-0.9) ;
\draw [thick] (12, -1.1) -- (12,-1.9) ;
\draw [thick] (12, -2.1) -- (12,-2.9) ;
\draw [thick] (11.6, 1) -- (12.4,1) ;
\draw [thick] (11.97, 0.05) -- (11.53,0.95) ;
\draw [thick] (12.03, 0.05) -- (12.47,0.95) ;  
\end{tikzpicture}
\ee  
If we keep activating FI deformations at the Abelian nodes we get, after folding, the quiver 
\be\label{quivk9L32} 
\begin{tikzpicture}
\filldraw[fill= white] (2,-6) circle [radius=0.1] node[below] {\scriptsize N-1};
\filldraw[fill= white] (3,-6) circle [radius=0.1] node[below] {\scriptsize 2N-1};
\filldraw[fill= white] (4,-6) circle [radius=0.1] node[below] {\scriptsize 3N-1};
\filldraw[fill= white] (5,-6) circle [radius=0.1] node[below] {\scriptsize 1};
\draw [thick] (3.1, -5.95) -- (3.9,-5.95) ;
\draw [thick] (3.1, -6) -- (3.9,-6) ;
\draw [thick] (3.1, -6.05) -- (3.9,-6.05) ;
\draw [thick] (2.1, -6) -- (2.9,-6) ;
\draw [thick] (3.4,-6) -- (3.6,-5.8);
\draw [thick] (3.4,-6) -- (3.6,-6.2);
\draw [thick] (4.1, -5.95) -- (4.9,-5.95);
\draw [thick] (4.1, -6) -- (4.9,-6);
\draw [thick] (4.1, -6.05) -- (4.9,-6.05) ;
\end{tikzpicture}
\ee 
As a consistency check of our proposal, we notice that for $N=1$ \eqref{quivk9L32} reduces to the magnetic quiver of a known rank-2 theory. 

In summary, our proposal for the mass deformation of the $SU(k)$ symmetry up to $k=3$ for $\ell=3$ SW theories is as follows: 
\be\label{massdefL3}
\begin{tikzpicture}
\filldraw[fill= white] (-1,-6) circle [radius=0.1] node[below] {\scriptsize 1};
\filldraw[fill= white] (0,-6) circle [radius=0.1] node[below] {\scriptsize 2};
\filldraw[fill= white] (1,-6) circle [radius=0.1] node[below] {\scriptsize 3};
\filldraw[fill= white] (2,-6) circle [radius=0.1] node[below] {\scriptsize A};
\filldraw[fill= white] (3,-6) circle [radius=0.1] node[below] {\scriptsize B};
\filldraw[fill= white] (4,-6) circle [radius=0.1] node[below] {\scriptsize C};
\draw [thick] (3.1, -5.95) -- (3.9,-5.95) ;
\draw [thick] (3.1, -6) -- (3.9,-6) ;
\draw [thick] (3.1, -6.05) -- (3.9,-6.05) ;
\draw [thick] (2.1, -6) -- (2.9,-6) ;
\draw [thick] (3.4,-6) -- (3.6,-5.8);
\draw [thick] (3.4,-6) -- (3.6,-6.2);
\draw [thick] (-0.1, -6) -- (-0.9,-6);
\draw [thick] (0.1, -6) -- (0.9,-6);
\draw [thick] (1.1, -6) -- (1.9,-6) ; 

\draw [->, thick] (2, -7) -- (2,-8) ; 

\filldraw[fill= white] (0,-9) circle [radius=0.1] node[below] {\scriptsize 1};
\filldraw[fill= white] (1,-9) circle [radius=0.1] node[below] {\scriptsize A-2};
\filldraw[fill= white] (2,-9) circle [radius=0.1] node[below] {\scriptsize B-2};
\filldraw[fill= white] (3,-9) circle [radius=0.1] node[below] {\scriptsize C-2};
\filldraw[fill= white] (4,-9) circle [radius=0.1] node[below] {\scriptsize 1};
\draw [thick] (2.1, -8.95) -- (2.9,-8.95) ;
\draw [thick] (2.1, -9) -- (2.9,-9) ;
\draw [thick] (2.1, -9.05) -- (2.9,-9.05) ;
\draw [thick] (2.4,-9) -- (2.6,-8.8);
\draw [thick] (2.4,-9) -- (2.6,-9.2);
\draw [thick] (3.1, -8.95) -- (3.9,-8.95);
\draw [thick] (3.1, -9.05) -- (3.9,-9.05);
\draw [thick] (0.1, -9) -- (0.9,-9);
\draw [thick] (1.1, -9) -- (1.9,-9) ;

\draw [->, thick] (2, -10) -- (2,-11) ; 

\filldraw[fill= white] (1,-12) circle [radius=0.1] node[below] {\scriptsize A-3};
\filldraw[fill= white] (2,-12) circle [radius=0.1] node[below] {\scriptsize B-3};
\filldraw[fill= white] (3,-12) circle [radius=0.1] node[below] {\scriptsize C-3};
\filldraw[fill= white] (4,-12) circle [radius=0.1] node[below] {\scriptsize 1};
\draw [thick] (2.1, -11.95) -- (2.9,-11.95) ;
\draw [thick] (2.1, -12) -- (2.9,-12) ;
\draw [thick] (2.1, -12.05) -- (2.9,-12.05) ;
\draw [thick] (1.1, -12) -- (1.9,-12) ;
\draw [thick] (2.4,-12) -- (2.6,-11.8);
\draw [thick] (2.4,-12) -- (2.6,-12.2);
\draw [thick] (3.1, -11.95) -- (3.9,-11.95);
\draw [thick] (3.1, -12) -- (3.9,-12);
\draw [thick] (3.1, -12.05) -- (3.9,-12.05) ;
\end{tikzpicture}
\ee 
Again, after ungauging at the Abelian node, we can notice that the number of flavors times the Dynkin label of node $C$ is equal to 2 and 3 for the second and third quivers in \eqref{massdefL3} respectively. This is perfectly consistent with equation \eqref{abelianrule}.

\subsubsection{The case $\ell=4$}

In this case we are dealing with the twisted affine $A_2$ Dynkin diagram 
\be\label{DynkinA2}
\begin{tikzpicture} 
 \filldraw[fill= white] (0,0) circle [radius=0.1] node[below] {\scriptsize A};
\filldraw[fill= white] (1,0) circle [radius=0.1] node[below] {\scriptsize B};

\draw [thick] (0.05,0.07)--(0.95,0.07);
\draw [thick] (0.07,0.02)--(0.93,0.02);
\draw [thick] (0.07,-0.02)--(0.93,-0.02);
\draw [thick] (0.05,-0.07)--(0.95,-0.07);
\draw [thick] (0.4,0) -- (0.6,-0.2);
\draw [thick] (0.4,0) -- (0.6,0.2);
\end{tikzpicture} 
\ee 
since we are combining the $\bbZ_4$ S-fold with a 7-brane of type $\mathcal{H}_2$. For illustration we consider the case $k=2$ with holonomy $n_4=2$. Starting from the magnetic quiver before the folding we activate the deformation 
\be\label{quivk=2L=4}
\begin{tikzpicture} 
\filldraw[fill= red] (1,-4.5) circle [radius=0.1] node[below] {\scriptsize 1};
\filldraw[fill= white] (2,-5) circle [radius=0.1] node[below] {\scriptsize 2};
\filldraw[fill= white] (3,-5.5) circle [radius=0.1] node[below] {\scriptsize N};
\filldraw[fill= white] (4,-6) circle [radius=0.1] node[below] {\scriptsize 2N};
\filldraw[fill= red] (1,-7.5) circle [radius=0.1] node[below] {\scriptsize 1};
\filldraw[fill= white] (2,-7) circle [radius=0.1] node[below] {\scriptsize 2};
\filldraw[fill= white] (3,-6.5) circle [radius=0.1] node[below] {\scriptsize N};
\filldraw[fill= blue] (7,-4.5) circle [radius=0.1] node[below] {\scriptsize 1};
\filldraw[fill= white] (6,-5) circle [radius=0.1] node[below] {\scriptsize 2};
\filldraw[fill= white] (5,-5.5) circle [radius=0.1] node[below] {\scriptsize N};
\filldraw[fill= blue] (7,-7.5) circle [radius=0.1] node[below] {\scriptsize 1};
\filldraw[fill= white] (6,-7) circle [radius=0.1] node[below] {\scriptsize 2};
\filldraw[fill= white] (5,-6.5) circle [radius=0.1] node[below] {\scriptsize N};
\draw [thick] (4.06, -5.97) -- (4.94,-5.52) ;
\draw [thick] (5.06, -5.47) -- (5.94,-5.02) ;
\draw [thick] (6.06, -4.97) -- (6.94,-4.52) ;
\draw [thick] (4.06, -6.03) -- (4.94,-6.47) ;
\draw [thick] (5.06, -6.53) -- (5.94,-6.97) ;
\draw [thick] (6.06, -7.03) -- (6.94,-7.47) ;
\draw [thick] (1.06, -4.53) -- (1.94,-4.97) ;
\draw [thick] (2.06, -5.03) -- (2.94,-5.47) ;
\draw [thick] (3.06, -5.53) -- (3.94,-5.97) ;
\draw [thick] (1.06, -7.47) -- (1.94,-7.03) ;
\draw [thick] (2.06, -6.97) -- (2.94,-6.53) ;
\draw [thick] (3.06, -6.47) -- (3.94,-6.03) ;

\draw[->, thick] (6.5,-6)--(7.5,-6); 

\filldraw[fill= white] (8,-5) circle [radius=0.1] node[below] {\scriptsize 1};
\filldraw[fill= white] (9,-5.5) circle [radius=0.1] node[below] {\scriptsize N-1};
\filldraw[fill= white] (10,-6) circle [radius=0.1] node[below] {\scriptsize 2N-1};
\filldraw[fill= white] (8,-7) circle [radius=0.1] node[below] {\scriptsize 1};
\filldraw[fill= white] (9,-6.5) circle [radius=0.1] node[below] {\scriptsize N-1};
\filldraw[fill= white] (12,-5) circle [radius=0.1] node[below] {\scriptsize 1};
\filldraw[fill= white] (11,-5.5) circle [radius=0.1] node[below] {\scriptsize N-1};
\filldraw[fill= white] (12,-7) circle [radius=0.1] node[below] {\scriptsize 1};
\filldraw[fill= white] (11,-6.5) circle [radius=0.1] node[below] {\scriptsize N-1};
\filldraw[fill= white] (10,-5) circle [radius=0.1] node[above] {\scriptsize 1};
\draw [thick] (10.06, -5.97) -- (10.94,-5.52) ;
\draw [thick] (11.06, -5.47) -- (11.94,-5.02) ;
\draw [thick] (10.06, -6.03) -- (10.94,-6.47) ;
\draw [thick] (11.06, -6.53) -- (11.94,-6.97) ;
\draw [thick] (8.06, -5.03) -- (8.94,-5.47) ;
\draw [thick] (9.06, -5.53) -- (9.94,-5.97) ;
\draw [thick] (8.06, -6.97) -- (8.94,-6.53) ;
\draw [thick] (9.06, -6.47) -- (9.94,-6.03) ; 
\draw [thick] (9.95,-5.95)--(9.95,-5.05);
\draw [thick] (10.05,-5.95)--(10.05,-5.05);
\end{tikzpicture} 
\ee 
If we now reiterate the deformation and subsequently turn on a FI at the last two surviving Abelian nodes we obtain the quiver 
\be\label{step2}
\begin{tikzpicture} 
\filldraw[fill= white] (9,-5.5) circle [radius=0.1] node[below] {\scriptsize N-2};
\filldraw[fill= white] (10,-6) circle [radius=0.1] node[below] {\scriptsize 2N-2};
\filldraw[fill= white] (9,-6.5) circle [radius=0.1] node[below] {\scriptsize N-2};
\filldraw[fill= white] (11,-5.5) circle [radius=0.1] node[below] {\scriptsize N-2};
\filldraw[fill= white] (11,-6.5) circle [radius=0.1] node[below] {\scriptsize N-2};
\filldraw[fill= white] (10,-5) circle [radius=0.1] node[above] {\scriptsize 1};
\draw [thick] (10.06, -5.97) -- (10.94,-5.52) ;
\draw [thick] (10.06, -6.03) -- (10.94,-6.47) ;
\draw [thick] (9.06, -5.53) -- (9.94,-5.97) ;
\draw [thick] (9.06, -6.47) -- (9.94,-6.03) ; 
\draw [thick] (9.93,-5.95)--(9.93,-5.05);
\draw [thick] (9.98,-5.93)--(9.98,-5.07);
\draw [thick] (10.02,-5.93)--(10.02,-5.07);
\draw [thick] (10.07,-5.95)--(10.07,-5.05);
\end{tikzpicture} 
\ee 
and after folding we get from \eqref{step2} 
\be\label{quivL=4}
\begin{tikzpicture} 
 \filldraw[fill= white] (0,0) circle [radius=0.1] node[below] {\scriptsize N-2};
\filldraw[fill= white] (1,0) circle [radius=0.1] node[below] {\scriptsize 2N-2};
\filldraw[fill= white] (2,0) circle [radius=0.1] node[below] {\scriptsize 1}; 

\draw [thick] (0.05,0.07)--(0.95,0.07);
\draw [thick] (0.07,0.02)--(0.93,0.02);
\draw [thick] (0.07,-0.02)--(0.93,-0.02);
\draw [thick] (0.05,-0.07)--(0.95,-0.07);
\draw [thick] (1.05,0.07)--(1.95,0.07);
\draw [thick] (1.07,0.02)--(1.93,0.02);
\draw [thick] (1.07,-0.02)--(1.93,-0.02);
\draw [thick] (1.05,-0.07)--(1.95,-0.07);
\draw [thick] (0.4,0) -- (0.6,-0.2);
\draw [thick] (0.4,0) -- (0.6,0.2);
\end{tikzpicture} 
\ee  
As in the previous cases we find the twisted Dynkin diagram \eqref{DynkinA2} with an Abelian node attached. 

As a check we can notice that for $N=2$ this reproduces the correct magnetic quiver for a known rank-2 SCFT. This is the model 55 discussed in \cite{Martone:2021ixp}, where it was pointed out that this theory can be higgsed to the rank-1 $\mathcal{N}=3$ SCFT labelled by the reflection group $G(4,1,1)$ with transverse slice $\mathbb{C}^2/\bbZ_2$. The $\mathcal{N}=3$ theory in turn has a Higgs branch isomorphic to $\mathbb{C}^2/\bbZ_4$. As a result, the quiver encoding this structure is\footnote{We thank Antoine Bourget and Julius Grimminger for discussions about this point.} 
$$\begin{tikzpicture} 
\filldraw[fill= white] (1,0) circle [radius=0.1] node[below] {\scriptsize 2};
\filldraw[fill= white] (2,0) circle [radius=0.1] node[below] {\scriptsize 1}; 

\draw [thick] (1.05,0.07)--(1.95,0.07);
\draw [thick] (1.07,0.02)--(1.93,0.02);
\draw [thick] (1.07,-0.02)--(1.93,-0.02);
\draw [thick] (1.05,-0.07)--(1.95,-0.07);
\end{tikzpicture}  $$
This agrees perfectly with \eqref{quivL=4} for $N=2$. 

For general choices of holonomy our proposal is therefore 
\be\label{defL=4}
 \begin{tikzpicture} 
\filldraw[fill= white] (-5,0) circle [radius=0.1] node[below] {\scriptsize 1}; 
\filldraw[fill= white] (-4,0) circle [radius=0.1] node[below] {\scriptsize 2}; 
 \filldraw[fill= white] (-3,0) circle [radius=0.1] node[below] {\scriptsize A};
\filldraw[fill= white] (-2,0) circle [radius=0.1] node[below] {\scriptsize B};

\draw [thick] (-4.9, 0) -- (-4.1,0) ; 
\draw [thick] (-3.9, 0) -- (-3.1,0) ; 
\draw [thick] (-2.05,0.07)--(-2.95,0.07);
\draw [thick] (-2.07,0.02)--(-2.93,0.02);
\draw [thick] (-2.07,-0.02)--(-2.93,-0.02);
\draw [thick] (-2.05,-0.07)--(-2.95,-0.07);
\draw [thick] (-2.6,0) -- (-2.4,-0.2);
\draw [thick] (-2.6,0) -- (-2.4,0.2);

\draw[->, thick] (-1.5,0)--(-0.5,0); 

 \filldraw[fill= white] (0,0) circle [radius=0.1] node[below] {\scriptsize A-2};
\filldraw[fill= white] (1,0) circle [radius=0.1] node[below] {\scriptsize B-2};
\filldraw[fill= white] (2,0) circle [radius=0.1] node[below] {\scriptsize 1}; 

\draw [thick] (0.05,0.07)--(0.95,0.07);
\draw [thick] (0.07,0.02)--(0.93,0.02);
\draw [thick] (0.07,-0.02)--(0.93,-0.02);
\draw [thick] (0.05,-0.07)--(0.95,-0.07);
\draw [thick] (1.05,0.07)--(1.95,0.07);
\draw [thick] (1.07,0.02)--(1.93,0.02);
\draw [thick] (1.07,-0.02)--(1.93,-0.02);
\draw [thick] (1.05,-0.07)--(1.95,-0.07);
\draw [thick] (0.4,0) -- (0.6,-0.2);
\draw [thick] (0.4,0) -- (0.6,0.2);
\end{tikzpicture} 
\ee
After ungauging at the Abelian node, we can notice that the number of flavors times the Dynkin label of node $B$ in the second quiver in \eqref{defL=4} is equal to $4$. This is consistent with equation \eqref{abelianrule} after the replacement 
\be\label{abelianrule2} \sum_{\text{nodes}}d_iF_i=2k\,.\ee

The difference in the r.h.s.~of \eqref{abelianrule} and \eqref{abelianrule2} suggests that the general rule for the attachment of the Abelian node to the Dynkin diagram is
\be\label{genabelianrule} 
\sum_{\text{nodes}}d_iF_i=\frac{M}{\ell'}\,,
\ee  
where $M=\ell k$ is the order of the automorphism and $\ell'$ is its order as an outer automorphism. The latter quantity is equal to $\ell$ for the 7-branes of type $E_6,D_4$ and to $2$ for the 7-brane of type $\mathcal{H}_2$ \cite{Giacomelli:2020gee} (see Table \eqref{xxxx}).\footnote{These values of $\ell'$ are for the maximal type of 7-brane compatible with the S-fold order. However the rule \eqref{genabelianrule} also applies to the cases with non-maximal 7-branes, which we will discuss in Section \ref{Sec:DecFlavors}. In these cases, $\ell'$ remains equal to $\ell$ for the 7-brane of type $D_4$, while it is equal to $1$ for the 7-branes of type $\mathcal{H}_2,\mathcal{H}_1$ and $\ell<4$ \cite{Giacomelli:2020gee} (see Table \eqref{xxxx}).}

\subsubsection{The cases $\ell=5,6$} \label{defell56}

In these cases the magnetic quivers describing the Higgs branch are very simple (see \eqref{eq:z5quiver} and \eqref{eq:z6quiver}) and are given by 
\be\label{ccell5} 
\begin{tikzpicture}
\draw [thick] (2.1, 0) -- (2.9,0) ;
\draw [thick] (2.4,0) -- (2.5,0.1);
\draw [thick] (2.4,0) -- (2.5,-0.1);
\draw [thick] (3.05, 0.05) to [out=45,in=315,looseness=12] (3.05,-0.05);
\filldraw[fill= gray] (2,0) circle [radius=0.1] node[below] {\scriptsize $k$};
\filldraw[fill= white] (3,0) circle [radius=0.1] node[below] {\scriptsize N};
\node[] at (2.5,0.25) {\scriptsize $\ell$}; 
\end{tikzpicture}
\ee 
where $\ell$ indeed denotes the multiplicity of the edge. If we turn on a mass term for a $SU(2)$ in $SU(k)$ the quiver \eqref{ccell5} can be shown to become 
\be\label{ccell52} 
\begin{tikzpicture}
\draw [thick] (2.1, 0) -- (2.9,0) ;
\draw [thick] (2.4,0) -- (2.5,0.1);
\draw [thick] (2.4,0) -- (2.5,-0.1);
\draw [thick] (3, 0.1) -- (3,0.9) ;
\draw [thick] (3.05, 0.05) to [out=45,in=315,looseness=12] (3.05,-0.05);
\filldraw[fill= gray] (2,0) circle [radius=0.1] node[below] {\scriptsize $k-2$};
\filldraw[fill= white] (3,0) circle [radius=0.1] node[below] {\scriptsize N-2};
\filldraw[fill= white] (3,1) circle [radius=0.1] node[left] {\scriptsize 1};
\node[] at (2.5,0.25) {\scriptsize $\ell$}; 
\node[] at (3.25,0.5) {\scriptsize $2\ell$}; 
\end{tikzpicture}
\ee 
The label $2\ell$ on top denotes the multiplicity of the unoriented edge, which is nothing but a bifundamental hypermultiplet.

\noindent To conclude our analysis of the various cases, let us discuss how the quiver changes if, for $k>2$, we proceed and deform away the $SU(k)$ symmetry completely. In this case the tail is removed and we land on 
\be\label{ccell53}
\begin{tikzpicture}
\draw [thick] (2.1, 0) -- (2.9,0) ;
\draw [thick] (3.05, 0.05) to [out=45,in=315,looseness=12] (3.05,-0.05);
\filldraw[fill= white] (2,0) circle [radius=0.1] node[below] {\scriptsize $1$};
\filldraw[fill= white] (3,0) circle [radius=0.1] node[below] {\scriptsize N-k};
\node[] at (2.5,0.2) {\scriptsize $k\ell$}; 
\end{tikzpicture}
\ee
where again $k\ell$ denotes the multiplicity of the bifundamental hypermultiplet.

\subsection{Theories on the $D3$-brane probes}\label{Sec:D3th}

In this section we will work out the other properties of the SCFTs obtained upon turning on a mass deformation for the global symmetry associated with the $T(SU(k))$ tail. We start by describing the Coulomb-branch spectrum of the resulting theory and then we proceed by determining the $a$, $c$, and the flavor central charges. We finally explain how the Hasse diagram of the theory is affected by the mass deformation.

\subsubsection{Coulomb-branch spectrum}\label{CBspectrumIIB}

Our task now is to determine the Coulomb-branch (CB) spectrum of the mass-deformed theories. Contrary to the undeformed case discussed in Section \ref{Sec:SWcompact}, we cannot use the results of \cite{Ohmori:2018ona} and therefore we need to find an alternative way to derive the answer. Our strategy is simple: We first consider theories with $\ell=1$ for which the mass deformation we are interested in was studied in detail in \cite{Giacomelli:2022drw}. These models were shown in    \cite{Giacomelli:2022drw} to correspond to class-$\mathcal S$ trinions (of type A untwisted) and therefore we can easily extract their Coulomb-branch spectrum using the techniques developed in \cite{Chacaltana:2010ks}. We will find a simple pattern for $\ell=1$ theories, and we will use it to guess the answer for $\ell>1$. We then test our proposal by comparing with the known spectrum of rank-2 theories, finding perfect agreement in all cases. 

Let us start by discussing the mass deformation for $\ell=1$ theories. We will look explicitly at a few examples. The first we consider is the $T^2$ compactification of the 6d orbi-instanton theory with $M=2$ and holonomy $n_{2'}=1$. As was discussed in \cite{Giacomelli:2022drw} the mass deformation is implemented at the level of the magnetic quiver as follows: 
\be\label{M2def}
\begin{tikzpicture}
\filldraw[fill= red] (-2,0) circle [radius=0.1] node[below] {\scriptsize 1};
\filldraw[fill= red] (-1,0) circle [radius=0.1] node[below] {\scriptsize 2};
\filldraw[fill= white] (0,0) circle [radius=0.1] node[below] {\scriptsize N+2};
\filldraw[fill= white] (1,0) circle [radius=0.1] node[below] {\scriptsize 2N+2};
\filldraw[fill= white] (2,0) circle [radius=0.1] node[below] {\scriptsize 3N+2};
\filldraw[fill= white] (3,0) circle [radius=0.1] node[below] {\scriptsize 4N+2};
\filldraw[fill= white] (4,0) circle [radius=0.1] node[below] {\scriptsize 5N+2};
\filldraw[fill= white] (5,0) circle [radius=0.1] node[below] {\scriptsize 6N+2};
\filldraw[fill= white] (6,0) circle [radius=0.1] node[below] {\scriptsize 4N+1};
\filldraw[fill= white] (5,1) circle [radius=0.1] node[above] {\scriptsize 3N+1};
\filldraw[fill= white] (7,0) circle [radius=0.1] node[below] {\scriptsize 2N};

\draw [thick] (-0.1, 0) -- (-0.9,0) ;
\draw [thick] (-1.1, 0) -- (-1.9,0) ;
\draw [thick] (0.1, 0) -- (0.9,0) ;
\draw [thick] (1.1, 0) -- (1.9,0) ;
\draw [thick] (2.1, 0) -- (2.9,0) ;
\draw [thick] (3.1, 0) -- (3.9,0) ;
\draw [thick] (4.1, 0) -- (4.9,0) ;
\draw [thick] (5.1, 0) -- (5.9,0) ;
\draw [thick] (5, 0.1) -- (5,0.9) ;
\draw [thick] (6.1, 0) -- (6.9,0) ;

\draw [->] (3,-0.6) -- (3,-2.6) ;

\filldraw[fill= white] (0,-3) circle [radius=0.1] node[below] {\scriptsize N};
\filldraw[fill= white] (1,-3) circle [radius=0.1] node[below] {\scriptsize 2N};
\filldraw[fill= white] (2,-3) circle [radius=0.1] node[below] {\scriptsize 3N};
\filldraw[fill= white] (3,-3) circle [radius=0.1] node[below] {\scriptsize 4N};
\filldraw[fill= white] (4,-3) circle [radius=0.1] node[below] {\scriptsize 5N};
\filldraw[fill= white] (5,-3) circle [radius=0.1] node[below] {\scriptsize 6N};
\filldraw[fill= white] (6,-3) circle [radius=0.1] node[below] {\scriptsize 4N};
\filldraw[fill= white] (5,-2) circle [radius=0.1] node[above] {\scriptsize 3N};
\filldraw[fill= white] (7,-3) circle [radius=0.1] node[below] {\scriptsize 2N};
\filldraw[fill= white] (8,-3) circle [radius=0.1] node[below] {\scriptsize 1};

\draw [thick] (7.1, -3) -- (7.9,-3) ;
\draw [thick] (0.1, -3) -- (0.9,-3) ;
\draw [thick] (1.1, -3) -- (1.9,-3) ;
\draw [thick] (2.1, -3) -- (2.9,-3) ;
\draw [thick] (3.1, -3) -- (3.9,-3) ;
\draw [thick] (4.1, -3) -- (4.9,-3) ;
\draw [thick] (5.1, -3) -- (5.9,-3) ;
\draw [thick] (5, -2.1) -- (5,-2.9) ;
\draw [thick] (6.1, -3) -- (6.9,-3) ;
\end{tikzpicture}
\ee
The Coulomb-branch spectrum of the theory in \eqref{M2def} before the mass deformation is known to be given by the sequence of integers
\be\label{beforedef} (6n, 6n+2)_{n=1,\dots, N}\,.\ee 
After the deformation we obtain instead, using the algorithm described in \cite{Chacaltana:2010ks}, the sequence 
\be\label{afterdef} (6n-2, 6n)_{n=1,\dots, N}\,.\ee 
This result was also given in \cite{Giacomelli:2022drw}. By comparing \eqref{beforedef} and \eqref{afterdef} we see that the effect of the mass deformation at the level of the Coulomb-branch spectrum is simply to uniformly shift the scaling dimension of all Coulomb-branch operators by two units.   

Let us confirm this pattern for a more complicated theory with a $T(SU(3))$ tail. We consider for definiteness the $T^2$ compactification of the orbi-instanton theory with $M=3$ and $n_{3'}=1$. The mass deformation of this theory was studied in \cite{Giacomelli:2022drw}:
\be
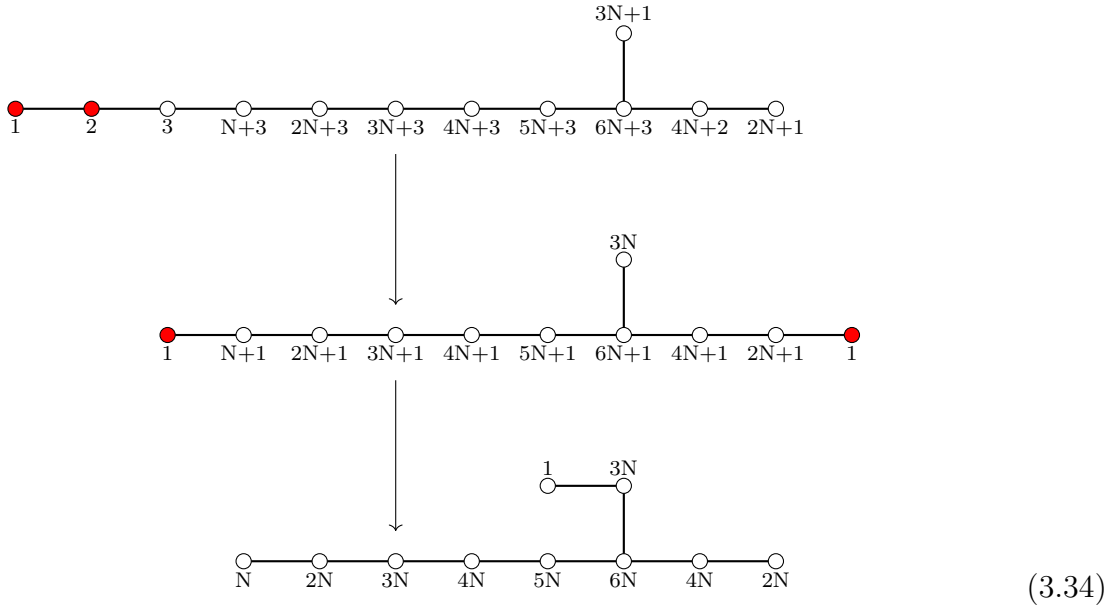
\label{M3def}
\begin{tikzpicture}
\filldraw[fill= red] (-3,0) circle [radius=0.1] node[below] {\scriptsize 1};
\filldraw[fill= red] (-2,0) circle [radius=0.1] node[below] {\scriptsize 2};
\filldraw[fill= white] (-1,0) circle [radius=0.1] node[below] {\scriptsize 3};
\filldraw[fill= white] (0,0) circle [radius=0.1] node[below] {\scriptsize N+3};
\filldraw[fill= white] (1,0) circle [radius=0.1] node[below] {\scriptsize 2N+3};
\filldraw[fill= white] (2,0) circle [radius=0.1] node[below] {\scriptsize 3N+3};
\filldraw[fill= white] (3,0) circle [radius=0.1] node[below] {\scriptsize 4N+3};
\filldraw[fill= white] (4,0) circle [radius=0.1] node[below] {\scriptsize 5N+3};
\filldraw[fill= white] (5,0) circle [radius=0.1] node[below] {\scriptsize 6N+3};
\filldraw[fill= white] (6,0) circle [radius=0.1] node[below] {\scriptsize 4N+2};
\filldraw[fill= white] (5,1) circle [radius=0.1] node[above] {\scriptsize 3N+1};
\filldraw[fill= white] (7,0) circle [radius=0.1] node[below] {\scriptsize 2N+1};

\draw [thick] (-0.1, 0) -- (-0.9,0) ;
\draw [thick] (-1.1, 0) -- (-1.9,0) ;
\draw [thick] (-2.1, 0) -- (-2.9,0) ;
\draw [thick] (0.1, 0) -- (0.9,0) ;
\draw [thick] (1.1, 0) -- (1.9,0) ;
\draw [thick] (2.1, 0) -- (2.9,0) ;
\draw [thick] (3.1, 0) -- (3.9,0) ;
\draw [thick] (4.1, 0) -- (4.9,0) ;
\draw [thick] (5.1, 0) -- (5.9,0) ;
\draw [thick] (5, 0.1) -- (5,0.9) ;
\draw [thick] (6.1, 0) -- (6.9,0) ;

\draw [->] (2,-0.6) -- (2,-2.6) ;

\filldraw[fill= red] (-1,-3) circle [radius=0.1] node[below] {\scriptsize 1};
\filldraw[fill= white] (0,-3) circle [radius=0.1] node[below] {\scriptsize N+1};
\filldraw[fill= white] (1,-3) circle [radius=0.1] node[below] {\scriptsize 2N+1};
\filldraw[fill= white] (2,-3) circle [radius=0.1] node[below] {\scriptsize 3N+1};
\filldraw[fill= white] (3,-3) circle [radius=0.1] node[below] {\scriptsize 4N+1};
\filldraw[fill= white] (4,-3) circle [radius=0.1] node[below] {\scriptsize 5N+1};
\filldraw[fill= white] (5,-3) circle [radius=0.1] node[below] {\scriptsize 6N+1};
\filldraw[fill= white] (6,-3) circle [radius=0.1] node[below] {\scriptsize 4N+1};
\filldraw[fill= white] (5,-2) circle [radius=0.1] node[above] {\scriptsize 3N};
\filldraw[fill= white] (7,-3) circle [radius=0.1] node[below] {\scriptsize 2N+1};
\filldraw[fill= red] (8,-3) circle [radius=0.1] node[below] {\scriptsize 1};

\draw [thick] (-0.1, -3) -- (-0.9,-3) ;
\draw [thick] (7.1, -3) -- (7.9,-3) ;
\draw [thick] (0.1, -3) -- (0.9,-3) ;
\draw [thick] (1.1, -3) -- (1.9,-3) ;
\draw [thick] (2.1, -3) -- (2.9,-3) ;
\draw [thick] (3.1, -3) -- (3.9,-3) ;
\draw [thick] (4.1, -3) -- (4.9,-3) ;
\draw [thick] (5.1, -3) -- (5.9,-3) ;
\draw [thick] (5, -2.1) -- (5,-2.9) ;
\draw [thick] (6.1, -3) -- (6.9,-3) ;

\draw [->] (2,-3.6) -- (2,-5.6) ;

\filldraw[fill= white] (0,-6) circle [radius=0.1] node[below] {\scriptsize N};
\filldraw[fill= white] (1,-6) circle [radius=0.1] node[below] {\scriptsize 2N};
\filldraw[fill= white] (2,-6) circle [radius=0.1] node[below] {\scriptsize 3N};
\filldraw[fill= white] (3,-6) circle [radius=0.1] node[below] {\scriptsize 4N};
\filldraw[fill= white] (4,-6) circle [radius=0.1] node[below] {\scriptsize 5N};
\filldraw[fill= white] (5,-6) circle [radius=0.1] node[below] {\scriptsize 6N};
\filldraw[fill= white] (6,-6) circle [radius=0.1] node[below] {\scriptsize 4N};
\filldraw[fill= white] (5,-5) circle [radius=0.1] node[above] {\scriptsize 3N};
\filldraw[fill= white] (7,-6) circle [radius=0.1] node[below] {\scriptsize 2N};
\filldraw[fill= white] (4,-5) circle [radius=0.1] node[above] {\scriptsize 1};

\draw [thick] (4.1, -5) -- (4.9,-5) ;
\draw [thick] (0.1, -6) -- (0.9,-6) ;
\draw [thick] (1.1, -6) -- (1.9,-6) ;
\draw [thick] (2.1, -6) -- (2.9,-6) ;
\draw [thick] (3.1, -6) -- (3.9,-6) ;
\draw [thick] (4.1, -6) -- (4.9,-6) ;
\draw [thick] (5.1, -6) -- (5.9,-6) ;
\draw [thick] (5, -5.1) -- (5,-5.9) ;
\draw [thick] (6.1, -6) -- (6.9,-6) ;
\end{tikzpicture}
\ee 
At the level of the Coulomb-branch spectrum the above deformation reads 
\be\label{def3M3} (6n, 6n+2, 6n+3)_{n=1,\dots, N}\rightarrow (6n-2, 6n, 6n+1)_{n=1,\dots, N}\rightarrow(6n-3, 6n-1, 6n)_{n=1,\dots, N}\,.\ee 
As we see from \eqref{def3M3} the effect of the first deformation is to decrease by two the scaling dimension of all Coulomb-branch operators, as in \eqref{afterdef}, while the second further lowers the scaling dimensions by one unit. By looking at a large number of examples we find the following uniform result: If the $T^2$ reduction of the orbi-instanton theory has a magnetic quiver with a $T(SU(k))$ tail, the CB spectrum of the SCFT's obtained by activating mass deformations for the $SU(k)$ symmetry can be derived by reducing the scaling dimension of all the CB operators of the original theory by the degree of one of the Casimir invariants of $SU(k)$. In the previous examples we have checked this for $k=2,3$. 

Our guess is that the same rule applies also to theories with $\ell>1$. The only ingredient which has no counterpart for $\ell=1$ is the presence of the operator whose dimension is divided by $\ell$ when the $-1$ curve in 6d supports a trivial or $USp(4m)$ gauge group. This happens for instance in the case $M=4$, $\ell=2$ and holonomy $n_{2'}=2$. For $N=1$ this corresponds to a known rank-2 theory with $USp(12)$ global symmetry and CB operators of dimension 4 and 6, or equivalently 6 and $8/2$. The mass deformation of this model leads to a SCFT with operators of dimension 3 and 4, which is also equal to $6-2$ and $(8-2)/2$. This suggests that we need to subtract the degree of the $SU(k)$ Casimir before dividing by $\ell$. In this case the $SU(2)$ symmetry enhances since it combines with the global symmetry arising from the $E_6$ 7-brane and therefore we can proceed further with the mass deformations. Our claim is that our rule for the spectrum of Coulomb-branch operators keeps working. Let us check this explicitly: If we proceed with another mass deformation we find a SCFT with operators of dimension $5/2$ and $3$, equivalent to $6-3$ and $(8-3)/2$. Finally, we predict a further deformation leading to a theory with spectrum $6-4$ and $(8-4)/2$, i.e. 2 and 2. Indeed such SCFTs are all known to exist as mass deformations of the $USp(12)$ theory \cite{Martone:2021drm}, thus confirming our guess.  

Our final proposal is therefore the following:
\begin{itemize}
\item Start from the spectrum of the undeformed theory $\{\Delta_r\}_{r=1,\dots,N}$.
\item Consider the shifted spectrum  $\{\Delta_r-j\}_{r=1,\dots,N}$ with $j=2,\dots k$, where we decrease by $j$ the dimension of every CB operator. The only exception is in the set $\Delta_1$ when the gauge group on the $-1$ curve is empty, in which case the operator of dimension $\frac{6N}{\ell}$ is replaced by $\frac{6N-j}{\ell}$, or when it is $USp(4m)$, in which case the operator of dimension $\frac{6N+2m}{\ell}$ is replaced by $\frac{6N+2m-j}{\ell}$. 
\item We have the mass deformation sequence:\\ 
$\{\Delta_r\}_{r=1,\dots,N}\rightarrow \{\Delta_r-2\}_{r=1,\dots,N}\rightarrow \{\Delta_r-3\}_{r=1,\dots,N}\rightarrow\dots$
\end{itemize}  
When we apply this algorithm to the well-understood case of rank-2 theories we find a perfect match with the known mass deformation trees \cite{Martone:2021drm} and we recover in this way almost all SCFT's with a unitary magnetic quiver listed in \cite{Bourget:2021csg}. We regard this as a highly non-trivial consistency check of our proposal. 

Finally, we would like to comment on the range of validity of the algorithm we have just illustrated. Clearly, for $k$ large enough we end up finding operators of dimension $2$ or $1$. In general, as long as the spectrum does not include operators of dimension $2$ or $1$, we expect the 4d theory to be superconformal and isolated, while the presence of operators of dimension $2$ implies that the theory can be written as a vector multiplet coupled to a matter sector (possibly strongly coupled). In this case the vanishing of the beta function is not implied just by the arguments we have provided, and we need to analyze the theory further to conclude that it is indeed conformal. An example of non conformality is given by the theory with $k=2$ and $n_{4'}=1$. For $N=1$ the theory has rank 2, with CB operators of dimension 6 and 8. Upon activating the mass deformation we have described above we end up with a theory with scaling dimensions 2 and 4. This theory is not conformal and actually it includes an infrared free $SU(2)$ sector. We will leave to future work the task of providing a precise conformality criterion when the spectrum includes operators of dimension 2.

\subsubsection{Central charges} 

In this section we continue the description of the theories on the $D3$-brane probes by deriving their $a$, $c$, and (non-Abelian) flavor-symmetry central charges. As is well known, the central charges are related to the following 't Hooft anomalies \cite{Anselmi:1997am, Anselmi:1997ys}:
\be \Tr\, rR^2=2(2a-c)\,;\qquad \Tr\, r^3=\Tr \,r=48(a-c)\,;\qquad k_{G}\delta^{ab}= -2\,\Tr\, r \,T_{G}^aT_{G}^b\,,\ee
where $r$ denotes the $U(1)_R$ generator of the superconformal algebra, $R$ the Cartan generator of the $SU(2)_R$ symmetry of the superconformal algebra, and $\{T^a_{G}\}_a$ indicate the generators of a simple factor $G$ of the global symmetry of the theory. Recall that the choice of holonomy for the gauge field supported on the 7-brane will break $G\to H$, and we are led to compute $k_H$ for each simple factor of the unbroken flavor symmetry.

Fortunately, using the near-horizon geometry of $D3$-branes probing an F-theory background, one can compute holographically all the central charges \cite{Aharony:2007dj,Apruzzi:2020pmv,Giacomelli:2020jel,Giacomelli:2020gee} (see \cite{Giacomelli:2022drw} for the extension to the case of 7-branes wrapped on Abelian orbifolds). We argue that, for the F-theory backgrounds considered in this paper, which combine 7-branes wrapped on $\mathbb{C}^2/\mathbb{Z}_k$ orbifolds with the presence of $\mathbb{C}^2/\mathbb{Z}_{\ell}$ S-folds, the holographic formulas for the central charges generalize to
\begin{eqnarray}\label{formulacc1} 
4(2a-c) & =& (M\Delta_7)N^2+(2M\Delta_7\,\epsilon+\Delta_7-1)N+\alpha N^0\,, \label{formulacc1}\\
24(c-a) & = & 6N(\Delta_7-1)+\beta N^0\,,\label{formulacc2}\\
k_H&=&2\Delta_7I_{H\hookrightarrow G}N+\gamma N^0\,, \label{formulacc3}
\end{eqnarray}
where $\Delta_7$ denotes the type of 7-brane,\footnote{This quantity is related to the deficit angle of the conical singularity created by the 7-brane in Type IIB string theory. It can take the following values: $\Delta_7=\tfrac65,\tfrac43,\tfrac32,2,3,4,6$, corresponding to the 7-branes of type $\mathcal{H}_0,\mathcal{H}_1,\mathcal{H}_2,D_4,E_6,E_7,E_8$ respectively. $\Delta_7=1$ corresponds to no 7-brane.} $N$ is the number of $D3$-brane probes, and $I_{H\hookrightarrow G}$ is the embedding index of $H$ in $G$. The quantity $\epsilon$ is the $D3$ charge of the background, which receives both a topological contribution proportional to the Euler character of the background and a contribution which depends on the choice of holonomy at infinity. Finally, $\alpha,\beta,\gamma$ denote the contribution at order $N^0$ to the central charges, and they are typically the hardest quantities to compute. However, as observed in \cite{Giacomelli:2022drw}, these $O(1)$ coefficients always vanish for the \emph{canonical} families, i.e.~the holonomy choice for which all the nodes of the magnetic quiver (except for the added Abelian node) are multiples of $N$.

By computing the CB spectrum with the prescription given in Section \ref{CBspectrumIIB}, and by using the Shapere-Tachikawa relation \eqref{STrelation}, we can immediately verify the $O(N^2)$ coefficient of \eqref{formulacc1} as well as compute $\epsilon$ and $\alpha$ for all the Type IIB orbi-S-fold theories with maximal 7-brane. Moreover, we observe that in every Hasse-diagram transition lowering the number of probes by one unit (see Section \ref{HasseMtheory}\footnote{As we will see in the next subsection, at large enough $N$ the Hasse diagrams for the Type IIB models coincide with the ones of their M-theory progenitors.}), the $\epsilon$ decreases by the quantity $1/M$. This nicely generalizes the analogous rule spotted for the pure S-fold \cite{Giacomelli:2020gee} and the pure orbifold \cite{Giacomelli:2022drw} cases, where the $\epsilon$ decreases by $1/\ell$ and $1/k$ respectively. It would be very nice to have a geometric derivation of this fact.

As for Eq.~\eqref{formulacc2}, we can easily verify its $O(N)$ coefficient as well as compute $\beta$ for all the fully-higgsable models: Indeed for them the quantity $24(c-a)$ is always equal to the quaternionic dimension of the Higgs branch $\text{dim}({\mathbb H})$, which can immediately be read off from the magnetic quiver (the sum of the ranks of all the nodes except the added Abelian node). We can check which models are fully higgsable by studying the corresponding Hasse diagram, which is the subject of the next subsection. We find that all the orbi-S-fold theories at S-fold order $\ell=2$ are fully higgsable as long as we turn on a mass deformation for a $SU(2)$ or $SU(3)$ symmetry, while orbi-S-fold theories at S-fold order $\ell=3$ are fully higgsable as long as we turn on a $SU(2)$ mass deformation. In all these cases the above applies. For the other values of $\ell$ and more general mass deformations one needs to check the structure of the Hasse diagram explicitly: In general $24(c-a)$ and $\text{dim}({\mathbb H})$ of a given theory differ by an $O(1)$ number. Hence, we can still verify the $O(N)$ coefficient of Eq.~\eqref{formulacc2} by looking at the magnetic quiver, whereas we can compute $\beta$ from the central charges of known rank-2 or rank-1 theories to which our models reduce for small enough $N$ via higgsing.

While we refer the reader to App.~\ref{App:IIBth} for the values of the central charges of all the orbi-S-fold theories with maximal 7-brane that we presented in this paper, let us discuss here explicitly the case $\ell=4,k=2$, which in fact features a mismatch between the quantity $24(c-a)$ and the Higgs-branch dimension. There are three families of SCFTs connected by higgsings as in \eqref{HasseM8l4} for large $N$. One of them has the magnetic quiver \eqref{quivL=4}, which gives $\text{dim}({\mathbb H}) = 3N-4$. The $O(N)$ coefficient indeed agrees with \eqref{formulacc2}, because here $\Delta_7=3/2$. The $N=2$ representative of this family happens to be the theory 55 of \cite{Martone:2021ixp,Martone:2021drm,Bourget:2021csg}, which has $\text{dim}({\mathbb H}) = 2$, but $24(c-a)=\text{dim}({\mathbb H})-1=1$. This fixes to $1$ the mismatch throughout the Hasse diagram. For instance, the family that immediately precedes \eqref{quivL=4} in the Hasse diagram will have $24(c-a)=3N-4$ but $\text{dim}({\mathbb H}) = 3N-3$. For $N=2$ this is a rank-3 SCFT with CB-operator dimensions $(\tfrac52,4,6)$ and magnetic quiver
\be
 \begin{tikzpicture}
 \filldraw[fill= white] (0,0) circle [radius=0.1] node[below] {\scriptsize 1};
\filldraw[fill= white] (1,0) circle [radius=0.1] node[below] {\scriptsize 2};
\filldraw[fill= white] (2,0) circle [radius=0.1] node[below] {\scriptsize 1}; 

\draw [thick] (0.05,0.07)--(0.95,0.07);
\draw [thick] (0.07,0.02)--(0.93,0.02);
\draw [thick] (0.07,-0.02)--(0.93,-0.02);
\draw [thick] (0.05,-0.07)--(0.95,-0.07);
\draw [thick] (1.05,0.07)--(1.95,0.07);
\draw [thick] (1.07,0.02)--(1.93,0.02);
\draw [thick] (1.07,-0.02)--(1.93,-0.02);
\draw [thick] (1.05,-0.07)--(1.95,-0.07);
\draw [thick] (0.4,0) -- (0.6,-0.2);
\draw [thick] (0.4,0) -- (0.6,0.2);
\end{tikzpicture} 
\ee
whose existence is a prediction of our construction.

Let us conclude this subsection by analyzing the flavor central charges.
Contrary to their M-theory progenitors of Section \ref{Sec:SWcompact}, 
the Type IIB orbi-S-fold theories do not have any extra non-Abelian flavor symmetry beyond the one carried by the 7-brane. The flavor central charges for all the M-theory models discussed in this paper are listed in App.~\ref{App:Mth}, and have been computed using Eqs.~\eqref{fccM1},\eqref{fccM2}. Given the holographic formula \eqref{formulacc3}, the deformation of the $T(SU(k))$ tail only affects the $O(1)$ coefficient $\gamma$, whereas the $O(N)$ coefficient remains unchanged in going from M-theory to Type IIB.

The values of the flavor central charges for all the Type IIB models we presented in this paper are listed in App.~\ref{App:IIBth}. Here we concentrate on how the number $\gamma$ in \eqref{formulacc3} is modified upon removing the tail. Again, by looking at known rank-2 or rank-1 theories to which our models reduce for small enough $N$, we can easily predict the change in $\gamma$ for all the orbi-S-fold theories we analyzed.\footnote{Recall that for $\ell=5,6$ there is no flavor symmetry associated to 7-branes.} We find that
\be
\gamma \;\stackrel{\rm def}{\longrightarrow}\; \gamma - \Delta\gamma\,,
\ee
with $\Delta\gamma$ given by
\be
\begin{array}{|c|c|}
 \hline 
\Delta\gamma & {\rm orbi}-{\rm S}-{\rm fold} \\ 
\hline
2I_{H\hookrightarrow G} & \ell=2,k=2 \\ 
\hline
3I_{H\hookrightarrow G} & \ell=2,k=3 \\ 
\hline
\frac{4}{3}I_{H\hookrightarrow G} & \ell=3,k=2 \\ 
\hline
I_{H\hookrightarrow G} & \ell=4,k=2 \\ 
\hline
\end{array}
\ee
The above result suggests that in general $\Delta\gamma=\frac{2k}{\ell}I_{H\hookrightarrow G}$ and it would be interesting to find a first-principle derivation of this result.

\subsubsection{Hasse Diagrams}\label{HasseIIB}

As we have seen in Section \ref{Sec:SWcompact}, we can interpolate between the various choices of holonomy via higgsing, and the knowledge of the magnetic quivers allows us to easily derive the corresponding Hasse diagram using quiver subtraction. Since all the quivers have the same tail attached to the Dynkin diagram, the result of the quiver subtraction only depends on the rank of the nodes in the Dynkin diagram. The tail simply disappears in the process. Since the mass deformations for the tail we have proposed in \eqref{defsell=2}, \eqref{massdefL3}, and \eqref{defL=4} modify the quivers always in the same way, regardless of the specific holonomy chosen, and crucially it does not affect the balancing of the nodes in the Dynkin diagram, the subtraction of two deformed quivers will lead to the same answer we get for the undeformed ones. We therefore find the same Hasse diagram with exactly the same transverse slices as in Section \ref{Sec:SWcompact}, and we can therefore simply refer the reader to Section \ref{sec:4dprop}. 

There is however an exception to this rule due to the fact that the statement of the previous paragraph holds true provided the number of probes $N$ is sufficiently large. This can be understood for example by looking at the last quiver in \eqref{defsell=2}: If one of the nodes $A$, $B$, $C$ or $D$ has rank less than 4, the quiver obtained upon mass deformation will formally have nodes with negative rank. This can indeed happen for $N$ low enough and reflects the fact that for those values of $N$ the mass deformation we are considering cannot actually be turned on. 

In order to explain how to deal with this problem and how it affects the Hasse diagram, let us go back to the parent 6d SCFT. For $N$ large enough, it is always the case that the last few $-2$ curves in the F-theory base host a $SU(M)$ gauge group, where $M$ is the order of the orbifold in M-theory. This ``plateau'' with gauge groups of constant rank in general follows a ramp with gauge groups of increasing rank. Let us call $\bar{N}$ the minimum value of $N$ such that the quiver on the 6d tensor branch includes a gauge group with $M$ colors; it can be either $SU(M)$ or $USp(M)$ (but the latter appears only in some cases with $\ell=1,2$ and $\bar{N}=1$). We have the following cases: 
\begin{itemize} 
\item {\bf For ${\bf N\geq\bar{N}}$} the Hasse diagram is as described above and the global symmetry is the one preserved by the holonomy (plus the symmetry from the tail if any). In general for $N=\bar{N}$ the global symmetry enhances. 
\item {\bf For ${\bf N=\bar{N}-1}$} some nodes in the quiver become ugly, meaning that the theory includes a free sector. This is a manifestation of the fact that the interacting part of the theory also has a realization with lower $k$. The actual value of $k$ is given by the number of colors of the last gauge node in the ramp divided by $\ell$. We can still turn on a mass for the $SU(k)$ symmetry although the $T(SU(k))$ tail and the Dynkin diagram in this case overlap inside the magnetic quiver which describes the interacting sector.
\item {\bf For ${\bf N<\bar{N}-1}$} some nodes in the quiver are bad, reflecting the fact that we need to describe the theory using the realization with lower $k$. In this case the tail is smaller and we cannot turn on all the mass deformations we have for higher values of $N$.
\end{itemize} 
As long as we are dealing with the undeformed theories studied in Section \ref{Sec:SWcompact}, the Hasse diagram is not affected by what we have just said and it is valid for every $N$, the only caveat being that for small $N$ some of the entries in the Hasse diagram need to be properly reinterpreted as theories with lower $k$. The story is however different for the mass-deformed theories.
  
As a result of the pattern we have just described, when $N\geq \text{Max}_i \{\bar{N}_i\}$ (where the maximum is taken over all possible choices of holonomy) the higgsing and mass deformations for the tail commute, and the Hasse diagram for the deformed theories is just the same as that for the undeformed ones detailed in Section \ref{sec:4dprop}. On the other hand, as we decrease $N$ below the value $\text{Max}_i \{\bar{N}_i\}$, some of the theories cannot be mass-deformed any more and should not appear in the Hasse diagram. Our proposal is that the Hasse diagram for the deformed theories is obtained simply by removing from the Hasse diagram for the undeformed models all the entries such that $\bar{N}-1>N$. With this prescription we can simply derive the desired result from the Hasse diagram of Section \ref{sec:4dprop}. 

\begin{figure}
\begin{tikzpicture}
\node[]  (a1) at (0,0) {\scriptsize $\mathring{\mathcal S}^{(3)}_{E_6,2} (0,0,0,1,1)$}; 
\node[]  (a2) at (-1.5,-1) {\scriptsize $\mathring{\mathcal S}^{(3)}_{E_6,2} (1,0,0,1,0)$}; 
\node[]  (a3) at (1.5,-1) {\scriptsize $\mathring{\mathcal T}^{(3)}_{E_6,2} (0,0,0,0,3)$}; 
\node[]  (a4) at (0,-2) {\scriptsize $\mathring{\mathcal T}^{(3)}_{E_6,2} (1,0,0,0,2)$}; 
\node[]  (a5) at (0,-3) {\scriptsize $\mathring{\mathcal T}^{(3)}_{E_6,2} (0,1,0,0,1)$}; 
\node[]  (a6) at (1.5,-4) {\scriptsize $\mathring{\mathcal T}^{(3)}_{E_6,2} (0,0,1,0,0)$}; 
\node[]  (a7) at (-1.5,-4) {\scriptsize $\mathring{\mathcal T}^{(3)}_{E_6,2} (2,0,0,0,1)$}; 
\node[]  (a8) at (0,-5) {\scriptsize $\mathring{\mathcal T}^{(3)}_{E_6,2} (1,1,0,0,0)$}; 
\node[]  (a9) at (0,-6) {{\color{blue}\scriptsize $\mathring{\mathcal T}^{(3)}_{E_6,2} (3,0,0,0,0)$}}; 
\draw[->, thick] (a1)--(a2); 
\draw[->, thick] (a1)--(a3); 
\draw[->, thick] (a2)--(a4); 
\draw[->, thick] (a3)--(a4); 
\draw[->, thick] (a4)--(a5); 
\draw[->, thick] (a5)--(a6); 
\draw[->, thick] (a5)--(a7); 
\draw[->, thick] (a7)--(a8);
\draw[->, thick] (a6)--(a8);  
\draw[->, thick] (a8)--(a9); 
\node[]  (b1) at (5,-4.5) {\scriptsize $\mathring{\mathcal S}^{(2)}_{E_6,2} (0,0,0,1,1)$}; 
\node[]  (b2) at (-1.5+5,-5.5) {\scriptsize $\mathring{\mathcal S}^{(2)}_{E_6,2} (1,0,0,1,0)$}; 
\node[]  (b3) at (1.5+5,-5.5) {\scriptsize $\mathring{\mathcal T}^{(2)}_{E_6,2} (0,0,0,0,3)$}; 
\node[]  (b4) at (5,-6.5) {\scriptsize $\mathring{\mathcal T}^{(2)}_{E_6,2} (1,0,0,0,2)$}; 
\node[]  (b5) at (5,-7.5) {\scriptsize $\mathring{\mathcal T}^{(2)}_{E_6,2} (0,1,0,0,1)$}; 
\node[]  (b6) at (1.5+5,-8.5) {\scriptsize $\mathring{\mathcal T}^{(2)}_{E_6,2} (0,0,1,0,0)$}; 
\node[]  (b7) at (-1.5+5,-8.5) {{\color{blue}\scriptsize $\mathring{\mathcal T}^{(2)}_{E_6,2} (2,0,0,0,1)$}}; 
\node[]  (b8) at (5,-9.5) {{\color{blue}\scriptsize $\mathring{\mathcal T}^{(2)}_{E_6,2} (1,1,0,0,0)$}}; 
\node[]  (b9) at (5,-10.5) {{\color{red}\scriptsize $\mathring{\mathcal T}^{(2)}_{E_6,2} (3,0,0,0,0)$}}; 
\draw[->, thick] (b1)--(b2); 
\draw[->, thick] (b1)--(b3); 
\draw[->, thick] (b2)--(b4); 
\draw[->, thick] (b3)--(b4); 
\draw[->, thick] (b4)--(b5); 
\draw[->, thick] (b5)--(b6); 
\draw[->, thick] (b5)--(b7); 
\draw[->, thick] (b7)--(b8);
\draw[->, thick] (b6)--(b8);  
\draw[->, thick] (b8)--(b9); 
\draw[->, thick] (a6)--(b1); 
\draw[->, thick] (a8)--(b2); 
\draw[->, thick] (a9)--(b7); 
\node[]  (c1) at (10,-9) {\scriptsize $\mathring{\mathcal S}^{(1)}_{E_6,2} (0,0,0,1,1)$}; 
\node[]  (c2) at (-1.5+10,-10) {{\color{blue}\scriptsize $\mathring{\mathcal S}^{(1)}_{E_6,2} (1,0,0,1,0)$}}; 
\node[]  (c3) at (1.5+10,-10) {\scriptsize $\mathring{\mathcal T}^{(1)}_{E_6,2} (0,0,0,0,3)$}; 
\node[]  (c4) at (10,-11) {{\color{blue}\scriptsize $\mathring{\mathcal T}^{(1)}_{E_6,2} (1,0,0,0,2)$}}; 
\node[]  (c5) at (10,-12) {{\color{blue}\scriptsize $\mathring{\mathcal T}^{(1)}_{E_6,2} (0,1,0,0,1)$}}; 
\node[]  (c6) at (1.5+10,-13) {{\color{blue}\scriptsize $\mathring{\mathcal T}^{(1)}_{E_6,2} (0,0,1,0,0)$}}; 
\node[]  (c7) at (-1.5+10,-13) {{\color{red} \scriptsize $\mathring{\mathcal T}^{(1)}_{E_6,2} (2,0,0,0,1)$}}; 
\node[]  (c8) at (10,-14) {{\color{red}\scriptsize $\mathring{\mathcal T}^{(1)}_{E_6,2} (1,1,0,0,0)$}}; 
\node[]  (c9) at (10,-15) {{\color{red}\scriptsize $\mathring{\mathcal T}^{(1)}_{E_6,2} (3,0,0,0,0)$}}; 
\draw[->, thick] (c1)--(c2); 
\draw[->, thick] (c1)--(c3); 
\draw[->, thick] (c2)--(c4); 
\draw[->, thick] (c3)--(c4); 
\draw[->, thick] (c4)--(c5); 
\draw[->, thick] (c5)--(c6); 
\draw[->, thick] (c5)--(c7); 
\draw[->, thick] (c7)--(c8);
\draw[->, thick] (c6)--(c8);  
\draw[->, thick] (c8)--(c9); 
\draw[->, thick] (b6)--(c1); 
\draw[->, thick] (b8)--(c2); 
\draw[->, thick] (b9)--(c7); 
\end{tikzpicture} 
\caption{Hasse diagram for $M=6$, $\ell=2$ and $N\leq3$. We color in blue theories such that $N=\bar{N}-1$. These describe a collection of free hypermultiplets plus an interacting sector which has another realization for $M<6$. We color in red theories for which $N<\bar{N}-1$. The Hasse diagram for the mass-deformed theories is obtained by removing the red entries.}\label{lowNHasse}
\end{figure}
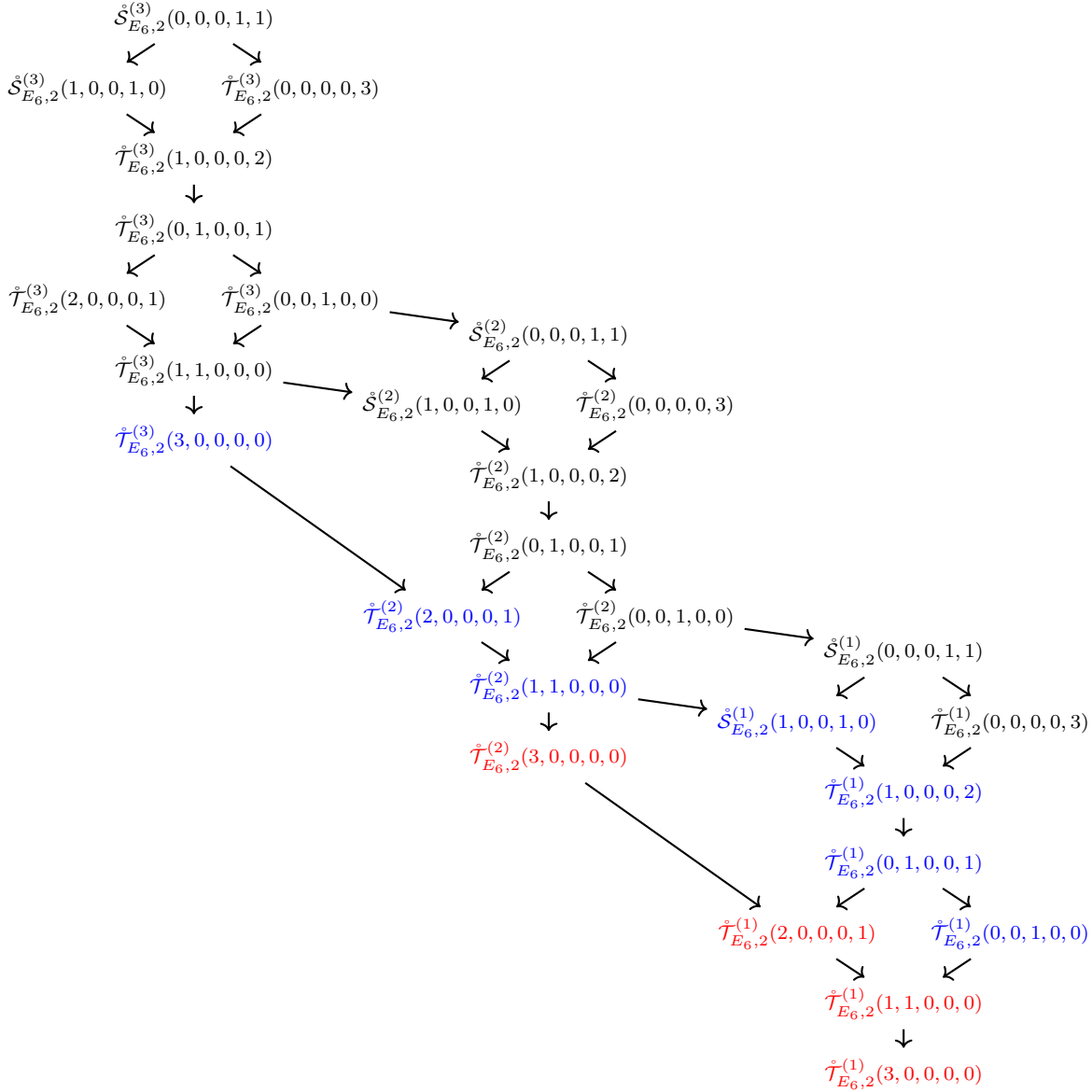

Let us illustrate the above discussion in the case $M=6$ and $\ell=2$ in which we have a $T(SU(3))$ tail. As we have seen, there are nine choices of holonomy corresponding to nine infinite sequences of theories. From the tensor branch description of the parent 6d theories we can easily determine the value of $\bar{N}$ for each one of them: 
\be\label{ttable}\begin{array}{|c|c|}
 \hline 
\text{Theory $S^{(N)}_{E_6,2}(a,b,c,d,e)$} & \text{Value of $\bar{N}$}\\ 
\hline
\mathcal S^{(N)}_{E_6,2} (0,0,0,1,1) & 1\\
\hline
\mathcal T^{(N)}_{E_6,2} (0,0,0,0,3) & 1\\
\hline
\mathcal S^{(N)}_{E_6,2} (1,0,0,1,0) & 2\\
\hline
\mathcal T^{(N)}_{E_6,2} (1,0,0,0,2) & 2\\
\hline
\mathcal T^{(N)}_{E_6,2} (0,1,0,0,1) & 2\\
\hline
\mathcal T^{(N)}_{E_6,2} (0,0,1,0,0) & 2\\
\hline
\mathcal T^{(N)}_{E_6,2} (2,0,0,0,1) & 3\\
\hline
\mathcal T^{(N)}_{E_6,2} (1,1,0,0,0) & 3\\
\hline
\mathcal T^{(N)}_{E_6,2} (3,0,0,0,0) & 4\\
\hline
\end{array}\ee 
As we can see from \eqref{ttable}, for $N\geq4$ we are not below any of the $\bar{N}$ values and therefore the Hasse diagram is the same for the undeformed and the $SU(3)$-deformed theories. The discussion is the same as in Section \ref{sec:4dprop}, and therefore we do not need to add anything. Let us focus instead on the case $N\leq3$. We report explicitly the Hasse diagram in Figure \ref{lowNHasse} for convenience. 

Let us start by discussing the interpretation of the various theories at low $N$. The most intricate case is that of theory $\mathcal T^{(N)}_{E_6,2} (3,0,0,0,0)$: 
\be\label{lowN1}  \mathcal T^{(3)}_{E_6,2} (3,0,0,0,0)\rightarrow \mathcal T^{(3)}_{E_6,2} (2,0,0,0,0)\;; \mathcal T^{(2)}_{E_6,2} (3,0,0,0,0)\rightarrow  \mathcal T^{(2)}_{E_6,2}\;; \mathcal T^{(1)}_{E_6,2} (3,0,0,0,0)\rightarrow \mathcal T^{(1)}_{E_6,2}\;\,\ee 
The theory for $N=3$ is equivalent to one of the models we found for $M=4$, for $N=2$ it reduces to the rank-2 $\mathcal{T}$ theory, and it becomes the $E_6$ Minahan-Nemeschansky theory for $N=1$. In all cases these interacting theories are accompanied by a collection of free hypermultiplets. 

\noindent We then have the models $\mathcal T^{(N)}_{E_6,2} (2,0,0,0,1)$ and $\mathcal T^{(N)}_{E_6,2} (1,1,0,0,0)$. For $N=2$ they both reduce to models which have a $M=4$ realization: 
\be\label{lowN2} \mathcal T^{(2)}_{E_6,2} (2,0,0,0,1)\rightarrow \mathcal T^{(2)}_{E_6,2} (1,0,0,0,1)\quad \mathcal T^{(2)}_{E_6,2} (1,1,0,0,0)\rightarrow \mathcal T^{(2)}_{E_6,2} (0,1,0,0,0)\,,\ee 
while for $N=1$ they reduce to rank-1 theories: 
 \be\label{lowN3} \mathcal T^{(1)}_{E_6,2} (2,0,0,0,1)\rightarrow \mathcal S^{(1)}_{E_6,2}\quad \mathcal T^{(1)}_{E_6,2} (1,1,0,0,0)\rightarrow \mathcal T^{(1)}_{E_6,2}\,.\ee 
Finally,  $\mathcal S^{(N)}_{E_6,2} (1,0,0,1,0)$ and  $\mathcal T^{(N)}_{E_6,2} (1,0,0,0,2)$ reduce to $M=4$ theories for $N=1$: 
\be\label{lowN4} \mathcal S^{(1)}_{E_6,2} (1,0,0,1,0)\rightarrow \mathcal S^{(1)}_{E_6,2} (0,0,0,1,0)\quad \mathcal T^{(1)}_{E_6,2} (1,0,0,0,2)\rightarrow \mathcal T^{(1)}_{E_6,2} (0,0,0,0,2)\,,\ee 
while $\mathcal T^{(N)}_{E_6,2} (0,1,0,0,1)$ and $\mathcal T^{(N)}_{E_6,2} (0,0,1,0,0)$ reduce to the same rank-1 model for $N=1$: 
 \be\label{lowN5} \mathcal T^{(1)}_{E_6,2} (0,1,0,0,1)\rightarrow \mathcal S^{(1)}_{E_6,2}\quad \mathcal T^{(1)}_{E_6,2} (0,0,1,0,0)\rightarrow \mathcal S^{(1)}_{E_6,2}\,.\ee 
Up to the replacements \eqref{lowN1}-\eqref{lowN5} the Hasse diagram in Figure \ref{lowNHasse} is perfectly valid for the undeformed theories without any modifications. There is one caveat though, namely that the transverse slices can be modified for low values of $N$. An example of this is provided by the theories in equation \eqref{lowN5}: For $N>1$ the higgsing between them has transverse slice $\mathbb{C}^2/\bbZ_2$ whereas for $N=1$ it is simply $\mathbb{C}^2$. 
This reflects the fact that for $N=1$ the interacting sector is the same and the two theories only differ in the number of free hypermultiplets accompanying it. The higgsing between them therefore trivializes in the sense that it just corresponds to activating a vev for the free scalars. 

The story becomes more interesting when we consider the mass-deformed SCFT's obtained by activating a mass deformation for the $SU(3)$ global symmetry carried by the $T(SU(3))$ tail for $N$ generic. As we have said, our proposal for the corresponding Hasse diagram is simply to remove the entries colored in red  in Figure \ref{lowNHasse}. The fact that we should remove the entry $\mathring{\mathcal T}^{(2)}_{E_6,2} (3,0,0,0,0)$ can be easily understood: From \eqref{lowN1} we know that this theory should be identified with the mass deformation of $\mathcal T^{(2)}_{E_6,2}$, whose global symmetry is known to be $F_4\times SO(4)$. The $F_4$ factor comes from the 7-brane and has nothing to do with the tail. The remaining $SO(4)$ simply does not contain a $SU(3)$ subgroup and therefore we cannot turn on a mass deformation for it. This is also reflected by the corresponding magnetic quiver which does not contain a $T(SU(3))$ tail (see \cite{Bourget:2020mez}). Furthermore, the mass-deformed version of the theory should be a descendant of the rank-2 $F_4\times U(1)$ theory (called $\widehat{\mathcal T}_{E_6,2}$ in \cite{Giacomelli:2020gee}) and no such SCFT is known to exist. This gives us even more confidence in claiming that the corresponding entry in the Hasse diagram should be removed. 

As a further check, let us consider the last portion of the Hasse diagram starting from the  $\mathcal T^{(2)}_{E_6,2} (1,1,0,0,0)$ theory. After the mass deformation we find a rank-3 SCFT whose Coulomb-branch operators have dimension $(3,9/2,5)$ and whose Higgs branch is described by the magnetic quiver 
$$
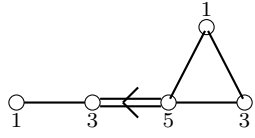
\begin{tikzpicture}
\filldraw[fill= white] (2,-4) circle [radius=0.1] node[below] {\scriptsize 1};
\filldraw[fill= white] (3,-4) circle [radius=0.1] node[below] {\scriptsize 3};
\filldraw[fill= white] (4,-4) circle [radius=0.1] node[below] {\scriptsize 5};
\filldraw[fill= white] (5,-4) circle [radius=0.1] node[below] {\scriptsize 3};
\filldraw[fill= white] (4.5,-3) circle [radius=0.1] node[above] {\scriptsize 1};
\draw [thick] (3.1, -3.95) -- (3.9,-3.95) ;
\draw [thick] (3.1, -4.05) -- (3.9,-4.05) ;
\draw [thick] (2.1, -4) -- (2.9,-4) ;
\draw [thick] (4.1, -4) -- (4.9,-4) ;
\draw [thick] (3.4,-4) -- (3.6,-3.8);
\draw [thick] (3.4,-4) -- (3.6,-4.2);
\draw [thick] (4.02, -3.95) -- (4.48,-3.05);
\draw [thick] (4.98, -3.95) -- (4.52,-3.05) ;
\end{tikzpicture}$$ 
The existence of this theory is a prediction of our construction. Following the Hasse diagram we find the mass deformation of $\mathcal S^{(1)}_{E_6,2} (1,0,0,1,0)$ which is a known rank-2 theory. From Figure \ref{lowNHasse} and the rule \eqref{defsell=2} we predict the following Hasse diagram for the mass-deformed theory: 
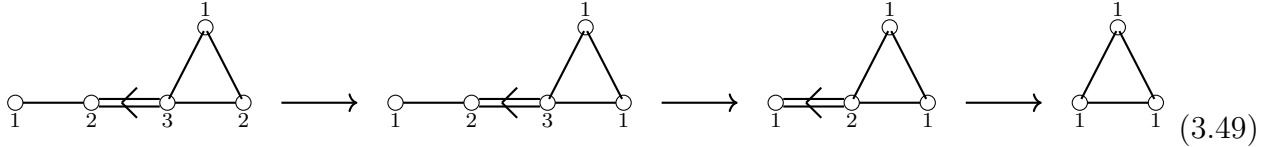
\be\label{Hassefin} \begin{tikzpicture} 
\filldraw[fill= white] (2,-4) circle [radius=0.1] node[below] {\scriptsize 1};
\filldraw[fill= white] (3,-4) circle [radius=0.1] node[below] {\scriptsize 2};
\filldraw[fill= white] (4,-4) circle [radius=0.1] node[below] {\scriptsize 3};
\filldraw[fill= white] (5,-4) circle [radius=0.1] node[below] {\scriptsize 2};
\filldraw[fill= white] (4.5,-3) circle [radius=0.1] node[above] {\scriptsize 1};
\draw [thick] (3.1, -3.95) -- (3.9,-3.95) ;
\draw [thick] (3.1, -4.05) -- (3.9,-4.05) ;
\draw [thick] (2.1, -4) -- (2.9,-4) ;
\draw [thick] (4.1, -4) -- (4.9,-4) ;
\draw [thick] (3.4,-4) -- (3.6,-3.8);
\draw [thick] (3.4,-4) -- (3.6,-4.2);
\draw [thick] (4.02, -3.95) -- (4.48,-3.05);
\draw [thick] (4.98, -3.95) -- (4.52,-3.05) ; 

\draw [->, thick] (5.5,-4)--(6.5,-4); 

\filldraw[fill= white] (7,-4) circle [radius=0.1] node[below] {\scriptsize 1};
\filldraw[fill= white] (8,-4) circle [radius=0.1] node[below] {\scriptsize 2};
\filldraw[fill= white] (9,-4) circle [radius=0.1] node[below] {\scriptsize 3};
\filldraw[fill= white] (10,-4) circle [radius=0.1] node[below] {\scriptsize 1};
\filldraw[fill= white] (9.5,-3) circle [radius=0.1] node[above] {\scriptsize 1};
\draw [thick] (8.1, -3.95) -- (8.9,-3.95) ;
\draw [thick] (8.1, -4.05) -- (8.9,-4.05) ;
\draw [thick] (7.1, -4) -- (7.9,-4) ;
\draw [thick] (9.1, -4) -- (9.9,-4) ;
\draw [thick] (8.4,-4) -- (8.6,-3.8);
\draw [thick] (8.4,-4) -- (8.6,-4.2);
\draw [thick] (9.02, -3.95) -- (9.48,-3.05);
\draw [thick] (9.98, -3.95) -- (9.52,-3.05) ; 

\draw [->, thick] (10.5,-4)--(11.5,-4); 

\filldraw[fill= white] (12,-4) circle [radius=0.1] node[below] {\scriptsize 1};
\filldraw[fill= white] (13,-4) circle [radius=0.1] node[below] {\scriptsize 2};
\filldraw[fill= white] (14,-4) circle [radius=0.1] node[below] {\scriptsize 1};
\filldraw[fill= white] (13.5,-3) circle [radius=0.1] node[above] {\scriptsize 1};
\draw [thick] (12.1, -3.95) -- (12.9,-3.95) ;
\draw [thick] (12.1, -4.05) -- (12.9,-4.05) ;
\draw [thick] (13.1, -4) -- (13.9,-4) ;
\draw [thick] (12.4,-4) -- (12.6,-3.8);
\draw [thick] (12.4,-4) -- (12.6,-4.2);
\draw [thick] (13.02, -3.95) -- (13.48,-3.05);
\draw [thick] (13.98, -3.95) -- (13.52,-3.05) ; 

\draw [->, thick] (14.5,-4)--(15.5,-4); 

\filldraw[fill= white] (16,-4) circle [radius=0.1] node[below] {\scriptsize 1};
\filldraw[fill= white] (17,-4) circle [radius=0.1] node[below] {\scriptsize 1};
\filldraw[fill= white] (16.5,-3) circle [radius=0.1] node[above] {\scriptsize 1};
\draw [thick] (16.1, -4) -- (16.9,-4) ;
\draw [thick] (16.02, -3.95) -- (16.48,-3.05);
\draw [thick] (16.98, -3.95) -- (16.52,-3.05) ; 
\end{tikzpicture}\ee 
where we have drawn the corresponding magnetic quivers. In \eqref{Hassefin} we recognize the magnetic quivers of the rank-1 theories $\mathcal S^{(1)}_{\mathcal{H}_2,2}$ and $\mathcal T^{(1)}_{\mathcal{H}_2,2}$, leading to the higgsing sequence 
$$\begin{tikzpicture} 
\node[]  (a) at (0,0) { $SU(3)\times SU(2)\times U(1)$}; 
\node[]  (b) at (5,0) { $Sp(3)\times U(1)$}; 
\node[]  (c) at (9,0) { $Sp(2)\times U(1)$}; 
\node[]  (d) at (12,0) { $SU(3)$};  

\draw[ thick, color=blue] (a)--(b); 
\draw[ thick, color=blue] (b)--(c); 
\draw[ thick, color=blue] (c)--(d); 
\end{tikzpicture}$$ 
This is indeed the known Hasse diagram of theory 36 of \cite{Martone:2021ixp}, confirming again that the red entries in Figure \ref{lowNHasse} should be dropped.

As a final remark, we would like to point out that the $SU(k)$ mass deformation for the theories with $N=\bar{N}-1$ (those colored in blue in Figure \ref{lowNHasse}) is a bit special: although the deformation is possible, the node $U(k)$ in this case is part of the graph of the Dynkin diagram and therefore the deformation cannot really be regarded as a removal of the tail. In particular the resulting theories cannot be higgsed to those corresponding to the flat 7-brane, as we would always expect for a Type IIB setup of the kind we are considering. Nevertheless, the mass-deformed SCFT's exist and our rules for determining the CB spectrum and magnetic quiver still work. We would also like to notice that when $N=\bar{N}=1$ the symmetry associated with the tail combines with the one coming from the 7-brane into a larger group. In this case the deformation of the tail and the deformation of the 7-brane (which we will consider in Section \ref{Sec:DecFlavors}) become indistinguishable: They give rise to the same mass deformation sequence while they differ for larger values of $N$. This is also confirmed at the level of the CB spectrum and at the level of the magnetic quiver.

Let us conclude by briefly discussing the Hasse diagram for the $\ell=3,4$ theories with $k=2$. In both cases it suffices to discuss the case $N\leq2$. 

For $\ell=3$ we only need to remove the theory $\mathring{\mathcal T}^{(1)}_{D_4,3} (2,0,0)$ according to our rule, as displayed in \eqref{ell3Nsmall}. The theory $\mathcal T^{(2)}_{D_4,3} (2,0,0)$ can be identified, modulo free hypers, with $\mathcal T^{(2)}_{D_4,3}$. Its global symmetry is known to be $G_2\times SU(2)$ and after the $SU(2)$ mass deformation we expect to find a model with $G_2$ global symmetry only. Indeed, by applying our rule \eqref{massdefL3} to the magnetic quiver of the $\mathcal T^{(2)}_{D_4,3}$ theory we get the magnetic quiver of another known rank-2 SCFT, namely theory  59 in \cite{Martone:2021ixp} whose global symmetry is precisely $G_2$. This model was called $\widehat{\mathcal T}_{D_4,3}$ in \cite{Giacomelli:2020gee}. Furthermore, the model $\mathcal S^{(1)}_{D_4,3} (1,0,1)$ is identified with the rank-1 theory $\mathcal S^{(1)}_{D_4,3}$ whose mass deformation is known to give $\mathcal S^{(1)}_{\mathcal{H}_1,3}$. This theory in turn is higgsed to the $(A_1,A_3)$ (a.k.a.~$\mathcal{H}_1$) Argyres-Douglas theory.
\be\label{ell3Nsmall}
\begin{tikzpicture}
\node[]  (a1) at (0,0) {\scriptsize $\mathring{\mathcal R}^{(2)}_{D_4,3} (0,0,2)$}; 
\node[]  (a2) at (0,-1) {\scriptsize $\mathring{\mathcal S}^{(2)}_{D_4,3} (1,0,1)$}; 
\node[]  (a3) at (0,-2) {\scriptsize $\mathring{\mathcal T}^{(2)}_{D_4,3} (0,1,0)$}; 
\node[]  (a4) at (0,-3) {{\color{blue} \scriptsize $\mathring{\mathcal T}^{(2)}_{D_4,3} (2,0,0)$}}; 
\draw[->, thick] (a1)--(a2); 
\draw[->, thick] (a2)--(a3); 
\draw[->, thick] (a3)--(a4); 
\node[]  (b1) at (4,-2.5) {\scriptsize $\mathring{\mathcal R}^{(1)}_{D_4,3} (0,0,2)$}; 
\node[]  (b2) at (4,-3.5) {{\color{blue} \scriptsize $\mathring{\mathcal S}^{(1)}_{D_4,3} (1,0,1)$}}; 
\node[]  (b3) at (4,-4.5) {{\color{blue} \scriptsize $\mathring{\mathcal T}^{(1)}_{D_4,3} (0,1,0)$}}; 
\node[]  (b4) at (4,-5.5) {{\color{red} \scriptsize $\mathring{\mathcal T}^{(1)}_{D_4,3} (2,0,0)$}}; 
\draw[->, thick] (b1)--(b2); 
\draw[->, thick] (b2)--(b3); 
\draw[->, thick] (b3)--(b4); 
\draw[->, thick] (a3)--(b1); 
\draw[->, thick] (a4)--(b2); 
\end{tikzpicture}
\ee 
As a result of the above discussion, we learn that all the theories in this series, like the mass-deformed theories with $M=6$ and $\ell=2$ we have discussed before, are completely higgsable, and therefore the dimension of the Higgs branch is always equal to $24(c-a)$. 

Notice also that the RG flow between $\mathcal T^{(2)}_{D_4,3}$ and $\widehat{\mathcal T}_{D_4,3}$ was dubbed non geometric in \cite{Giacomelli:2020gee}, since it is not described by the deformation of the underlying 7-brane in the S-fold setup considered in \cite{Giacomelli:2020gee}. It does however fit naturally in the more general framework we are developing in the present work. 

Finally, in the case $\ell=4$ we find that the $\mathcal T^{(2)}_{\mathcal{H}_2,4} (2,0)$ theory can be identified with the $S$-fold model $\mathcal{T}^{(2)}_{A_2,4}$ and, as we have seen in the previous subsection, its mass-deformed version becomes the theory 55 of \cite{Martone:2021ixp}. This model was called $\widehat{\mathcal T}_{A_2,4}$ in \cite{Giacomelli:2020gee}. According to our Hasse diagram \eqref{ell4Nsmall} this can be higgsed to the mass deformation of  $\mathcal{S}^{(1)}_{\mathcal{H}_2,4} (1,1)= \mathcal{S}^{(1)}_{A_2,4}$. This is a known rank-1 theory whose mass deformation leads to a $\mathcal{N}=3$ theory with Coulomb-branch operator of dimension 4. 
\be\label{ell4Nsmall}
\begin{tikzpicture}
\node[]  (a2) at (0,-1) {\scriptsize $\mathring{\mathcal T}^{(2)}_{\mathcal{H}_2,4} (0,2)$}; 
\node[]  (a1) at (0,0) {\scriptsize $\mathring{\mathcal S}^{(2)}_{\mathcal{H}_2,4} (1,1)$}; 
\node[]  (a3) at (0,-2) {{\color{blue}\scriptsize $\mathring{\mathcal T}^{(2)}_{\mathcal{H}_2,4} (2,0)$}}; 
\draw[->, thick] (a1)--(a2); 
\draw[->, thick] (a2)--(a3); 
\node[]  (b2) at (4,-2-1.5) {{\color{blue}\scriptsize $\mathring{\mathcal T}^{(1)}_{\mathcal{H}_2,4} (0,2)$}}; 
\node[]  (b1) at (4,-1-1.5) {{\color{blue}\scriptsize $\mathring{\mathcal S}^{(1)}_{\mathcal{H}_2,4} (1,1)$}}; 
\node[]  (b3) at (4,-3-1.5) {{\color{red}\scriptsize $\mathring{\mathcal T}^{(1)}_{\mathcal{H}_2,4} (2,0)$}}; 
\draw[->, thick] (b1)--(b2); 
\draw[->, thick] (b2)--(b3);  
\draw[->, thick] (a3)--(b1); 
\end{tikzpicture}
\ee
We therefore conclude that theory  55 of \cite{Martone:2021ixp} can be higgsed to a $\mathcal N=3$ theory, in perfect agreement with the findings of \cite{Martone:2021ixp}. This in particular implies, as we have already mentioned, that for this entire series of theories $24(c-a)$ is equal to the dimension of the Higgs branch minus one. Finally, the $\mathcal N=3$ rank-1 theory is higgsed to $\mathring{\mathcal T}^{(1)}_{\mathcal{H}_2,4} (0,2)$, which is simply a free hyper plus a free Abelian vector multiplet (or equivalently the $\mathcal N=4$ Abelian Super Yang-Mills theory). Again we find that the RG flow between $\mathcal T^{(2)}_{A_2,4}$ and $\widehat{\mathcal T}_{A_2,4}$ becomes geometric in our generalized orbi-S-fold setup, unlike in \cite{Giacomelli:2020gee}.

\subsubsection{Two predictions about rank-2 theories}

The goal of this subsection is to propose a S-fold (and therefore the first stringy) realization of theory 69 in \cite{Martone:2021ixp}, namely the $USp(4)$ gauge theory with a half-hypermultiplet in the {\bf 16}, and suggest a new rank-2 theory. In order to do this let us recall that, as we have seen, the RG flows from $\mathcal T_{G,\ell}^{(2)}$ to $\widehat{\mathcal T}_{G,\ell}^{(2)}$ theories, which were dubbed as non-geometric in \cite{Giacomelli:2020gee}, fit naturally in our construction as special examples of the mass deformations we have discussed throughout this section. We only consider the case of maximal 7-brane for each value of $\ell$.

The models $\mathcal T_{G,\ell}^{(2)}$ are SW compactifications of the 6d theories
\begin{align}
[\mathfrak{e}_8] \, \, 1 \, \, \overset{\mathfrak{su}_{\ell}}{2}[\mathfrak{su}_{2\ell}]\,,
\end{align}
and exist for every $\ell\leq 6$. In \cite{Giacomelli:2020gee} only the cases $\ell\geq2$ were considered, but we can also take $\ell=1$, in which case the 4d theory is simply the rank-2 $E_8$ Minahan-Nemeschansky theory. In this case the $SU(2)$ mass deformation leads to the $D_2^{20}(E_8)$ theory studied in \cite{Giacomelli:2017ckh}, which can be regarded as the $\widehat{\mathcal T}_{G,1}^{(2)}$.

The defining data of the $\widehat{\mathcal T}_{G,\ell}^{(2)}$ theories for $\ell\leq4$ were worked out in \cite{Giacomelli:2020gee}. In particular the Coulomb-branch operators are known to have dimension $4$ and $\frac{10}{\ell}$, which is compatible with the prescription given in Section \ref{CBspectrumIIB}. The combination of central charges $24(c-a)$ satisfies the formula \eqref{formulacc2} with $N=2$ and $\Delta_7=6/\ell$. Actually, both $\mathcal T_{G,\ell}^{(2)}$ and $\widehat{\mathcal T}_{G,\ell}^{(2)}$ satisfy \eqref{formulacc2}, but with different values of the parameter $\beta$. We do not know how to compute $\beta$ from first principles but we notice that the difference $\Delta\beta$ between the parameter $\beta$ in the UV (namely for $\mathcal T_{G,\ell}^{(2)}$ theories) and in the IR (namely for $\widehat{\mathcal T}_{G,\ell}^{(2)}$ theories) for $\ell\leq4$ has a very simple expression: 
\be\label{conj1} \Delta\beta=2(\Delta_7+1).\ee 
If we also assume the validity of \eqref{conj1} for $\ell=5$ we find that the putative  $\widehat{\mathcal T}_{\varnothing,5}^{(2)}$ theory has Coulomb-branch operators of dimension $2$ and $4$ and satisfies the relation $24(c-a)=-2$. Furthermore, from \eqref{ccell52} (or also \eqref{ccell53}) we see that for $N=2$ the magnetic quiver becomes trivial after the mass deformation, indicating that $\widehat{\mathcal T}_{\varnothing,5}^{(2)}$ has trivial Higgs branch and global symmetry. Overall, these are the data of theory 69 in \cite{Martone:2021ixp}. This observation remarkably provides a stringy realization of that theory, by connecting it to the S-fold construction. 

Let us now move to the case $\ell=6$. The theory $\mathcal T_{\varnothing,6}^{(2)}$ is known to be $\mathcal N=4$ Super Yang-Mills with gauge group $G_2$ \cite{Giacomelli:2020gee}. By turning on a mass for the $SU(2)$ global symmetry we get the massive $\mathcal N=2^*$ theory. Our candidate $\widehat{\mathcal T}_{\varnothing,6}^{(2)}$ theory has again trivial Higgs branch (see \eqref{ccell52}) and no global symmetry. The Coulomb-branch operators have dimension $\frac{5}{3}$ and $4$ and from \eqref{conj1} we find $24(c-a)=-4$. The scaling dimensions $(\frac{5}{3},4)$ constitute an allowed pair at rank 2 according to the criterion of \cite{Caorsi:2018zsq}, but to our knowledge this rank-2 theory has never been discussed before and constitutes a prediction of our construction. Our argument in particular suggests that this theory arises at a singular point of the Coulomb branch of the $\mathcal N=2^*$ $G_2$ theory. It would definitely be interesting to investigate this possibility in detail by studying the curve of the massive $G_2$ theory, perhaps using the findings of \cite{Bordner:1998xs, Bordner:1998sw}\footnote{We would like to thank Philip Argyres for suggesting these references.} (see also \cite{Argyres:2023eij}).

\section{7-brane deformations}\label{Sec:MassDef}

In this section we will discuss in detail the second family of mass deformations of SW theories we consider in this paper. More specifically, we will explain how the properties of the 4d SCFT change as we deform the 7-brane, leaving the $SU(k)$ symmetry inherited from M-theory intact. We will start by explaining in Section \ref{Sec:DecFlavors} how the magnetic quivers and the Coulomb-branch spectrum change upon considering non-maximal 7-branes. It turns out that the result can be written explicitly in terms of the holonomy data, giving full access to the moduli space of the field theory. We will then focus on the study of the case in which the 7-brane is removed completely, as the resulting 4d theories have special properties. In Section \ref{Sec:Pert} we focus on the case $\ell=2$ with trivial 7-brane, which lead to perturbative setups in Type IIB once the B-field is turned on. We will explain with a few examples how to incorporate the B-field and connect our construction with the well-known worldsheet analysis, which provides a standard lagrangian description for the theory on the worldvolume of the $D3$-branes. Finally, in Section \ref{Sec:1form} we will discuss 1-form symmetries for our 4d theories. These are absent whenever the F-theory background includes 7-branes, but can emerge once the 7-branes have been removed completely. We focus for simplicity on the case $\ell=2$ which is simpler to analyze. We also briefly comment on the case $\ell>2$, which is the most natural playground for investigating isolated and strongly-coupled $\mathcal{N}=2$ SCFTs with non-trivial 1-form symmetries, generalizing the understood case of theories with  $\mathcal{N}\geq 3$ supersymmetry.

\subsection{Decoupling flavors}\label{Sec:DecFlavors}

All the theories we have been discussing so far had a common feature: The probed background includes the maximal type of 7-brane\footnote{We sloppily adopt this terminology to refer to the M-theory nine-brane wall too.} allowed by the S-fold, which, for a given S-fold order $\ell$, is such that $\ell\Delta_7=6$. However, $\ell\Delta_7$ is also allowed to be another integer, as long as it is strictly smaller than $5$. Clearly, for some S-fold backgrounds, this opens up the possibility of accommodating different types of 7-branes. In this section we would like to derive the probe field theories associated to the non-maximal choice of 7-brane for all the orbi-S-fold geometries discussed in Section \ref{Sec:SWcompact}. More precisely, starting from the case of the maximal 7-brane presented there, we will reach the minimal case by successive iterations of mass deformations, which will progressively break the flavor symmetry of the 7-brane. As usual, we will implement this process at the level of the magnetic quivers by turning on FI parameters at specific nodes, as described at length in the previous sections.

Let us remark that the deformations we will discuss here only involve the flavor symmetry carried by the 7-brane, and are thus somewhat ``complementary'' to those discussed in Section \ref{Sec:IIBSfolds}, which instead concerned exclusively the geometric $SU(k)$ symmetry of the M-theory background. As was the case for the latter, we will now show that deforming the 7-brane while leaving the $T(SU(k))$ tail intact will also lead to a regular changing pattern of the field theories, in particular when it comes to their CB spectrum.

\subsubsection{The case {\bf$\ell=2$}}
We start by considering orbi-S-fold geometries with $\ell=2$ and any value of $k$. In this case, the maximal 7-brane has $\Delta_7=3$ (the 7-brane of type $E_6$), and the rule to derive the field theory for any choice of the 6d holonomy is summarized in the magnetic quiver \eqref{eq:ell2quiv}. It is useful to divide all possible holonomy choices into two groups according to the parity of $n_{4'}$. Defining the quantity
\be
\bar{n}_{4'}\equiv\left\{\begin{array}{ll}0\quad& n_{4'} \;\;{\rm even}\\ 1\quad& n_{4'} \;\;{\rm odd}\end{array}\right.
\ee
the magnetic quiver for generic $k$ can be written as follows
\begin{align}\label{eq:ell2maxgen}
\begin{tikzpicture}
\filldraw[fill= gray] (1,0) circle [radius=0.1] node[below] {\scriptsize {$k$}};
\filldraw[fill= white] (2,0) circle [radius=0.1] node[below] {\scriptsize {A}};
\filldraw[fill= white] (3,0) circle [radius=0.1] node[below] {\scriptsize{B}};
\filldraw[fill= white] (4,0) circle [radius=0.1] node[below] {\scriptsize{C}};
\filldraw[fill= white] (5,0) circle [radius=0.1] node[below] {\scriptsize{D}};
\filldraw[fill= white] (6,0) circle [radius=0.1] node[below] {\scriptsize{E}};
\draw [thick] (1.1, 0) -- (1.9,0) ;
\draw [thick] (2.1, 0) -- (2.9,0) ;
\draw [thick] (4.1, 0.05) -- (4.9,0.05) ;
\draw [thick] (4.1, -0.05) -- (4.9,-0.05) ;
\draw [thick] (3.1, 0) -- (3.9,0) ;
\draw [thick] (5.1, 0) -- (5.9,0) ;
\draw [thick] (4.4,0) -- (4.6,0.2);
\draw [thick] (4.4,0) -- (4.6,-0.2);
\end{tikzpicture} 
\end{align}
where
\begin{eqnarray}\label{EtichetteNodiell2}
{\rm A}&=&N+n_{2'}+n_4+\frac{3n_{4'}+\bar{n}_{4'}}{2}+2n_6\,,\nonumber\\
{\rm B}&=&2N+n_{2'}+n_{4'}+\bar{n}_{4'}+n_6\,,\nonumber\\
{\rm C}&=&3N+n_{2'}+\frac{n_{4'}+3\bar{n}_{4'}}{2}\,,\nonumber\\
{\rm D}&=&4N+n_{2'}+2\bar{n}_{4'}\,,\nonumber\\
{\rm E}&=&2N+\bar{n}_{4'}\,.
\end{eqnarray}

We now want to deform the 7-brane and descend to the next-to-maximal choice, i.e.~the 7-brane of type $D_4$ ($\Delta_7=2$). Such a deformation is very easy to describe for all the holonomy configurations where $n_6=0$.\footnote{The cases with $n_6\neq0$ must be deformed in a different fashion, which however leads to a different changing pattern of the CB spectrum as well as to a network of higgsings not simply inherited from the one before the deformation. Therefore we refrain from presenting them here.} It simply amounts to turn on FI terms at the B and C nodes\footnote{Strictly speaking this works for large enough $N$, such that the rank of the C-node does not exceed the one of the D-node. The precise condition is $N> \left\lfloor \frac{n_{4'}}{2} \right\rfloor$.}
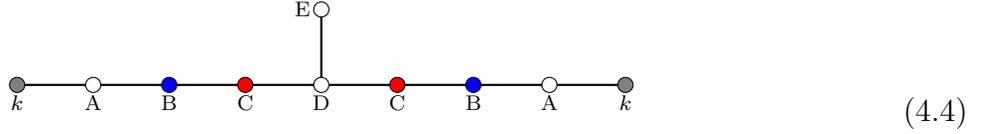
\begin{align}
\begin{tikzpicture}
\filldraw[fill= gray] (1,0) circle [radius=0.1] node[below] {\scriptsize $k$};
\filldraw[fill= white] (2,0) circle [radius=0.1] node[below] {\scriptsize A};
\filldraw[fill= blue] (3,0) circle [radius=0.1] node[below] {\scriptsize B};
\filldraw[fill= red] (4,0) circle [radius=0.1] node[below] {\scriptsize C};
\filldraw[fill= white] (5,0) circle [radius=0.1] node[below] {\scriptsize D};
\filldraw[fill= red] (6,0) circle [radius=0.1] node[below] {\scriptsize C};
\filldraw[fill= blue] (7,0) circle [radius=0.1] node[below] {\scriptsize B};
\filldraw[fill= white] (8,0) circle [radius=0.1] node[below] {\scriptsize A};
\filldraw[fill= gray] (9,0) circle [radius=0.1] node[below] {\scriptsize $k$};
\filldraw[fill= white] (5,1) circle [radius=0.1] node[left] {\scriptsize E};
\draw [thick] (1.1, 0) -- (1.9,0) ;
\draw [thick] (2.1, 0) -- (2.9,0) ;
\draw [thick] (3.1, 0) -- (3.9,0) ;
\draw [thick] (4.1, 0) -- (4.9,0) ;
\draw [thick] (5.1, 0) -- (5.9,0) ;
\draw [thick] (6.1, 0) -- (6.9,0) ;
\draw [thick] (7.1, 0) -- (7.9,0) ;
\draw [thick] (8.1, 0) -- (8.9,0) ;
\draw [thick] (5, 0.1) -- (5,0.9) ;
\end{tikzpicture} 
\end{align}
resulting in the following quiver
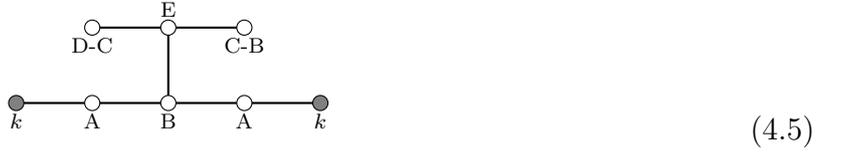
\begin{align}\label{ell2Nmaxgen}
\begin{tikzpicture}
\filldraw[fill= white] (4,1) circle [radius=0.1] node[below] {\scriptsize D-C};
\filldraw[fill= white] (6,1) circle [radius=0.1] node[below] {\scriptsize C-B};
\filldraw[fill= gray] (3,0) circle [radius=0.1] node[below] {\scriptsize $k$};
\filldraw[fill= white] (4,0) circle [radius=0.1] node[below] {\scriptsize A};
\filldraw[fill= white] (5,0) circle [radius=0.1] node[below] {\scriptsize B};
\filldraw[fill= white] (6,0) circle [radius=0.1] node[below] {\scriptsize A};
\filldraw[fill= gray] (7,0) circle [radius=0.1] node[below] {\scriptsize $k$};
\filldraw[fill= white] (5,1) circle [radius=0.1] node[above] {\scriptsize E};
\draw [thick] (3.1, 0) -- (3.9,0) ;
\draw [thick] (4.1, 0) -- (4.9,0) ;
\draw [thick] (4.1, 1) -- (4.9,1) ;
\draw [thick] (5.1, 0) -- (5.9,0) ;
\draw [thick] (5.1, 1) -- (5.9,1) ;
\draw [thick] (6.1, 0) -- (6.9,0) ;
\draw [thick] (5, 0.1) -- (5,0.9) ;
\end{tikzpicture} 
\end{align}
Crucially, as can be seen from \eqref{EtichetteNodiell2}, ${\rm D}-{\rm C}={\rm C}-{\rm B}$ iff $n_6=0$, and so the above quiver can be folded, yielding
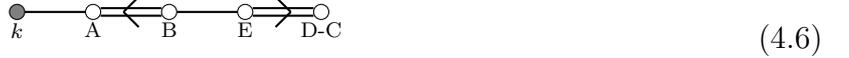
\begin{equation}\label{eq:ell2Nmaxgen}
\begin{tikzpicture}
\filldraw[fill= gray] (2,0) circle [radius=0.1] node[below] {\scriptsize $k$};
\filldraw[fill= white] (3,0) circle [radius=0.1] node[below] {\scriptsize A};
\filldraw[fill= white] (4,0) circle [radius=0.1] node[below] {\scriptsize B};
\filldraw[fill= white] (5,0) circle [radius=0.1] node[below] {\scriptsize E};
\filldraw[fill= white] (6,0) circle [radius=0.1] node[below] {\scriptsize  D-C};
\draw [thick] (2.1, 0) -- (2.9,0) ;
\draw [thick] (3.1, 0.05) -- (3.9,0.05) ;
\draw [thick] (3.1, -0.05) -- (3.9,-0.05) ;
\draw [thick] (4.1, 0) -- (4.9,0) ;
\draw [thick] (5.1, 0.05) -- (5.9,0.05) ;
\draw [thick] (5.1, -0.05) -- (5.9,-0.05) ;
\draw [thick] (3.4,0) -- (3.6,0.2);
\draw [thick] (3.4,0) -- (3.6,-0.2);
\draw [thick] (5.6,0) -- (5.4,0.2);
\draw [thick] (5.6,0) -- (5.4,-0.2);
\end{tikzpicture} 
\end{equation} 
which indeed has the shape of the $2$-twisted affine Dynkin diagram of $D_4$ connected to the $T(SU(k))$ tail. Formula \eqref{eq:ell2Nmaxgen} collectively indicates the magnetic quivers associated to all $\ell=2$ orbi-S-fold theories with a $D_4$-type 7-brane and vanishing holonomy label $n_6$.

We now proceed to the next deformation, converting the 7-brane of type $D_4$ to the minimal 7-brane for $\ell=2$, i.e.~the one of type $\mathcal{H}_2$ ($\Delta_7=\tfrac32$). To this end we make a further restriction on the choice of 6d holonomy that we consider: We take configurations where $n_4=n_6=0$.  After turning FI terms at the A nodes of \eqref{ell2Nmaxgen}, we get
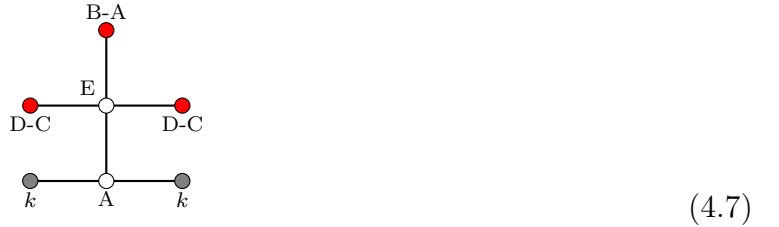
\begin{align}
\begin{tikzpicture}
\filldraw[fill= red] (4,1) circle [radius=0.1] node[below] {\scriptsize D-C};
\filldraw[fill= red] (6,1) circle [radius=0.1] node[below] {\scriptsize D-C};
\filldraw[fill= gray] (4,0) circle [radius=0.1] node[below] {\scriptsize $k$};
\filldraw[fill= white] (5,0) circle [radius=0.1] node[below] {\scriptsize A};
\filldraw[fill= red] (5,2) circle [radius=0.1] node[above] {\scriptsize B-A};
\filldraw[fill= gray] (6,0) circle [radius=0.1] node[below] {\scriptsize $k$};
\filldraw[fill= white] (5,1) circle [radius=0.1] node[above left] {\scriptsize E};
\draw [thick] (4.1, 0) -- (4.9,0) ;
\draw [thick] (4.1, 1) -- (4.9,1) ;
\draw [thick] (5.1, 0) -- (5.9,0) ;
\draw [thick] (5.1, 1) -- (5.9,1) ;
\draw [thick] (5, 0.1) -- (5,0.9) ;
\draw [thick] (5, 1.1) -- (5,1.9) ;
\end{tikzpicture} 
\end{align}
Due to our restrictions on the holonomy, we notice from \eqref{EtichetteNodiell2} that ${\rm D}-{\rm C}={\rm B}-{\rm A}$. Therefore we can turn on FI terms at the red nodes in the above quiver and get, after two steps,
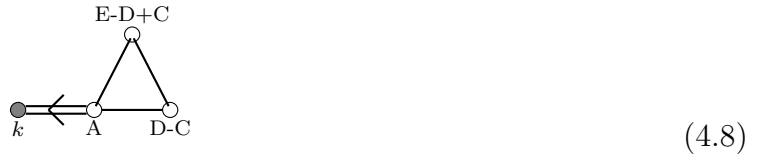
\begin{equation}\label{eq:ell2mingen}
\begin{tikzpicture}
\filldraw[fill= gray] (3,-4) circle [radius=0.1] node[below] {\scriptsize $k$};
\filldraw[fill= white] (4,-4) circle [radius=0.1] node[below] {\scriptsize A};
\filldraw[fill= white] (5,-4) circle [radius=0.1] node[below] {\scriptsize D-C};
\filldraw[fill= white] (4.5,-3) circle [radius=0.1] node[above] {\scriptsize E-D+C};
\draw [thick] (3.1, -3.95) -- (3.9,-3.95) ;
\draw [thick] (3.1, -4.05) -- (3.9,-4.05) ;
\draw [thick] (4.1, -4) -- (4.9,-4) ;
\draw [thick] (3.4,-4) -- (3.6,-3.8);
\draw [thick] (3.4,-4) -- (3.6,-4.2);
\draw [thick] (4.02, -3.95) -- (4.48,-3.05);
\draw [thick] (4.98, -3.95) -- (4.52,-3.05) ;
\end{tikzpicture}
\end{equation}
which indeed has the shape of the affine Dynkin diagram of $A_2$ connected twice to the $T(SU(k))$ tail. Formula \eqref{eq:ell2mingen} collectively indicates the magnetic quivers associated to all $\ell=2$ orbi-S-fold theories with a $\mathcal{H}_2$-type 7-brane and vanishing holonomy labels $n_4,n_6$.

\subsubsection{The case {\bf$\ell=3$}}
The only other S-fold order for which deforming the maximal 7-brane to a lower one is possible is $\ell=3$. Here the maximal one is the 7-brane of type $D_4$, and we can only deform it to the minimal one, i.e.~the 7-brane of type $\mathcal{H}_1$ ($\Delta_7=\tfrac43$).

The starting point is the quiver \eqref{eq:ell3quiv}, which summarizes the rule to derive the field theory corresponding to a generic $\ell=3$ orbi-S-fold for any choice of the 6d holonomy. In this case it is useful to organize the various holonomy configurations into three groups according to the value of $n_{3'}$ modulo $3$. Therefore, if we define the quantity
\be
\bar{n}_{3'}\equiv\left\{\begin{array}{ll}0\quad& n_{3'}=0\;{\rm mod}\,3\\ 1\quad& n_{3'}=1\;{\rm mod}\,3\\ 2\quad& n_{3'}=2\;{\rm mod}\,3 \end{array}\right.
\ee
the magnetic quiver for generic $k$ can be written as follows
\begin{align}\label{eq:ell3maxgen}
\begin{tikzpicture}
\filldraw[fill= gray] (1,0) circle [radius=0.1] node[below] {\scriptsize $k$};
\filldraw[fill= white] (2,0) circle [radius=0.1] node[below] {\scriptsize A};
\filldraw[fill= white] (3,0) circle [radius=0.1] node[below] {\scriptsize B};
\filldraw[fill= white] (4,0) circle [radius=0.1] node[below] {\scriptsize C};
\draw [thick] (1.1, 0) -- (1.9,0) ;
\draw [thick] (2.1, 0) -- (2.9,0) ;
\draw [thick] (3.1, 0.07) -- (3.9,0.07) ;
\draw [thick] (3.1, -0.07) -- (3.9,-0.07) ;
\draw [thick] (3.1, 0) -- (3.9,0) ;
\draw [thick] (3.4,0) -- (3.6,0.2);
\draw [thick] (3.4,0) -- (3.6,-0.2);
\end{tikzpicture}\,.
\end{align}
where
\begin{eqnarray}\label{EtichetteNodiell3}
{\rm A}&=&N+\frac{2n_{3'}+\bar{n}_{3'}}{3}+n_6\,,\nonumber\\
{\rm B}&=&2N+\frac{n_{3'}+2\bar{n}_{3'}}{3}\,,\nonumber\\
{\rm C}&=&3N+\bar{n}_{3'}\,.
\end{eqnarray}

Let us now perform the deformation of the 7-brane. Again, a very simple one exists upon restricting to a subset of holonomy configurations: In this case we select those with $n_6=0$. We simply turn FI terms at the A and B nodes\footnote{Strictly speaking this works for large enough $N$, such that the rank of the B-node does not exceed the one of the C-node. The precise condition is $N> \left\lfloor \frac{n_{3'}}{3} \right\rfloor$.}
\begin{align}
\begin{tikzpicture}
\filldraw[fill= gray] (2,0) circle [radius=0.1] node[below] {\scriptsize $k$};
\filldraw[fill= blue] (3,0) circle [radius=0.1] node[below] {\scriptsize A};
\filldraw[fill= red] (4,0) circle [radius=0.1] node[below] {\scriptsize B};
\filldraw[fill= white] (5,0) circle [radius=0.1] node[below] {\scriptsize C};
\filldraw[fill= red] (6,0) circle [radius=0.1] node[below] {\scriptsize B};
\filldraw[fill= blue] (7,0) circle [radius=0.1] node[below] {\scriptsize A};
\filldraw[fill= gray] (8,0) circle [radius=0.1] node[below] {\scriptsize $k$};
\filldraw[fill= red] (5,1) circle [radius=0.1] node[left] {\scriptsize B};
\filldraw[fill= blue] (5,2) circle [radius=0.1] node[left] {\scriptsize A};
\filldraw[fill= gray] (5,3) circle [radius=0.1] node[left] {\scriptsize $k$};
\draw [thick] (2.1, 0) -- (2.9,0) ;
\draw [thick] (3.1, 0) -- (3.9,0) ;
\draw [thick] (4.1, 0) -- (4.9,0) ;
\draw [thick] (5.1, 0) -- (5.9,0) ;
\draw [thick] (6.1, 0) -- (6.9,0) ;
\draw [thick] (7.1, 0) -- (7.9,0) ;
\draw [thick] (5, 0.1) -- (5,0.9) ;
\draw [thick] (5, 1.1) -- (5,1.9) ;
\draw [thick] (5, 2.1) -- (5,2.9) ;
\end{tikzpicture} 
\end{align}
and, in the process, make use of the fact that ${\rm C}-{\rm B}={\rm B}-{\rm A}$ iff $n_6=0$, as can be seen from \eqref{EtichetteNodiell3}. After a few simple passages, we land on the following quiver
\begin{align}\label{eq:ell3mingen}
\begin{tikzpicture}
\filldraw[fill= gray] (2,-9) circle [radius=0.1] node[below] {\scriptsize $k$};
\filldraw[fill= white] (3,-9) circle [radius=0.1] node[below] {\scriptsize A};
\filldraw[fill= white] (4,-9) circle [radius=0.1] node[below] {\scriptsize C-B};
\draw [thick] (2.1, -8.95) -- (2.9,-8.95) ;
\draw [thick] (2.1, -9) -- (2.9,-9) ;
\draw [thick] (2.1, -9.05) -- (2.9,-9.05) ;
\draw [thick] (2.4,-9) -- (2.6,-8.8);
\draw [thick] (2.4,-9) -- (2.6,-9.2);
\draw [thick] (3.1, -8.95) -- (3.9,-8.95);
\draw [thick] (3.1, -9.05) -- (3.9,-9.05);
\end{tikzpicture} 
\end{align}
which indeed has the shape of the affine Dynkin diagram of $A_1$ connected thrice to the $T(SU(k))$ tail. Formula \eqref{eq:ell3mingen} collectively indicates the magnetic quivers associated to all $\ell=3$ orbi-S-fold theories with a $\mathcal{H}_1$-type 7-brane and vanishing holonomy label $n_6$.

\subsubsection{Coulomb-branch spectrum}\label{CB7branedef}
The deformations described above all lead to SCFTs whose CB spectrum can immediately be deduced from the parent theories with the maximal 7-brane. For the latter theories (i.e.~when $\ell\Delta_7=6$), we summarized the general rule to get the CB spectrum in Section \ref{CBspectrumMaxM}. For the non-maximal 7-branes, all what it takes to do is to substitute the $6$ appearing in the formulas for the conformal dimensions with the actual value of $\ell\Delta_7$. Therefore we have $N-1$ blocks of operators with dimension
\begin{align}
\Delta_r =  \{ \ell\Delta_7 r \}\,  \cup\, \{  \ell\Delta_7 r + d \, | \, d= 2, \dots, p_r\} \qquad r=2,\ldots,N\,,
\end{align}
associated to the (-2)-curves carrying gauge algebra $\mathfrak{su}_{\ell p_r}$ in 6d. The last block has operators of dimensions that depend on the gauge algebra carried by the (-1)-curve:
\begin{align}
\mathfrak g = \varnothing\,, \quad &\rightarrow \quad \Delta_1 = \left\{\Delta_7 N\right\}\,,\nonumber \\
\mathfrak g = \mathfrak{su}_{\ell p_1} \,, \quad &\rightarrow \quad \Delta_1 =  \{ \ell\Delta_7 N \}\,  \cup\, \{ \ell\Delta_7 N + d \, | \, d= 2, \dots, p_1\}\,, \nonumber \\
\mathfrak g = \mathfrak{usp}_{4m+2} \,, \quad &\rightarrow \quad \Delta_1 =  \{ \ell\Delta_7 N \}\,  \cup\, \{ \ell\Delta_7 N + d \, | \, d= 2, \dots, 2m\}\,, \nonumber \\
\mathfrak g = \mathfrak{usp}_{4m} \,, \quad &\rightarrow \quad \Delta_1 =  \{ \ell\Delta_7 N \}\,  \cup\, \{ \ell\Delta_7 N + d \, | \, d= 2, \dots, 2m-2\} \, \cup \, \{\Delta_7 N+\frac{2m}{\ell}\}\,.
\end{align}

We would now like to observe a curious fact relating the deformation of the 7-brane which we just discussed with the deformation of the tail described in Section \ref{Sec:IIBSfolds}. It turns out that, when $N=1$, and only then, the two procedures are equivalent and lead to the same theory. Let us see how this works in a couple of examples.

Take $\ell=2$ and a generic orbifold order $k$, and consider the theory $\mathcal T^{(1)}_{E_6,2} (0,0,0,0,k)$, whose magnetic quiver is (see Eq.~\eqref{eq:ell2maxgen})
\begin{align}
\begin{tikzpicture}
\filldraw[fill= gray] (1,0) circle [radius=0.1] node[below] {\scriptsize $k$};
\filldraw[fill= white] (2,0) circle [radius=0.1] node[below] {\scriptsize $k+1$};
\filldraw[fill= white] (3,0) circle [radius=0.1] node[below] {\scriptsize $k+2$};
\filldraw[fill= white] (4,0) circle [radius=0.1] node[below] {\scriptsize $k+3$};
\filldraw[fill= white] (5,0) circle [radius=0.1] node[below] {\scriptsize $k+4$};
\filldraw[fill= white] (6,0) circle [radius=0.1] node[below] {\scriptsize  $2$};
\draw [thick] (1.1, 0) -- (1.9,0) ;
\draw [thick] (2.1, 0) -- (2.9,0) ;
\draw [thick] (4.1, 0.05) -- (4.9,0.05) ;
\draw [thick] (4.1, -0.05) -- (4.9,-0.05) ;
\draw [thick] (3.1, 0) -- (3.9,0) ;
\draw [thick] (5.1, 0) -- (5.9,0) ;
\draw [thick] (4.4,0) -- (4.6,0.2);
\draw [thick] (4.4,0) -- (4.6,-0.2);
\end{tikzpicture} \,.
\end{align}
We now perform a (partial) deformation of the $T(SU(k))$ tail by means of the passages described explicitly in Eq.~\eqref{quiv1}.\footnote{If $k=2$ this amounts to getting rid of the tail and breaking the geometric $SU(2)$ completely.} The result is the following quiver
\begin{equation}
\begin{tikzpicture}
\filldraw[fill= gray] (2,0) circle [radius=0.1] node[below] {\scriptsize $k$};
\filldraw[fill= white] (3,0) circle [radius=0.1] node[below] {\scriptsize $k+1$};
\filldraw[fill= white] (4,0) circle [radius=0.1] node[below] {\scriptsize $k+2$};
\filldraw[fill= white] (5,0) circle [radius=0.1] node[below] {\scriptsize $2$};
\filldraw[fill= white] (6,0) circle [radius=0.1] node[below] {\scriptsize  $1$};
\draw [thick] (2.1, 0) -- (2.9,0) ;
\draw [thick] (3.1, 0.05) -- (3.9,0.05) ;
\draw [thick] (3.1, -0.05) -- (3.9,-0.05) ;
\draw [thick] (4.1, 0) -- (4.9,0) ;
\draw [thick] (5.1, 0.05) -- (5.9,0.05) ;
\draw [thick] (5.1, -0.05) -- (5.9,-0.05) ;
\draw [thick] (3.4,0) -- (3.6,0.2);
\draw [thick] (3.4,0) -- (3.6,-0.2);
\draw [thick] (5.6,0) -- (5.4,0.2);
\draw [thick] (5.6,0) -- (5.4,-0.2);
\end{tikzpicture} 
\end{equation} 
which coincides with \eqref{eq:ell2Nmaxgen} for the holonomy configuration we chose, i.e.~$n_{2'}=k$ being the only non-vanishing label. We have found the theory associated to the $D_4$ 7-brane (the next-to-maximal one) on the same orbi-S-fold, but with the $T(SU(k))$ tail intact.

Let us go one step further in the deformation of the $T(SU(k))$ tail, as described explicitly in Eq.~\eqref{quiv2}.\footnote{If $k=3$ this extra step destroys the tail and breaks the geometric $SU(3)$ completely.} The result is the following quiver
\begin{equation}
\begin{tikzpicture}
\filldraw[fill= gray] (3,-4) circle [radius=0.1] node[below] {\scriptsize $k$};
\filldraw[fill= white] (4,-4) circle [radius=0.1] node[below] {\scriptsize $k+1$};
\filldraw[fill= white] (5,-4) circle [radius=0.1] node[below] {\scriptsize $1$};
\filldraw[fill= white] (4.5,-3) circle [radius=0.1] node[above] {\scriptsize $1$};
\draw [thick] (3.1, -3.95) -- (3.9,-3.95) ;
\draw [thick] (3.1, -4.05) -- (3.9,-4.05) ;
\draw [thick] (4.1, -4) -- (4.9,-4) ;
\draw [thick] (3.4,-4) -- (3.6,-3.8);
\draw [thick] (3.4,-4) -- (3.6,-4.2);
\draw [thick] (4.02, -3.95) -- (4.48,-3.05);
\draw [thick] (4.98, -3.95) -- (4.52,-3.05) ;
\end{tikzpicture}
\end{equation}
which coincides with \eqref{eq:ell2mingen} for the holonomy configuration we chose, i.e.~$n_{2'}=k$ being the only non-vanishing label. We have found the theory associated to the $\mathcal{H}_2$ 7-brane (the minimal one) on the same orbi-S-fold, but with the $T(SU(k))$ tail intact. The magnetic quivers in this specific example were already discussed in \cite{Bourget:2020asf} and the results presented in that paper agree perfectly with ours.

Take now $\ell=3, k=2$, and consider the theory $\mathcal R^{(1)}_{D_4,3} (0,0,2)$, whose magnetic quiver is (see Eq.~\eqref{eq:ell3maxgen})
\begin{align}
\begin{tikzpicture}
\filldraw[fill= gray] (1,0) circle [radius=0.1] node[below] {\scriptsize $2$};
\filldraw[fill= white] (2,0) circle [radius=0.1] node[below] {\scriptsize $3$};
\filldraw[fill= white] (3,0) circle [radius=0.1] node[below] {\scriptsize $4$};
\filldraw[fill= white] (4,0) circle [radius=0.1] node[below] {\scriptsize $5$};
\draw [thick] (1.1, 0) -- (1.9,0) ;
\draw [thick] (2.1, 0) -- (2.9,0) ;
\draw [thick] (3.1, 0.07) -- (3.9,0.07) ;
\draw [thick] (3.1, -0.07) -- (3.9,-0.07) ;
\draw [thick] (3.1, 0) -- (3.9,0) ;
\draw [thick] (3.4,0) -- (3.6,0.2);
\draw [thick] (3.4,0) -- (3.6,-0.2);
\end{tikzpicture}\,.
\end{align}
We now want to deform the $T(SU(2))$ tail by following the procedure outlined in Eq.~\eqref{defL=3}. This yields the quiver
\begin{align}
\begin{tikzpicture}
\filldraw[fill= gray] (2,-9) circle [radius=0.1] node[below] {\scriptsize $2$};
\filldraw[fill= white] (3,-9) circle [radius=0.1] node[below] {\scriptsize $3$};
\filldraw[fill= white] (4,-9) circle [radius=0.1] node[below] {\scriptsize $1$};
\draw [thick] (2.1, -8.95) -- (2.9,-8.95) ;
\draw [thick] (2.1, -9) -- (2.9,-9) ;
\draw [thick] (2.1, -9.05) -- (2.9,-9.05) ;
\draw [thick] (2.4,-9) -- (2.6,-8.8);
\draw [thick] (2.4,-9) -- (2.6,-9.2);
\draw [thick] (3.1, -8.95) -- (3.9,-8.95);
\draw [thick] (3.1, -9.05) -- (3.9,-9.05);
\end{tikzpicture} 
\end{align}
which coincides with \eqref{eq:ell3mingen} for the holonomy configuration we chose, i.e.~$n_{3'}=2$ being the only non-vanishing label. We have found the theory associated to the $\mathcal{H}_1$ 7-brane (the minimal one) on the same orbi-S-fold, but with the $T(SU(2))$ tail intact. Also in this case we find perfect agreement with \cite{Bourget:2020asf}.

 Let us conclude with a couple of remarks concerning the Hasse diagram and 't Hooft anomalies of the SCFTs discussed in this section. 

\paragraph{Hasse diagrams for general choices of 7-brane.}

In contrast to the situation we have encountered in Section \ref{HasseIIB} concerning the tail deformation, here we are activating a mass deformation for the same symmetry that is involved in the higgsings encoded by the Hasse diagrams and this a priori can affect the Hasse diagram in a non-trivial way. Nevertheless, one can easily show using quiver subtraction that the Hasse diagrams one finds for the mass-deformed theories associated with non-maximal 7-branes are almost identical to those for the maximal 7-branes (see Section \ref{HasseMtheory}), although transverse slices may change in general. The main difference we observe is that one needs to remove from the Hasse diagram of the undeformed theories all the entries corresponding to holonomy choices such that the mass deformation cannot be activated (like e.g. theories with $n_4\neq0$ in the case of 7-brane of type $\mathcal H_2$ and $\ell=2$), and include new transitions by concatenating the arrows passing through the removed entries. Specifically, the Hasse diagram for the case $\ell=k=2$ and 7-brane of type $\mathcal{H}_2$ is obtained from \eqref{E6ell2k2Hasse} by eliminating the family $\mathcal T^{(N)}_{E_6,2} (0,1,0,0,0)$, which is indeed the only one characterized by a non-vanishing $n_4$. As a result, two new elementary transitions appear, both with a one-dimensional transverse slice: $\mathcal T^{(N)}_{E_6,2} (1,0,0,0,1)\to\mathcal T^{(N)}_{E_6,2} (2,0,0,0,0)$ with transverse slice $\mathbb{C}^2/\mathbb{Z}_4$ and $\mathcal T^{(N)}_{E_6,2} (1,0,0,0,1)\to\mathcal S^{(N-1)}_{E_6,2} (0,0,0,1,0)$ with transverse slice $\mathbb{C}^2/\mathbb{Z}_3$.

\paragraph{Central charges of theories associated with non maximal 7-brane.}

The second remark is about the change of the central charges upon 7-brane deformations. Formula \eqref{formulacc3} for the central charge of the global symmetry carried by the 7-brane tells us directly how this quantity varies when we deform the 7-brane from the maximal one: By looking at known examples,\footnote{We have checked the validity of our proposal for theories with $\ell=1$ (no S-fold) and S-fold theories with $k=1$ (no orbifold), for which the central charges are known for arbitrary $N$. We have also tested it in the case of orbi-S-fold theories at rank 2.} we argue that one just needs to put the correct value of $\Delta_7$ in the $N$-linear term, whereas the $N$-independent term remains the same. As for the central charge of the symmetry associated to the $T(SU(k))$ tail, this quantity does not depend on $N$, but does depend on $\Delta_7$. By inspecting the pure-orbifold SCFTs (i.e.~$\ell=1$), which are all theories of class $\mathcal{S}$, we found that $k_{SU(k)}=2(\Delta_7+k)$. By using the formulas for the central charges of section \ref{sec:4dprop}, we readily conclude that, for $\ell>1$, the latter expression generalizes to
\be
k_{SU(k)}=2(\ell\Delta_7+k)\,,
\ee
which immediately tells us how the flavor central charge associated to the tail symmetry changes when we deform the 7-brane. Finally, we found an empirical rule for the modification of the quantity $24(c-a)$ upon 7-brane deformations: Assuming to know its variation at the first step (the difference between the value for the maximal 7-brane and the one for the next-to-maximal 7-brane), call it $\delta$, the difference in all the other steps is $\delta/2$, including the final step that gets rid of the 7-brane, as long as it corresponds to a deformation that does not pass through non-conformal theories.

\subsection{Perturbative families}\label{Sec:Pert}

In the case of orbi-S-folds with $\ell=2$ and no 7-branes we obtain a perturbative Type IIB background formed by combining an $O3$-plane and an orbifold singularity. Since there are well-established techniques to read off a lagrangian description of the theory on the $D3$ probes,\footnote{See e.g.~\cite{Giacomelli:2023qyc} where these techniques are reviewed and the corresponding quiver gauge theories are described.} a natural question is how these are related to the tools we have developed in this paper. We would now like to illustrate with a few examples how this works. It turns out that the connection we have discussed in Section \ref{Sec:IIBSfolds} between the B-field in string theory and the mass deformation for the tail symmetry in the 4d field theory plays a crucial role.

The first example we consider is the theory with $M=2k=4$ and holonomy $n_{2'}=2$. Starting from the magnetic quiver \eqref{eq:ell2mingen}, which corresponds to the minimal type of 7-brane, we can activate an extra mass deformation to get rid of the 7-brane completely. This amounts to turn on FI terms as follows
\begin{equation}\label{FI1modelpert}
\begin{tikzpicture}
\filldraw[fill= white] (2.5,-2.5) circle [radius=0.1] node[below] {\scriptsize $1$};
\filldraw[fill= white] (3.2,-3.2) circle [radius=0.1] node[below] {\scriptsize $2$};
\filldraw[fill= white] (2,-4) circle [radius=0.1] node[below] {\scriptsize $1$};
\filldraw[fill= white] (3,-4) circle [radius=0.1] node[below] {\scriptsize $2$};
\filldraw[fill= white] (4,-4) circle [radius=0.1] node[below] {\scriptsize N+2};
\filldraw[fill= red] (5,-4) circle [radius=0.1] node[below] {\scriptsize N};
\filldraw[fill= red] (4.5,-3) circle [radius=0.1] node[above] {\scriptsize N};
\draw [thick] (2.1, -3.95) -- (2.9,-3.95) ;
\draw [thick] (3.1, -3.95) -- (3.9,-3.95) ;
\draw [thick] (4.1, -4) -- (4.9,-4) ;
\draw [thick] (4.02, -3.95) -- (4.48,-3.05);
\draw [thick] (4.98, -3.95) -- (4.52,-3.05) ;
\draw [thick] (3.98, -3.95) -- (3.3,-3.25) ;
\draw [thick] (3.18, -3.15) -- (2.6,-2.55) ;

\draw [->] (5.5,-3.5) -- (6.5,-3.5) ;

\filldraw[fill= white] (7,-3) circle [radius=0.1] node[above] {\scriptsize $1$};
\filldraw[fill= white] (8,-3) circle [radius=0.1] node[above] {\scriptsize $2$};
\filldraw[fill= white] (9,-3) circle [radius=0.1] node[above] {\scriptsize $2$};
\filldraw[fill= white] (9,-4) circle [radius=0.1] node[right] {\scriptsize $2$};
\filldraw[fill= white] (9,-5) circle [radius=0.1] node[right] {\scriptsize $1$};
\filldraw[fill= lime] (8,-4) circle [radius=0.1] node[below] {\scriptsize N};

\draw [thick] (7.1, -3) -- (7.9,-3) ;
\draw [thick] (8.1, -3) -- (8.9,-3) ;
\draw [thick] (9, -3.1) -- (9,-3.9) ;
\draw [thick] (9, -4.1) -- (9,-4.9) ;
\draw [thick] (8.1, -4) -- (8.9,-4) ;
\draw [thick] (8, -3.9) -- (8,-3.1) ;
\draw [thick] (7.95, -3.95) to [out=120,in=200,looseness=15] (7.95, -4.05);

\end{tikzpicture}
\end{equation}
where we have used \eqref{EtichetteNodiell2} for our choice of holonomy. By condensing the bifundamental fields between the $U(2)$ nodes, we arrive at
\begin{align}
\begin{tikzpicture}\label{k2l2nobrane}
\filldraw[fill= white] (2,-9) circle [radius=0.1] node[below] {\scriptsize 1};
\filldraw[fill= white] (3,-9) circle [radius=0.1] node[below] {\scriptsize 2};
\filldraw[fill= white] (4,-9) circle [radius=0.1] node[below] {\scriptsize N};
\draw [thick] (2.1, -8.95) -- (2.9,-8.95) ;
\draw [thick] (2.1, -9.05) -- (2.9,-9.05) ;
\draw [thick] (2.4,-9) -- (2.6,-8.8);
\draw [thick] (2.4,-9) -- (2.6,-9.2);
\draw [thick] (3.1, -8.95) -- (3.9,-8.95);
\draw [thick] (3.1, -9.05) -- (3.9,-9.05);
\draw [thick] (4.05, -8.95) to [out=45,in=315,looseness=15] (4.05,-9.05);
\end{tikzpicture} 
\end{align}
Note that, in this process, we have not deformed the $T(SU(2))$ tail, whose associated $SU(2)$ symmetry remains thus intact: The tail just fused to the rest of the diagram. We can therefore apply the usual rule of Section \ref{CB7branedef} to deduce the CB spectrum of the SCFT family \eqref{k2l2nobrane}, finding
\be\label{cbex1}
(2r,2r+2)_{r=1,\dots,N-1};\;\; (2N,N+1)\,.
\ee
For $N=1$ this is just SQCD with gauge group $SO(4)$ and $2$ fundamental hypermultiplets.\footnote{Such fundamental matter may be unexpected, given the absence of 7-branes. But recall that we are still in the regime of vanishing B-field, where the Type IIB picture of (fractional) $D3$ probes breaks down, while the dual (SW-reduced) 6d picture from M-theory is more appropriate.} For higher $N$, however, the theory is not lagrangian and is rather a conformal gauging of a strongly-coupled theory. We would like to understand what this theory is. A hint is given by \eqref{formulacc1} which tells us that the combination $24(c-a)$ does not depend on $N$. Since its value for $N=1$ is $2$, we conclude that for the entire family \eqref{k2l2nobrane}
\be\label{ccex1} 
24(c-a)=2\,.
\ee 
We can then notice that \eqref{cbex1} always includes the Casimirs of $SO(2N+2)$, and therefore this group is our guess for the gauge group. For $N=2$, for instance, this means that we have a $SO(6)$ gauging of a rank-1 theory with a CB operator of dimension $4$. From \eqref{ccex1} we also conclude that this putative rank-1 theory must have $24(c-a)$ equal to $2+15=17$. These are the data of the $E_7$ Minahan-Nemeschansky theory, and hence we are led to consider a $SO(6)$ gauging of such a SCFT. Back to generic $N$, remarkably we find in class $\mathcal{S}$ a series of theories which has the correct properties to be interpreted as the matter sector that we want to gauge with $SO(2N+2)$. These are three-punctured spheres of type A$_{2N-1}$, with punctures specified by the following partitions: 
\be\label{trinion1} 
p_1=(1^{2N})\,;\quad p_2=(N,N)\,;\quad p_3=(N-1,N-1,1,1)\,.
\ee
The magnetic quiver corresponding to this matter sector is
\be
\begin{tikzpicture}\label{trinion1mq}
\filldraw[fill= white] (-2,0) circle [radius=0.1] node[below] {\scriptsize 1};
\filldraw[fill= white] (-1,0) circle [radius=0.1] node[below] {\scriptsize 2};
\filldraw[fill= white] (0,0) circle [radius=0.1] node[below] {\scriptsize N+1};
\filldraw[fill= white] (1,0) circle [radius=0.1] node[below] {\scriptsize 2N};
\filldraw[fill= white] (2,0) circle [radius=0.1] node[below] {\scriptsize 2N-1};
\node[] at (3,0) {\dots};
\filldraw[fill= white] (4,0) circle [radius=0.1] node[below] {\scriptsize 2};
\filldraw[fill= white] (5,0) circle [radius=0.1] node[below] {\scriptsize 1};
\filldraw[fill= white] (1,1) circle [radius=0.1] node[above] {\scriptsize N};

\draw [thick] (-0.1, 0) -- (-0.9,0) ;
\draw [thick] (-1.1, 0) -- (-1.9,0) ;
\draw [thick] (0.1, 0) -- (0.9,0) ;
\draw [thick] (1.1, 0) -- (1.9,0) ;
\draw [thick] (2.1, 0) -- (2.7,0) ;
\draw [thick] (3.3, 0) -- (3.9,0) ;
\draw [thick] (4.1, 0) -- (4.9,0) ;
\draw [thick] (1, 0.1) -- (1,0.9) ;
\end{tikzpicture}
\ee
Using the standard rules for class-$\mathcal{S}$ theories, one can check that once the $SO(2N+2)$ vector multiplet is added we recover for every $N$ the CB spectrum \eqref{cbex1}, and furthermore the relation \eqref{ccex1} holds. The global symmetry of the theory \eqref{trinion1mq} for $N>2$ is $$SO(4N+4)_{4N}\times SU(2)_8\,,$$ while for $N=1$ it describes the free theory of eight hypermultiplets, and for $N=2$ it reduces to the $E_7$ Minahan-Nemeschansky theory, as expected. The $SO(2N+2)$ gauging  is conformal if we gauge the diagonal combination of $SO(2N+2)\times SO(2N+2)\subset SO(4N+4)$. The $SU(2)$ global symmetry surviving the gauging can be interpreted as the symmetry inherited from M-theory, carried by the $TSU(2))$ tail of \eqref{trinion1mq}, and, as usual (see App.~\ref{App:Mth}), its level does not depend on $N$. As we have explained in Section \ref{Sec:IIBSfolds}, turning on a mass for it corresponds to turning on the B-field in Type IIB string theory.

In summary, we propose that the theory \eqref{trinion1mq}, coupled to a $SO(2N+2)$ vector multiplet, is the worldvolume theory on $D3$-branes probing an $O3$-plane combined with a $\bbZ_2$ orbifold, at least for some choice of the orientifold projection and distribution of the fractional branes. The fact that the theory is strongly coupled reflects the absence of the B-field. Only once the B-field is turned on, does the theory reduce to one of the ordinary quivers predicted by the worldsheet analysis. We therefore expect that, by activating a mass deformation for the $SU(2)_8$ symmetry, we land on a orbi-orientifold quiver. Indeed, by switching on said mass term (or FI term in the magnetic quiver) we find: 
\be\label{MMdef}
\begin{tikzpicture}
\filldraw[fill= red] (-2,0) circle [radius=0.1] node[below] {\scriptsize 1};
\filldraw[fill= red] (-1,0) circle [radius=0.1] node[below] {\scriptsize 2};
\filldraw[fill= white] (0,0) circle [radius=0.1] node[below] {\scriptsize N+1};
\filldraw[fill= white] (1,0) circle [radius=0.1] node[below] {\scriptsize 2N};
\filldraw[fill= white] (2,0) circle [radius=0.1] node[below] {\scriptsize 2N-1};
\node[] at (3,0) {\dots};
\filldraw[fill= white] (4,0) circle [radius=0.1] node[below] {\scriptsize 2};
\filldraw[fill= white] (5,0) circle [radius=0.1] node[below] {\scriptsize 1};
\filldraw[fill= white] (1,1) circle [radius=0.1] node[above] {\scriptsize N};

\draw [thick] (-0.1, 0) -- (-0.9,0) ;
\draw [thick] (-1.1, 0) -- (-1.9,0) ;
\draw [thick] (0.1, 0) -- (0.9,0) ;
\draw [thick] (1.1, 0) -- (1.9,0) ;
\draw [thick] (2.1, 0) -- (2.7,0) ;
\draw [thick] (3.3, 0) -- (3.9,0) ;
\draw [thick] (4.1, 0) -- (4.9,0) ;
\draw [thick] (1, 0.1) -- (1,0.9) ;

\draw [->] (2,-0.6) -- (2,-2.2) ;

\filldraw[fill= white] (0,-3) circle [radius=0.1] node[below] {\scriptsize N-1};
\filldraw[fill= white] (1,-3) circle [radius=0.1] node[below] {\scriptsize 2N-2};
\filldraw[fill= white] (2,-3) circle [radius=0.1] node[below] {\scriptsize 2N-2};
\filldraw[fill= white] (3,-3) circle [radius=0.1] node[below] {\scriptsize 2N-2};
\filldraw[fill= white] (4,-3) circle [radius=0.1] node[below] {\scriptsize 2N-3};
\node[] at (5,-3) {\dots};
\filldraw[fill= white] (6,-3) circle [radius=0.1] node[below] {\scriptsize 2};
\filldraw[fill= white] (1,-2) circle [radius=0.1] node[above] {\scriptsize N-1};
\filldraw[fill= white] (7,-3) circle [radius=0.1] node[below] {\scriptsize 1};
\filldraw[fill= white] (3,-2) circle [radius=0.1] node[above] {\scriptsize 1};

\draw [thick] (0.1, -3) -- (0.9,-3) ;
\draw [thick] (1.1, -3) -- (1.9,-3) ;
\draw [thick] (2.1, -3) -- (2.9,-3) ;
\draw [thick] (3.1, -3) -- (3.9,-3) ;
\draw [thick] (4.1, -3) -- (4.7,-3) ;
\draw [thick] (5.3, -3) -- (5.9,-3) ;
\draw [thick] (1, -2.1) -- (1,-2.9) ;
\draw [thick] (3, -2.1) -- (3,-2.9) ;
\draw [thick] (6.1, -3) -- (6.9,-3) ;
\end{tikzpicture}
\ee 
In the second quiver of \eqref{MMdef} we recognize the mirror dual of $USp(2N-2)$ SQCD with $2N+2$ fundamental hypermultiplets. The theory is now lagrangian, although no longer conformal. Upon the $SO(2N+2)$ gauging, we finally find the (electric) quiver 
\be\label{lagg1}
\begin{tikzpicture}[thick, scale=0.5]
  \node(L1) at (6,-4){$USp(2N-2)$};
  \node (L2) at (14,-4){$SO(2N+2)$ ,};
 \path[every node/.style={font=\sffamily\small,
  		fill=white,inner sep=1pt}]
(L1) edge  (L2);
\end{tikzpicture}
\ee 
which is indeed one of the two families of theories we can get in weakly-coupled Type IIB string theory with a $\bbZ_2$ orbi-orientifold setup \cite{Giacomelli:2023qyc}. This represents a strong check of the picture we have proposed.

Let us now discuss one example with a $\bbZ_3$ orbifold, focusing on the theory with $M=2k=6$ and holonomy $n_{2'}=3$. Again we can start from the magnetic quiver \eqref{eq:ell2mingen} and activate an extra mass deformation to get rid of the 7-brane completely. This amounts to turn on FI terms in a way exactly analogous to \eqref{FI1modelpert}, which brings us to
\begin{align}
\begin{tikzpicture}\label{k3l2nobrane}
\filldraw[fill= white] (1,-9) circle [radius=0.1] node[below] {\scriptsize 1};
\filldraw[fill= white] (2,-9) circle [radius=0.1] node[below] {\scriptsize 2};
\filldraw[fill= white] (3,-9) circle [radius=0.1] node[below] {\scriptsize 3};
\filldraw[fill= white] (4,-9) circle [radius=0.1] node[below] {\scriptsize N};
\draw [thick] (1.1, -9) -- (1.9,-9) ;
\draw [thick] (2.1, -8.95) -- (2.9,-8.95) ;
\draw [thick] (2.1, -9.05) -- (2.9,-9.05) ;
\draw [thick] (2.4,-9) -- (2.6,-8.8);
\draw [thick] (2.4,-9) -- (2.6,-9.2);
\draw [thick] (3.1, -8.95) -- (3.9,-8.95);
\draw [thick] (3.1, -9.05) -- (3.9,-9.05);
\draw [thick] (4.05, -8.95) to [out=45,in=315,looseness=15] (4.05,-9.05);
\end{tikzpicture} 
\end{align}
According to the rules of Section \ref{CB7branedef}, the CB spectrum of the SCFT family \eqref{k3l2nobrane} is 
\be\label{cbex2}
(2r,2r+2,2r+3)_{r=1,\dots,N-1};\;\; (2N,2N+2)\,.
\ee 
For $N=1$ this is exactly the lagrangian rank-2 theory $SO(5)$ SQCD with three fundamental hypermultiplets \cite{Martone:2021drm}, whereby the $SU(3)$ symmetry we expect from the $\mathbb{C}^2/\bbZ_3$ orbifold in M-theory gets enhanced to $USp(6)$. From this case we learn that the $SU(3)$ subgroup, which is there for any $N$, has level $10$. For $N=2$ we have, instead, a rank-5 theory with CB operators of dimension $(2,4,4,5,6)$. Since all Lie groups with a Casimir operator of degree 5 also have one of degree 3, we conclude that the CB operator of dimension 5 arises from the matter sector, which must be a SCFT of rank at least 2 (since 5 is not an admissible scaling dimension at rank one). A good candidate for the matter sector is the rank-2 theory with symmetry $SU(10)_{10}$, which clearly has the $SU(3)_{10}$ subgroup we expect, and CB operators of dimension $(4,5)$. This hints at a gauge group with Casimirs $(2,4,6)$, which can be either $USp(6)$ or $SO(7)$. The $SO(7)$ option looks more appealing since gauging a $SO(7)\subset SU(7)\subset SU(10)_{10}$ makes the beta function precisely vanish and leaves as a commutant the $SU(3)_{10}$ we expect. As in the previous example, we can generalize the above to $N>2$ by considering as matter sector the class-$\mathcal{S}$ family of trinions labelled by the punctures 
 \be\label{trinion2}
 p_1=(1^{2N+1})\,;\quad p_2=(N+1,N)\,;\quad p_3=(N-1,N-1,1,1,1)\,,
\ee 
whose magnetic quiver is
\be\label{MMdef2}
\begin{tikzpicture}
\filldraw[fill= white] (-3,0) circle [radius=0.1] node[below] {\scriptsize 1};
\filldraw[fill= white] (-2,0) circle [radius=0.1] node[below] {\scriptsize 2};
\filldraw[fill= white] (-1,0) circle [radius=0.1] node[below] {\scriptsize 3};
\filldraw[fill= white] (0,0) circle [radius=0.1] node[below] {\scriptsize N+2};
\filldraw[fill= white] (1,0) circle [radius=0.1] node[below] {\scriptsize 2N+1};
\filldraw[fill= white] (2,0) circle [radius=0.1] node[below] {\scriptsize 2N};
\node[] at (3,0) {\dots};
\filldraw[fill= white] (4,0) circle [radius=0.1] node[below] {\scriptsize 2};
\filldraw[fill= white] (5,0) circle [radius=0.1] node[below] {\scriptsize 1};
\filldraw[fill= white] (1,1) circle [radius=0.1] node[above] {\scriptsize N};

\draw [thick] (-2.1, 0) -- (-2.9,0) ;
\draw [thick] (-0.1, 0) -- (-0.9,0) ;
\draw [thick] (-1.1, 0) -- (-1.9,0) ;
\draw [thick] (0.1, 0) -- (0.9,0) ;
\draw [thick] (1.1, 0) -- (1.9,0) ;
\draw [thick] (2.1, 0) -- (2.7,0) ;
\draw [thick] (3.3, 0) -- (3.9,0) ;
\draw [thick] (4.1, 0) -- (4.9,0) ;
\draw [thick] (1, 0.1) -- (1,0.9) ;
\end{tikzpicture}
\ee 
The global symmetry and the CB spectrum are 
\be
SU(2N+3)_{4N+2}\times SU(3)_{10}\times U(1)\,,\qquad\quad (4,5,6,\dots ,2N,2N+1)\,.
\ee
We introduce a $SO(2N+3)$ vector multiplet that gauges the $SO(2N+3)_{8N+4}\subset SU(2N+3)_{4N+2}$ symmetry of the matter sector. The gauging is conformal, the CB spectrum is precisely as in \eqref{cbex2}, we have the expected $SU(3)_{10}$ symmetry and the combination $24(c-a)$ is equal to 5 for every $N$, in agreement with \eqref{formulacc1}. 

In conclusion, our proposal for the worldvolume theory on the $D3$-branes in the absence of B-field is a $SO(2N+3)$ gauging of the SCFT \eqref{MMdef2}. If we now want to understand the effect of the B-field, we should turn on a mass as in \eqref{MMdef}, but this time for the whole $SU(3)$ symmetry carried by the $T(SU(3))$ tail on the left of \eqref{MMdef2}. By implementing the deformation as explained in \cite{Giacomelli:2022drw}, we find that the theory \eqref{MMdef2} flows in the infrared to a $U(2N-2)$ gauge theory with a hypermultiplet in the antisymmetric and $2N+3$ hypermultiplets in the fundamental representation. Upon the $SO(2N+3)$ gauging, we finally land on the lagrangian, although no longer conformal, family
\be\label{lagg2}
\begin{tikzpicture}[thick, scale=0.5]
  \node(L1) at (5,-4){$SO(2N+3)$};
  \node (L2) at (12,-4){$U(2N-2)$};
 \path[every node/.style={font=\sffamily\small,
  		fill=white,inner sep=1pt}]
		(L2) edge [loop, in=10, out=35, looseness=4] node[above=1mm] {$A$}  (L2)
(L1) edge  (L2);
\end{tikzpicture}
\ee  
We indeed recognize in \eqref{lagg2} one of the families of gauge theories we can construct by probing with $D3$-branes an $O3$-plane combined with a $\mathbb{C}^2/\bbZ_3$ orbifold in weakly-coupled Type IIB string theory \cite{Giacomelli:2023qyc}. 

As the two examples above clearly show, by activating the mass deformation which corresponds to turning on the B-field, we recover theories which naturally arise in the context of perturbative type IIB (whenever the axio-dilaton is not frozen at a value of order one). The limit of vanishing B-field leads instead to field theories with $SU(k)$ global symmetry which are in general strongly-coupled and are best described in the context of M-theory.

\subsection{$1$-form symmetries of orbi-S-fold theories}\label{Sec:1form}

In this section we would like to mention a few things regarding the possible $1$-form symmetries \cite{Gaiotto:2014kfa} of the theories arising from probing the orbi-S-fold geometries discussed in this paper. We will focus on backgrounds without 7-branes, because the presence of the latter introduces flavors which typically screen all the line operators, thus preventing the appearance of $1$-form symmetries. 

Theories with $\ell=2$ correspond to orbi-orientifold backgrounds (with $O3$-planes) and hence can be discussed very explicitly: A complete list of them up to orbifold order $k=4$ can be found in \cite{Giacomelli:2023qyc}, and we are now going to analyze their $1$-form symmetries. In particular, we will study the fate of such symmetries under higgsing processes which reduce the orbifold order: While such mechanisms can never lead to any breaking of these symmetries, because they are triggered by vevs of uncharged fields, we will observe that there may appear $1$-form-symmetry enhancements along the RG flow: More precisely, this happens whenever hypermultiplets transforming in the symmetric representation of unitary gauge groups condense, whereas vevs for all other types of fields (bifundamental, antisymmetric) leave the 1-form symmetry unchanged. 

As is well known, we have three families of $\mathcal{N}=4$ theories corresponding to pure $\ell=2$ backgrounds with no orbifold, depending on the type of $O3$ projection: They have gauge group $USp(2N), SO(2N+1), SO(2N)$. The first two families have a $\mathbb{Z}_2$ 1-form symmetry, whereas the third family (for $N>1$) has either a $\mathbb{Z}_2^2$ or a $\mathbb{Z}_4$ 1-form symmetry, according to whether $N$ is even or odd respectively.

Introducing the orbifold, as usual, produces quiver gauge theories with $\mathcal{N}=2$ supersymmetry, which can be analyzed with the technique of fractional branes (see \cite{Giacomelli:2023qyc} for the details). On $\mathbb{C}^2/\mathbb{Z}_2$ we have two inequivalent orientifold projections. They give the following quivers
\begin{center}
\begin{tikzpicture}[thick, scale=0.5]
  \node(L1) at (6,-4){$USp(2N)$};
  \node (L2) at (12,-4){$SO(M)$};
 \path[every node/.style={font=\sffamily\small,
  		fill=white,inner sep=1pt}]
(L1) edge  (L2);
\end{tikzpicture}\vspace{-0.9cm}
\end{center}\be\label{ZZ2O3} \ee\\
and
\begin{center}
\begin{tikzpicture}[thick, scale=0.8]
  \node(L1) at (10,-2){$U(N)$};
   \path[every node/.style={font=\sffamily\small,
  		fill=white,inner sep=1pt}]
(L1) edge [loop, in=10, out=40, looseness=4] node[above=2mm] {$A$}  (L1)
(L1) edge [loop, in=-40, out=-70, looseness=4] node[below=1.5mm] {$S$}  (L1);
\end{tikzpicture}\vspace{-1cm}\be\label{Z2MixedQuiver2}\ee
\end{center}
where in the second quiver the two loops indicate hypermultiplets in the 2-index symmetric and antisymmetric representations of $SU(N)$. It is worth pointing out that, while the second family, decoupling the central $U(1)$, is conformal for any value of $N$, the first quiver gives a family of SCFTs iff $M=2N+2$. This configuration corresponds to having an ``unbalanced'' quantity of fractional $D3$-branes associated to the two irreducible representations of $\mathbb{Z}_2$. This means, in turn, that by giving vev to the bifundamental hypermultiplet, all the fractional branes pair up into $2N$ integral branes, except for two of them which stay unbounded, and therefore remain stuck at the singularity, which cannot be resolved. The field theory one obtains at the end of the RG flow will contain, besides the $\mathcal{N}=4$ sector, a free $\mathcal{N}=2$ vector. Consequently the 1-form symmetry will become continuous, after all massive degrees of freedom have been integrated out. To avoid the appearance of such free vectors, from now on we will stick to the families characterized by an homogeneous distribution of fractional branes, at the price of losing conformality.

Consider the quiver \eqref{ZZ2O3} with $M=2N$. The 1-form symmetry of this family is $\mathbb{Z}_2^2$ or $\mathbb{Z}_4$ if $N$ is even or odd respectively, and it is generated by $\omega_{USp}\otimes\omega_{SO}$, with the two factors being the generators of the centers $Z(USp(2N))$ and $Z(SO(2N))$ respectively.\footnote{It is easy to see that the bifundamental hypermultiplet is neutral under $\omega_{USp}\otimes\omega_{SO}$.} Now we give vev to the bifundamental field: A quick count of degrees of freedom says that we flow to an $\mathcal{N}=4$ family (we have resolved the orbifold), which could be either the $USp(2N)$ or the $SO(2N)$ gauge theory. However, the first family has only a $\mathbb{Z}_2$ 1-form symmetry, contradicting the fact that the higgsing process does not break the original 1-form symmetry. Therefore, we conclude that, by resolving the orbifold, we land on the $SO(2N)$ gauge theory, which has exactly the same 1-form symmetry we started with.

The quiver \eqref{Z2MixedQuiver2} describes a family of theories with $\mathbb{Z}_2$ 1-form symmetry. Both the antisymmetric and the symmetric fields originate from open strings connecting the two fractional stacks of $D3$-branes, which for this particular projection happen to be the orientifold image of one another. Therefore vevs for these fields are responsible for resolving the orbifold. In particular, giving a vev to the hypermultiplet in the symmetric representation breaks $U(N)\to SO(N)$ and the antisymmetric hypermultiplet becomes the adjoint needed to enhance to $\mathcal{N}=4$. Hence, while for odd $N$ the 1-form symmetry remains the same, for even $N$ it enhances to $\mathbb{Z}_2^2$ if $N$ is a multiple of $4$ and to $\mathbb{Z}_4$ otherwise. Moreover, when $N$ is even, a vev for the antisymmetric field instead breaks $U(N)\to USp(N)$, leading to no enhancement of the 1-form symmetry.

On $\mathbb{C}^2/\mathbb{Z}_3$ we also have two inequivalent projections, producing the following quivers
\begin{center}
\begin{tikzpicture}[thick, scale=0.5]
  \node(L1) at (6,-4){$USp(2N)$};
  \node (L2) at (12,-4){$U(2N)$};
 \path[every node/.style={font=\sffamily\small,
  		fill=white,inner sep=1pt}]
		(L2) edge [loop, in=10, out=40, looseness=4] node[above=1mm] {$S$}  (L2) 
(L1) edge  (L2);
\end{tikzpicture}\vspace{-0.9cm}
\end{center}\be\label{ZZ3O3a} \ee\\
and
\begin{center}
\begin{tikzpicture}[thick, scale=0.5]
  \node(L1) at (6,-4){$SO(2N)$};
  \node (L2) at (12,-4){$U(2N)$};
 \path[every node/.style={font=\sffamily\small,
  		fill=white,inner sep=1pt}]
		(L2) edge [loop, in=10, out=40, looseness=4] node[above=1mm] {$A$}  (L2)
(L1) edge  (L2);
\end{tikzpicture}\vspace{-0.9cm}
\end{center}\be\label{ZZ3O3b} \ee\\
where for simplicity we chose the homogeneous distribution of fractional branes and neglected the analog of \eqref{ZZ3O3b} with odd number of colors. The family \eqref{ZZ3O3a} has a $\mathbb{Z}_2$ 1-form symmetry for any $N$, whereas the family \eqref{ZZ3O3b} has $\mathbb{Z}_2^2$, $\mathbb{Z}_4$ for $N$ even, odd respectively.

On the one hand, as expected, giving vev to the symmetric in \eqref{ZZ3O3a} and to the antisymmetric in \eqref{ZZ3O3b} has the same effect of partially resolving the orbifold, producing the quiver \eqref{ZZ2O3} with $M=2N$. Once again, we observe that vevs for the antisymmetric leaves the 1-form symmetry intact, whereas vevs for the symmetric enhances it. 

On the other hand, giving vev to the bifundamental field resolves the orbifold completely and produces the $\mathcal{N}=4$ theories associated to $2N$ integral $D3$-branes free to move in the orientifold background, i.e.~$USp(2N)$, $SO(2N)$ gauge theory for quiver \eqref{ZZ3O3a}, \eqref{ZZ3O3b} respectively. Under this operation, the 1-form symmetry stays unchanged.

Let us end this discussion by quickly looking at the $\mathbb{C}^2/\mathbb{Z}_4$ orbifold. Here we also have two inequivalent projections, producing the quivers
\begin{center}
\begin{tikzpicture}[thick, scale=0.5]
  \node(L1) at (6,-4){$USp(2N)$};
  \node (L2) at (12,-4){$U(2N)$};
  \node (L3) at (18,-4){$SO(2N)$};
 \path[every node/.style={font=\sffamily\small,
  		fill=white,inner sep=1pt}]
(L1) edge  (L2)
(L2) edge  (L3);
\end{tikzpicture}\vspace{-0.9cm}
\end{center}\be\label{ZZ4O3a} \ee\\
and
\begin{center}
\begin{tikzpicture}[thick, scale=0.8]
  \node(L2) at (12,-2){$U(2N)$};
  \node(L1) at (8,-2){$U(2N)$};
   \path[every node/.style={font=\sffamily\small,
  		fill=white,inner sep=1pt}]
	(L1) edge  (L2)	
(L2) edge [loop, in=10, out=40, looseness=4] node[above=2mm] {$S$}  (L2)
(L1) edge [loop, in=130, out=160, looseness=4] node[above=1.5mm] {$A$}  (L1);
\end{tikzpicture}\vspace{-1cm}\be\label{ZZ4O3b}\ee
\end{center}
where again, for simplicity, we restricted to the non-conformal configuration of equal numbers of fractional branes and neglected the analog of \eqref{ZZ4O3b} with odd number of colors.

The 1-form symmetry of the family \eqref{ZZ4O3a} is $\mathbb{Z}_2^2$, $\mathbb{Z}_4$ for $N$ even, odd respectively, and it is generated by the ``diagonal'' combination $\omega_{USp}\otimes\omega_{U}\otimes\omega_{SO}$, with the three factors being the generators of the centers $Z(USp(2N))$, $Z(U(2N))$, $Z(SO(2N))$ respectively. Giving vev to either of the two bifundamental fields lowers the orbifold order to $2$ and sends us back to \eqref{ZZ2O3}
with $M=2N$, thus leaving the 1-form symmetry unchanged.

The 1-form symmetry of the family \eqref{ZZ4O3b}, instead, is only $\mathbb{Z}_2$ for any $N$. Giving vev to the antisymmetric field partially resolves the orbifold and gives \eqref{ZZ3O3a}, thus leaving the 1-form symmetry intact. Giving vev to the symmetric field still lowers the orbifold order to $3$, but sends us to \eqref{ZZ3O3b}, hence giving an enhancement of the 1-form symmetry to $\mathbb{Z}_2^2$, $\mathbb{Z}_4$ for $N$ even, odd respectively. Finally, condensing the bifundamental hypermultiplet identifies the two gauge groups and produces the quiver \eqref{Z2MixedQuiver2}, corresponding to a $\mathbb{C}^2/\mathbb{Z}_2$ orbifold, still with the same $\mathbb{Z}_2$ 1-form symmetry.

Let us conclude this section by making a first rough estimate of the $1$-form symmetries of $\ell>2$ orbi-S-fold theories at generic orbifold order. The 1-form symmetries of the pure S-fold $\mathcal{N}=3$ theories are known to be $\mathbb{Z}_3,\mathbb{Z}_2,\mathbbm{1}$ for $\ell=3,4,6$ respectively \cite{Etheredge:2023ler}. Based on what we learned in the $\ell=2$ case, orbi-S-fold theories with $\ell=3$ and any $k$ enjoy either a $\mathbb{Z}_3$ or no 1-form symmetry, those with with $\ell=4$ and any $k$ either $\mathbb{Z}_2$ or none, and finally orbi-S-fold theories with $\ell=6$ at any orbifold order have no non-trivial 1-form symmetry. It would be interesting to verify these statements and actually compute the 1-form symmetries of the cases $\ell=3,4$ using techniques analogous to those employed in \cite{Etheredge:2023ler}.

\newpage

\section{A list of 4d SCFTs of rank $\bf{3}$}\label{Sec:Rank3}

The purpose of this section is to illustrate the effectiveness of the tools we have developed by showing what we can say about the landscape of rank-3 4d  $\mathcal{N}=2$ SCFTs, for which, at present, we only have scattered results in the literature (see e.g.~\cite{Caorsi:2018zsq, Cecotti:2021ouq, Cecotti:2023ksl, Argyres:2020nrr, Argyres:2020wmq}). In order to identify all possible theories sitting at the top of the mass-deformation trees, we need to consider 6d orbi-instantons with non-trivial $\sigma$ (see Section \ref{Sec:OrbiInst}). These are obtained from higher-rank models by higgsing the $SU(k)$ symmetry supported on the tail. We have not discussed these explicitly so far, but our analysis can be easily generalized to incorporate them. We will do it in this section for the cases relevant for the problem at hand.

We organize our findings by discussing each value of $\ell$ separately. We start by presenting a diagram showing how the various theories are organized into mass-deformation trees. We mostly label theories using their global symmetry,\footnote{To read off the global symmetry, we use the well-known criterion of balanced nodes of the magnetic quiver \cite{Gaiotto:2008ak}. However, in some cases, this criterion is known to deliver only a proper subgroup of the actual global symmetry (see e.g.~\cite{Gledhill:2021cbe}). In all cases the enhancement is known to occur, we will indicate the actual global symmetry, but there may be other cases of enhancement.} unless they have a name commonly used in the literature (such as $T_N$ or $(G,G')$), in which case we use the standard name. We draw in black the mass deformations we have discussed in detail in the previous sections and in blue the others. We color in red lagrangian theories. We then proceed by providing tables collecting all the relevant data, such as central charges, the scaling dimension of Coulomb-branch operators, which we denote $(\Delta_1,\Delta_2,\Delta_3)$, and the magnetic quiver, from which the Higgs-branch Hasse diagram can be derived. As for the $a$ and $c$ central charges, all the theories we discuss satisfy the Shapere-Tachikawa relation \cite{Shapere:2008zf}, and therefore the combination $2a-c$ is known once the Coulomb-branch spectrum is given: 
\be 8a-4c=2(\Delta_1+\Delta_2+\Delta_3)-3\,.\ee 
The quantity $24(c-a)$ is often, but not always, equal to the dimension of the Higgs branch, which can be read off from the magnetic quiver. We include this quantity explicitly in the tables. We also provide explicitly for lagrangian theories the gauge group and matter content. 

We do not perform a systematic scan of Argyres-Douglas theories and only include a few which we know how to reach via mass deformations. The data of these theories have been worked out in several references (see for instance \cite{Cecotti:2010fi, Cecotti:2012jx, Cecotti:2013lda, Xie:2012hs, Xie:2013jc, Wang:2015mra, Wang:2018gvb, Giacomelli:2017ckh}). Some well-known AD theories, such as $(A_3,A_3)$, $(A_1,E_6)$, and $(A_1,E_7)$, are not included in our list.

Overall we find 115 theories. Perhaps an indication of how much we capture of the rank-3 landscape is given by considering the lagrangian subclass: We find 17 lagrangian models out of the 31 we have at rank 3 \cite{Bhardwaj:2013qia}.

\subsection{Theories with $\ell=1$} 

These are the theories we find at rank 3 by combining orbifolds and 7-branes, without introducing S-folds. As was found in \cite{Giacomelli:2022drw} this leads to class-$\mathcal{S}$ theories (of type A untwisted) on the sphere, whose properties are well understood and discussed in detail in \cite{Chacaltana:2010ks}. The theories at the top of the deformation trees are the $T^2$ compactifications of 6d SCFTs with one curve supporting either a $SU(3)$ or $USp(4)$ gauge group, two curves one of which supports a $SU(2)$ gauge group, or three curves all with trivial gauge group. 

For this class of theories the associated magnetic quiver is simply-laced and therefore it is relatively easy to identify possible mass deformation by studying FI deformations of the quiver, which we analyze thoroughly. By looking at the diagram \eqref{diagramell=1} we can notice that, by restricting to the class of mass deformations we have discussed, we capture only half of the possible theories. However, we expect this to improve significantly for $\ell>1$. This intuition is based on the knowledge of the rank-2 landscape (see\cite{Martone:2021drm, Bourget:2021csg}): Also in that case for $\ell=1$ we reach only half of the theories by restricting to the mass deformations we have studied in the previous sections, but when we look at theories with $\ell>1$ we miss only 2 out of 40.  

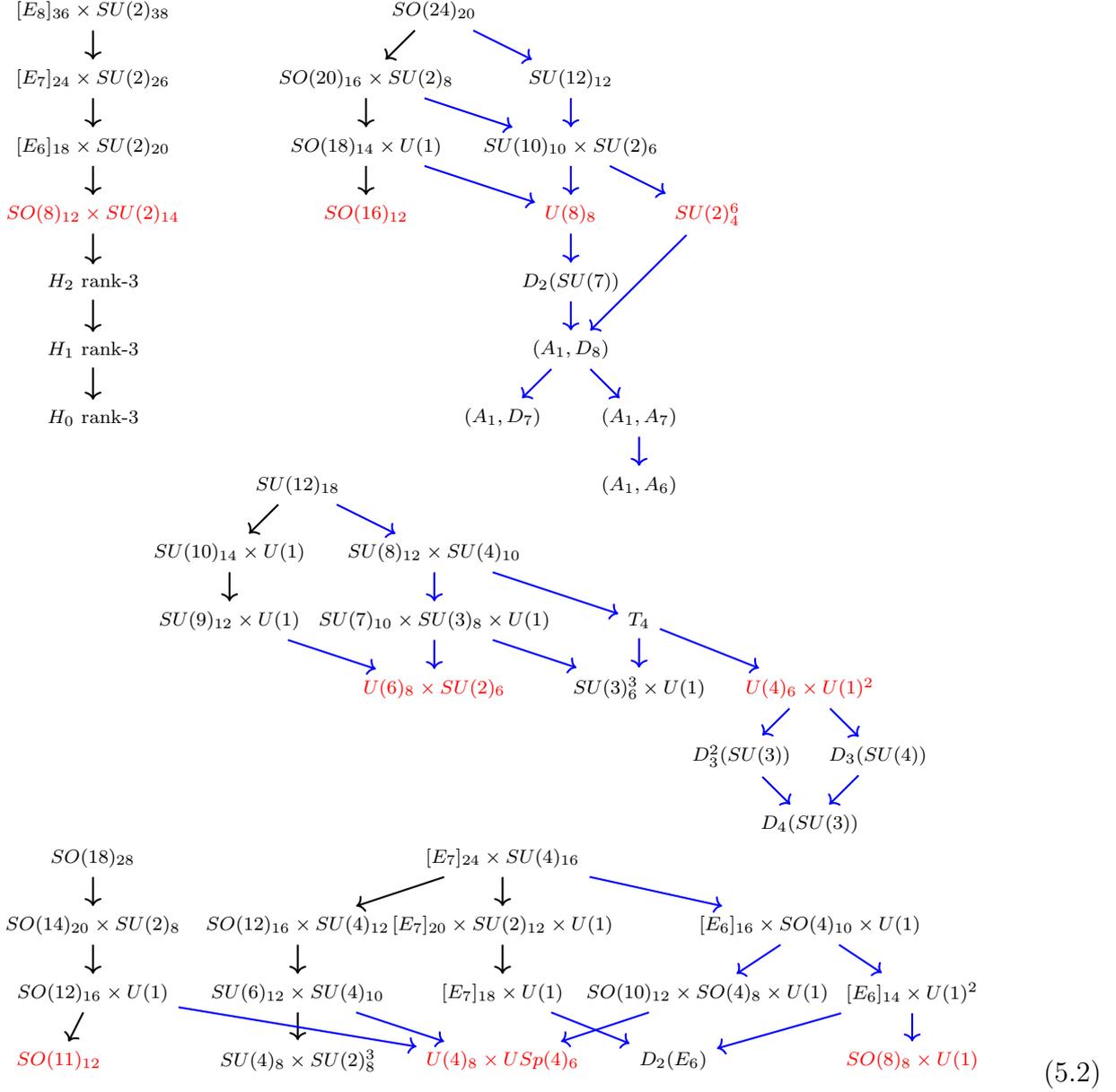
\begin{equation}\label{diagramell=1} 
\begin{tikzpicture} 
\node[] (a1) at (0,0) {\scriptsize $[E_8]_{36}\times SU(2)_{38}$}; 
\node[] (a2) at (0,-1) {\scriptsize $[E_7]_{24}\times SU(2)_{26}$}; 
\node[] (a3) at (0,-2) {\scriptsize $[E_6]_{18}\times SU(2)_{20}$}; 
\node[] (a4) at (0,-3) {\scriptsize {\color{red} $SO(8)_{12}\times SU(2)_{14}$}}; 
\node[] (a5) at (0,-4) {\scriptsize $H_2$ rank-3}; 
\node[] (a6) at (0,-5) {\scriptsize $H_1$ rank-3};
\node[] (a7) at (0,-6) {\scriptsize $H_0$ rank-3};   

\node[] (b1) at (3,-7) {\scriptsize $SU(12)_{18}$}; 
\node[] (b2) at (2,-8) {\scriptsize $SU(10)_{14}\times U(1)$}; 
\node[] (b3) at (2,-9) {\scriptsize $SU(9)_{12}\times U(1)$}; 
\node[] (b4) at (5,-10) {\scriptsize {\color{red} $U(6)_{8}\times SU(2)_6$}}; 
\node[] (B2) at (5,-8) {\scriptsize $S_6$}; 
\node[] (B3) at (8,-9) {\scriptsize $T_4$};
\node[] (B4) at (10.5,-10) {\scriptsize {\color{red}$U(4)_6\times U(1)^2$}};
\node[] (B5) at (5,-9) {\scriptsize $R_{1,5}$}; %
\node[] (B6) at (8,-10) {\scriptsize $SU(3)_{6}^3\times U(1)$}; 
\node[] (Bb) at (10.5,-12) {\scriptsize $D_4(SU(3))$};
\node[] (Bbb) at (9.5,-11) {\scriptsize $D_3^2(SU(3))$};
\node[] (BB) at (11.5,-11) {\scriptsize $D_3(SU(4))$}; 

\node[] (c1) at (5,0) {\scriptsize $SO(24)_{20}$}; 
\node[] (c2) at (4,-1) {\scriptsize $SO(20)_{16}\times SU(2)_8$};
\node[] (c3) at (4,-2) {\scriptsize $R_{2,7}$}; 
\node[] (c4) at (4,-3) {\scriptsize {\color{red} $SO(16)_{12}$}}; 
\node[] (C2) at (7,-1) {\scriptsize $SU(12)_{12}$};
\node[] (C3) at (7,-2) {\scriptsize $R_{0,5}$ }; %$SU(10)_{10}\times SU(2)_6$ 
\node[] (C4) at (7,-3) {\scriptsize {\color{red}$U(8)_8$}};
\node[] (C10) at (9,-3) {\scriptsize {\color{red}$SU(2)_4^6$}};
\node[] (C5) at (7,-4) {\scriptsize $D_2(SU(7))$};
\node[] (C6) at (7,-5) {\scriptsize $(A_1,D_8)$}; 
\node[] (C7) at (6,-6) {\scriptsize $(A_1,D_7)$};
\node[] (C8) at (8,-6) {\scriptsize $(A_1,A_7)$};
\node[] (C9) at (8,-7) {\scriptsize $(A_1,A_6)$};

\node[] (d1) at (6,-12.5) {\scriptsize $[E_7]_{24}\times SO(7)_{16}$}; 
\node[] (d2) at (3,-13.5) {\scriptsize $SO(12)_{16}\times SO(7)_{12}$}; 
\node[] (d3) at (3,-14.5) {\scriptsize $SU(6)_{12}\times SO(7)_{10}$}; 
\node[] (d4) at (3,-15.5) {\scriptsize $SU(4)_8\times SU(2)_8^3$}; 
\node[] (d5) at (6,-15.5) {\scriptsize {\color{red}$U(4)_8\times USp(4)_6$}};
\node[] (D2) at (6,-13.5) {\scriptsize $[E_7]_{20}\times SU(2)_{12}\times U(1)$}; 
\node[] (D3) at (6,-14.5) {\scriptsize $[E_7]_{18}\times U(1)$};
\node[] (D4) at (8.5,-15.5) {\scriptsize $D_2(E_6)$};
\node[] (E2) at (10.5,-13.5) {\scriptsize $[E_6]_{16}\times SO(4)_{10}\times U(1)$};
\node[] (E3) at (12,-14.5) {\scriptsize $[E_6]_{14}\times U(1)^2$}; 
\node[] (E4) at (12,-15.5) {\scriptsize {\color{red}$SO(8)_8\times U(1)$}}; 
\node[] (E5) at (9,-14.5) {\scriptsize $SO(10)_{12}\times SO(4)_8\times U(1)$}; 
\node[] (e1) at (0,-12.5) {\scriptsize $SO(19)_{28}$}; 
\node[] (e2) at (0,-13.5) {\scriptsize $SO(15)_{20}\times SU(2)_{8}$}; 
\node[] (e3) at (0,-14.5) {\scriptsize $SO(13)_{16}\times U(1)$}; 
\node[] (e4) at (-0.5,-15.5) {\scriptsize {\color{red}$SO(11)_{12}$}};

\draw[->, thick] (a1)--(a2); 
\draw[->, thick] (a2)--(a3); 
\draw[->, thick] (a3)--(a4); 
\draw[->, thick] (a4)--(a5); 
\draw[->, thick] (a5)--(a6); 
\draw[->, thick] (a6)--(a7); 
\draw[->, thick] (b1)--(b2); 
\draw[->, thick] (b2)--(b3); 
\draw[->, thick, color=blue] (b3)--(b4);
\draw[->, thick, color=blue] (b1)--(B2);
\draw[->, thick, color=blue] (B2)--(B3);
\draw[->, thick, color=blue] (B3)--(B4);
\draw[->, thick, color=blue] (B2)--(B5);
\draw[->, thick, color=blue] (B5)--(B6);
\draw[->, thick, color=blue] (B3)--(B6);
\draw[->, thick, color=blue] (B5)--(b4);
\draw[->, thick, color=blue] (Bbb)--(Bb);
\draw[->, thick, color=blue] (BB)--(Bb);
\draw[->, thick, color=blue] (B4)--(Bbb);
\draw[->, thick, color=blue] (B4)--(BB);
\draw[->, thick] (c1)--(c2); 
\draw[->, thick] (c2)--(c3); 
\draw[->, thick] (c3)--(c4);
\draw[->, thick, color=blue] (c1)--(C2);
\draw[->, thick, color=blue] (C2)--(C3);
\draw[->, thick, color=blue] (C3)--(C4);
\draw[->, thick, color=blue] (C4)--(C5);
\draw[->, thick, color=blue] (C5)--(C6);
\draw[->, thick, color=blue] (C6)--(C7);
\draw[->, thick, color=blue] (C6)--(C8);
\draw[->, thick, color=blue] (C8)--(C9);
\draw[->, thick, color=blue] (C3)--(C10);
\draw[->, thick, color=blue] (C10)--(C6);
\draw[->, thick, color=blue] (c2)--(C3);
\draw[->, thick, color=blue] (c3)--(C4);
\draw[->, thick] (d1)--(d2); 
\draw[->, thick] (d2)--(d3); 
\draw[->, thick] (d3)--(d4);
\draw[->, thick, color=blue] (d3)--(d5);
\draw[->, thick] (d1)--(D2);
\draw[->, thick] (D2)--(D3);
\draw[->, thick, color=blue] (D3)--(D4);
\draw[->, thick, color=blue] (d1)--(E2);
\draw[->, thick, color=blue] (E2)--(E3);
\draw[->, thick, color=blue] (E3)--(E4);
\draw[->, thick, color=blue] (E3)--(D4);
\draw[->, thick, color=blue] (E2)--(E5); 
\draw[->, thick, color=blue] (E5)--(d5);
\draw[->, thick] (e1)--(e2); 
\draw[->, thick] (e2)--(e3); 
\draw[->, thick] (e3)--(e4);
\draw[->, thick, color=blue] (e3)--(d5);
\end{tikzpicture} 
\end{equation} 
\vspace{10pt}
\begin{itemize} 
\item $[E_8]_{36}\times SU(2)_{38}$: $T^2$ reduction of the rank-3 E-string theory. 
\item $SO(24)_{20}$: $T^2$ reduction of the $\bbZ_4$ orbi-instanton with $(n_{2'}=2)$ and one tensor.
\item $SU(12)_{18}$: $T^2$ reduction of the $\bbZ_3$ orbi-instanton with $(n_{3'}=1)$ and one tensor.
\item $[E_7]_{24}\times SO(7)_{16}$: $T^2$ reduction of the $\bbZ_2$ orbi-instanton with $(n_{2}=1)$ and two tensors.
\item  $SO(19)_{28}$: $T^2$ reduction of the $\bbZ_2$ orbi-instanton with $(n_{2'}=1)$, two tensors and non-trivial $\sigma=(2,2)$.
\end{itemize}

\paragraph{Descendants of the $[E_8]_{36}\times SU(2)_{38}$ theory} 

$$\begin{array} {|c|c|c|c|} 
\hline 
\text{Theory}& (\Delta_1, \Delta_2, \Delta_3)& 24(c-a)& \text{Magnetic Quiver}\\ 
\hline 
[E_8]_{36}\times SU(2)_{39} & (6, 12, 18)& 90& \begin{tikzpicture}
 \filldraw[fill= white] (0,0) circle [radius=0.1] node[below] {\scriptsize 1};
\filldraw[fill= white] (1,0) circle [radius=0.1] node[below] {\scriptsize 3};
\filldraw[fill= white] (2,0) circle [radius=0.1] node[below] {\scriptsize 6}; 
 \filldraw[fill= white] (3,0) circle [radius=0.1] node[below] {\scriptsize 9};
\filldraw[fill= white] (4,0) circle [radius=0.1] node[below] {\scriptsize 12};
\filldraw[fill= white] (5,0) circle [radius=0.1] node[below] {\scriptsize 15}; 
 \filldraw[fill= white] (6,0) circle [radius=0.1] node[below] {\scriptsize 18};
\filldraw[fill= white] (7,0) circle [radius=0.1] node[below] {\scriptsize 12};
\filldraw[fill= white] (8,0) circle [radius=0.1] node[below] {\scriptsize 6}; 
\filldraw[fill= white] (6,1) circle [radius=0.1] node[left] {\scriptsize 9}; 
\draw [thick] (0.1,0)--(0.9,0);
\draw [thick] (1.1,0)--(1.9,0);
\draw [thick] (2.1,0)--(2.9,0);
\draw [thick] (3.1,0)--(3.9,0);
\draw [thick] (4.1,0)--(4.9,0);
\draw [thick] (5.1,0)--(5.9,0);
\draw [thick] (6.1,0)--(6.9,0);
\draw [thick] (7.1,0)--(7.9,0); 
\draw [thick] (6,0.1)--(6,0.9);
\end{tikzpicture}\\ 
\hline 
[E_7]_{24}\times SU(2)_{27} & (4, 8, 12)& 54& \begin{tikzpicture}
 \filldraw[fill= white] (0,0) circle [radius=0.1] node[below] {\scriptsize 1};
\filldraw[fill= white] (1,0) circle [radius=0.1] node[below] {\scriptsize 3};
\filldraw[fill= white] (2,0) circle [radius=0.1] node[below] {\scriptsize 6}; 
 \filldraw[fill= white] (3,0) circle [radius=0.1] node[below] {\scriptsize 9};
\filldraw[fill= white] (4,0) circle [radius=0.1] node[below] {\scriptsize 12};
\filldraw[fill= white] (5,0) circle [radius=0.1] node[below] {\scriptsize 9}; 
 \filldraw[fill= white] (6,0) circle [radius=0.1] node[below] {\scriptsize 6};
\filldraw[fill= white] (7,0) circle [radius=0.1] node[below] {\scriptsize 3};
\filldraw[fill= white] (4,1) circle [radius=0.1] node[left] {\scriptsize 6}; 
\draw [thick] (0.1,0)--(0.9,0);
\draw [thick] (1.1,0)--(1.9,0);
\draw [thick] (2.1,0)--(2.9,0);
\draw [thick] (3.1,0)--(3.9,0);
\draw [thick] (4.1,0)--(4.9,0);
\draw [thick] (5.1,0)--(5.9,0);
\draw [thick] (6.1,0)--(6.9,0);
\draw [thick] (4,0.1)--(4,0.9);
\end{tikzpicture}\\ 
\hline 
[E_6]_{18}\times SU(2)_{21} & (3, 6, 9)& 36& \begin{tikzpicture}
 \filldraw[fill= white] (0,0) circle [radius=0.1] node[below] {\scriptsize 1};
\filldraw[fill= white] (1,0) circle [radius=0.1] node[below] {\scriptsize 3};
\filldraw[fill= white] (2,0) circle [radius=0.1] node[below] {\scriptsize 6}; 
 \filldraw[fill= white] (3,0) circle [radius=0.1] node[below] {\scriptsize 9};
\filldraw[fill= white] (4,0) circle [radius=0.1] node[below] {\scriptsize 6};
\filldraw[fill= white] (5,0) circle [radius=0.1] node[below] {\scriptsize 3}; 
 \filldraw[fill= white] (3,1) circle [radius=0.1] node[left] {\scriptsize 6};
\filldraw[fill= white] (4,1) circle [radius=0.1] node[right] {\scriptsize 3}; 
\draw [thick] (0.1,0)--(0.9,0);
\draw [thick] (1.1,0)--(1.9,0);
\draw [thick] (2.1,0)--(2.9,0);
\draw [thick] (3.1,0)--(3.9,0);
\draw [thick] (4.1,0)--(4.9,0);
\draw [thick] (3.1,1)--(3.9,1);
\draw [thick] (3,0.1)--(3,0.9);
\end{tikzpicture}\\ 
\hline 
SO(8)_{12}\times SU(2)_{15} & (2, 4, 6)& 18& \begin{tikzpicture}
 \filldraw[fill= white] (0,0) circle [radius=0.1] node[below] {\scriptsize 1};
\filldraw[fill= white] (1,0) circle [radius=0.1] node[below] {\scriptsize 3};
\filldraw[fill= white] (2,0) circle [radius=0.1] node[below] {\scriptsize 6}; 
 \filldraw[fill= white] (3,0) circle [radius=0.1] node[below] {\scriptsize 3}; 
 \filldraw[fill= white] (1,1) circle [radius=0.1] node[left] {\scriptsize 3};
\filldraw[fill= white] (3,1) circle [radius=0.1] node[right] {\scriptsize 3}; 
\draw [thick] (0.1,0)--(0.9,0);
\draw [thick] (1.1,0)--(1.9,0);
\draw [thick] (2.1,0)--(2.9,0);
\draw [thick] (2.05,0.05)--(2.95,0.95);
\draw [thick] (1.95,0.05)--(1.05,0.95);
\end{tikzpicture}\\ 
\hline 
\text{$H_2$ rank-3} & (\frac{3}{2}, 3, \frac{9}{2})& 9& \begin{tikzpicture}
 \filldraw[fill= white] (0,0) circle [radius=0.1] node[below] {\scriptsize 1};
\filldraw[fill= white] (1,0) circle [radius=0.1] node[below] {\scriptsize 3};
\filldraw[fill= white] (2,0) circle [radius=0.1] node[below] {\scriptsize 3}; 
 \filldraw[fill= white] (2,1) circle [radius=0.1] node[right] {\scriptsize 3};  
\draw [thick] (0.1,0)--(0.9,0);
\draw [thick] (1.1,0)--(1.9,0);
\draw [thick] (1.05,0.05)--(1.95,0.95);
\draw [thick] (2,0.1)--(2,0.9);
\end{tikzpicture}\\ 
\hline 
\text{$H_1$ rank-3} & (\frac{4}{3}, \frac{8}{3}, 4)& 6& \begin{tikzpicture}
 \filldraw[fill= white] (0,0) circle [radius=0.1] node[below] {\scriptsize 1};
\filldraw[fill= white] (1,0) circle [radius=0.1] node[below] {\scriptsize 3};
\filldraw[fill= white] (2,0) circle [radius=0.1] node[below] {\scriptsize 3};   
\node [] at (0.5,0.5) {};
\draw [thick] (0.1,0)--(0.9,0);
\draw [thick] (1.05,0.05)--(1.95,0.05);
\draw [thick] (1.05,-0.05)--(1.95,-0.05);
\end{tikzpicture}\\ 
\hline 
\text{$H_0$ rank-3} & (\frac{6}{5}, \frac{12}{5}, \frac{18}{5})& \frac{18}{5}& \begin{tikzpicture}
 \filldraw[fill= white] (0,0) circle [radius=0.1] node[below] {\scriptsize 1};
\filldraw[fill= white] (1,0) circle [radius=0.1] node[below] {\scriptsize 3};
\draw [thick] (0.1,0)--(0.9,0);
\draw [thick] (1.05, 0.05) to [out=45,in=315,looseness=15] (1.05,-0.05);
\end{tikzpicture}\\ 
\hline 
\end{array}$$ 
This family is a bit special since the Type IIB setup, which involves a flat 7-brane, always includes a free hypermultiplet. This is also included in the magnetic quivers displayed above. We decided to provide in the table the central charges for the theory with the free hypermultiplet included. The central charges of the interacting part only can be obtained by subtracting one from $24(c-a)$ and by decreasing by one unit the level of the $SU(2)$ factor of the global symmetry. These data were first computed in \cite{Aharony:2007dj}.

\paragraph{Descendants of the $SO(24)_{20}$ theory} 

$$\begin{array} {|c|c|c|c|} 
\hline 
\text{Theory}& (\Delta_1, \Delta_2, \Delta_3)& 24(c-a)& \text{Magnetic Quiver}\\ 
\hline 
SO(24)_{20} & (6, 8, 10)& 67& \begin{tikzpicture}[scale=0.80]
\filldraw[fill= white] (-3,0) circle [radius=0.1] node[below] {\scriptsize 1};
\filldraw[fill= white] (-2,0) circle [radius=0.1] node[below] {\scriptsize 2};
\filldraw[fill= white] (-1,0) circle [radius=0.1] node[below] {\scriptsize 3};
\filldraw[fill= white] (0,0) circle [radius=0.1] node[below] {\scriptsize 4};
\filldraw[fill= white] (1,0) circle [radius=0.1] node[below] {\scriptsize 5};
\filldraw[fill= white] (2,0) circle [radius=0.1] node[below] {\scriptsize 6}; 
 \filldraw[fill= white] (3,0) circle [radius=0.1] node[below] {\scriptsize 7};
\filldraw[fill= white] (4,0) circle [radius=0.1] node[below] {\scriptsize 8};
\filldraw[fill= white] (5,0) circle [radius=0.1] node[below] {\scriptsize 9}; 
 \filldraw[fill= white] (6,0) circle [radius=0.1] node[below] {\scriptsize 10};
\filldraw[fill= white] (7,0) circle [radius=0.1] node[below] {\scriptsize 6};
\filldraw[fill= white] (8,0) circle [radius=0.1] node[below] {\scriptsize 2}; 
\filldraw[fill= white] (6,1) circle [radius=0.1] node[left] {\scriptsize 5}; 
\draw [thick] (-0.1,0)--(-0.9,0);
\draw [thick] (-1.1,0)--(-1.9,0);
\draw [thick] (-2.1,0)--(-2.9,0);
\draw [thick] (0.1,0)--(0.9,0);
\draw [thick] (1.1,0)--(1.9,0);
\draw [thick] (2.1,0)--(2.9,0);
\draw [thick] (3.1,0)--(3.9,0);
\draw [thick] (4.1,0)--(4.9,0);
\draw [thick] (5.1,0)--(5.9,0);
\draw [thick] (6.1,0)--(6.9,0);
\draw [thick] (7.1,0)--(7.9,0); 
\draw [thick] (6,0.1)--(6,0.9);
\end{tikzpicture}\\ 
\hline 
SO(20)_{16}\times SU(2)_{8} & (4, 6, 8)& 47& \begin{tikzpicture}[scale=0.80]
\filldraw[fill= white] (-3,0) circle [radius=0.1] node[below] {\scriptsize 1};
\filldraw[fill= white] (-2,0) circle [radius=0.1] node[below] {\scriptsize 2};
\filldraw[fill= white] (-1,0) circle [radius=0.1] node[below] {\scriptsize 3};
 \filldraw[fill= white] (0,0) circle [radius=0.1] node[below] {\scriptsize 4};
\filldraw[fill= white] (1,0) circle [radius=0.1] node[below] {\scriptsize 5};
\filldraw[fill= white] (2,0) circle [radius=0.1] node[below] {\scriptsize 6}; 
 \filldraw[fill= white] (3,0) circle [radius=0.1] node[below] {\scriptsize 7};
\filldraw[fill= white] (4,0) circle [radius=0.1] node[below] {\scriptsize 8};
\filldraw[fill= white] (5,0) circle [radius=0.1] node[below] {\scriptsize 5}; 
 \filldraw[fill= white] (6,0) circle [radius=0.1] node[below] {\scriptsize 2};
\filldraw[fill= white] (7,0) circle [radius=0.1] node[below] {\scriptsize 1};
\filldraw[fill= white] (4,1) circle [radius=0.1] node[left] {\scriptsize 4}; 
\draw [thick] (-0.1,0)--(-0.9,0);
\draw [thick] (-1.1,0)--(-1.9,0);
\draw [thick] (-2.1,0)--(-2.9,0);
\draw [thick] (0.1,0)--(0.9,0);
\draw [thick] (1.1,0)--(1.9,0);
\draw [thick] (2.1,0)--(2.9,0);
\draw [thick] (3.1,0)--(3.9,0);
\draw [thick] (4.1,0)--(4.9,0);
\draw [thick] (5.1,0)--(5.9,0);
\draw [thick] (6.1,0)--(6.9,0);
\draw [thick] (4,0.1)--(4,0.9);
\end{tikzpicture}\\ 
\hline 
SU(12)_{12} & (4, 5, 6)& 37& \begin{tikzpicture}[scale=0.80]
\filldraw[fill= white] (-3,0) circle [radius=0.1] node[below] {\scriptsize 1};
\filldraw[fill= white] (-2,0) circle [radius=0.1] node[below] {\scriptsize 2};
\filldraw[fill= white] (-1,0) circle [radius=0.1] node[below] {\scriptsize 3};
 \filldraw[fill= white] (0,0) circle [radius=0.1] node[below] {\scriptsize 4};
\filldraw[fill= white] (1,0) circle [radius=0.1] node[below] {\scriptsize 5};
\filldraw[fill= white] (2,0) circle [radius=0.1] node[below] {\scriptsize 6}; 
 \filldraw[fill= white] (3,0) circle [radius=0.1] node[below] {\scriptsize 5};
\filldraw[fill= white] (4,0) circle [radius=0.1] node[below] {\scriptsize 4};
\filldraw[fill= white] (5,0) circle [radius=0.1] node[below] {\scriptsize 3}; 
 \filldraw[fill= white] (6,0) circle [radius=0.1] node[below] {\scriptsize 2};
\filldraw[fill= white] (7,0) circle [radius=0.1] node[below] {\scriptsize 1};
\filldraw[fill= white] (2,1) circle [radius=0.1] node[left] {\scriptsize 2}; 
\draw [thick] (-0.1,0)--(-0.9,0);
\draw [thick] (-1.1,0)--(-1.9,0);
\draw [thick] (-2.1,0)--(-2.9,0);
\draw [thick] (0.1,0)--(0.9,0);
\draw [thick] (1.1,0)--(1.9,0);
\draw [thick] (2.1,0)--(2.9,0);
\draw [thick] (3.1,0)--(3.9,0);
\draw [thick] (4.1,0)--(4.9,0);
\draw [thick] (5.1,0)--(5.9,0);
\draw [thick] (6.1,0)--(6.9,0);
\draw [thick] (2,0.1)--(2,0.9);
\end{tikzpicture}\\ 
\hline 
R_{2,7} & (3, 5, 7)& 37& \begin{tikzpicture}
\filldraw[fill= white] (-3,0) circle [radius=0.1] node[below] {\scriptsize 1};
\filldraw[fill= white] (-2,0) circle [radius=0.1] node[below] {\scriptsize 2};
\filldraw[fill= white] (-1,0) circle [radius=0.1] node[below] {\scriptsize 3};
 \filldraw[fill= white] (0,0) circle [radius=0.1] node[below] {\scriptsize 4};
\filldraw[fill= white] (1,0) circle [radius=0.1] node[below] {\scriptsize 5};
\filldraw[fill= white] (2,0) circle [radius=0.1] node[below] {\scriptsize 6}; 
 \filldraw[fill= white] (3,0) circle [radius=0.1] node[below] {\scriptsize 7};
\filldraw[fill= white] (4,0) circle [radius=0.1] node[below] {\scriptsize 4};
\filldraw[fill= white] (5,0) circle [radius=0.1] node[below] {\scriptsize 1}; 
 \filldraw[fill= white] (3,1) circle [radius=0.1] node[left] {\scriptsize 4};
\filldraw[fill= white] (4,1) circle [radius=0.1] node[right] {\scriptsize 1}; 
\draw [thick] (-0.1,0)--(-0.9,0);
\draw [thick] (-1.1,0)--(-1.9,0);
\draw [thick] (-2.1,0)--(-2.9,0);
\draw [thick] (0.1,0)--(0.9,0);
\draw [thick] (1.1,0)--(1.9,0);
\draw [thick] (2.1,0)--(2.9,0);
\draw [thick] (3.1,0)--(3.9,0);
\draw [thick] (4.1,0)--(4.9,0);
\draw [thick] (3.1,1)--(3.9,1);
\draw [thick] (3,0.1)--(3,0.9);
\end{tikzpicture}\\ 
\hline 
R_{0,5} & (3, 4, 5)& 27& \begin{tikzpicture}
\filldraw[fill= white] (-2,0) circle [radius=0.1] node[below] {\scriptsize 1};
\filldraw[fill= white] (-1,0) circle [radius=0.1] node[below] {\scriptsize 2};
 \filldraw[fill= white] (0,0) circle [radius=0.1] node[below] {\scriptsize 3};
\filldraw[fill= white] (1,0) circle [radius=0.1] node[below] {\scriptsize 4};
\filldraw[fill= white] (2,0) circle [radius=0.1] node[below] {\scriptsize 5}; 
 \filldraw[fill= white] (3,0) circle [radius=0.1] node[below] {\scriptsize 4};
\filldraw[fill= white] (4,0) circle [radius=0.1] node[below] {\scriptsize 3};
\filldraw[fill= white] (5,0) circle [radius=0.1] node[below] {\scriptsize 2}; 
 \filldraw[fill= white] (6,0) circle [radius=0.1] node[below] {\scriptsize 1};
\filldraw[fill= white] (3,1) circle [radius=0.1] node[right] {\scriptsize 1};
\filldraw[fill= white] (2,1) circle [radius=0.1] node[left] {\scriptsize 2}; 
\draw [thick] (-0.1,0)--(-0.9,0);
\draw [thick] (-1.1,0)--(-1.9,0);
\draw [thick] (0.1,0)--(0.9,0);
\draw [thick] (1.1,0)--(1.9,0);
\draw [thick] (2.1,0)--(2.9,0);
\draw [thick] (3.1,0)--(3.9,0);
\draw [thick] (4.1,0)--(4.9,0);
\draw [thick] (5.1,0)--(5.9,0);
\draw [thick] (2.1,1)--(2.9,1);
\draw [thick] (2,0.1)--(2,0.9);
\end{tikzpicture}\\  %SU(10)_{10}\times SU(2)_6
\hline 
SO(16)_{12} & (2, 4, 6)& 27& \begin{tikzpicture}
\filldraw[fill= white] (-3,0) circle [radius=0.1] node[below] {\scriptsize 1};
\filldraw[fill= white] (-2,0) circle [radius=0.1] node[below] {\scriptsize 2};
 \filldraw[fill= white] (-1,0) circle [radius=0.1] node[below] {\scriptsize 3};
 \filldraw[fill= white] (0,0) circle [radius=0.1] node[below] {\scriptsize 4};
\filldraw[fill= white] (1,0) circle [radius=0.1] node[below] {\scriptsize 5};
\filldraw[fill= white] (2,0) circle [radius=0.1] node[below] {\scriptsize 6}; 
 \filldraw[fill= white] (3,0) circle [radius=0.1] node[below] {\scriptsize 3}; 
 \filldraw[fill= white] (1,1) circle [radius=0.1] node[left] {\scriptsize 1};
\filldraw[fill= white] (3,1) circle [radius=0.1] node[right] {\scriptsize 3}; 
\draw [thick] (-0.1,0)--(-0.9,0);
\draw [thick] (-1.1,0)--(-1.9,0);
\draw [thick] (-2.1,0)--(-2.9,0);
\draw [thick] (0.1,0)--(0.9,0);
\draw [thick] (1.1,0)--(1.9,0);
\draw [thick] (2.1,0)--(2.9,0);
\draw [thick] (2.05,0.05)--(2.95,0.95);
\draw [thick] (1.95,0.05)--(1.05,0.95);
\end{tikzpicture}\\ 
\hline 
U(8)_{8} & (2, 3, 4)& 17& \begin{tikzpicture}
\filldraw[fill= white] (-3,0) circle [radius=0.1] node[below] {\scriptsize 1};
\filldraw[fill= white] (-2,0) circle [radius=0.1] node[below] {\scriptsize 2};
 \filldraw[fill= white] (-1,0) circle [radius=0.1] node[below] {\scriptsize 3};
 \filldraw[fill= white] (0,0) circle [radius=0.1] node[below] {\scriptsize 4};
\filldraw[fill= white] (1,0) circle [radius=0.1] node[below] {\scriptsize 3};
\filldraw[fill= white] (2,0) circle [radius=0.1] node[below] {\scriptsize 2}; 
 \filldraw[fill= white] (3,0) circle [radius=0.1] node[below] {\scriptsize 1}; 
 \filldraw[fill= white] (-1,1) circle [radius=0.1] node[left] {\scriptsize 1};
\filldraw[fill= white] (1,1) circle [radius=0.1] node[right] {\scriptsize 1}; 
\draw [thick] (-0.1,0)--(-0.9,0);
\draw [thick] (-1.1,0)--(-1.9,0);
\draw [thick] (-2.1,0)--(-2.9,0);
\draw [thick] (0.1,0)--(0.9,0);
\draw [thick] (1.1,0)--(1.9,0);
\draw [thick] (2.1,0)--(2.9,0);
\draw [thick] (0.05,0.05)--(0.95,0.95);
\draw [thick] (-0.05,0.05)--(-0.95,0.95);
\end{tikzpicture}\\ 
\hline 
SU(2)_{4}^6 & (2, 2, 2)& 7& \begin{tikzpicture}
\filldraw[fill= white] (-0.5,1) circle [radius=0.1] node[left] {\scriptsize 1};
 \filldraw[fill= white] (-1,0) circle [radius=0.1] node[below] {\scriptsize 1};
 \filldraw[fill= white] (0,0) circle [radius=0.1] node[below] {\scriptsize 2};
\filldraw[fill= white] (1,0) circle [radius=0.1] node[below] {\scriptsize 1};
\filldraw[fill= white] (0.5,1) circle [radius=0.1] node[right] {\scriptsize 1};  
 \filldraw[fill= white] (-1,1) circle [radius=0.1] node[left] {\scriptsize 1};
\filldraw[fill= white] (1,1) circle [radius=0.1] node[right] {\scriptsize 1}; 
\draw [thick] (-0.1,0)--(-0.9,0);
\draw [thick] (0.03,0.07)--(0.47,0.93);
\draw [thick] (-0.03,0.07)--(-0.47,0.93);
\draw [thick] (0.1,0)--(0.9,0);
\draw [thick] (0.05,0.05)--(0.95,0.95);
\draw [thick] (-0.05,0.05)--(-0.95,0.95);
\end{tikzpicture}\\ 
\hline 
D_2(SU(7)) & (\frac{3}{2}, \frac{5}{2}, \frac{7}{2})& 12& \begin{tikzpicture}
 \filldraw[fill= white] (-1,0) circle [radius=0.1] node[below] {\scriptsize 1};
 \filldraw[fill= white] (0,0) circle [radius=0.1] node[below] {\scriptsize 2};
\filldraw[fill= white] (1,0) circle [radius=0.1] node[below] {\scriptsize 3};
\filldraw[fill= white] (2,0) circle [radius=0.1] node[below] {\scriptsize 3}; 
 \filldraw[fill= white] (3,0) circle [radius=0.1] node[below] {\scriptsize 2};
 \filldraw[fill= white] (4,0) circle [radius=0.1] node[below] {\scriptsize 1};
 \filldraw[fill= white] (2,1) circle [radius=0.1] node[right] {\scriptsize 1};  
\draw [thick] (-0.1,0)--(-0.9,0);
\draw [thick] (0.1,0)--(0.9,0);
\draw [thick] (1.1,0)--(1.9,0);
\draw [thick] (2.1,0)--(2.9,0);
\draw [thick] (3.1,0)--(3.9,0);
\draw [thick] (1.05,0.05)--(1.95,0.95);
\draw [thick] (2,0.1)--(2,0.9);
\end{tikzpicture}\\ 
\hline
(A_1,D_8) & (\frac{5}{4}, \frac{6}{4}, \frac{7}{4}) & 2 & \begin{tikzpicture} 
\filldraw[fill= white] (1,0) circle [radius=0.1] node[above] {\scriptsize 1};
\filldraw[fill= white] (2,0) circle [radius=0.1] node[above] {\scriptsize 1}; 
\filldraw[fill= white] (1.5,-1) circle [radius=0.1] node[left] {\scriptsize 1};   
\draw [thick] (1.05,0.05)--(1.95,0.05);
\draw [thick] (1.05,0)--(1.95,0);
\draw [thick] (1.05,-0.05)--(1.95,-0.05); 
\draw [thick] (1.53,-0.93)--(1.97,-0.07); 
\draw [thick] (1.47,-0.93)--(1.03,-0.07);
\end{tikzpicture}\\ 
\hline
(A_1,D_7) & (\frac{8}{7}, \frac{10}{7}, \frac{12}{7})& \frac{9}{7}& \begin{tikzpicture}
\filldraw[fill= white] (1,0) circle [radius=0.1] node[below] {\scriptsize 1};
\filldraw[fill= white] (2,0) circle [radius=0.1] node[below] {\scriptsize 1};   
\node [] at (3,0) {+}; 
\node [] at (5,0) {\scriptsize \text{2 Free Hypermultiplets}};
\node [] at (0.5,0.5) {};
\draw [thick] (1.05,0.05)--(1.95,0.05);
\draw [thick] (1.05,-0.05)--(1.95,-0.05);
\end{tikzpicture}\\ 
\hline  
(A_1,A_7) & (\frac{6}{5}, \frac{7}{5}, \frac{8}{5})& 1& \begin{tikzpicture}
\filldraw[fill= white] (1,0) circle [radius=0.1] node[below] {\scriptsize 1};
\filldraw[fill= white] (2,0) circle [radius=0.1] node[below] {\scriptsize 1};   
\node [] at (1.5,0.5) {};
\draw [thick] (1.03,0.07)--(1.97,0.07);
\draw [thick] (1.07,0.02)--(1.93,0.02);
\draw [thick] (1.07,-0.02)--(1.93,-0.02);
\draw [thick] (1.03,-0.07)--(1.97,-0.07);
\end{tikzpicture}\\ 
\hline 
(A_1,A_6) & (\frac{10}{9}, \frac{12}{9}, \frac{14}{9})& \frac{1}{3}& {\scriptsize \text{3 Free Hypermultiplets}}\\ 
\hline 
\end{array}$$

\paragraph{Descendants of the $SU(12)_{18}$ theory} 

$$\begin{array} {|c|c|c|c|} 
\hline 
\text{Theory}& (\Delta_1, \Delta_2, \Delta_3)& 24(c-a)& \text{Magnetic Quiver}\\ 
\hline 
SU(12)_{18} & (6, 8, 9)& 57& \begin{tikzpicture}[scale=0.80]
\filldraw[fill= white] (-2,0) circle [radius=0.1] node[below] {\scriptsize 1};
\filldraw[fill= white] (-1,0) circle [radius=0.1] node[below] {\scriptsize 2};
\filldraw[fill= white] (0,0) circle [radius=0.1] node[below] {\scriptsize 3};
\filldraw[fill= white] (1,0) circle [radius=0.1] node[below] {\scriptsize 4};
\filldraw[fill= white] (2,0) circle [radius=0.1] node[below] {\scriptsize 5}; 
 \filldraw[fill= white] (3,0) circle [radius=0.1] node[below] {\scriptsize 6};
\filldraw[fill= white] (4,0) circle [radius=0.1] node[below] {\scriptsize 7};
\filldraw[fill= white] (5,0) circle [radius=0.1] node[below] {\scriptsize 8}; 
 \filldraw[fill= white] (6,0) circle [radius=0.1] node[below] {\scriptsize 9};
\filldraw[fill= white] (7,0) circle [radius=0.1] node[below] {\scriptsize 6};
\filldraw[fill= white] (8,0) circle [radius=0.1] node[below] {\scriptsize 3}; 
\filldraw[fill= white] (6,1) circle [radius=0.1] node[left] {\scriptsize 4}; 
\draw [thick] (-0.1,0)--(-0.9,0);
\draw [thick] (-1.1,0)--(-1.9,0);
\draw [thick] (0.1,0)--(0.9,0);
\draw [thick] (1.1,0)--(1.9,0);
\draw [thick] (2.1,0)--(2.9,0);
\draw [thick] (3.1,0)--(3.9,0);
\draw [thick] (4.1,0)--(4.9,0);
\draw [thick] (5.1,0)--(5.9,0);
\draw [thick] (6.1,0)--(6.9,0);
\draw [thick] (7.1,0)--(7.9,0); 
\draw [thick] (6,0.1)--(6,0.9);
\end{tikzpicture}\\ 
\hline 
SU(10)_{14}\times U(1) & (4, 6, 7)& 39& \begin{tikzpicture}[scale=0.80]
\filldraw[fill= white] (-2,0) circle [radius=0.1] node[below] {\scriptsize 1};
\filldraw[fill= white] (-1,0) circle [radius=0.1] node[below] {\scriptsize 2};
 \filldraw[fill= white] (0,0) circle [radius=0.1] node[below] {\scriptsize 3};
\filldraw[fill= white] (1,0) circle [radius=0.1] node[below] {\scriptsize 4};
\filldraw[fill= white] (2,0) circle [radius=0.1] node[below] {\scriptsize 5}; 
 \filldraw[fill= white] (3,0) circle [radius=0.1] node[below] {\scriptsize 6};
\filldraw[fill= white] (4,0) circle [radius=0.1] node[below] {\scriptsize 7};
\filldraw[fill= white] (5,0) circle [radius=0.1] node[below] {\scriptsize 5}; 
 \filldraw[fill= white] (6,0) circle [radius=0.1] node[below] {\scriptsize 3};
\filldraw[fill= white] (7,0) circle [radius=0.1] node[below] {\scriptsize 1};
\filldraw[fill= white] (4,1) circle [radius=0.1] node[left] {\scriptsize 3}; 
\draw [thick] (-0.1,0)--(-0.9,0);
\draw [thick] (-1.1,0)--(-1.9,0);
\draw [thick] (0.1,0)--(0.9,0);
\draw [thick] (1.1,0)--(1.9,0);
\draw [thick] (2.1,0)--(2.9,0);
\draw [thick] (3.1,0)--(3.9,0);
\draw [thick] (4.1,0)--(4.9,0);
\draw [thick] (5.1,0)--(5.9,0);
\draw [thick] (6.1,0)--(6.9,0);
\draw [thick] (4,0.1)--(4,0.9);
\end{tikzpicture}\\ 
\hline 
S_6 & (4, 5, 6)& 33& \begin{tikzpicture}[scale=0.80]
\filldraw[fill= white] (-3,0) circle [radius=0.1] node[below] {\scriptsize 1};
\filldraw[fill= white] (-2,0) circle [radius=0.1] node[below] {\scriptsize 2};
\filldraw[fill= white] (-1,0) circle [radius=0.1] node[below] {\scriptsize 3};
 \filldraw[fill= white] (0,0) circle [radius=0.1] node[below] {\scriptsize 4};
\filldraw[fill= white] (1,0) circle [radius=0.1] node[below] {\scriptsize 5};
\filldraw[fill= white] (2,0) circle [radius=0.1] node[below] {\scriptsize 6}; 
 \filldraw[fill= white] (3,0) circle [radius=0.1] node[below] {\scriptsize 4};
\filldraw[fill= white] (4,0) circle [radius=0.1] node[below] {\scriptsize 3};
\filldraw[fill= white] (5,0) circle [radius=0.1] node[below] {\scriptsize 2}; 
 \filldraw[fill= white] (6,0) circle [radius=0.1] node[below] {\scriptsize 1};
\filldraw[fill= white] (2,1) circle [radius=0.1] node[left] {\scriptsize 3}; 
\draw [thick] (-0.1,0)--(-0.9,0);
\draw [thick] (-1.1,0)--(-1.9,0);
\draw [thick] (-2.1,0)--(-2.9,0);
\draw [thick] (0.1,0)--(0.9,0);
\draw [thick] (1.1,0)--(1.9,0);
\draw [thick] (2.1,0)--(2.9,0);
\draw [thick] (3.1,0)--(3.9,0);
\draw [thick] (4.1,0)--(4.9,0);
\draw [thick] (5.1,0)--(5.9,0);
\draw [thick] (2,0.1)--(2,0.9);
\end{tikzpicture}\\ 
\hline 
SU(9)_{12}\times U(1) & (3, 5, 6)& 30& \begin{tikzpicture}
\filldraw[fill= white] (-2,0) circle [radius=0.1] node[below] {\scriptsize 1};
\filldraw[fill= white] (-1,0) circle [radius=0.1] node[below] {\scriptsize 2};
 \filldraw[fill= white] (0,0) circle [radius=0.1] node[below] {\scriptsize 3};
\filldraw[fill= white] (1,0) circle [radius=0.1] node[below] {\scriptsize 4};
\filldraw[fill= white] (2,0) circle [radius=0.1] node[below] {\scriptsize 5}; 
 \filldraw[fill= white] (3,0) circle [radius=0.1] node[below] {\scriptsize 6};
\filldraw[fill= white] (4,0) circle [radius=0.1] node[below] {\scriptsize 4};
\filldraw[fill= white] (5,0) circle [radius=0.1] node[below] {\scriptsize 2}; 
 \filldraw[fill= white] (3,1) circle [radius=0.1] node[left] {\scriptsize 3};
\filldraw[fill= white] (4,1) circle [radius=0.1] node[right] {\scriptsize 1}; 
\draw [thick] (-0.1,0)--(-0.9,0);
\draw [thick] (-1.1,0)--(-1.9,0);
\draw [thick] (0.1,0)--(0.9,0);
\draw [thick] (1.1,0)--(1.9,0);
\draw [thick] (2.1,0)--(2.9,0);
\draw [thick] (3.1,0)--(3.9,0);
\draw [thick] (4.1,0)--(4.9,0);
\draw [thick] (3.1,1)--(3.9,1);
\draw [thick] (3,0.1)--(3,0.9);
\end{tikzpicture}\\ 
\hline 
 R_{1,5}& (3, 4, 5)& 24& \begin{tikzpicture}
\filldraw[fill= white] (-2,0) circle [radius=0.1] node[below] {\scriptsize 1};
\filldraw[fill= white] (-1,0) circle [radius=0.1] node[below] {\scriptsize 2};
 \filldraw[fill= white] (0,0) circle [radius=0.1] node[below] {\scriptsize 3};
\filldraw[fill= white] (1,0) circle [radius=0.1] node[below] {\scriptsize 4};
\filldraw[fill= white] (2,0) circle [radius=0.1] node[below] {\scriptsize 5}; 
 \filldraw[fill= white] (3,0) circle [radius=0.1] node[below] {\scriptsize 3};
\filldraw[fill= white] (4,0) circle [radius=0.1] node[below] {\scriptsize 2};
\filldraw[fill= white] (5,0) circle [radius=0.1] node[below] {\scriptsize 1}; 
\filldraw[fill= white] (3,1) circle [radius=0.1] node[right] {\scriptsize 1};
\filldraw[fill= white] (2,1) circle [radius=0.1] node[left] {\scriptsize 3}; 
\draw [thick] (-0.1,0)--(-0.9,0);
\draw [thick] (-1.1,0)--(-1.9,0);
\draw [thick] (0.1,0)--(0.9,0);
\draw [thick] (1.1,0)--(1.9,0);
\draw [thick] (2.1,0)--(2.9,0);
\draw [thick] (3.1,0)--(3.9,0);
\draw [thick] (4.1,0)--(4.9,0);
\draw [thick] (2.1,1)--(2.9,1);
\draw [thick] (2,0.1)--(2,0.9);
\end{tikzpicture}\\  %SU(7)_{10}\times SU(3)_8\times U(1)
\hline 
T_4\;\;\; (SU(4)_{8}^3)& (3, 4, 4)& 21& \begin{tikzpicture}
\filldraw[fill= white] (3,1) circle [radius=0.1] node[above] {\scriptsize 2};
\filldraw[fill= white] (-1,0) circle [radius=0.1] node[below] {\scriptsize 1};
 \filldraw[fill= white] (0,0) circle [radius=0.1] node[below] {\scriptsize 2};
\filldraw[fill= white] (1,0) circle [radius=0.1] node[below] {\scriptsize 3};
\filldraw[fill= white] (2,0) circle [radius=0.1] node[below] {\scriptsize 4}; 
 \filldraw[fill= white] (3,0) circle [radius=0.1] node[below] {\scriptsize 3};
\filldraw[fill= white] (4,0) circle [radius=0.1] node[below] {\scriptsize 2};
\filldraw[fill= white] (5,0) circle [radius=0.1] node[below] {\scriptsize 1}; 
\filldraw[fill= white] (4,1) circle [radius=0.1] node[right] {\scriptsize 1};
\filldraw[fill= white] (2,1) circle [radius=0.1] node[left] {\scriptsize 3}; 
\draw [thick] (-0.1,0)--(-0.9,0);
\draw [thick] (3.1,1)--(3.9,1);
\draw [thick] (0.1,0)--(0.9,0);
\draw [thick] (1.1,0)--(1.9,0);
\draw [thick] (2.1,0)--(2.9,0);
\draw [thick] (3.1,0)--(3.9,0);
\draw [thick] (4.1,0)--(4.9,0);
\draw [thick] (2.1,1)--(2.9,1);
\draw [thick] (2,0.1)--(2,0.9);
\end{tikzpicture}\\ 
\hline 
U(6)_{8}\times SU(2)_6 & (2, 3, 4)& 15& \begin{tikzpicture}
\filldraw[fill= white] (-2,0) circle [radius=0.1] node[below] {\scriptsize 1};
 \filldraw[fill= white] (-1,0) circle [radius=0.1] node[below] {\scriptsize 2};
 \filldraw[fill= white] (0,0) circle [radius=0.1] node[below] {\scriptsize 3};
\filldraw[fill= white] (1,0) circle [radius=0.1] node[below] {\scriptsize 4};
\filldraw[fill= white] (2,0) circle [radius=0.1] node[below] {\scriptsize 2}; 
 \filldraw[fill= white] (3,0) circle [radius=0.1] node[below] {\scriptsize 1}; 
 \filldraw[fill= white] (0,1) circle [radius=0.1] node[left] {\scriptsize 1};
\filldraw[fill= white] (2,1) circle [radius=0.1] node[right] {\scriptsize 2}; 
\draw [thick] (-0.1,0)--(-0.9,0);
\draw [thick] (-1.1,0)--(-1.9,0);
\draw [thick] (0.1,0)--(0.9,0);
\draw [thick] (1.1,0)--(1.9,0);
\draw [thick] (2.1,0)--(2.9,0);
\draw [thick] (1.05,0.05)--(1.95,0.95);
\draw [thick] (0.95,0.05)--(0.05,0.95);
\end{tikzpicture}\\ 
\hline 
SU(3)_{6}^3\times U(1) & (2, 3, 3)& 12& \begin{tikzpicture}
\filldraw[fill= white] (-2,0) circle [radius=0.1] node[below] {\scriptsize 1};
 \filldraw[fill= white] (-1,0) circle [radius=0.1] node[below] {\scriptsize 2};
 \filldraw[fill= white] (0,0) circle [radius=0.1] node[below] {\scriptsize 3};
\filldraw[fill= white] (1,0) circle [radius=0.1] node[below] {\scriptsize 2};
\filldraw[fill= white] (2,0) circle [radius=0.1] node[below] {\scriptsize 1}; 
 \filldraw[fill= white] (0,1) circle [radius=0.1] node[left] {\scriptsize 2}; 
 \filldraw[fill= white] (-1,1) circle [radius=0.1] node[left] {\scriptsize 1};
\filldraw[fill= white] (1,1) circle [radius=0.1] node[right] {\scriptsize 1}; 
\draw [thick] (-0.1,0)--(-0.9,0);
\draw [thick] (-1.1,0)--(-1.9,0);
\draw [thick] (0.1,0)--(0.9,0);
\draw [thick] (1.1,0)--(1.9,0);
\draw [thick] (0,0.1)--(0,0.9); 
\draw [thick] (0.1,1)--(0.9,1);
\draw [thick] (-0.05,0.05)--(-0.95,0.95);
\end{tikzpicture}\\ 
\hline 
U(4)_{6}\times U(1)^2 & (2, 2, 3)& 9& \begin{tikzpicture}
\filldraw[fill= white] (-2,0) circle [radius=0.1] node[below] {\scriptsize 1};
 \filldraw[fill= white] (-1,0) circle [radius=0.1] node[below] {\scriptsize 2};
 \filldraw[fill= white] (0,0) circle [radius=0.1] node[below] {\scriptsize 3};
\filldraw[fill= white] (1,0) circle [radius=0.1] node[below] {\scriptsize 1};
\filldraw[fill= white] (0.5,1) circle [radius=0.1] node[right] {\scriptsize 1};  
 \filldraw[fill= white] (-0.5,1) circle [radius=0.1] node[left] {\scriptsize 1};
\filldraw[fill= white] (1,1) circle [radius=0.1] node[right] {\scriptsize 1}; 
\draw [thick] (-1.1,0)--(-1.9,0);
\draw [thick] (-0.1,0)--(-0.9,0);
\draw [thick] (0.03,0.07)--(0.47,0.93);
\draw [thick] (-0.03,0.07)--(-0.47,0.93);
\draw [thick] (0.1,0)--(0.9,0);
\draw [thick] (0.05,0.05)--(0.95,0.95);
\end{tikzpicture}\\ 
\hline 
D_3(SU(4)) & (\frac{4}{3}, \frac{5}{3}, \frac{8}{3})& 5& \begin{tikzpicture}
 \filldraw[fill= white] (0,0) circle [radius=0.1] node[below] {\scriptsize 1};
\filldraw[fill= white] (1,0) circle [radius=0.1] node[below] {\scriptsize 2};
\filldraw[fill= white] (2,0) circle [radius=0.1] node[below] {\scriptsize 2}; 
 \filldraw[fill= white] (2,1) circle [radius=0.1] node[right] {\scriptsize 1};  
\draw [thick] (0.1,0)--(0.9,0);
\draw [thick] (1.1,0)--(1.9,0);
\draw [thick] (1.05,0.05)--(1.95,0.95);
\draw [thick] (1.95,0.05)--(1.95,0.95);
\draw [thick] (2.05,0.05)--(2.05,0.95);
\end{tikzpicture}\\ 
\hline
D_3^2(SU(3)) & (\frac{4}{3}, \frac{5}{3}, \frac{7}{3})& 4& \begin{tikzpicture}
 \filldraw[fill= white] (3,0) circle [radius=0.1] node[below] {\scriptsize 1};
\filldraw[fill= white] (1,0) circle [radius=0.1] node[below] {\scriptsize 1};
\filldraw[fill= white] (2,0) circle [radius=0.1] node[below] {\scriptsize 2}; 
 \filldraw[fill= white] (2,1) circle [radius=0.1] node[right] {\scriptsize 1};  
\draw [thick] (2.1,0)--(2.9,0);
\draw [thick] (1.1,0)--(1.9,0);
\draw [thick] (2.95,0.05)--(2.05,0.95);
\draw [thick] (1.95,0.05)--(1.95,0.95);
\draw [thick] (2.05,0.05)--(2.05,0.95);
\end{tikzpicture}\\ 
\hline
D_4(SU(3)) & (\frac{5}{4}, \frac{6}{4}, \frac{9}{4})& 3& \begin{tikzpicture}
\filldraw[fill= white] (0,0) circle [radius=0.1] node[below] {\scriptsize 1};  
\filldraw[fill= white] (1,0) circle [radius=0.1] node[below] {\scriptsize 2};
\filldraw[fill= white] (2,0) circle [radius=0.1] node[below] {\scriptsize 1};   
\node [] at (0.5,0.5) {};
\draw [thick] (0.1,0)--(0.9,0);
\draw [thick] (1.05,0.05)--(1.95,0.05);
\draw [thick] (1.1,0)--(1.9,0);
\draw [thick] (1.05,-0.05)--(1.95,-0.05);
\end{tikzpicture}\\ 
\hline  
\end{array}$$

\paragraph{Descendants of the $[E_7]_{24}\times SO(7)_{16}$ theory} 

$$\begin{array} {|c|c|c|c|} 
\hline 
\text{Theory}& (\Delta_1, \Delta_2, \Delta_3)& 24(c-a)& \text{Magnetic Quiver}\\ 
\hline 
[E_7]_{24}\times SO(7)_{16} & (6, 8, 12)& 63& \begin{tikzpicture}[scale=0.80]
\filldraw[fill= white] (-1,0) circle [radius=0.1] node[below] {\scriptsize 1};
\filldraw[fill= white] (0,0) circle [radius=0.1] node[below] {\scriptsize 2};
\filldraw[fill= white] (1,0) circle [radius=0.1] node[below] {\scriptsize 3};
\filldraw[fill= white] (2,0) circle [radius=0.1] node[below] {\scriptsize 4}; 
 \filldraw[fill= white] (3,0) circle [radius=0.1] node[below] {\scriptsize 6};
\filldraw[fill= white] (4,0) circle [radius=0.1] node[below] {\scriptsize 8};
\filldraw[fill= white] (5,0) circle [radius=0.1] node[below] {\scriptsize 10}; 
 \filldraw[fill= white] (6,0) circle [radius=0.1] node[below] {\scriptsize 12};
\filldraw[fill= white] (7,0) circle [radius=0.1] node[below] {\scriptsize 8};
\filldraw[fill= white] (8,0) circle [radius=0.1] node[below] {\scriptsize 4}; 
\filldraw[fill= white] (6,1) circle [radius=0.1] node[left] {\scriptsize 6}; 
\draw [thick] (-0.1,0)--(-0.9,0);
\draw [thick] (0.1,0)--(0.9,0);
\draw [thick] (1.1,0)--(1.9,0);
\draw [thick] (2.1,0)--(2.9,0);
\draw [thick] (3.1,0)--(3.9,0);
\draw [thick] (4.1,0)--(4.9,0);
\draw [thick] (5.1,0)--(5.9,0);
\draw [thick] (6.1,0)--(6.9,0);
\draw [thick] (7.1,0)--(7.9,0); 
\draw [thick] (6,0.1)--(6,0.9);
\end{tikzpicture}\\ 
\hline 
SO(12)_{16}\times SO(7)_{12} & (4, 6, 8)& 39& \begin{tikzpicture}[scale=0.80]
\filldraw[fill= white] (-1,0) circle [radius=0.1] node[below] {\scriptsize 1};
 \filldraw[fill= white] (0,0) circle [radius=0.1] node[below] {\scriptsize 2};
\filldraw[fill= white] (1,0) circle [radius=0.1] node[below] {\scriptsize 3};
\filldraw[fill= white] (2,0) circle [radius=0.1] node[below] {\scriptsize 4}; 
 \filldraw[fill= white] (3,0) circle [radius=0.1] node[below] {\scriptsize 6};
\filldraw[fill= white] (4,0) circle [radius=0.1] node[below] {\scriptsize 8};
\filldraw[fill= white] (5,0) circle [radius=0.1] node[below] {\scriptsize 6}; 
 \filldraw[fill= white] (6,0) circle [radius=0.1] node[below] {\scriptsize 4};
\filldraw[fill= white] (7,0) circle [radius=0.1] node[below] {\scriptsize 2};
\filldraw[fill= white] (4,1) circle [radius=0.1] node[left] {\scriptsize 4}; 
\draw [thick] (-0.1,0)--(-0.9,0);
\draw [thick] (0.1,0)--(0.9,0);
\draw [thick] (1.1,0)--(1.9,0);
\draw [thick] (2.1,0)--(2.9,0);
\draw [thick] (3.1,0)--(3.9,0);
\draw [thick] (4.1,0)--(4.9,0);
\draw [thick] (5.1,0)--(5.9,0);
\draw [thick] (6.1,0)--(6.9,0);
\draw [thick] (4,0.1)--(4,0.9);
\end{tikzpicture}\\ 
\hline 
[E_7]_{20}\times SU(2)_{12}\times U(1) & (4, 6, 10)& 47& \begin{tikzpicture}[scale=0.80]
\filldraw[fill= white] (-2,0) circle [radius=0.1] node[below] {\scriptsize 1};
\filldraw[fill= white] (-1,0) circle [radius=0.1] node[below] {\scriptsize 2};
 \filldraw[fill= white] (0,0) circle [radius=0.1] node[below] {\scriptsize 4};
\filldraw[fill= white] (1,0) circle [radius=0.1] node[below] {\scriptsize 6};
\filldraw[fill= white] (2,0) circle [radius=0.1] node[below] {\scriptsize 8}; 
 \filldraw[fill= white] (3,0) circle [radius=0.1] node[below] {\scriptsize 10};
\filldraw[fill= white] (4,0) circle [radius=0.1] node[below] {\scriptsize 7};
\filldraw[fill= white] (5,0) circle [radius=0.1] node[below] {\scriptsize 4}; 
 \filldraw[fill= white] (6,0) circle [radius=0.1] node[below] {\scriptsize 1};
\filldraw[fill= white] (3,1) circle [radius=0.1] node[left] {\scriptsize 5}; 
\draw [thick] (-0.1,0)--(-0.9,0);
\draw [thick] (-1.1,0)--(-1.9,0);
\draw [thick] (0.1,0)--(0.9,0);
\draw [thick] (1.1,0)--(1.9,0);
\draw [thick] (2.1,0)--(2.9,0);
\draw [thick] (3.1,0)--(3.9,0);
\draw [thick] (4.1,0)--(4.9,0);
\draw [thick] (5.1,0)--(5.9,0);
\draw [thick] (3,0.1)--(3,0.9);
\end{tikzpicture}\\ 
\hline 
[E_6]_{16}\times SO(4)_{10}\times U(1) & (4, 5, 8)& 37& \begin{tikzpicture}[scale=0.80]
\filldraw[fill= white] (-2,0) circle [radius=0.1] node[below] {\scriptsize 1};
\filldraw[fill= white] (-1,0) circle [radius=0.1] node[below] {\scriptsize 2};
 \filldraw[fill= white] (0,0) circle [radius=0.1] node[below] {\scriptsize 4};
\filldraw[fill= white] (1,0) circle [radius=0.1] node[below] {\scriptsize 6};
\filldraw[fill= white] (2,0) circle [radius=0.1] node[below] {\scriptsize 8}; 
 \filldraw[fill= white] (3,0) circle [radius=0.1] node[below] {\scriptsize 6};
\filldraw[fill= white] (4,0) circle [radius=0.1] node[below] {\scriptsize 4};
\filldraw[fill= white] (5,0) circle [radius=0.1] node[below] {\scriptsize 2}; 
 \filldraw[fill= white] (6,0) circle [radius=0.1] node[below] {\scriptsize 1};
\filldraw[fill= white] (2,1) circle [radius=0.1] node[left] {\scriptsize 4}; 
\draw [thick] (-0.1,0)--(-0.9,0);
\draw [thick] (-1.1,0)--(-1.9,0);
\draw [thick] (0.1,0)--(0.9,0);
\draw [thick] (1.1,0)--(1.9,0);
\draw [thick] (2.1,0)--(2.9,0);
\draw [thick] (3.1,0)--(3.9,0);
\draw [thick] (4.1,0)--(4.9,0);
\draw [thick] (5.1,0)--(5.9,0);
\draw [thick] (2,0.1)--(2,0.9);
\end{tikzpicture}\\ 
\hline 
SU(6)_{12}\times SO(7)_{10} & (3, 5, 6)& 27& \begin{tikzpicture}
\filldraw[fill= white] (-1,0) circle [radius=0.1] node[below] {\scriptsize 1};
 \filldraw[fill= white] (0,0) circle [radius=0.1] node[below] {\scriptsize 2};
\filldraw[fill= white] (1,0) circle [radius=0.1] node[below] {\scriptsize 3};
\filldraw[fill= white] (2,0) circle [radius=0.1] node[below] {\scriptsize 4}; 
 \filldraw[fill= white] (3,0) circle [radius=0.1] node[below] {\scriptsize 6};
\filldraw[fill= white] (4,0) circle [radius=0.1] node[below] {\scriptsize 4};
\filldraw[fill= white] (5,0) circle [radius=0.1] node[below] {\scriptsize 2}; 
 \filldraw[fill= white] (3,1) circle [radius=0.1] node[left] {\scriptsize 4};
\filldraw[fill= white] (4,1) circle [radius=0.1] node[right] {\scriptsize 2}; 
\draw [thick] (-0.1,0)--(-0.9,0);
\draw [thick] (0.1,0)--(0.9,0);
\draw [thick] (1.1,0)--(1.9,0);
\draw [thick] (2.1,0)--(2.9,0);
\draw [thick] (3.1,0)--(3.9,0);
\draw [thick] (4.1,0)--(4.9,0);
\draw [thick] (3.1,1)--(3.9,1);
\draw [thick] (3,0.1)--(3,0.9);
\end{tikzpicture}\\ 
\hline 
[E_7]_{18}\times U(1) & (3, 5, 9)& 39& \begin{tikzpicture}
\filldraw[fill= white] (-2,0) circle [radius=0.1] node[below] {\scriptsize 1};
\filldraw[fill= white] (-1,0) circle [radius=0.1] node[below] {\scriptsize 3};
 \filldraw[fill= white] (0,0) circle [radius=0.1] node[below] {\scriptsize 5};
\filldraw[fill= white] (1,0) circle [radius=0.1] node[below] {\scriptsize 7};
\filldraw[fill= white] (2,0) circle [radius=0.1] node[below] {\scriptsize 9}; 
 \filldraw[fill= white] (3,0) circle [radius=0.1] node[below] {\scriptsize 6};
\filldraw[fill= white] (4,0) circle [radius=0.1] node[below] {\scriptsize 3}; 
\filldraw[fill= white] (3,1) circle [radius=0.1] node[right] {\scriptsize 1};
\filldraw[fill= white] (2,1) circle [radius=0.1] node[left] {\scriptsize 5}; 
\draw [thick] (-0.1,0)--(-0.9,0);
\draw [thick] (-1.1,0)--(-1.9,0);
\draw [thick] (0.1,0)--(0.9,0);
\draw [thick] (1.1,0)--(1.9,0);
\draw [thick] (2.1,0)--(2.9,0);
\draw [thick] (3.1,0)--(3.9,0);
\draw [thick] (2.1,1)--(2.9,1);
\draw [thick] (2,0.1)--(2,0.9);
\end{tikzpicture}\\ 
\hline 
SO(10)_{12}\times SO(4)_8\times U(1) & (3, 4, 6)& 25& \begin{tikzpicture}
\filldraw[fill= white] (-2,0) circle [radius=0.1] node[below] {\scriptsize 1};
\filldraw[fill= white] (-1,0) circle [radius=0.1] node[below] {\scriptsize 2};
 \filldraw[fill= white] (0,0) circle [radius=0.1] node[below] {\scriptsize 4};
\filldraw[fill= white] (1,0) circle [radius=0.1] node[below] {\scriptsize 6};
\filldraw[fill= white] (2,0) circle [radius=0.1] node[below] {\scriptsize 4}; 
 \filldraw[fill= white] (3,0) circle [radius=0.1] node[below] {\scriptsize 2};
\filldraw[fill= white] (4,0) circle [radius=0.1] node[below] {\scriptsize 1}; 
\filldraw[fill= white] (2,1) circle [radius=0.1] node[right] {\scriptsize 2};
\filldraw[fill= white] (1,1) circle [radius=0.1] node[left] {\scriptsize 4}; 
\draw [thick] (-0.1,0)--(-0.9,0);
\draw [thick] (-1.1,0)--(-1.9,0);
\draw [thick] (0.1,0)--(0.9,0);
\draw [thick] (1.1,0)--(1.9,0);
\draw [thick] (2.1,0)--(2.9,0);
\draw [thick] (3.1,0)--(3.9,0);
\draw [thick] (1.1,1)--(1.9,1);
\draw [thick] (1,0.1)--(1,0.9);
\end{tikzpicture}\\ 
\hline 
[E_6]_{14}\times U(1)^2& (3, 4, 7)& 29& \begin{tikzpicture}
\filldraw[fill= white] (3,1) circle [radius=0.1] node[right] {\scriptsize 1};
\filldraw[fill= white] (-1,0) circle [radius=0.1] node[below] {\scriptsize 1};
 \filldraw[fill= white] (0,0) circle [radius=0.1] node[below] {\scriptsize 3};
\filldraw[fill= white] (1,0) circle [radius=0.1] node[below] {\scriptsize 5};
\filldraw[fill= white] (2,0) circle [radius=0.1] node[below] {\scriptsize 7}; 
 \filldraw[fill= white] (3,0) circle [radius=0.1] node[below] {\scriptsize 5};
\filldraw[fill= white] (4,0) circle [radius=0.1] node[below] {\scriptsize 3};
\filldraw[fill= white] (5,0) circle [radius=0.1] node[below] {\scriptsize 1}; 
\filldraw[fill= white] (2,1) circle [radius=0.1] node[left] {\scriptsize 4}; 
\draw [thick] (-0.1,0)--(-0.9,0);
\draw [thick] (0.1,0)--(0.9,0);
\draw [thick] (1.1,0)--(1.9,0);
\draw [thick] (2.1,0)--(2.9,0);
\draw [thick] (3.1,0)--(3.9,0);
\draw [thick] (4.1,0)--(4.9,0);
\draw [thick] (2.1,1)--(2.9,1);
\draw [thick] (2,0.1)--(2,0.9);
\end{tikzpicture}\\ 
\hline 
U(4)_{8}\times USp(4)_6 & (2, 3, 4)& 13& \begin{tikzpicture}
 \filldraw[fill= white] (-1,0) circle [radius=0.1] node[below] {\scriptsize 1};
 \filldraw[fill= white] (0,0) circle [radius=0.1] node[below] {\scriptsize 2};
\filldraw[fill= white] (1,0) circle [radius=0.1] node[below] {\scriptsize 4};
\filldraw[fill= white] (2,0) circle [radius=0.1] node[below] {\scriptsize 2}; 
 \filldraw[fill= white] (3,0) circle [radius=0.1] node[below] {\scriptsize 1}; 
 \filldraw[fill= white] (0,1) circle [radius=0.1] node[left] {\scriptsize 2};
\filldraw[fill= white] (2,1) circle [radius=0.1] node[right] {\scriptsize 2}; 
\draw [thick] (-0.1,0)--(-0.9,0);
\draw [thick] (0.1,0)--(0.9,0);
\draw [thick] (1.1,0)--(1.9,0);
\draw [thick] (2.1,0)--(2.9,0);
\draw [thick] (1.05,0.05)--(1.95,0.95);
\draw [thick] (0.95,0.05)--(0.05,0.95);
\end{tikzpicture}\\ 
\hline 
SU(4)_{8}\times SU(2)_8^3 & (2, 4, 4)& 15& \begin{tikzpicture}
\filldraw[fill= white] (-2,0) circle [radius=0.1] node[below] {\scriptsize 1};
 \filldraw[fill= white] (-1,0) circle [radius=0.1] node[below] {\scriptsize 2};
 \filldraw[fill= white] (0,0) circle [radius=0.1] node[below] {\scriptsize 3};
\filldraw[fill= white] (1,0) circle [radius=0.1] node[below] {\scriptsize 4};
\filldraw[fill= white] (2,0) circle [radius=0.1] node[below] {\scriptsize 2}; 
 \filldraw[fill= white] (2,1) circle [radius=0.1] node[right] {\scriptsize 2}; 
\filldraw[fill= white] (1,1) circle [radius=0.1] node[left] {\scriptsize 2}; 
\draw [thick] (-0.1,0)--(-0.9,0);
\draw [thick] (-1.1,0)--(-1.9,0);
\draw [thick] (0.1,0)--(0.9,0);
\draw [thick] (1.1,0)--(1.9,0);
\draw [thick] (1,0.1)--(1,0.9); 
\draw [thick] (1.05,0.05)--(1.95,0.95);
\end{tikzpicture}\\ 
\hline 
D_2(E_6) & (2, 3, 6)& 21& \begin{tikzpicture}
\filldraw[fill= white] (-2,0) circle [radius=0.1] node[below] {\scriptsize 2};
 \filldraw[fill= white] (-1,0) circle [radius=0.1] node[below] {\scriptsize 4};
 \filldraw[fill= white] (0,0) circle [radius=0.1] node[below] {\scriptsize 6};
\filldraw[fill= white] (1,0) circle [radius=0.1] node[below] {\scriptsize 4};
\filldraw[fill= white] (2,0) circle [radius=0.1] node[below] {\scriptsize 2};  
 \filldraw[fill= white] (0,1) circle [radius=0.1] node[left] {\scriptsize 3};
\filldraw[fill= white] (1,1) circle [radius=0.1] node[right] {\scriptsize 1}; 
\draw [thick] (-1.1,0)--(-1.9,0);
\draw [thick] (-0.1,0)--(-0.9,0);
\draw [thick] (0.1,0)--(0.9,0);
\draw [thick] (1.1,0)--(1.9,0);
\draw [thick] (0.05,0.05)--(0.95,0.95);
\draw [thick] (0,0.1)--(0,0.9);
\end{tikzpicture}\\ 
\hline 
SO(8)_{8}\times U(1) & (2, 2, 4)& 11& \begin{tikzpicture}
\filldraw[fill= white] (0,1) circle [radius=0.1] node[left] {\scriptsize 1};
 \filldraw[fill= white] (0,0) circle [radius=0.1] node[below] {\scriptsize 2};
\filldraw[fill= white] (1,0) circle [radius=0.1] node[below] {\scriptsize 4};
\filldraw[fill= white] (2,0) circle [radius=0.1] node[below] {\scriptsize 2}; 
 \filldraw[fill= white] (2,1) circle [radius=0.1] node[right] {\scriptsize 2}; 
\filldraw[fill= white] (1,1) circle [radius=0.1] node[left] {\scriptsize 1}; 
\draw [thick] (0.95,0.05)--(0.05,0.95);
\draw [thick] (0.1,0)--(0.9,0);
\draw [thick] (1.1,0)--(1.9,0);
\draw [thick] (1,0.1)--(1,0.9); 
\draw [thick] (1.05,0.05)--(1.95,0.95);
\end{tikzpicture}\\ 
\hline 
\end{array}$$

\paragraph{Descendants of the $SO(19)_{28}$ theory} 

$$\begin{array} {|c|c|c|c|} 
\hline 
\text{Theory}& (\Delta_1, \Delta_2, \Delta_3)& 24(c-a)& \text{Magnetic Quiver}\\ 
\hline 
SO(19)_{28} & (6, 12, 14)& 75& \begin{tikzpicture}
 \filldraw[fill= white] (0,0) circle [radius=0.1] node[below] {\scriptsize 2};
\filldraw[fill= white] (1,0) circle [radius=0.1] node[below] {\scriptsize 4};
\filldraw[fill= white] (2,0) circle [radius=0.1] node[below] {\scriptsize 6}; 
 \filldraw[fill= white] (3,0) circle [radius=0.1] node[below] {\scriptsize 8};
\filldraw[fill= white] (4,0) circle [radius=0.1] node[below] {\scriptsize 10};
\filldraw[fill= white] (5,0) circle [radius=0.1] node[below] {\scriptsize 12}; 
 \filldraw[fill= white] (6,0) circle [radius=0.1] node[below] {\scriptsize 14};
\filldraw[fill= white] (7,0) circle [radius=0.1] node[below] {\scriptsize 9};
\filldraw[fill= white] (8,0) circle [radius=0.1] node[below] {\scriptsize 4}; 
\filldraw[fill= white] (6,1) circle [radius=0.1] node[left] {\scriptsize 7}; 
\draw [thick] (0.1,0)--(0.9,0);
\draw [thick] (1.1,0)--(1.9,0);
\draw [thick] (2.1,0)--(2.9,0);
\draw [thick] (3.1,0)--(3.9,0);
\draw [thick] (4.1,0)--(4.9,0);
\draw [thick] (5.1,0)--(5.9,0);
\draw [thick] (6.1,0)--(6.9,0);
\draw [thick] (7.1,0)--(7.9,0); 
\draw [thick] (6,0.1)--(6,0.9);
\end{tikzpicture}\\ 
\hline 
SO(15)_{20}\times SU(2)_{16} & (4, 8, 10)& 47& \begin{tikzpicture}
 \filldraw[fill= white] (0,0) circle [radius=0.1] node[below] {\scriptsize 2};
\filldraw[fill= white] (1,0) circle [radius=0.1] node[below] {\scriptsize 4};
\filldraw[fill= white] (2,0) circle [radius=0.1] node[below] {\scriptsize 6}; 
 \filldraw[fill= white] (3,0) circle [radius=0.1] node[below] {\scriptsize 8};
\filldraw[fill= white] (4,0) circle [radius=0.1] node[below] {\scriptsize 10};
\filldraw[fill= white] (5,0) circle [radius=0.1] node[below] {\scriptsize 7}; 
 \filldraw[fill= white] (6,0) circle [radius=0.1] node[below] {\scriptsize 4};
\filldraw[fill= white] (7,0) circle [radius=0.1] node[below] {\scriptsize 2};
\filldraw[fill= white] (4,1) circle [radius=0.1] node[left] {\scriptsize 5}; 
\draw [thick] (0.1,0)--(0.9,0);
\draw [thick] (1.1,0)--(1.9,0);
\draw [thick] (2.1,0)--(2.9,0);
\draw [thick] (3.1,0)--(3.9,0);
\draw [thick] (4.1,0)--(4.9,0);
\draw [thick] (5.1,0)--(5.9,0);
\draw [thick] (6.1,0)--(6.9,0);
\draw [thick] (4,0.1)--(4,0.9);
\end{tikzpicture}\\ 
\hline 
SO(13)_{16}\times U(1) & (3, 6, 8)& 33& \begin{tikzpicture}
 \filldraw[fill= white] (0,0) circle [radius=0.1] node[below] {\scriptsize 2};
\filldraw[fill= white] (1,0) circle [radius=0.1] node[below] {\scriptsize 4};
\filldraw[fill= white] (2,0) circle [radius=0.1] node[below] {\scriptsize 6}; 
 \filldraw[fill= white] (3,0) circle [radius=0.1] node[below] {\scriptsize 8};
\filldraw[fill= white] (4,0) circle [radius=0.1] node[below] {\scriptsize 5};
\filldraw[fill= white] (5,0) circle [radius=0.1] node[below] {\scriptsize 2}; 
 \filldraw[fill= white] (3,1) circle [radius=0.1] node[left] {\scriptsize 5};
\filldraw[fill= white] (4,1) circle [radius=0.1] node[right] {\scriptsize 2}; 
\draw [thick] (0.1,0)--(0.9,0);
\draw [thick] (1.1,0)--(1.9,0);
\draw [thick] (2.1,0)--(2.9,0);
\draw [thick] (3.1,0)--(3.9,0);
\draw [thick] (4.1,0)--(4.9,0);
\draw [thick] (3.1,1)--(3.9,1);
\draw [thick] (3,0.1)--(3,0.9);
\end{tikzpicture}\\ 
\hline 
SO(11)_{12} & (2, 4, 6)& 19& \begin{tikzpicture}
 \filldraw[fill= white] (0,0) circle [radius=0.1] node[below] {\scriptsize 2};
\filldraw[fill= white] (1,0) circle [radius=0.1] node[below] {\scriptsize 4};
\filldraw[fill= white] (2,0) circle [radius=0.1] node[below] {\scriptsize 6}; 
 \filldraw[fill= white] (3,0) circle [radius=0.1] node[below] {\scriptsize 3}; 
 \filldraw[fill= white] (1,1) circle [radius=0.1] node[left] {\scriptsize 2};
\filldraw[fill= white] (3,1) circle [radius=0.1] node[right] {\scriptsize 3}; 
\draw [thick] (0.1,0)--(0.9,0);
\draw [thick] (1.1,0)--(1.9,0);
\draw [thick] (2.1,0)--(2.9,0);
\draw [thick] (2.05,0.05)--(2.95,0.95);
\draw [thick] (1.95,0.05)--(1.05,0.95);
\end{tikzpicture}\\ 
\hline 
\end{array}$$ 
Notice that, although the $U(4)_8\times USp(4)_6$ theory is a descendant of the $SO(13)_{16}\times U(1)$ model, we did not include it here since it already appears in the table of descendants of the $[E_7]_{24}\times SO(7)_{16}$ theory.

\paragraph{Lagrangian theories} 

Within the $\ell=1$ class of theories we find 9 lagrangian models and we now specify for all of them the gauge group and matter content. 
\begin{enumerate} 
\item ${\bf SO(8)_{12}\times SU(2)_{14}:}$ $USp(6)$ gauge theory with 4 hypermultiplets in the {\bf 6} and one in the {\bf 14} (rank-2 antisymmetric). 
\item ${\bf SO(16)_{12}:}$ $USp(6)$ gauge theory with 8 hypermultiplets in the {\bf 6} ($USp(6)$ SQCD).
\item ${\bf U(8)_8:}$ $SU(4)$ gauge theory with 8 hypermultiplets in the {\bf 4} ($SU(4)$ SQCD).
 \item ${\bf SU(2)_4^6:}$ $SU(2)^3$ linear quiver ($\boxed{2}-SU(2)-SU(2)-SU(2)-\boxed{2}$).
\item ${\bf U(6)_8\times SU(2)_6:}$ $SU(4)$ gauge theory with 6 hypermultiplets in the {\bf 4} and one hypermultiplet in the {\bf 6}.
\item ${\bf U(4)_6\times U(1)^2:}$ $SU(2)\times SU(3)$ linear quiver ($\boxed{1}-SU(2)-SU(3)-\boxed{4}$).
\item ${\bf SO(11)_{12}:}$ $USp(6)$ gauge theory with 11 half-hypermultiplets in the {\bf 6} and a half-hypermultiplet in the {\bf 14'} (rank-3 antisymmetric).
\item ${\bf U(4)_8\times USp(4)_6:}$ $SU(4)$ gauge theory with 4 hypermultiplets in the {\bf 4} and two hypermultiplets in the {\bf 6}. 
\item ${\bf SO(8)_8\times U(1):}$ $SU(2)\times USp(4)$ linear quiver ($SU(2)-USp(4)-\boxed{4}$).
\end{enumerate}

\subsection{Theories with $\ell=2$} 

For $\ell=2$ we have nine possible starting points and two of them have $k>3$, which we have not discussed explicitly in detail so far. In this case we need to consider two 6d orbi-instanton theories with non-trivial $\sigma$. A common feature for $\ell>1$ is that, with the methods developed in this paper alone, we do not gain a precise understanding of models with trivial 7-brane. We analyze them one by one and we omit the theory whenever we are unable to determine most of its data. We color in blue RG flows which correspond to a sequence of two or more consecutive mass deformations.

\begin{equation}\label{diagramell=2} 
\begin{tikzpicture} 
\node[] (a1) at (0,0) {\scriptsize $USp(8)_{19}\times SU(2)_{57}$}; 
\node[] (a2) at (0,-1) {\scriptsize $USp(4)_{13}\times SU(2)_{24}\times SU(2)_{39}$}; 
\node[] (a3) at (0,-2) {\scriptsize $SU(2)_{10}\times SU(2)_{30}\times U(1)$}; 
\node[] (a4) at (0,-3) {\scriptsize {\color{red} $SU(2)_{21}$}}; 

\node[] (b1) at (4,0) {\scriptsize $[F_4]_{18}\times SU(2)_{39}$}; 
\node[] (b2) at (4,-1) {\scriptsize $SO(7)_{12}\times SU(2)_{27}$}; 
\node[] (b3) at (4,-2) {\scriptsize $SU(3)_{9}\times SU(2)_{21}$}; 
\node[] (b4) at (4,-3) {\scriptsize {\color{red} $SU(2)_{15}$}}; 

\node[] (c1) at (7,0) {\scriptsize $USp(18)_{11}$}; 
\node[] (c2) at (7,-1) {\scriptsize $USp(14)_{9}\times SU(2)_{8}$}; 
\node[] (c3) at (7,-2) {\scriptsize $USp(12)_{8}\times U(1)$}; 
\node[] (c4) at (7,-3) {\scriptsize {\color{red} $USp(10)_{7}$}}; 

\node[] (d1) at (10,0) {\scriptsize $USp(16)_{10}$}; 
\node[] (d2) at (10,-1) {\scriptsize $USp(12)_{8}\times SU(2)_{8}$}; 
\node[] (d3) at (10,-2) {\scriptsize $USp(10)_{7}\times U(1)$}; 
\node[] (d4) at (10,-3) {\scriptsize {\color{red} $USp(8)_{6}$}}; 

\node[] (e1) at (-.5,-4) {\scriptsize $SU(7)_{18}\times U(1)$}; 
\node[] (e2) at (-.5,-5) {\scriptsize $SU(5)_{14}\times U(1)^2$}; 
\node[] (e3) at (-.5,-6) {\scriptsize $SU(4)_{12}\times U(1)^2$}; 
\node[] (e4) at (-.5,-7) {\scriptsize {\color{red} $U(2)_{8}\times U(1)$}};

\node[] (f1) at (3.5,-4) {\scriptsize $SU(5)_{28}\times SU(2)_{15}$}; 
\node[] (f2) at (3.5,-5) {\scriptsize $SU(3)_{20}\times SU(2)_{11}\times U(1)$}; 
\node[] (f3) at (3.5,-6) {\scriptsize $SU(2)_{16}\times SU(2)_{9}\times U(1)$};

\node[] (g1) at (1.5,-8) {\scriptsize $USp(10)_{14}\times U(1)$}; 
\node[] (g2) at (1.5,-9) {\scriptsize $USp(6)_{10}\times SU(2)_{16} \times U(1)$}; 
\node[] (g3) at (1.5,-10) {\scriptsize $USp(4)_{8}\times U(1)^2 $}; 
\node[] (g4) at (4,-11) {\scriptsize { {\color{red} $SU(2)_{6}\times U(1)$}}};

\node[] (h1) at (8.5,-8) {\scriptsize $USp(6)_{13}\times SU(3)_{16}  \times U(1)$}; 
\node[] (h2) at (6,-9) {\scriptsize $SU(2)_{9}\times SU(2)_{?}\times SU(3)_{12} \times U(1) $};
\node[] (h3) at (6,-10) {\scriptsize $SU(3)_{10}\times U(1)^2 $};  
\node[] (H2) at (10.2,-9) {\scriptsize $USp(6)_{11}\times U(1)^2 $};
\node[] (H3) at (10.2,-10) {\scriptsize $USp(6)_{10}\times U(1) $};

\node[] (i1) at (8.5,-4) {\scriptsize $SO(7)_{12}\times SU(4)_{16}$}; 
\node[] (i2) at (7.2,-5) {\scriptsize $SO(5)_{8}\times SU(4)_{8} $};
\node[] (i3) at (10.7,-5) {\scriptsize $SO(7)_{10}\times SU(2)_{12} \times U(1)$};  
\node[] (i4) at (10.7,-6) {\scriptsize { $SO(7)_{9}\times U(1) $}};  
\node[] (i5) at (11.5,-7) {\scriptsize $SU(3)_6$};
\node[] (i6) at (10,-7) {\scriptsize{\color{red}  $U(1)$}};

\draw[->, thick, color=blue] (h3)--(g4);
\draw[->, thick, color=blue] (i4)--(i6);
\draw[->, thick] (i1)--(i2);
\draw[->, thick] (i1)--(i3);
\draw[->, thick] (i3)--(i4);
\draw[->, thick, color=blue] (i4)--(i5);
\draw[->, thick] (H2)--(H3);
\draw[->, thick] (h1)--(H2);
\draw[->, thick] (h2)--(h3);
\draw[->, thick] (h1)--(h2);
\draw[->, thick] (g2)--(g3);
\draw[->, thick] (g3)--(g4);
\draw[->, thick] (g1)--(g2);
\draw[->, thick] (f2)--(f3);
\draw[->, thick] (f1)--(f2);
\draw[->, thick] (e2)--(e3);
\draw[->, thick, color=blue] (e3)--(e4);
\draw[->, thick] (e1)--(e2);
\draw[->, thick] (d2)--(d3);
\draw[->, thick] (d3)--(d4);
\draw[->, thick] (d1)--(d2);
\draw[->, thick] (c2)--(c3);
\draw[->, thick] (c3)--(c4);
\draw[->, thick] (c1)--(c2); 
\draw[->, thick] (a1)--(a2); 
\draw[->, thick] (a2)--(a3);
\draw[->, thick] (a3)--(a4);
\draw[->, thick] (b1)--(b2); 
\draw[->, thick] (b2)--(b3);
\draw[->, thick] (b3)--(b4);
\end{tikzpicture}
\end{equation}

\begin{itemize} 
\item $[F_4]_{18}\times SU(2)_{39}$: SW compactification of the $\bbZ_2$ orbi-instanton with $(n_2=1)$ and three tensors. 
\item  $USp(8)_{19}\times SU(2)_{57}$: SW compactification of the $\bbZ_2$ orbi-instanton with $(n_{2'}=1)$ and three tensors.
\item $USp(6)_{13}\times SU(3)_{16}  \times U(1)$: SW compactification of the $\bbZ_4$ orbi-instanton with $(n_2=n_{2'}=1)$ and two tensors.
\item $SO(7)_{12}\times SU(4)_{16}$: SW compactification of the $\bbZ_4$ orbi-instanton with $(n_{4}=1)$ and two tensors. 
\item  $USp(10)_{14}\times U(1)$: SW compactification of the $\bbZ_4$ orbi-instanton with $(n_{2'}=2)$, two tensors and non-trivial $\sigma=(2,2)$. 
\item $SU(5)_{28}\times SU(2)_{15}$: SW compactification of the $\bbZ_4$ orbi-instanton with $(n_{4'}=1)$, two tensors and non-trivial $\sigma=(2,2)$. 
\item $SU(7)_{18}\times U(1)$: SW compactification of the $\bbZ_6$ orbi-instanton with $(n_{2'}=n_{4'}=1)$ and one tensor.
\item  $USp(16)_{10}$: SW compactification of the $\bbZ_8$ orbi-instanton with $(n_{2'}=4)$ and one tensor.
\item $USp(18)_{11}$: SW compactification of the $\bbZ_{10}$ orbi-instanton with $(n_{2'}=5)$ and one tensor.
\end{itemize}

\paragraph{Descendants of the $USp(8)_{19}\times SU(2)_{57}$ theory} 

$$\begin{array} {|c|c|c|c|} 
\hline 
\text{Theory}& (\Delta_1, \Delta_2, \Delta_3)& 24(c-a)& \text{Magnetic Quiver}\\ 
\hline 
USp(8)_{19}\times SU(2)_{57} & (6, 12, 18)& 40& \begin{tikzpicture}
% \filldraw[fill= white] (-2,0) circle [radius=0.1] node[below] {\scriptsize 1};
% \filldraw[fill= white] (-1,0) circle [radius=0.1] node[below] {\scriptsize 2};
%\filldraw[fill= white] (0,0) circle [radius=0.1] node[below] {\scriptsize 3};
\filldraw[fill= white] (1,0) circle [radius=0.1] node[below] {\scriptsize 1};
\filldraw[fill= white] (2,0) circle [radius=0.1] node[below] {\scriptsize 4};
\filldraw[fill= white] (3,0) circle [radius=0.1] node[below] {\scriptsize 7};
\filldraw[fill= white] (4,0) circle [radius=0.1] node[below] {\scriptsize 10};
\filldraw[fill= white] (5,0) circle [radius=0.1] node[below] {\scriptsize 13};
\filldraw[fill= white] (6,0) circle [radius=0.1] node[below] {\scriptsize 6}; 
\node[] at (2.5,0.5) {};

%\draw [thick] (-1.1, 0) -- (-1.9,0) ;
%\draw [thick] (-0.1, 0) -- (-0.9,0) ;
%\draw [thick] (0.1, 0) -- (0.9,0) ;
\draw [thick] (1.1, 0) -- (1.9,0) ;
\draw [thick] (2.1, 0) -- (2.9,0) ;
\draw [thick] (4.1, 0.05) -- (4.9,0.05) ;
\draw [thick] (4.1, -0.05) -- (4.9,-0.05) ;
\draw [thick] (3.1, 0) -- (3.9,0) ;
\draw [thick] (5.1, 0) -- (5.9,0) ;
\draw [thick] (4.4,0) -- (4.6,0.2);
\draw [thick] (4.4,0) -- (4.6,-0.2); 
\end{tikzpicture}\\ 
\hline 
USp(4)_{13}\times SU(2)_{24}\times SU(2)_{39} & (4, 8, 12)& 20& \begin{tikzpicture}
%\filldraw[fill= white] (-1,-2) circle [radius=0.1] node[below] {\scriptsize  1};
%\filldraw[fill= white] (0,-2) circle [radius=0.1] node[below] {\scriptsize 2};
%\filldraw[fill= white] (1,-2) circle [radius=0.1] node[below] {\scriptsize A-2};
\filldraw[fill= white] (2,-2) circle [radius=0.1] node[below] {\scriptsize 1};
\filldraw[fill= white] (3,-2) circle [radius=0.1] node[below] {\scriptsize 4};
\filldraw[fill= white] (4,-2) circle [radius=0.1] node[below] {\scriptsize 7};
\filldraw[fill= white] (5,-2) circle [radius=0.1] node[below] {\scriptsize 6};
\filldraw[fill= white] (6,-2) circle [radius=0.1] node[below] {\scriptsize 3};
\node[] at (2.5,-1.5) {};

%\draw [thick] (-0.1, -2) -- (-0.9,-2) ;
%\draw [thick] (0.1, -2) -- (0.9,-2) ;
%\draw [thick] (1.1, -2) -- (1.9,-2) ;
\draw [thick] (3.1, -1.95) -- (3.9,-1.95) ;
\draw [thick] (3.1, -2.05) -- (3.9,-2.05) ;
\draw [thick] (2.1, -2) -- (2.9,-2) ;
\draw [thick] (4.1, -2) -- (4.9,-2) ;
\draw [thick] (5.1, -1.95) -- (5.9,-1.95) ;
\draw [thick] (5.1, -2.05) -- (5.9,-2.05) ;
\draw [thick] (3.4,-2) -- (3.6,-1.8);
\draw [thick] (3.4,-2) -- (3.6,-2.2);
\draw [thick] (5.6,-2) -- (5.4,-1.8);
\draw [thick] (5.6,-2) -- (5.4,-2.2);

\end{tikzpicture}\\ 
\hline 
SU(2)_{10}\times SU(2)_{30}\times U(1) & (3, 6, 9)& 10& \begin{tikzpicture}
%\filldraw[fill= white] (0,-4) circle [radius=0.1] node[below] {\scriptsize  1};
%\filldraw[fill= white] (1,-4) circle [radius=0.1] node[below] {\scriptsize A-3};
%\filldraw[fill= white] (2,-4) circle [radius=0.1] node[below] {\scriptsize B-3};
\filldraw[fill= white] (3,-4) circle [radius=0.1] node[below] {\scriptsize 1};
\filldraw[fill= white] (4,-4) circle [radius=0.1] node[below] {\scriptsize 4};
\filldraw[fill= white] (5,-4) circle [radius=0.1] node[below] {\scriptsize 3};
\filldraw[fill= white] (4.5,-3) circle [radius=0.1] node[above] {\scriptsize 3};
%\draw [thick] (0.1, -4) -- (0.9,-4) ;
%\draw [thick] (1.1, -4) -- (1.9,-4) ;
\draw [thick] (3.1, -3.95) -- (3.9,-3.95) ;
\draw [thick] (3.1, -4.05) -- (3.9,-4.05) ;
%\draw [thick] (2.1, -4) -- (2.9,-4) ;
\draw [thick] (4.1, -4) -- (4.9,-4) ;
\draw [thick] (3.4,-4) -- (3.6,-3.8);
\draw [thick] (3.4,-4) -- (3.6,-4.2);
\draw [thick] (4.02, -3.95) -- (4.48,-3.05);
\draw [thick] (4.98, -3.95) -- (4.52,-3.05) ;
\end{tikzpicture}\\ 
\hline 
SU(2)_{21} & (2, 4, 6)& 0&\begin{tikzpicture}
 \filldraw[fill= white] (0,0) circle [radius=0.1] node[below] {\scriptsize 1};
\filldraw[fill= white] (1,0) circle [radius=0.1] node[below] {\scriptsize 3};
\draw [thick] (0.1,0.05)--(0.9,0.05);
\draw [thick] (0.1,-0.05)--(0.9,-0.05);
%\draw [thick] (0.4,0) -- (0.6,0.2);
%\draw [thick] (0.4,0) -- (0.6,-0.2);
\draw [thick] (1.05, 0.05) to [out=45,in=315,looseness=15] (1.05,-0.05);
\end{tikzpicture}\\ 
\hline 
\end{array}$$ 

\paragraph{Descendants of the $[F_4]_{18}\times SU(2)_{39}$ theory} 

$$\begin{array} {|c|c|c|c|} 
\hline 
\text{Theory}& (\Delta_1, \Delta_2, \Delta_3)& 24(c-a)& \text{Magnetic Quiver}\\ 
\hline 
[F_4]_{18}\times SU(2)_{39} & (6, 9, 12)& 36& \begin{tikzpicture}
% \filldraw[fill= white] (-2,0) circle [radius=0.1] node[below] {\scriptsize 1};
% \filldraw[fill= white] (-1,0) circle [radius=0.1] node[below] {\scriptsize 2};
%\filldraw[fill= white] (0,0) circle [radius=0.1] node[below] {\scriptsize 3};
\filldraw[fill= white] (1,0) circle [radius=0.1] node[below] {\scriptsize 1};
\filldraw[fill= white] (2,0) circle [radius=0.1] node[below] {\scriptsize 3};
\filldraw[fill= white] (3,0) circle [radius=0.1] node[below] {\scriptsize 6};
\filldraw[fill= white] (4,0) circle [radius=0.1] node[below] {\scriptsize 9};
\filldraw[fill= white] (5,0) circle [radius=0.1] node[below] {\scriptsize 12};
\filldraw[fill= white] (6,0) circle [radius=0.1] node[below] {\scriptsize 6}; 
\node[] at (2.5,0.5) {};

%\draw [thick] (-1.1, 0) -- (-1.9,0) ;
%\draw [thick] (-0.1, 0) -- (-0.9,0) ;
%\draw [thick] (0.1, 0) -- (0.9,0) ;
\draw [thick] (1.1, 0) -- (1.9,0) ;
\draw [thick] (2.1, 0) -- (2.9,0) ;
\draw [thick] (4.1, 0.05) -- (4.9,0.05) ;
\draw [thick] (4.1, -0.05) -- (4.9,-0.05) ;
\draw [thick] (3.1, 0) -- (3.9,0) ;
\draw [thick] (5.1, 0) -- (5.9,0) ;
\draw [thick] (4.4,0) -- (4.6,0.2);
\draw [thick] (4.4,0) -- (4.6,-0.2); 
\end{tikzpicture}\\ 
\hline 
SO(7)_{12}\times SU(2)_{27} & (4, 6, 8)& 18& \begin{tikzpicture}
%\filldraw[fill= white] (-1,-2) circle [radius=0.1] node[below] {\scriptsize  1};
%\filldraw[fill= white] (0,-2) circle [radius=0.1] node[below] {\scriptsize 2};
%\filldraw[fill= white] (1,-2) circle [radius=0.1] node[below] {\scriptsize A-2};
\filldraw[fill= white] (2,-2) circle [radius=0.1] node[below] {\scriptsize 1};
\filldraw[fill= white] (3,-2) circle [radius=0.1] node[below] {\scriptsize 3};
\filldraw[fill= white] (4,-2) circle [radius=0.1] node[below] {\scriptsize 6};
\filldraw[fill= white] (5,-2) circle [radius=0.1] node[below] {\scriptsize 6};
\filldraw[fill= white] (6,-2) circle [radius=0.1] node[below] {\scriptsize 3};
\node[] at (2.5,-1.5) {};

%\draw [thick] (-0.1, -2) -- (-0.9,-2) ;
%\draw [thick] (0.1, -2) -- (0.9,-2) ;
%\draw [thick] (1.1, -2) -- (1.9,-2) ;
\draw [thick] (3.1, -1.95) -- (3.9,-1.95) ;
\draw [thick] (3.1, -2.05) -- (3.9,-2.05) ;
\draw [thick] (2.1, -2) -- (2.9,-2) ;
\draw [thick] (4.1, -2) -- (4.9,-2) ;
\draw [thick] (5.1, -1.95) -- (5.9,-1.95) ;
\draw [thick] (5.1, -2.05) -- (5.9,-2.05) ;
\draw [thick] (3.4,-2) -- (3.6,-1.8);
\draw [thick] (3.4,-2) -- (3.6,-2.2);
\draw [thick] (5.6,-2) -- (5.4,-1.8);
\draw [thick] (5.6,-2) -- (5.4,-2.2);

\end{tikzpicture}\\ 
\hline 
SU(3)_{9}\times SU(2)_{21} & (3, \frac{9}{2}, 6)& 9& \begin{tikzpicture}
%\filldraw[fill= white] (0,-4) circle [radius=0.1] node[below] {\scriptsize  1};
%\filldraw[fill= white] (1,-4) circle [radius=0.1] node[below] {\scriptsize A-3};
%\filldraw[fill= white] (2,-4) circle [radius=0.1] node[below] {\scriptsize B-3};
\filldraw[fill= white] (3,-4) circle [radius=0.1] node[below] {\scriptsize 1};
\filldraw[fill= white] (4,-4) circle [radius=0.1] node[below] {\scriptsize 3};
\filldraw[fill= white] (5,-4) circle [radius=0.1] node[below] {\scriptsize 3};
\filldraw[fill= white] (4.5,-3) circle [radius=0.1] node[above] {\scriptsize 3};
%\draw [thick] (0.1, -4) -- (0.9,-4) ;
%\draw [thick] (1.1, -4) -- (1.9,-4) ;
\draw [thick] (3.1, -3.95) -- (3.9,-3.95) ;
\draw [thick] (3.1, -4.05) -- (3.9,-4.05) ;
%\draw [thick] (2.1, -4) -- (2.9,-4) ;
\draw [thick] (4.1, -4) -- (4.9,-4) ;
\draw [thick] (3.4,-4) -- (3.6,-3.8);
\draw [thick] (3.4,-4) -- (3.6,-4.2);
\draw [thick] (4.02, -3.95) -- (4.48,-3.05);
\draw [thick] (4.98, -3.95) -- (4.52,-3.05) ;
\end{tikzpicture}\\ 
\hline 
SU(2)_{15} & (2, 3, 4)& 0& \begin{tikzpicture}
 \filldraw[fill= white] (0,0) circle [radius=0.1] node[below] {\scriptsize 1};
\filldraw[fill= white] (1,0) circle [radius=0.1] node[below] {\scriptsize 3};
\draw [thick] (0.1,0.05)--(0.9,0.05);
\draw [thick] (0.1,-0.05)--(0.9,-0.05);
\draw [thick] (0.4,0) -- (0.6,0.2);
\draw [thick] (0.4,0) -- (0.6,-0.2);
\draw [thick] (1.05, 0.05) to [out=45,in=315,looseness=15] (1.05,-0.05);
\end{tikzpicture}\\ 
\hline 
\end{array}$$

\paragraph{Descendants of the $USp(6)_{13}\times SU(3)_{16}  \times U(1)$ theory} 

$$\begin{array} {|c|c|c|c|} 
\hline 
\text{Theory}& (\Delta_1, \Delta_2, \Delta_3)& 24(c-a)& \text{Magnetic Quiver}\\ 
\hline 
USp(6)_{13}\times SU(3)_{16} \times U(1)& (6, 8, 12)& 30& \begin{tikzpicture}
\filldraw[fill= white] (0,0) circle [radius=0.1] node[below] {\scriptsize 1};
\filldraw[fill= white] (1,0) circle [radius=0.1] node[below] {\scriptsize 2};
\filldraw[fill= white] (2,0) circle [radius=0.1] node[below] {\scriptsize 3};
\filldraw[fill= white] (3,0) circle [radius=0.1] node[below] {\scriptsize 5};
\filldraw[fill= white] (4,0) circle [radius=0.1] node[below] {\scriptsize 7};
\filldraw[fill= white] (5,0) circle [radius=0.1] node[below] {\scriptsize 9};
\filldraw[fill= white] (6,0) circle [radius=0.1] node[below] {\scriptsize 4}; 
\node[] at (2.5,0.5) {};

\draw [thick] (0.1, 0) -- (0.9,0) ;
\draw [thick] (1.1, 0) -- (1.9,0) ;
\draw [thick] (2.1, 0) -- (2.9,0) ;
\draw [thick] (4.1, 0.05) -- (4.9,0.05) ;
\draw [thick] (4.1, -0.05) -- (4.9,-0.05) ;
\draw [thick] (3.1, 0) -- (3.9,0) ;
\draw [thick] (5.1, 0) -- (5.9,0) ;
\draw [thick] (4.4,0) -- (4.6,0.2);
\draw [thick] (4.4,0) -- (4.6,-0.2); 
\end{tikzpicture}\\ 
\hline 
USp(6)_{11}\times U(1)^2 & (4, 6, 10)& 20& \begin{tikzpicture}
\filldraw[fill= white] (1,-2) circle [radius=0.1] node[below] {\scriptsize 1};
\filldraw[fill= white] (2,-2) circle [radius=0.1] node[below] {\scriptsize 3};
\filldraw[fill= white] (3,-2) circle [radius=0.1] node[below] {\scriptsize 5};
\filldraw[fill= white] (4,-2) circle [radius=0.1] node[below] {\scriptsize 7};
\filldraw[fill= white] (5,-2) circle [radius=0.1] node[below] {\scriptsize 4};
\filldraw[fill= white] (6,-2) circle [radius=0.1] node[below] {\scriptsize 1};
\node[] at (2.5,-1.5) {};

\draw [thick] (1.1, -2) -- (1.9,-2) ;
\draw [thick] (3.1, -1.95) -- (3.9,-1.95) ;
\draw [thick] (3.1, -2.05) -- (3.9,-2.05) ;
\draw [thick] (2.1, -2) -- (2.9,-2) ;
\draw [thick] (4.1, -2) -- (4.9,-2) ;
\draw [thick] (5.1, -1.95) -- (5.9,-1.95) ;
\draw [thick] (5.1, -2.05) -- (5.9,-2.05) ;
\draw [thick] (3.4,-2) -- (3.6,-1.8);
\draw [thick] (3.4,-2) -- (3.6,-2.2);
\draw [thick] (5.6,-2) -- (5.4,-1.8);
\draw [thick] (5.6,-2) -- (5.4,-2.2);

\end{tikzpicture}\\ 
\hline 
SU(2)_{9}\times SU(2)_{?}\times SU(3)_{12} \times U(1) & (4, 6, 8)& 16& \begin{tikzpicture}
\filldraw[fill= white] (1,-2) circle [radius=0.1] node[below] {\scriptsize 1};
\filldraw[fill= white] (2,-2) circle [radius=0.1] node[below] {\scriptsize 2};
\filldraw[fill= white] (3,-2) circle [radius=0.1] node[below] {\scriptsize 3};
\filldraw[fill= white] (4,-2) circle [radius=0.1] node[below] {\scriptsize 5};
\filldraw[fill= white] (5,-2) circle [radius=0.1] node[below] {\scriptsize 4};
\filldraw[fill= white] (6,-2) circle [radius=0.1] node[below] {\scriptsize 2};
\node[] at (2.5,-1.5) {};

\draw [thick] (1.1, -2) -- (1.9,-2) ;
\draw [thick] (3.1, -1.95) -- (3.9,-1.95) ;
\draw [thick] (3.1, -2.05) -- (3.9,-2.05) ;
\draw [thick] (2.1, -2) -- (2.9,-2) ;
\draw [thick] (4.1, -2) -- (4.9,-2) ;
\draw [thick] (5.1, -1.95) -- (5.9,-1.95) ;
\draw [thick] (5.1, -2.05) -- (5.9,-2.05) ;
\draw [thick] (3.4,-2) -- (3.6,-1.8);
\draw [thick] (3.4,-2) -- (3.6,-2.2);
\draw [thick] (5.6,-2) -- (5.4,-1.8);
\draw [thick] (5.6,-2) -- (5.4,-2.2);

\end{tikzpicture}\\ 
\hline 
USp(6)_{10}\times U(1) & (3, 5, 9)& 15& \begin{tikzpicture}
\filldraw[fill= white] (2,-4) circle [radius=0.1] node[below] {\scriptsize 2};
\filldraw[fill= white] (3,-4) circle [radius=0.1] node[below] {\scriptsize 4};
\filldraw[fill= white] (4,-4) circle [radius=0.1] node[below] {\scriptsize 6};
\filldraw[fill= white] (5,-4) circle [radius=0.1] node[below] {\scriptsize 3};
\filldraw[fill= white] (4.5,-3) circle [radius=0.1] node[above] {\scriptsize 1};

\draw [thick] (3.1, -3.95) -- (3.9,-3.95) ;
\draw [thick] (3.1, -4.05) -- (3.9,-4.05) ;
\draw [thick] (2.1, -4) -- (2.9,-4) ;
\draw [thick] (4.1, -4) -- (4.9,-4) ;
\draw [thick] (3.4,-4) -- (3.6,-3.8);
\draw [thick] (3.4,-4) -- (3.6,-4.2);
\draw [thick] (4.02, -3.95) -- (4.48,-3.05);
\draw [thick] (4.98, -3.95) -- (4.52,-3.05) ;
\end{tikzpicture}\\ 
\hline 
SU(3)_{10}\times U(1)^2 & (3, 5, 6)& 9& \begin{tikzpicture}
\filldraw[fill= white] (2,-4) circle [radius=0.1] node[below] {\scriptsize 1};
\filldraw[fill= white] (3,-4) circle [radius=0.1] node[below] {\scriptsize 2};
\filldraw[fill= white] (4,-4) circle [radius=0.1] node[below] {\scriptsize 3};
\filldraw[fill= white] (5,-4) circle [radius=0.1] node[below] {\scriptsize 2};
\filldraw[fill= white] (4.5,-3) circle [radius=0.1] node[above] {\scriptsize 2};

\draw [thick] (3.1, -3.95) -- (3.9,-3.95) ;
\draw [thick] (3.1, -4.05) -- (3.9,-4.05) ;
\draw [thick] (2.1, -4) -- (2.9,-4) ;
\draw [thick] (4.1, -4) -- (4.9,-4) ;
\draw [thick] (3.4,-4) -- (3.6,-3.8);
\draw [thick] (3.4,-4) -- (3.6,-4.2);
\draw [thick] (4.02, -3.95) -- (4.48,-3.05);
\draw [thick] (4.98, -3.95) -- (4.52,-3.05) ;
\end{tikzpicture}\\ 
\hline 
SU(2)_6 \times U(1)& (2, 3, 4)& 1 & ? \\ 
\hline 
\end{array}$$ 
For the model with spectrum $(4,6,8)$ in the above table, we are unable to determine the flavor central charge of one of the $SU(2)$ factors coming from $USp(6)$ of the parent theory since we do not know its embedding index. The theory in the last line associated with trivial 7-brane is supposed to have spectrum $(2,4,4)$ and symmetry which includes a $SU(2)_8$ according to our rules. From these data we can identify it with a non conformal $USp(4)$ gauging of the $E_7$ rank-1 theory. If we now turn on a mass deformation for the latter we get a $USp(4)$ gauging of the $E_6$ Minahan-Nemeschansky theory, which is conformal and is discussed in the class-$\mathcal{S}$ context in \cite{Chacaltana:2012ch}. It corresponds to the $A_3$ theory on a torus with a minimal puncture and a twist line along one of the cycles of the torus. This theory is equivalent to the $SU(4)$ gauge theory with hypermultiplets in the symmetric and antisymmetric rank-2 representations. We include the latter in the table. We will find another occurrence of this theory below.

\paragraph{Descendants of the $SO(7)_{12}\times SU(4)_{16}$ theory} 

$$\begin{array} {|c|c|c|c|} 
\hline 
\text{Theory}& (\Delta_1, \Delta_2, \Delta_3)& 24(c-a)& \text{Magnetic Quiver}\\ 
\hline 
SO(7)_{12}\times SU(4)_{16}& (6, 6, 8)& 27& \begin{tikzpicture}
\filldraw[fill= white] (0,0) circle [radius=0.1] node[below] {\scriptsize 1};
\filldraw[fill= white] (1,0) circle [radius=0.1] node[below] {\scriptsize 2};
\filldraw[fill= white] (2,0) circle [radius=0.1] node[below] {\scriptsize 3};
\filldraw[fill= white] (3,0) circle [radius=0.1] node[below] {\scriptsize 4};
\filldraw[fill= white] (4,0) circle [radius=0.1] node[below] {\scriptsize 6};
\filldraw[fill= white] (5,0) circle [radius=0.1] node[below] {\scriptsize 8};
\filldraw[fill= white] (6,0) circle [radius=0.1] node[below] {\scriptsize 4}; 
\node[] at (2.5,0.5) {};

\draw [thick] (0.1, 0) -- (0.9,0) ;
\draw [thick] (1.1, 0) -- (1.9,0) ;
\draw [thick] (2.1, 0) -- (2.9,0) ;
\draw [thick] (4.1, 0.05) -- (4.9,0.05) ;
\draw [thick] (4.1, -0.05) -- (4.9,-0.05) ;
\draw [thick] (3.1, 0) -- (3.9,0) ;
\draw [thick] (5.1, 0) -- (5.9,0) ;
\draw [thick] (4.4,0) -- (4.6,0.2);
\draw [thick] (4.4,0) -- (4.6,-0.2); 
\end{tikzpicture}\\ 
\hline 
SO(7)_{10}\times SU(2)_{12} \times U(1) & (4, 5, 6)& 17& \begin{tikzpicture}
\filldraw[fill= white] (1,-2) circle [radius=0.1] node[below] {\scriptsize 1};
\filldraw[fill= white] (2,-2) circle [radius=0.1] node[below] {\scriptsize 2};
\filldraw[fill= white] (3,-2) circle [radius=0.1] node[below] {\scriptsize 4};
\filldraw[fill= white] (4,-2) circle [radius=0.1] node[below] {\scriptsize 6};
\filldraw[fill= white] (5,-2) circle [radius=0.1] node[below] {\scriptsize 4};
\filldraw[fill= white] (6,-2) circle [radius=0.1] node[below] {\scriptsize 1};
\node[] at (2.5,-1.5) {};

\draw [thick] (1.1, -2) -- (1.9,-2) ;
\draw [thick] (3.1, -1.95) -- (3.9,-1.95) ;
\draw [thick] (3.1, -2.05) -- (3.9,-2.05) ;
\draw [thick] (2.1, -2) -- (2.9,-2) ;
\draw [thick] (4.1, -2) -- (4.9,-2) ;
\draw [thick] (5.1, -1.95) -- (5.9,-1.95) ;
\draw [thick] (5.1, -2.05) -- (5.9,-2.05) ;
\draw [thick] (3.4,-2) -- (3.6,-1.8);
\draw [thick] (3.4,-2) -- (3.6,-2.2);
\draw [thick] (5.6,-2) -- (5.4,-1.8);
\draw [thick] (5.6,-2) -- (5.4,-2.2);

\end{tikzpicture}\\ 
\hline 
SO(5)_{8}\times SU(4)_{8} & (4, 4, 6)& 15& \begin{tikzpicture}
\filldraw[fill= white] (1,-2) circle [radius=0.1] node[below] {\scriptsize 1};
\filldraw[fill= white] (2,-2) circle [radius=0.1] node[below] {\scriptsize 2};
\filldraw[fill= white] (3,-2) circle [radius=0.1] node[below] {\scriptsize 3};
\filldraw[fill= white] (4,-2) circle [radius=0.1] node[below] {\scriptsize 4};
\filldraw[fill= white] (5,-2) circle [radius=0.1] node[below] {\scriptsize 4};
\filldraw[fill= white] (6,-2) circle [radius=0.1] node[below] {\scriptsize 2};
\node[] at (2.5,-1.5) {};

\draw [thick] (1.1, -2) -- (1.9,-2) ;
\draw [thick] (3.1, -1.95) -- (3.9,-1.95) ;
\draw [thick] (3.1, -2.05) -- (3.9,-2.05) ;
\draw [thick] (2.1, -2) -- (2.9,-2) ;
\draw [thick] (4.1, -2) -- (4.9,-2) ;
\draw [thick] (5.1, -1.95) -- (5.9,-1.95) ;
\draw [thick] (5.1, -2.05) -- (5.9,-2.05) ;
\draw [thick] (3.4,-2) -- (3.6,-1.8);
\draw [thick] (3.4,-2) -- (3.6,-2.2);
\draw [thick] (5.6,-2) -- (5.4,-1.8);
\draw [thick] (5.6,-2) -- (5.4,-2.2);

\end{tikzpicture}\\ 
\hline 
SO(7)_{9}\times U(1) & (3, \frac{9}{2}, 5)& 12& \begin{tikzpicture}
\filldraw[fill= white] (2,-4) circle [radius=0.1] node[below] {\scriptsize 1};
\filldraw[fill= white] (3,-4) circle [radius=0.1] node[below] {\scriptsize 3};
\filldraw[fill= white] (4,-4) circle [radius=0.1] node[below] {\scriptsize 5};
\filldraw[fill= white] (5,-4) circle [radius=0.1] node[below] {\scriptsize 3};
\filldraw[fill= white] (4.5,-3) circle [radius=0.1] node[above] {\scriptsize 1};

\draw [thick] (3.1, -3.95) -- (3.9,-3.95) ;
\draw [thick] (3.1, -4.05) -- (3.9,-4.05) ;
\draw [thick] (2.1, -4) -- (2.9,-4) ;
\draw [thick] (4.1, -4) -- (4.9,-4) ;
\draw [thick] (3.4,-4) -- (3.6,-3.8);
\draw [thick] (3.4,-4) -- (3.6,-4.2);
\draw [thick] (4.02, -3.95) -- (4.48,-3.05);
\draw [thick] (4.98, -3.95) -- (4.52,-3.05) ;
\end{tikzpicture}\\ 
\hline 
SU(3)_{6} & (2, 3, 3)& 3 & \begin{tikzpicture}
 \filldraw[fill= white] (0,0) circle [radius=0.1] node[below] {\scriptsize 1};
\filldraw[fill= white] (1,0) circle [radius=0.1] node[below] {\scriptsize 2};
\filldraw[fill= white] (0,1) circle [radius=0.1] node[left] {\scriptsize 2};
\draw [thick] (0.1,0.05)--(0.9,0.05);
\draw [thick] (0.1,-0.05)--(0.9,-0.05);
\draw [thick] (0.05, 0.1)--(0.05, 0.9);
\draw [thick] (-0.05,0.1)--(-0.05, 0.9);
\draw [thick] (0.05,0.95)--(0.95,0.05);
\end{tikzpicture}\\ \hline
 U(1) & (2,2,2) & -1&  \begin{tikzpicture}
\filldraw[fill= white] (1,0) circle [radius=0.1] node[below] {\scriptsize 2};
\draw [thick] (1.05, 0.05) to [out=45,in=315,looseness=15] (1.05,-0.05);
\draw [thick] (0.95, 0.05) to [out=135,in=225,looseness=15] (0.95,-0.05);
\end{tikzpicture}\\
\hline 

\end{array}$$ 
In this case we include only the theory associated with the 7-brane of type $D_4$ and we do not proceed to $\mathcal H_2$ since we did not find such a deformation when $n_4\neq 0$. As a result of the $SU(4)$ mass deformation discussed in Section \ref{Sec:IIBSfolds} we land on a theory with CB spectrum $(2,4,4)$ and symmetry $SO(7)_8$. We can readily identify it with a (non-conformal) $SU(2)$ gauging of the rank-2 $SO(7)_8\times SU(2)_5^2$ theory. The latter can in turn be deformed to two different theories: Either the theory with $SU(3)_6\times SU(2)_4^2$ symmetry and CB spectrum $(3,3)$ or $SO(4)$ SQCD with two vectors \cite{Martone:2021drm}. In both cases the $SU(2)$ gauging is conformal and we therefore include these two theories. The first option leads to the non-lagrangian SCFT with symmetry $SU(3)_6$ while the second is lagrangian and describes a $SU(2)\times SO(4)$ circular quiver. This is also equivalent to the $A_1$ class-$\mathcal{S}$ theory on a genus-2 surface without punctures. A class-$\mathcal{S}$ description of the $SU(3)_6$ model is given by the $A_2$ theory on a torus with a full puncture and a $\bbZ_2$ twist line along one cycle of the torus (see \cite{Chacaltana:2012ch}). We add our proposal for its magnetic quiver.

\paragraph{Descendants of the $USp(10)_{14}\times U(1)$ theory} 

$$\begin{array} {|c|c|c|c|} 
\hline 
\text{Theory}& (\Delta_1, \Delta_2, \Delta_3)& 24(c-a)& \text{Magnetic Quiver}\\ 
\hline 
USp(10)_{14}\times U(1) & (6, 7, 12)& 33& \begin{tikzpicture}
% \filldraw[fill= white] (-2,0) circle [radius=0.1] node[below] {\scriptsize 1};
% \filldraw[fill= white] (-1,0) circle [radius=0.1] node[below] {\scriptsize 2};
%\filldraw[fill= white] (0,0) circle [radius=0.1] node[below] {\scriptsize 3};
\filldraw[fill= white] (1,0) circle [radius=0.1] node[below] {\scriptsize 2};
\filldraw[fill= white] (2,0) circle [radius=0.1] node[below] {\scriptsize 4};
\filldraw[fill= white] (3,0) circle [radius=0.1] node[below] {\scriptsize 6};
\filldraw[fill= white] (4,0) circle [radius=0.1] node[below] {\scriptsize 8};
\filldraw[fill= white] (5,0) circle [radius=0.1] node[below] {\scriptsize 10};
\filldraw[fill= white] (6,0) circle [radius=0.1] node[below] {\scriptsize 4}; 
\node[] at (2.5,0.5) {};

%\draw [thick] (-1.1, 0) -- (-1.9,0) ;
%\draw [thick] (-0.1, 0) -- (-0.9,0) ;
%\draw [thick] (0.1, 0) -- (0.9,0) ;
\draw [thick] (1.1, 0) -- (1.9,0) ;
\draw [thick] (2.1, 0) -- (2.9,0) ;
\draw [thick] (4.1, 0.05) -- (4.9,0.05) ;
\draw [thick] (4.1, -0.05) -- (4.9,-0.05) ;
\draw [thick] (3.1, 0) -- (3.9,0) ;
\draw [thick] (5.1, 0) -- (5.9,0) ;
\draw [thick] (4.4,0) -- (4.6,0.2);
\draw [thick] (4.4,0) -- (4.6,-0.2); 
\end{tikzpicture}\\ 
\hline 
USp(6)_{10}\times SU(2)_{16} \times U(1)& (4, 5, 8)& 17& \begin{tikzpicture}
%\filldraw[fill= white] (-1,-2) circle [radius=0.1] node[below] {\scriptsize  1};
%\filldraw[fill= white] (0,-2) circle [radius=0.1] node[below] {\scriptsize 2};
%\filldraw[fill= white] (1,-2) circle [radius=0.1] node[below] {\scriptsize A-2};
\filldraw[fill= white] (2,-2) circle [radius=0.1] node[below] {\scriptsize 2};
\filldraw[fill= white] (3,-2) circle [radius=0.1] node[below] {\scriptsize 4};
\filldraw[fill= white] (4,-2) circle [radius=0.1] node[below] {\scriptsize 6};
\filldraw[fill= white] (5,-2) circle [radius=0.1] node[below] {\scriptsize 4};
\filldraw[fill= white] (6,-2) circle [radius=0.1] node[below] {\scriptsize 2};
\node[] at (2.5,-1.5) {};

%\draw [thick] (-0.1, -2) -- (-0.9,-2) ;
%\draw [thick] (0.1, -2) -- (0.9,-2) ;
%\draw [thick] (1.1, -2) -- (1.9,-2) ;
\draw [thick] (3.1, -1.95) -- (3.9,-1.95) ;
\draw [thick] (3.1, -2.05) -- (3.9,-2.05) ;
\draw [thick] (2.1, -2) -- (2.9,-2) ;
\draw [thick] (4.1, -2) -- (4.9,-2) ;
\draw [thick] (5.1, -1.95) -- (5.9,-1.95) ;
\draw [thick] (5.1, -2.05) -- (5.9,-2.05) ;
\draw [thick] (3.4,-2) -- (3.6,-1.8);
\draw [thick] (3.4,-2) -- (3.6,-2.2);
\draw [thick] (5.6,-2) -- (5.4,-1.8);
\draw [thick] (5.6,-2) -- (5.4,-2.2);

\end{tikzpicture}\\ 
\hline 
USp(4)_{8}\times U(1)^2 & (3, 4, 6)& 9& \begin{tikzpicture}
%\filldraw[fill= white] (0,-4) circle [radius=0.1] node[below] {\scriptsize  1};
%\filldraw[fill= white] (1,-4) circle [radius=0.1] node[below] {\scriptsize A-3};
%\filldraw[fill= white] (2,-4) circle [radius=0.1] node[below] {\scriptsize B-3};
\filldraw[fill= white] (3,-4) circle [radius=0.1] node[below] {\scriptsize 2};
\filldraw[fill= white] (4,-4) circle [radius=0.1] node[below] {\scriptsize 4};
\filldraw[fill= white] (5,-4) circle [radius=0.1] node[below] {\scriptsize 2};
\filldraw[fill= white] (4.5,-3) circle [radius=0.1] node[above] {\scriptsize 2};
%\draw [thick] (0.1, -4) -- (0.9,-4) ;
%\draw [thick] (1.1, -4) -- (1.9,-4) ;
\draw [thick] (3.1, -3.95) -- (3.9,-3.95) ;
\draw [thick] (3.1, -4.05) -- (3.9,-4.05) ;
%\draw [thick] (2.1, -4) -- (2.9,-4) ;
\draw [thick] (4.1, -4) -- (4.9,-4) ;
\draw [thick] (3.4,-4) -- (3.6,-3.8);
\draw [thick] (3.4,-4) -- (3.6,-4.2);
\draw [thick] (4.02, -3.95) -- (4.48,-3.05);
\draw [thick] (4.98, -3.95) -- (4.52,-3.05) ;
\end{tikzpicture}\\ 
\hline 
SU(2)_6 \times U(1)& (2, 3, 4)& 1 & ? \\ 
\hline 
\end{array}$$ 
The level $16$ for $SU(2)$ in the second line of the above table does not follow from our rules and is our guess: We observe that the global symmetry is the same as in the pure S-fold ($k=1$) $\mathcal{S}$ series at rank 1, but all ranks in the magnetic quivers and all other flavor central charges are doubled. We therefore guess that also the $SU(2)$ flavor central charge is doubled with respect to its rank-1 counterpart. Regarding the model with spectrum $(2,3,4)$, which is again the $SU(4)$ gauge theory with symmetric and antisymmetric matter, we find directly the defining data of this theory from our algorithm, without further deformations. Finally, the extra $U(1)$ factor in the global symmetry of the first three theories is inherited from the parent 6d model and is preserved by the SW twist.\footnote{We thank the referee for pointing this out.}

 %We can also confirm this result more indirectly as follows: Start from the orbi-instanton theory with trivial $\sigma$. The theory associated with the trivial 7-brane was studied in Section \ref{Sec:Pert}, where it was found that for $N=2$ we have a $SU(4)$ vector multiplet coupled to the $E_7$ Minahan-Nemeschansky theory. We then turn on $\sigma$ in this theory by higgsing the $E_7$ theory. This turns the theory into 16 free hypermultiplets organized into symmetric and antisymmetric rank-2 tensors. 

\paragraph{Descendants of the $SU(5)_{28}\times SU(2)_{15}$ theory} 

$$\begin{array} {|c|c|c|c|} 
\hline 
\text{Theory}& (\Delta_1, \Delta_2, \Delta_3)& 24(c-a)& \text{Magnetic Quiver}\\ 
\hline 
SU(5)_{28}\times SU(2)_{15} & (6, 12, 14)& 34& \begin{tikzpicture}
% \filldraw[fill= white] (-2,0) circle [radius=0.1] node[below] {\scriptsize 1};
% \filldraw[fill= white] (-1,0) circle [radius=0.1] node[below] {\scriptsize 2};
%\filldraw[fill= white] (0,0) circle [radius=0.1] node[below] {\scriptsize 3};
\filldraw[fill= white] (1,0) circle [radius=0.1] node[below] {\scriptsize 2};
\filldraw[fill= white] (2,0) circle [radius=0.1] node[below] {\scriptsize 4};
\filldraw[fill= white] (3,0) circle [radius=0.1] node[below] {\scriptsize 6};
\filldraw[fill= white] (4,0) circle [radius=0.1] node[below] {\scriptsize 8};
\filldraw[fill= white] (5,0) circle [radius=0.1] node[below] {\scriptsize 10};
\filldraw[fill= white] (6,0) circle [radius=0.1] node[below] {\scriptsize 5}; 
\node[] at (2.5,0.5) {};

%\draw [thick] (-1.1, 0) -- (-1.9,0) ;
%\draw [thick] (-0.1, 0) -- (-0.9,0) ;
%\draw [thick] (0.1, 0) -- (0.9,0) ;
\draw [thick] (1.1, 0) -- (1.9,0) ;
\draw [thick] (2.1, 0) -- (2.9,0) ;
\draw [thick] (4.1, 0.05) -- (4.9,0.05) ;
\draw [thick] (4.1, -0.05) -- (4.9,-0.05) ;
\draw [thick] (3.1, 0) -- (3.9,0) ;
\draw [thick] (5.1, 0) -- (5.9,0) ;
\draw [thick] (4.4,0) -- (4.6,0.2);
\draw [thick] (4.4,0) -- (4.6,-0.2); 
\end{tikzpicture}\\ 
\hline 
SU(3)_{20}\times SU(2)_{11}\times U(1) & (4, 8, 10)& 18& \begin{tikzpicture}
%\filldraw[fill= white] (-1,-2) circle [radius=0.1] node[below] {\scriptsize  1};
%\filldraw[fill= white] (0,-2) circle [radius=0.1] node[below] {\scriptsize 2};
%\filldraw[fill= white] (1,-2) circle [radius=0.1] node[below] {\scriptsize A-2};
\filldraw[fill= white] (2,-2) circle [radius=0.1] node[below] {\scriptsize 2};
\filldraw[fill= white] (3,-2) circle [radius=0.1] node[below] {\scriptsize 4};
\filldraw[fill= white] (4,-2) circle [radius=0.1] node[below] {\scriptsize 6};
\filldraw[fill= white] (5,-2) circle [radius=0.1] node[below] {\scriptsize 5};
\filldraw[fill= white] (6,-2) circle [radius=0.1] node[below] {\scriptsize 2};
\node[] at (2.5,-1.5) {};

%\draw [thick] (-0.1, -2) -- (-0.9,-2) ;
%\draw [thick] (0.1, -2) -- (0.9,-2) ;
%\draw [thick] (1.1, -2) -- (1.9,-2) ;
\draw [thick] (3.1, -1.95) -- (3.9,-1.95) ;
\draw [thick] (3.1, -2.05) -- (3.9,-2.05) ;
\draw [thick] (2.1, -2) -- (2.9,-2) ;
\draw [thick] (4.1, -2) -- (4.9,-2) ;
\draw [thick] (5.1, -1.95) -- (5.9,-1.95) ;
\draw [thick] (5.1, -2.05) -- (5.9,-2.05) ;
\draw [thick] (3.4,-2) -- (3.6,-1.8);
\draw [thick] (3.4,-2) -- (3.6,-2.2);
\draw [thick] (5.6,-2) -- (5.4,-1.8);
\draw [thick] (5.6,-2) -- (5.4,-2.2);

\end{tikzpicture}\\ 
\hline 
SU(2)_{16}\times SU(2)_{9}\times U(1) & (3, 6, 8)& 10& \begin{tikzpicture}
%\filldraw[fill= white] (0,-4) circle [radius=0.1] node[below] {\scriptsize  1};
%\filldraw[fill= white] (1,-4) circle [radius=0.1] node[below] {\scriptsize A-3};
%\filldraw[fill= white] (2,-4) circle [radius=0.1] node[below] {\scriptsize B-3};
\filldraw[fill= white] (3,-4) circle [radius=0.1] node[below] {\scriptsize 2};
\filldraw[fill= white] (4,-4) circle [radius=0.1] node[below] {\scriptsize 4};
\filldraw[fill= white] (5,-4) circle [radius=0.1] node[below] {\scriptsize 2};
\filldraw[fill= white] (4.5,-3) circle [radius=0.1] node[above] {\scriptsize 3};
%\draw [thick] (0.1, -4) -- (0.9,-4) ;
%\draw [thick] (1.1, -4) -- (1.9,-4) ;
\draw [thick] (3.1, -3.95) -- (3.9,-3.95) ;
\draw [thick] (3.1, -4.05) -- (3.9,-4.05) ;
%\draw [thick] (2.1, -4) -- (2.9,-4) ;
\draw [thick] (4.1, -4) -- (4.9,-4) ;
\draw [thick] (3.4,-4) -- (3.6,-3.8);
\draw [thick] (3.4,-4) -- (3.6,-4.2);
\draw [thick] (4.02, -3.95) -- (4.48,-3.05);
\draw [thick] (4.98, -3.95) -- (4.52,-3.05) ;
\end{tikzpicture}\\ 
\hline  
\end{array}$$ 
The model in the last row above is expected to flow, upon mass deformation, to a theory with CB spectrum $(2,4,6)$. We currently do not have a precise understanding of this theory and therefore we refrain from including it.

\paragraph{Descendants of the $SU(7)_{18}\times U(1)$ theory} 

$$\begin{array} {|c|c|c|c|} 
\hline 
\text{Theory}& (\Delta_1, \Delta_2, \Delta_3)& 24(c-a)& \text{Magnetic Quiver}\\ 
\hline 
SU(7)_{18}\times U(1) & (6, 8, 9)& 30& \begin{tikzpicture}
 \filldraw[fill= white] (-1,0) circle [radius=0.1] node[below] {\scriptsize 1};
\filldraw[fill= white] (0,0) circle [radius=0.1] node[below] {\scriptsize 2};
\filldraw[fill= white] (1,0) circle [radius=0.1] node[below] {\scriptsize 3};
\filldraw[fill= white] (2,0) circle [radius=0.1] node[below] {\scriptsize 4};
\filldraw[fill= white] (3,0) circle [radius=0.1] node[below] {\scriptsize 5};
\filldraw[fill= white] (4,0) circle [radius=0.1] node[below] {\scriptsize 6};
\filldraw[fill= white] (5,0) circle [radius=0.1] node[below] {\scriptsize 7};
\filldraw[fill= white] (6,0) circle [radius=0.1] node[below] {\scriptsize 3}; 
\node[] at (2.5,0.5) {};

\draw [thick] (-0.1, 0) -- (-0.9,0) ;
\draw [thick] (0.1, 0) -- (0.9,0) ;
\draw [thick] (1.1, 0) -- (1.9,0) ;
\draw [thick] (2.1, 0) -- (2.9,0) ;
\draw [thick] (4.1, 0.05) -- (4.9,0.05) ;
\draw [thick] (4.1, -0.05) -- (4.9,-0.05) ;
\draw [thick] (3.1, 0) -- (3.9,0) ;
\draw [thick] (5.1, 0) -- (5.9,0) ;
\draw [thick] (4.4,0) -- (4.6,0.2);
\draw [thick] (4.4,0) -- (4.6,-0.2); 
\end{tikzpicture}\\ 
\hline 
SU(5)_{14}\times U(1)^2 & (4, 6, 7)& 18& \begin{tikzpicture}
\filldraw[fill= white] (0,-2) circle [radius=0.1] node[below] {\scriptsize 1};
\filldraw[fill= white] (1,-2) circle [radius=0.1] node[below] {\scriptsize 2};
\filldraw[fill= white] (2,-2) circle [radius=0.1] node[below] {\scriptsize 3};
\filldraw[fill= white] (3,-2) circle [radius=0.1] node[below] {\scriptsize 4};
\filldraw[fill= white] (4,-2) circle [radius=0.1] node[below] {\scriptsize 5};
\filldraw[fill= white] (5,-2) circle [radius=0.1] node[below] {\scriptsize 3};
\filldraw[fill= white] (6,-2) circle [radius=0.1] node[below] {\scriptsize 1};
\node[] at (2.5,-1.5) {};

\draw [thick] (0.1, -2) -- (0.9,-2) ;
\draw [thick] (1.1, -2) -- (1.9,-2) ;
\draw [thick] (3.1, -1.95) -- (3.9,-1.95) ;
\draw [thick] (3.1, -2.05) -- (3.9,-2.05) ;
\draw [thick] (2.1, -2) -- (2.9,-2) ;
\draw [thick] (4.1, -2) -- (4.9,-2) ;
\draw [thick] (5.1, -1.95) -- (5.9,-1.95) ;
\draw [thick] (5.1, -2.05) -- (5.9,-2.05) ;
\draw [thick] (3.4,-2) -- (3.6,-1.8);
\draw [thick] (3.4,-2) -- (3.6,-2.2);
\draw [thick] (5.6,-2) -- (5.4,-1.8);
\draw [thick] (5.6,-2) -- (5.4,-2.2);

\end{tikzpicture}\\ 
\hline 
SU(4)_{12}\times U(1)^2 & (3, 5, 6)& 12& \begin{tikzpicture}
\filldraw[fill= white] (1,-4) circle [radius=0.1] node[below] {\scriptsize 1};
\filldraw[fill= white] (2,-4) circle [radius=0.1] node[below] {\scriptsize 2};
\filldraw[fill= white] (3,-4) circle [radius=0.1] node[below] {\scriptsize 3};
\filldraw[fill= white] (4,-4) circle [radius=0.1] node[below] {\scriptsize 4};
\filldraw[fill= white] (5,-4) circle [radius=0.1] node[below] {\scriptsize 2};
\filldraw[fill= white] (4.5,-3) circle [radius=0.1] node[above] {\scriptsize 1};

\draw [thick] (1.1, -4) -- (1.9,-4) ;
\draw [thick] (3.1, -3.95) -- (3.9,-3.95) ;
\draw [thick] (3.1, -4.05) -- (3.9,-4.05) ;
\draw [thick] (2.1, -4) -- (2.9,-4) ;
\draw [thick] (4.1, -4) -- (4.9,-4) ;
\draw [thick] (3.4,-4) -- (3.6,-3.8);
\draw [thick] (3.4,-4) -- (3.6,-4.2);
\draw [thick] (4.02, -3.95) -- (4.48,-3.05);
\draw [thick] (4.98, -3.95) -- (4.52,-3.05) ;
\end{tikzpicture}\\ 
\hline 
U(2)_{8}\times U(1) & (2, 3, 4)& 3 &?\\ 
\hline 
\end{array}$$ 
The last line of the above table deserves some comments. According to our algorithm, the theory should have spectrum $(2,4,5)$ and global symmetry which includes a $SU(3)_{10}$ factor. From these data we can interpret the theory as a $SU(2)$ gauging of model 35 in \cite{Martone:2021drm}. The gauging however is not conformal. Theory 35, in turn, can be mass-deformed to model 37 and, after this mass deformation, the $SU(2)$ gauging becomes conformal. We therefore include this SCFT rather than the $(2,4,5)$ model in the table. We can now notice that all the data of the SCFT match those of the $SU(4)$ gauge theory with two flavors and one hypermultiplet in the symmetric and therefore we identify our SCFT with the gauge theory. This is a special case of the Argyres-Seiberg-like duality \cite{Argyres:2007cn} observed in \cite{Zafrir:2016wkk}. An alternative dual description of the same gauge theory is given by example 14 of \cite{Argyres:2007tq}.

\paragraph{Descendants of the $USp(16)_{10}$ theory} 

$$\begin{array} {|c|c|c|c|} 
\hline 
\text{Theory}& (\Delta_1, \Delta_2, \Delta_3)& 24(c-a)& \text{Magnetic Quiver}\\ 
\hline 
USp(16)_{10} & (5, 6, 8)& 37& \begin{tikzpicture}
\filldraw[fill= white] (-2,0) circle [radius=0.1] node[below] {\scriptsize 1};
 \filldraw[fill= white] (-1,0) circle [radius=0.1] node[below] {\scriptsize 2};
\filldraw[fill= white] (0,0) circle [radius=0.1] node[below] {\scriptsize 3};
\filldraw[fill= white] (1,0) circle [radius=0.1] node[below] {\scriptsize 4};
\filldraw[fill= white] (2,0) circle [radius=0.1] node[below] {\scriptsize 5};
\filldraw[fill= white] (3,0) circle [radius=0.1] node[below] {\scriptsize 6};
\filldraw[fill= white] (4,0) circle [radius=0.1] node[below] {\scriptsize 7};
\filldraw[fill= white] (5,0) circle [radius=0.1] node[below] {\scriptsize 8};
\filldraw[fill= white] (6,0) circle [radius=0.1] node[below] {\scriptsize 2}; 
\node[] at (2.5,0.5) {};

\draw [thick] (-1.1, 0) -- (-1.9,0) ;
\draw [thick] (-0.1, 0) -- (-0.9,0) ;
\draw [thick] (0.1, 0) -- (0.9,0) ;
\draw [thick] (1.1, 0) -- (1.9,0) ;
\draw [thick] (2.1, 0) -- (2.9,0) ;
\draw [thick] (4.1, 0.05) -- (4.9,0.05) ;
\draw [thick] (4.1, -0.05) -- (4.9,-0.05) ;
\draw [thick] (3.1, 0) -- (3.9,0) ;
\draw [thick] (5.1, 0) -- (5.9,0) ;
\draw [thick] (4.4,0) -- (4.6,0.2);
\draw [thick] (4.4,0) -- (4.6,-0.2); 
\end{tikzpicture}\\ 
\hline 
USp(12)_{8}\times SU(2)_{8} & (4, 4, 6)& 23& \begin{tikzpicture}
\filldraw[fill= white] (-1,-2) circle [radius=0.1] node[below] {\scriptsize  1};
\filldraw[fill= white] (0,-2) circle [radius=0.1] node[below] {\scriptsize 2};
\filldraw[fill= white] (1,-2) circle [radius=0.1] node[below] {\scriptsize 3};
\filldraw[fill= white] (2,-2) circle [radius=0.1] node[below] {\scriptsize 4};
\filldraw[fill= white] (3,-2) circle [radius=0.1] node[below] {\scriptsize 5};
\filldraw[fill= white] (4,-2) circle [radius=0.1] node[below] {\scriptsize 6};
\filldraw[fill= white] (5,-2) circle [radius=0.1] node[below] {\scriptsize 2};
\filldraw[fill= white] (6,-2) circle [radius=0.1] node[below] {\scriptsize 1};
\node[] at (2.5,-1.5) {};

\draw [thick] (-0.1, -2) -- (-0.9,-2) ;
\draw [thick] (0.1, -2) -- (0.9,-2) ;
\draw [thick] (1.1, -2) -- (1.9,-2) ;
\draw [thick] (3.1, -1.95) -- (3.9,-1.95) ;
\draw [thick] (3.1, -2.05) -- (3.9,-2.05) ;
\draw [thick] (2.1, -2) -- (2.9,-2) ;
\draw [thick] (4.1, -2) -- (4.9,-2) ;
\draw [thick] (5.1, -1.95) -- (5.9,-1.95) ;
\draw [thick] (5.1, -2.05) -- (5.9,-2.05) ;
\draw [thick] (3.4,-2) -- (3.6,-1.8);
\draw [thick] (3.4,-2) -- (3.6,-2.2);
\draw [thick] (5.6,-2) -- (5.4,-1.8);
\draw [thick] (5.6,-2) -- (5.4,-2.2);

\end{tikzpicture}\\ 
\hline 
USp(10)_{7}\times U(1) & (3, \frac{7}{2}, 5)& 16& \begin{tikzpicture}
\filldraw[fill= white] (0,-4) circle [radius=0.1] node[below] {\scriptsize  1};
\filldraw[fill= white] (1,-4) circle [radius=0.1] node[below] {\scriptsize 2};
\filldraw[fill= white] (2,-4) circle [radius=0.1] node[below] {\scriptsize 3};
\filldraw[fill= white] (3,-4) circle [radius=0.1] node[below] {\scriptsize 4};
\filldraw[fill= white] (4,-4) circle [radius=0.1] node[below] {\scriptsize 5};
\filldraw[fill= white] (5,-4) circle [radius=0.1] node[below] {\scriptsize 1};
\filldraw[fill= white] (4.5,-3) circle [radius=0.1] node[above] {\scriptsize 1};

\draw [thick] (0.1, -4) -- (0.9,-4) ;
\draw [thick] (1.1, -4) -- (1.9,-4) ;
\draw [thick] (3.1, -3.95) -- (3.9,-3.95) ;
\draw [thick] (3.1, -4.05) -- (3.9,-4.05) ;
\draw [thick] (2.1, -4) -- (2.9,-4) ;
\draw [thick] (4.1, -4) -- (4.9,-4) ;
\draw [thick] (3.4,-4) -- (3.6,-3.8);
\draw [thick] (3.4,-4) -- (3.6,-4.2);
\draw [thick] (4.02, -3.95) -- (4.48,-3.05);
\draw [thick] (4.98, -3.95) -- (4.52,-3.05) ;
\end{tikzpicture}\\ 
\hline 
USp(8)_{6} & (2, 3, 4)& 9&\begin{tikzpicture}
\filldraw[fill= white] (1,-6) circle [radius=0.1] node[below] {\scriptsize 1};
\filldraw[fill= white] (2,-6) circle [radius=0.1] node[below] {\scriptsize 2};
\filldraw[fill= white] (3,-6) circle [radius=0.1] node[below] {\scriptsize 3};
\filldraw[fill= white] (4,-6) circle [radius=0.1] node[below] {\scriptsize 4};
\filldraw[fill= white] (4,-5) circle [radius=0.1] node[above] {\scriptsize 1};

\draw [thick] (1.1, -6) -- (1.9,-6) ;
\draw [thick] (3.1, -5.95) -- (3.9,-5.95) ;
\draw [thick] (3.1, -6.05) -- (3.9,-6.05) ;
\draw [thick] (2.1, -6) -- (2.9,-6) ;
\draw [thick] (3.4,-6) -- (3.6,-5.8);
\draw [thick] (3.4,-6) -- (3.6,-6.2);
\draw [thick] (4.05, -5.95) -- (4.05,-5.05);
\draw [thick] (3.95, -5.95) -- (3.95,-5.05) ;
\end{tikzpicture}\\ 
\hline 
\end{array}$$

\paragraph{Descendants of the $USp(18)_{11}$ theory} 

$$\begin{array} {|c|c|c|c|} 
\hline 
\text{Theory}& (\Delta_1, \Delta_2, \Delta_3)& 24(c-a)& \text{Magnetic Quiver}\\ 
\hline 
USp(18)_{11} & (6, 8, 10)& 46& \begin{tikzpicture}
\filldraw[fill= white] (-3,0) circle [radius=0.1] node[below] {\scriptsize 1};
\filldraw[fill= white] (-2,0) circle [radius=0.1] node[below] {\scriptsize 2};
 \filldraw[fill= white] (-1,0) circle [radius=0.1] node[below] {\scriptsize 3};
\filldraw[fill= white] (0,0) circle [radius=0.1] node[below] {\scriptsize 4};
\filldraw[fill= white] (1,0) circle [radius=0.1] node[below] {\scriptsize 5};
\filldraw[fill= white] (2,0) circle [radius=0.1] node[below] {\scriptsize 6};
\filldraw[fill= white] (3,0) circle [radius=0.1] node[below] {\scriptsize 7};
\filldraw[fill= white] (4,0) circle [radius=0.1] node[below] {\scriptsize 8};
\filldraw[fill= white] (5,0) circle [radius=0.1] node[below] {\scriptsize 9};
\filldraw[fill= white] (6,0) circle [radius=0.1] node[below] {\scriptsize 2}; 
\node[] at (2.5,0.5) {};

\draw [thick] (-2.1, 0) -- (-2.9,0) ;
\draw [thick] (-1.1, 0) -- (-1.9,0) ;
\draw [thick] (-0.1, 0) -- (-0.9,0) ;
\draw [thick] (0.1, 0) -- (0.9,0) ;
\draw [thick] (1.1, 0) -- (1.9,0) ;
\draw [thick] (2.1, 0) -- (2.9,0) ;
\draw [thick] (4.1, 0.05) -- (4.9,0.05) ;
\draw [thick] (4.1, -0.05) -- (4.9,-0.05) ;
\draw [thick] (3.1, 0) -- (3.9,0) ;
\draw [thick] (5.1, 0) -- (5.9,0) ;
\draw [thick] (4.4,0) -- (4.6,0.2);
\draw [thick] (4.4,0) -- (4.6,-0.2); 
\end{tikzpicture}\\ 
\hline 
USp(14)_{9}\times SU(2)_{8} & (4, 6, 8)& 30& \begin{tikzpicture}
\filldraw[fill= white] (-2,-2) circle [radius=0.1] node[below] {\scriptsize  1};
\filldraw[fill= white] (-1,-2) circle [radius=0.1] node[below] {\scriptsize  2};
\filldraw[fill= white] (0,-2) circle [radius=0.1] node[below] {\scriptsize 3};
\filldraw[fill= white] (1,-2) circle [radius=0.1] node[below] {\scriptsize 4};
\filldraw[fill= white] (2,-2) circle [radius=0.1] node[below] {\scriptsize 5};
\filldraw[fill= white] (3,-2) circle [radius=0.1] node[below] {\scriptsize 6};
\filldraw[fill= white] (4,-2) circle [radius=0.1] node[below] {\scriptsize 7};
\filldraw[fill= white] (5,-2) circle [radius=0.1] node[below] {\scriptsize 2};
\filldraw[fill= white] (6,-2) circle [radius=0.1] node[below] {\scriptsize 1};
\node[] at (2.5,-1.5) {};

\draw [thick] (-1.1, -2) -- (-1.9,-2) ;
\draw [thick] (-0.1, -2) -- (-0.9,-2) ;
\draw [thick] (0.1, -2) -- (0.9,-2) ;
\draw [thick] (1.1, -2) -- (1.9,-2) ;
\draw [thick] (3.1, -1.95) -- (3.9,-1.95) ;
\draw [thick] (3.1, -2.05) -- (3.9,-2.05) ;
\draw [thick] (2.1, -2) -- (2.9,-2) ;
\draw [thick] (4.1, -2) -- (4.9,-2) ;
\draw [thick] (5.1, -1.95) -- (5.9,-1.95) ;
\draw [thick] (5.1, -2.05) -- (5.9,-2.05) ;
\draw [thick] (3.4,-2) -- (3.6,-1.8);
\draw [thick] (3.4,-2) -- (3.6,-2.2);
\draw [thick] (5.6,-2) -- (5.4,-1.8);
\draw [thick] (5.6,-2) -- (5.4,-2.2);

\end{tikzpicture}\\ 
\hline 
USp(12)_{8}\times U(1) & (3, 5, 7)& 22& \begin{tikzpicture}
\filldraw[fill= white] (-1,-4) circle [radius=0.1] node[below] {\scriptsize  1};
\filldraw[fill= white] (0,-4) circle [radius=0.1] node[below] {\scriptsize  2};
\filldraw[fill= white] (1,-4) circle [radius=0.1] node[below] {\scriptsize 3};
\filldraw[fill= white] (2,-4) circle [radius=0.1] node[below] {\scriptsize 4};
\filldraw[fill= white] (3,-4) circle [radius=0.1] node[below] {\scriptsize 5};
\filldraw[fill= white] (4,-4) circle [radius=0.1] node[below] {\scriptsize 6};
\filldraw[fill= white] (5,-4) circle [radius=0.1] node[below] {\scriptsize 1};
\filldraw[fill= white] (4.5,-3) circle [radius=0.1] node[above] {\scriptsize 1};

\draw [thick] (-0.1, -4) -- (-0.9,-4) ;
\draw [thick] (0.1, -4) -- (0.9,-4) ;
\draw [thick] (1.1, -4) -- (1.9,-4) ;
\draw [thick] (3.1, -3.95) -- (3.9,-3.95) ;
\draw [thick] (3.1, -4.05) -- (3.9,-4.05) ;
\draw [thick] (2.1, -4) -- (2.9,-4) ;
\draw [thick] (4.1, -4) -- (4.9,-4) ;
\draw [thick] (3.4,-4) -- (3.6,-3.8);
\draw [thick] (3.4,-4) -- (3.6,-4.2);
\draw [thick] (4.02, -3.95) -- (4.48,-3.05);
\draw [thick] (4.98, -3.95) -- (4.52,-3.05) ;
\end{tikzpicture}\\ 
\hline 
USp(10)_{7} & (2, 4, 6)& 14&\begin{tikzpicture}
\filldraw[fill= white] (0,-6) circle [radius=0.1] node[below] {\scriptsize 1};
\filldraw[fill= white] (1,-6) circle [radius=0.1] node[below] {\scriptsize 2};
\filldraw[fill= white] (2,-6) circle [radius=0.1] node[below] {\scriptsize 3};
\filldraw[fill= white] (3,-6) circle [radius=0.1] node[below] {\scriptsize 4};
\filldraw[fill= white] (4,-6) circle [radius=0.1] node[below] {\scriptsize 5};
\filldraw[fill= white] (4,-5) circle [radius=0.1] node[above] {\scriptsize 1};

\draw [thick] (.1, -6) -- (.9,-6) ;
\draw [thick] (1.1, -6) -- (1.9,-6) ;
\draw [thick] (3.1, -5.95) -- (3.9,-5.95) ;
\draw [thick] (3.1, -6.05) -- (3.9,-6.05) ;
\draw [thick] (2.1, -6) -- (2.9,-6) ;
\draw [thick] (3.4,-6) -- (3.6,-5.8);
\draw [thick] (3.4,-6) -- (3.6,-6.2);
\draw [thick] (4.05, -5.95) -- (4.05,-5.05);
\draw [thick] (3.95, -5.95) -- (3.95,-5.05) ;
\end{tikzpicture}\\ 
\hline 
\end{array}$$

\paragraph{Lagrangian theories} 

Within the $\ell=2$ class of theories we find 7 lagrangian models and we now specify for all of them the gauge group and matter content. 
\begin{enumerate} 
\item ${\bf SU(2)_{21}:}$ Either $USp(6)$ or $SO(7)$ $\mathcal N=4$ Super Yang-Mills. 
\item ${\bf SU(2)_{15}:}$ $SU(4)$ $\mathcal N=4$ Super Yang-Mills. 
\item ${\bf SU(2)_{6}\times U(1):}$ $SU(4)$ gauge theory with 1 hypermultiplet in the symmetric and 1 hypermultiplets in the antisymmetric.
\item ${\bf U(2)_{8}\times U(1):}$ $SU(4)$ gauge theory with 1 hypermultiplet in the symmetric and 2 hypermultiplets in the fundamental.
\item ${\bf USp(8)_{6}:}$ $SO(6)$ SQCD with 4 hypermultiplets in the fundamental.
\item ${\bf USp(10)_{7}:}$ $SO(7)$ SQCD with 5 hypermultiplets in the fundamental.
\item ${\bf U(1):}$ $USp(2)\times SO(4)$ gauge theory with 1 hypermultiplet in the bifundamental.
\end{enumerate}

\subsection{Theories with $\ell=3$} 

For $\ell=3$ we find five orbi-instanton starting points, all with $k\leq 2$, which we have already analyzed in detail. In one case we need to consider a theory with non-trivial $\sigma$. We find one lagrangian model in this sequence, which however involves activating two mass deformations starting from the parent SCFT. We denote the corresponding RG flow with a blue arrow. In one case we do not provide any 7-brane mass deformations since the corresponding holonomy ($n_6\neq0$) does not satisfy the constraint discussed in Section \ref{Sec:MassDef} establishing the existence of the mass deformation. In some cases we do not know how to construct the magnetic quiver for the theory with trivial 7-brane.

\begin{equation}\label{diagramell=3} 
\begin{tikzpicture} 
\node[] (a1) at (0,0) {\scriptsize $SU(3)_{38}\times U(1)$}; 
\node[] (a2) at (0,-1) {\scriptsize $U(1)\times U(1)$}; 
\node[] (a3) at (0,-2) {\scriptsize $G(3,1,3)$}; 

\node[] (b1) at (3,0) {\scriptsize $[G_2]_{12}\times U(1)$}; 
\node[] (b2) at (3,-1) {\scriptsize $SU(2)_{8}\times U(1)$}; 
\node[] (b3) at (3,-2) {\scriptsize $G(3,3,3)$}; 

\node[] (c1) at (7,0) {\scriptsize $SU(3)_{16}\times SU(2)_{26}\times U(1)$}; 
\node[] (c2) at (6,-1) {\scriptsize $SU(3)_{\frac{32}{3}}\times U(1)$};
\node[] (c3) at (6,-2) {\scriptsize $SU(2)_{8}$};  
\node[] (C2) at (8.5,-1) {\scriptsize $SU(2)_{22}\times U(1)^2$};
\node[] (C3) at (8.5,-2) {\scriptsize $SU(2)_{20}\times U(1)$};

\node[] (d1) at (12,0) {\scriptsize $SU(4)_{16} \times SU(2)_{8}$}; 
\node[] (d2) at (12,-1) {\scriptsize $SU(2)_{12}\times SU(2)_{\frac{20}{3}} $}; 
\node[] (d3) at (12,-2) {\scriptsize $SU(2)_{\frac{18}{3}} \times U(1)$}; 
\node[] (d4) at (12,-3) {\scriptsize {\color{red}$\varnothing$}}; 

\node[] (e1) at (15,0) {\scriptsize $SU(4)_{28}$}; 
\node[] (e2) at (15,-1) {\scriptsize $SU(2)_{20}\times U(1)$}; 
\node[] (e3) at (15,-2) {\scriptsize $U(1)$}; 

\draw[->, thick] (e1)--(e2); 
\draw[->, thick] (e2)--(e3); 
\draw[->, thick, color=blue] (d3)--(d4); 
\draw[->, thick] (d2)--(d3); 
\draw[->, thick] (d1)--(d2); 
\draw[->, thick] (c1)--(c2); 
\draw[->, thick] (c2)--(c3); 
\draw[->, thick] (c1)--(C2);
\draw[->, thick] (C2)--(C3);
\draw[->, thick] (a1)--(a2); 
\draw[->, thick] (a2)--(a3);
\draw[->, thick] (b1)--(b2); 
\draw[->, thick] (b2)--(b3);
\end{tikzpicture}
\end{equation}

\begin{itemize} 
\item $[G_2]_{12}\times U(1)$: SW compactification of the $\bbZ_3$ orbi-instanton with $(n_3=1)$ and three tensors. 
\item $SU(3)_{38}\times U(1)$: SW compactification of the $\bbZ_3$ orbi-instanton with $(n_{3'}=1)$ and three tensors.
\item $SU(3)_{16}\times SU(2)_{26}\times U(1)$: SW compactification of the $\bbZ_6$ orbi-instanton with $(n_3=n_{3'}=1)$ and two tensors.
\item $SU(4)_{16} \times SU(2)_{8}$: SW compactification of the $\bbZ_6$ orbi-instanton with $(n_{6}=1)$ and two tensors. 
\item  $SU(4)_{28}$: SW compactification of the $\bbZ_6$ orbi-instanton with $(n_{3'}=2)$, two tensors and non-trivial $\sigma=(2,2,2)$.
\end{itemize}

\paragraph{Descendants of the $SU(3)_{38}\times U(1)$ theory} 

$$\begin{array} {|c|c|c|c|} 
\hline 
\text{Theory}& (\Delta_1, \Delta_2, \Delta_3)& 24(c-a)& \text{Magnetic Quiver}\\ 
\hline 
SU(3)_{38}\times U(1) & (6, 12, 18)& 21& \begin{tikzpicture}
%\filldraw[fill= white] (-1,-6) circle [radius=0.1] node[below] {\scriptsize 1};
%\filldraw[fill= white] (0,-6) circle [radius=0.1] node[below] {\scriptsize 2};
\filldraw[fill= white] (1,-6) circle [radius=0.1] node[below] {\scriptsize 1};
\filldraw[fill= white] (2,-6) circle [radius=0.1] node[below] {\scriptsize 4};
\filldraw[fill= white] (3,-6) circle [radius=0.1] node[below] {\scriptsize 7};
\filldraw[fill= white] (4,-6) circle [radius=0.1] node[below] {\scriptsize 10};
\node[] at (2.5,-5.5) {};
\draw [thick] (3.1, -5.95) -- (3.9,-5.95) ;
\draw [thick] (3.1, -6) -- (3.9,-6) ;
\draw [thick] (3.1, -6.05) -- (3.9,-6.05) ;
\draw [thick] (2.1, -6) -- (2.9,-6) ;
\draw [thick] (3.4,-6) -- (3.6,-5.8);
\draw [thick] (3.4,-6) -- (3.6,-6.2);
%\draw [thick] (-0.1, -6) -- (-0.9,-6);
%\draw [thick] (0.1, -6) -- (0.9,-6);
\draw [thick] (1.1, -6) -- (1.9,-6) ; 

\end{tikzpicture}\\ 
\hline 
U(1)\times U(1) & (4, 8, 12)& 7& \begin{tikzpicture}
%\filldraw[fill= white] (0,-9) circle [radius=0.1] node[below] {\scriptsize 1};
%\filldraw[fill= white] (1,-9) circle [radius=0.1] node[below] {\scriptsize A-2};
\filldraw[fill= white] (2,-9) circle [radius=0.1] node[below] {\scriptsize 1};
\filldraw[fill= white] (3,-9) circle [radius=0.1] node[below] {\scriptsize 4};
\filldraw[fill= white] (4,-9) circle [radius=0.1] node[below] {\scriptsize 3};
\node[] at (2.5,-8.5) {};
\draw [thick] (2.1, -8.95) -- (2.9,-8.95) ;
\draw [thick] (2.1, -9) -- (2.9,-9) ;
\draw [thick] (2.1, -9.05) -- (2.9,-9.05) ;
\draw [thick] (2.4,-9) -- (2.6,-8.8);
\draw [thick] (2.4,-9) -- (2.6,-9.2);
\draw [thick] (3.1, -8.95) -- (3.9,-8.95);
\draw [thick] (3.1, -9.05) -- (3.9,-9.05);
%\draw [thick] (0.1, -9) -- (0.9,-9);
%\draw [thick] (1.1, -9) -- (1.9,-9) ;
\end{tikzpicture}\\ 
\hline 
G(3,1,3) & (3, 6, 9)& 0& \begin{tikzpicture}
 \filldraw[fill= white] (0,0) circle [radius=0.1] node[below] {\scriptsize 1};
\filldraw[fill= white] (1,0) circle [radius=0.1] node[below] {\scriptsize 3};
\draw [thick] (0.1,0)--(0.9,0);
\draw [thick] (0.1,0.05)--(0.9,0.05);
\draw [thick] (0.1,-0.05)--(0.9,-0.05);
\draw [thick] (1.05, 0.05) to [out=45,in=315,looseness=15] (1.05,-0.05);
\end{tikzpicture}\\ 
\hline  
\end{array}$$ 

\paragraph{Descendants of the  $[G_2]_{12}\times U(1)$ theory} 

$$\begin{array} {|c|c|c|c|} 
\hline 
\text{Theory}& (\Delta_1, \Delta_2, \Delta_3)& 24(c-a)& \text{Magnetic Quiver}\\ 
\hline 
 [G_2]_{12}\times U(1) & (6, 6, 12)& 18& \begin{tikzpicture}
%\filldraw[fill= white] (-1,-6) circle [radius=0.1] node[below] {\scriptsize 1};
%\filldraw[fill= white] (0,-6) circle [radius=0.1] node[below] {\scriptsize 2};
\filldraw[fill= white] (1,-6) circle [radius=0.1] node[below] {\scriptsize 1};
\filldraw[fill= white] (2,-6) circle [radius=0.1] node[below] {\scriptsize 3};
\filldraw[fill= white] (3,-6) circle [radius=0.1] node[below] {\scriptsize 6};
\filldraw[fill= white] (4,-6) circle [radius=0.1] node[below] {\scriptsize 9};
\node[] at (2.5,-5.5) {};
\draw [thick] (3.1, -5.95) -- (3.9,-5.95) ;
\draw [thick] (3.1, -6) -- (3.9,-6) ;
\draw [thick] (3.1, -6.05) -- (3.9,-6.05) ;
\draw [thick] (2.1, -6) -- (2.9,-6) ;
\draw [thick] (3.4,-6) -- (3.6,-5.8);
\draw [thick] (3.4,-6) -- (3.6,-6.2);
%\draw [thick] (-0.1, -6) -- (-0.9,-6);
%\draw [thick] (0.1, -6) -- (0.9,-6);
\draw [thick] (1.1, -6) -- (1.9,-6) ; 

\end{tikzpicture}\\ 
\hline 
SU(2)_{8}\times U(1) & (4, 4, 8)& 6& \begin{tikzpicture}
%\filldraw[fill= white] (0,-9) circle [radius=0.1] node[below] {\scriptsize 1};
%\filldraw[fill= white] (1,-9) circle [radius=0.1] node[below] {\scriptsize A-2};
\filldraw[fill= white] (2,-9) circle [radius=0.1] node[below] {\scriptsize 1};
\filldraw[fill= white] (3,-9) circle [radius=0.1] node[below] {\scriptsize 3};
\filldraw[fill= white] (4,-9) circle [radius=0.1] node[below] {\scriptsize 3};
\node[] at (2.5,-8.5) {};
\draw [thick] (2.1, -8.95) -- (2.9,-8.95) ;
\draw [thick] (2.1, -9) -- (2.9,-9) ;
\draw [thick] (2.1, -9.05) -- (2.9,-9.05) ;
\draw [thick] (2.4,-9) -- (2.6,-8.8);
\draw [thick] (2.4,-9) -- (2.6,-9.2);
\draw [thick] (3.1, -8.95) -- (3.9,-8.95);
\draw [thick] (3.1, -9.05) -- (3.9,-9.05);
%\draw [thick] (0.1, -9) -- (0.9,-9);
%\draw [thick] (1.1, -9) -- (1.9,-9) ;
\end{tikzpicture}\\ 
\hline 
G(3,3,3) & (3, 3, 6)& 0& \begin{tikzpicture}
 \filldraw[fill= white] (0,0) circle [radius=0.1] node[below] {\scriptsize 1};
\filldraw[fill= white] (1,0) circle [radius=0.1] node[below] {\scriptsize 3};
\draw [thick] (0.1,0)--(0.9,0);
\draw [thick] (0.1,0.05)--(0.9,0.05);
\draw [thick] (0.1,-0.05)--(0.9,-0.05);
\draw [thick] (0.4,0) -- (0.6,0.2);
\draw [thick] (0.4,0) -- (0.6,-0.2);
\draw [thick] (1.05, 0.05) to [out=45,in=315,looseness=15] (1.05,-0.05);
\end{tikzpicture}\\ 
\hline  
\end{array}$$

\paragraph{Descendants of the  $SU(3)_{16}\times SU(2)_{26}\times U(1)$ theory} 

$$\begin{array} {|c|c|c|c|} 
\hline 
\text{Theory}& (\Delta_1, \Delta_2, \Delta_3)& 24(c-a)& \text{Magnetic Quiver}\\ 
\hline 
SU(3)_{16}\times SU(2)_{26}\times U(1) & (6, 8, 12)& 17& \begin{tikzpicture}
%\filldraw[fill= white] (-1,-6) circle [radius=0.1] node[below] {\scriptsize 1};
\filldraw[fill= white] (0,-6) circle [radius=0.1] node[below] {\scriptsize 1};
\filldraw[fill= white] (1,-6) circle [radius=0.1] node[below] {\scriptsize 2};
\filldraw[fill= white] (2,-6) circle [radius=0.1] node[below] {\scriptsize 3};
\filldraw[fill= white] (3,-6) circle [radius=0.1] node[below] {\scriptsize 5};
\filldraw[fill= white] (4,-6) circle [radius=0.1] node[below] {\scriptsize 7};
\node[] at (2.5,-5.5) {};
\draw [thick] (3.1, -5.95) -- (3.9,-5.95) ;
\draw [thick] (3.1, -6) -- (3.9,-6) ;
\draw [thick] (3.1, -6.05) -- (3.9,-6.05) ;
\draw [thick] (2.1, -6) -- (2.9,-6) ;
\draw [thick] (3.4,-6) -- (3.6,-5.8);
\draw [thick] (3.4,-6) -- (3.6,-6.2);
%\draw [thick] (-0.1, -6) -- (-0.9,-6);
\draw [thick] (0.1, -6) -- (0.9,-6);
\draw [thick] (1.1, -6) -- (1.9,-6) ; 

\end{tikzpicture}\\ 
\hline 
SU(2)_{22}\times U(1)^2  & (4, 6, 10)& 9& \begin{tikzpicture}
%\filldraw[fill= white] (0,-9) circle [radius=0.1] node[below] {\scriptsize 1};
\filldraw[fill= white] (1,-9) circle [radius=0.1] node[below] {\scriptsize 1};
\filldraw[fill= white] (2,-9) circle [radius=0.1] node[below] {\scriptsize 3};
\filldraw[fill= white] (3,-9) circle [radius=0.1] node[below] {\scriptsize 5};
\filldraw[fill= white] (4,-9) circle [radius=0.1] node[below] {\scriptsize 1};
\node[] at (2.5,-8.5) {};
\draw [thick] (2.1, -8.95) -- (2.9,-8.95) ;
\draw [thick] (2.1, -9) -- (2.9,-9) ;
\draw [thick] (2.1, -9.05) -- (2.9,-9.05) ;
\draw [thick] (2.4,-9) -- (2.6,-8.8);
\draw [thick] (2.4,-9) -- (2.6,-9.2);
\draw [thick] (3.1, -8.95) -- (3.9,-8.95);
\draw [thick] (3.1, -9.05) -- (3.9,-9.05);
%\draw [thick] (0.1, -9) -- (0.9,-9);
\draw [thick] (1.1, -9) -- (1.9,-9) ;
\end{tikzpicture}\\ 
\hline 
SU(3)_{\frac{32}{3}}\times U(1)  & (4, 6, 8)& 7& \begin{tikzpicture}
%\filldraw[fill= white] (0,-9) circle [radius=0.1] node[below] {\scriptsize 1};
\filldraw[fill= white] (1,-9) circle [radius=0.1] node[below] {\scriptsize 1};
\filldraw[fill= white] (2,-9) circle [radius=0.1] node[below] {\scriptsize 2};
\filldraw[fill= white] (3,-9) circle [radius=0.1] node[below] {\scriptsize 3};
\filldraw[fill= white] (4,-9) circle [radius=0.1] node[below] {\scriptsize 2};
\node[] at (2.5,-8.5) {};
\draw [thick] (2.1, -8.95) -- (2.9,-8.95) ;
\draw [thick] (2.1, -9) -- (2.9,-9) ;
\draw [thick] (2.1, -9.05) -- (2.9,-9.05) ;
\draw [thick] (2.4,-9) -- (2.6,-8.8);
\draw [thick] (2.4,-9) -- (2.6,-9.2);
\draw [thick] (3.1, -8.95) -- (3.9,-8.95);
\draw [thick] (3.1, -9.05) -- (3.9,-9.05);
%\draw [thick] (0.1, -9) -- (0.9,-9);
\draw [thick] (1.1, -9) -- (1.9,-9) ;
\end{tikzpicture}\\ 
\hline  
 SU(2)_{20}\times U(1) & (3, 5, 9)& 6& \begin{tikzpicture}
%\filldraw[fill= white] (1,-12) circle [radius=0.1] node[below] {\scriptsize A-3};
\filldraw[fill= white] (2,-12) circle [radius=0.1] node[below] {\scriptsize 2};
\filldraw[fill= white] (3,-12) circle [radius=0.1] node[below] {\scriptsize 4};
\filldraw[fill= white] (4,-12) circle [radius=0.1] node[below] {\scriptsize 1};
\node[] at (2.5,-11.5) {};
\draw [thick] (2.1, -11.95) -- (2.9,-11.95) ;
\draw [thick] (2.1, -12) -- (2.9,-12) ;
\draw [thick] (2.1, -12.05) -- (2.9,-12.05) ;
%\draw [thick] (1.1, -12) -- (1.9,-12) ;
\draw [thick] (2.4,-12) -- (2.6,-11.8);
\draw [thick] (2.4,-12) -- (2.6,-12.2);
\draw [thick] (3.1, -11.95) -- (3.9,-11.95);
\draw [thick] (3.1, -12) -- (3.9,-12);
\draw [thick] (3.1, -12.05) -- (3.9,-12.05) ;
\end{tikzpicture}\\
\hline
 SU(2)_{8} & (3, 5, 6)& 2&?\\
\hline
\end{array}$$

\paragraph{Descendants of the  $SU(4)_{16} \times SU(2)_{8}$ theory} 

$$\begin{array} {|c|c|c|c|} 
\hline 
\text{Theory}& (\Delta_1, \Delta_2, \Delta_3)& 24(c-a)& \text{Magnetic Quiver}\\ 
\hline 
SU(4)_{16} \times SU(2)_{8} & (4, 6, 8)& 15& \begin{tikzpicture}
%\filldraw[fill= white] (-1,-6) circle [radius=0.1] node[below] {\scriptsize 1};
\filldraw[fill= white] (0,-6) circle [radius=0.1] node[below] {\scriptsize 1};
\filldraw[fill= white] (1,-6) circle [radius=0.1] node[below] {\scriptsize 2};
\filldraw[fill= white] (2,-6) circle [radius=0.1] node[below] {\scriptsize 3};
\filldraw[fill= white] (3,-6) circle [radius=0.1] node[below] {\scriptsize 4};
\filldraw[fill= white] (4,-6) circle [radius=0.1] node[below] {\scriptsize 6};
\node[] at (2.5,-5.5) {};
\draw [thick] (3.1, -5.95) -- (3.9,-5.95) ;
\draw [thick] (3.1, -6) -- (3.9,-6) ;
\draw [thick] (3.1, -6.05) -- (3.9,-6.05) ;
\draw [thick] (2.1, -6) -- (2.9,-6) ;
\draw [thick] (3.4,-6) -- (3.6,-5.8);
\draw [thick] (3.4,-6) -- (3.6,-6.2);
%\draw [thick] (-0.1, -6) -- (-0.9,-6);
\draw [thick] (0.1, -6) -- (0.9,-6);
\draw [thick] (1.1, -6) -- (1.9,-6) ; 

\end{tikzpicture}\\ 
\hline 
SU(2)_{12}\times SU(2)_{\frac{20}{3}}  & (\frac{10}{3}, 4, 6)& 7& \begin{tikzpicture}
%\filldraw[fill= white] (0,-9) circle [radius=0.1] node[below] {\scriptsize 1};
\filldraw[fill= white] (1,-9) circle [radius=0.1] node[below] {\scriptsize 1};
\filldraw[fill= white] (2,-9) circle [radius=0.1] node[below] {\scriptsize 2};
\filldraw[fill= white] (3,-9) circle [radius=0.1] node[below] {\scriptsize 4};
\filldraw[fill= white] (4,-9) circle [radius=0.1] node[below] {\scriptsize 1};
\node[] at (2.5,-8.5) {};
\draw [thick] (2.1, -8.95) -- (2.9,-8.95) ;
\draw [thick] (2.1, -9) -- (2.9,-9) ;
\draw [thick] (2.1, -9.05) -- (2.9,-9.05) ;
\draw [thick] (2.4,-9) -- (2.6,-8.8);
\draw [thick] (2.4,-9) -- (2.6,-9.2);
\draw [thick] (3.1, -8.95) -- (3.9,-8.95);
\draw [thick] (3.1, -9.05) -- (3.9,-9.05);
%\draw [thick] (0.1, -9) -- (0.9,-9);
\draw [thick] (1.1, -9) -- (1.9,-9) ;
\end{tikzpicture}\\ 
\hline 
SU(2)_{\frac{18}{3}} \times U(1) & (3,3,5)& 4& \begin{tikzpicture}
%\filldraw[fill= white] (1,-12) circle [radius=0.1] node[below] {\scriptsize A-3};
\filldraw[fill= white] (2,-12) circle [radius=0.1] node[below] {\scriptsize 1};
\filldraw[fill= white] (3,-12) circle [radius=0.1] node[below] {\scriptsize 3};
\filldraw[fill= white] (4,-12) circle [radius=0.1] node[below] {\scriptsize 1};
\node[] at (2.5,-11.5) {};
\draw [thick] (2.1, -11.95) -- (2.9,-11.95) ;
\draw [thick] (2.1, -12) -- (2.9,-12) ;
\draw [thick] (2.1, -12.05) -- (2.9,-12.05) ;
%\draw [thick] (1.1, -12) -- (1.9,-12) ;
\draw [thick] (2.4,-12) -- (2.6,-11.8);
\draw [thick] (2.4,-12) -- (2.6,-12.2);
\draw [thick] (3.1, -11.95) -- (3.9,-11.95);
\draw [thick] (3.1, -12) -- (3.9,-12);
\draw [thick] (3.1, -12.05) -- (3.9,-12.05) ;
\end{tikzpicture}\\ 
\hline
 \varnothing & (2, 2, 3)& -3&\varnothing \\
\hline
\end{array}$$ 
According to our algorithm, in the last line of the above table we should have a theory with CB spectrum $(2,8/3,4)$. This can be immediately identified with a (non-conformal) $SU(2)$ gauging of model 58 in \cite{Martone:2021drm}. The latter can flow upon a mass deformation to $SU(3)$ $\mathcal N=4$ Super Yang-Mills, and once this is done the gauging becomes conformal. We include this lagrangian theory in the table. Since the theory has no Higgs branch and no global symmetry, the corresponding magnetic quiver is trivial.

\paragraph{Descendants of the  $SU(4)_{28}$ theory} 

$$\begin{array} {|c|c|c|c|} 
\hline 
\text{Theory}& (\Delta_1, \Delta_2, \Delta_3)& 24(c-a)& \text{Magnetic Quiver}\\ 
\hline 
SU(4)_{28} & (6, 12, 14)& 19& \begin{tikzpicture}
%\filldraw[fill= white] (-1,-6) circle [radius=0.1] node[below] {\scriptsize 1};
%\filldraw[fill= white] (0,-6) circle [radius=0.1] node[below] {\scriptsize 2};
\filldraw[fill= white] (1,-6) circle [radius=0.1] node[below] {\scriptsize 2};
\filldraw[fill= white] (2,-6) circle [radius=0.1] node[below] {\scriptsize 4};
\filldraw[fill= white] (3,-6) circle [radius=0.1] node[below] {\scriptsize 6};
\filldraw[fill= white] (4,-6) circle [radius=0.1] node[below] {\scriptsize 8};
\node[] at (2.5,-5.5) {};
\draw [thick] (3.1, -5.95) -- (3.9,-5.95) ;
\draw [thick] (3.1, -6) -- (3.9,-6) ;
\draw [thick] (3.1, -6.05) -- (3.9,-6.05) ;
\draw [thick] (2.1, -6) -- (2.9,-6) ;
\draw [thick] (3.4,-6) -- (3.6,-5.8);
\draw [thick] (3.4,-6) -- (3.6,-6.2);
%\draw [thick] (-0.1, -6) -- (-0.9,-6);
%\draw [thick] (0.1, -6) -- (0.9,-6);
\draw [thick] (1.1, -6) -- (1.9,-6) ; 

\end{tikzpicture}\\ 
\hline 
SU(2)_{20}\times U(1) & (4, 8, 10)& 7& \begin{tikzpicture}
%\filldraw[fill= white] (0,-9) circle [radius=0.1] node[below] {\scriptsize 1};
%\filldraw[fill= white] (1,-9) circle [radius=0.1] node[below] {\scriptsize A-2};
\filldraw[fill= white] (2,-9) circle [radius=0.1] node[below] {\scriptsize 2};
\filldraw[fill= white] (3,-9) circle [radius=0.1] node[below] {\scriptsize 4};
\filldraw[fill= white] (4,-9) circle [radius=0.1] node[below] {\scriptsize 2};
\node[] at (2.5,-8.5) {};
\draw [thick] (2.1, -8.95) -- (2.9,-8.95) ;
\draw [thick] (2.1, -9) -- (2.9,-9) ;
\draw [thick] (2.1, -9.05) -- (2.9,-9.05) ;
\draw [thick] (2.4,-9) -- (2.6,-8.8);
\draw [thick] (2.4,-9) -- (2.6,-9.2);
\draw [thick] (3.1, -8.95) -- (3.9,-8.95);
\draw [thick] (3.1, -9.05) -- (3.9,-9.05);
%\draw [thick] (0.1, -9) -- (0.9,-9);
%\draw [thick] (1.1, -9) -- (1.9,-9) ;
\end{tikzpicture}\\ 
\hline 
U(1) & (3, 6, 8)& 1 &?\\ 
\hline  
\end{array}$$ 

\paragraph{Lagrangian theories} 

Within the $\ell=3$ series of theories we find only one lagrangian model, namely 

${\bf \varnothing: }$ $SU(2)\times SU(3)$ gauge theory with a half-hypermultiplet in the ({\bf 2},{\bf 8}).

\subsection{Theories with $\ell=4$} 

For $\ell=4$ we find 4 orbi-instanton starting points, all at $k\leq2$. In one case we do not know how to compute the 't Hooft anomaly $24(c-a)$ and the magnetic quiver for the theory with trivial 7-brane.

\begin{equation}\label{diagramell=4} 
\begin{tikzpicture} 
\node[] (a1) at (0,0) {\scriptsize $SU(2)_{38}\times U(1)$}; 
\node[] (a2) at (0,-1) {\scriptsize $G(4,1,3)$}; 

\node[] (b1) at (2.5,0) {\scriptsize $SU(2)_{9}\times U(1)$}; 
\node[] (b2) at (2.5,-1) {\scriptsize $G(4,4,3)$}; 

\node[] (c1) at (6,0) {\scriptsize $SU(3)_{16}\times U(1)$}; 
\node[] (c2) at (5,-1) {\scriptsize $SU(2)_{12}$}; 
\node[] (C2) at (7,-1) {\scriptsize $U(1)\times U(1)$};

\node[] (d1) at (10.5,0) {\scriptsize $SU(4)_{16}$}; 
\node[] (D2) at (10.5,-1) {\scriptsize $SU(2)_{12}\times U(1)$};

\draw[->, thick] (d1)--(D2); 
\draw[->, thick] (c1)--(c2); 
\draw[->, thick] (c1)--(C2); 
\draw[->, thick] (a1)--(a2); 
\draw[->, thick] (b1)--(b2); 
\end{tikzpicture}
\end{equation}

\begin{itemize} 
\item $SU(2)_{9}\times U(1)$: SW compactification of the $\bbZ_4$ orbi-instanton with $(n_4=1)$ and three tensors. 
\item $SU(2)_{38}\times U(1)$: SW compactification of the $\bbZ_4$ orbi-instanton with $(n_{4'}=1)$ and three tensors.
\item $SU(3)_{16}\times U(1)$: SW compactification of the $\bbZ_8$ orbi-instanton with $(n_4=n_{4'}=1)$ and two tensors.
\item $SU(4)_{16}$: SW compactification of the $\bbZ_8$ orbi-instanton with $(n_{4'}=2)$ and two tensors. 
\end{itemize}

\paragraph{Descendants of the $SU(2)_{38}\times U(1)$ theory} 

$$\begin{array} {|c|c|c|c|} 
\hline 
\text{Theory}& (\Delta_1, \Delta_2, \Delta_3)& 24(c-a)& \text{Magnetic Quiver}\\ 
\hline 
SU(2)_{38}\times U(1) & (6, 12, 18)& 11& \begin{tikzpicture}
%\filldraw[fill= white] (-5,0) circle [radius=0.1] node[below] {\scriptsize 1}; 
\filldraw[fill= white] (-4,0) circle [radius=0.1] node[below] {\scriptsize 1}; 
 \filldraw[fill= white] (-3,0) circle [radius=0.1] node[below] {\scriptsize 4};
\filldraw[fill= white] (-2,0) circle [radius=0.1] node[below] {\scriptsize 7};
\node[] at (-3.5,0.5) {};
%\draw [thick] (-4.9, 0) -- (-4.1,0) ; 
\draw [thick] (-3.9, 0) -- (-3.1,0) ; 
\draw [thick] (-2.05,0.07)--(-2.95,0.07);
\draw [thick] (-2.07,0.02)--(-2.93,0.02);
\draw [thick] (-2.07,-0.02)--(-2.93,-0.02);
\draw [thick] (-2.05,-0.07)--(-2.95,-0.07);
\draw [thick] (-2.6,0) -- (-2.4,-0.2);
\draw [thick] (-2.6,0) -- (-2.4,0.2);
\end{tikzpicture}\\ 
\hline 
G(4,1,3) & (4, 8, 12)& 0& \begin{tikzpicture}
 \filldraw[fill= white] (0,0) circle [radius=0.1] node[below] {\scriptsize 1};
\filldraw[fill= white] (1,0) circle [radius=0.1] node[below] {\scriptsize 3};
\draw [thick] (0.07,0.02)--(0.93,0.02);
\draw [thick] (0.07,-0.02)--(0.93,-0.02);
\draw [thick] (0.05,0.07)--(0.95,0.07);
\draw [thick] (0.05,-0.07)--(0.95,-0.07);
%\draw [thick] (0.4,0) -- (0.6,0.2);
%\draw [thick] (0.4,0) -- (0.6,-0.2);
\draw [thick] (1.05, 0.05) to [out=45,in=315,looseness=15] (1.05,-0.05);
\end{tikzpicture}\\ 
\hline  
\end{array}$$ 

\paragraph{Descendants of the $SU(2)_{9}\times U(1)$ theory} 

$$\begin{array} {|c|c|c|c|} 
\hline 
\text{Theory}& (\Delta_1, \Delta_2, \Delta_3)& 24(c-a)& \text{Magnetic Quiver}\\ 
\hline 
SU(2)_{9}\times U(1) & (\frac{9}{2}, 6, 12)& 9& \begin{tikzpicture}
%\filldraw[fill= white] (-5,0) circle [radius=0.1] node[below] {\scriptsize 1}; 
\filldraw[fill= white] (-4,0) circle [radius=0.1] node[below] {\scriptsize 1}; 
 \filldraw[fill= white] (-3,0) circle [radius=0.1] node[below] {\scriptsize 3};
\filldraw[fill= white] (-2,0) circle [radius=0.1] node[below] {\scriptsize 6};
\node[] at (-3.5,0.5) {};
%\draw [thick] (-4.9, 0) -- (-4.1,0) ; 
\draw [thick] (-3.9, 0) -- (-3.1,0) ; 
\draw [thick] (-2.05,0.07)--(-2.95,0.07);
\draw [thick] (-2.07,0.02)--(-2.93,0.02);
\draw [thick] (-2.07,-0.02)--(-2.93,-0.02);
\draw [thick] (-2.05,-0.07)--(-2.95,-0.07);
\draw [thick] (-2.6,0) -- (-2.4,-0.2);
\draw [thick] (-2.6,0) -- (-2.4,0.2);
\end{tikzpicture}\\ 
\hline 
G(4,4,3) & (3, 4, 8)& 0& \begin{tikzpicture}
 \filldraw[fill= white] (0,0) circle [radius=0.1] node[below] {\scriptsize 1};
\filldraw[fill= white] (1,0) circle [radius=0.1] node[below] {\scriptsize 3};
\draw [thick] (0.07,0.02)--(0.93,0.02);
\draw [thick] (0.07,-0.02)--(0.93,-0.02);
\draw [thick] (0.05,0.07)--(0.95,0.07);
\draw [thick] (0.05,-0.07)--(0.95,-0.07);
\draw [thick] (0.4,0) -- (0.6,0.2);
\draw [thick] (0.4,0) -- (0.6,-0.2);
\draw [thick] (1.05, 0.05) to [out=45,in=315,looseness=15] (1.05,-0.05);
\end{tikzpicture}\\ 
\hline  
\end{array}$$

\paragraph{Descendants of the $SU(3)_{16}\times U(1)$} 

$$\begin{array} {|c|c|c|c|} 
\hline 
\text{Theory}& (\Delta_1, \Delta_2, \Delta_3)& 24(c-a)& \text{Magnetic Quiver}\\ 
\hline 
SU(3)_{16}\times U(1) & (6, 8, 12)& 10& \begin{tikzpicture}
\filldraw[fill= white] (-5,0) circle [radius=0.1] node[below] {\scriptsize 1}; 
\filldraw[fill= white] (-4,0) circle [radius=0.1] node[below] {\scriptsize 2}; 
 \filldraw[fill= white] (-3,0) circle [radius=0.1] node[below] {\scriptsize 3};
\filldraw[fill= white] (-2,0) circle [radius=0.1] node[below] {\scriptsize 5};
\node[] at (-3.5,0.5) {};
\draw [thick] (-4.9, 0) -- (-4.1,0) ; 
\draw [thick] (-3.9, 0) -- (-3.1,0) ; 
\draw [thick] (-2.05,0.07)--(-2.95,0.07);
\draw [thick] (-2.07,0.02)--(-2.93,0.02);
\draw [thick] (-2.07,-0.02)--(-2.93,-0.02);
\draw [thick] (-2.05,-0.07)--(-2.95,-0.07);
\draw [thick] (-2.6,0) -- (-2.4,-0.2);
\draw [thick] (-2.6,0) -- (-2.4,0.2);
\end{tikzpicture}\\ 
\hline 
U(1)\times U(1) & (4, 6, 10)& 3& \begin{tikzpicture}
 \filldraw[fill= white] (0,0) circle [radius=0.1] node[below] {\scriptsize 1};
\filldraw[fill= white] (1,0) circle [radius=0.1] node[below] {\scriptsize 3};
\filldraw[fill= white] (2,0) circle [radius=0.1] node[below] {\scriptsize 1}; 
\node[] at (0.5,0.5) {};
\draw [thick] (0.05,0.07)--(0.95,0.07);
\draw [thick] (0.07,0.02)--(0.93,0.02);
\draw [thick] (0.07,-0.02)--(0.93,-0.02);
\draw [thick] (0.05,-0.07)--(0.95,-0.07);
\draw [thick] (1.05,0.07)--(1.95,0.07);
\draw [thick] (1.07,0.02)--(1.93,0.02);
\draw [thick] (1.07,-0.02)--(1.93,-0.02);
\draw [thick] (1.05,-0.07)--(1.95,-0.07);
\draw [thick] (0.4,0) -- (0.6,-0.2);
\draw [thick] (0.4,0) -- (0.6,0.2);
\end{tikzpicture}\\ 
\hline  
SU(2)_{12} & (4, 6, 8)& ? & ? \\
\hline  
\end{array}$$

\paragraph{Descendants of the $SU(4)_{16}$} 

$$\begin{array} {|c|c|c|c|} 
\hline 
\text{Theory}& (\Delta_1, \Delta_2, \Delta_3)& 24(c-a)& \text{Magnetic Quiver}\\ 
\hline 
SU(4)_{16} & (3, 6, 8)& 9& \begin{tikzpicture}
\filldraw[fill= white] (-5,0) circle [radius=0.1] node[below] {\scriptsize 1}; 
\filldraw[fill= white] (-4,0) circle [radius=0.1] node[below] {\scriptsize 2}; 
 \filldraw[fill= white] (-3,0) circle [radius=0.1] node[below] {\scriptsize 3};
\filldraw[fill= white] (-2,0) circle [radius=0.1] node[below] {\scriptsize 4};
\node[] at (-3.5,0.5) {};
\draw [thick] (-4.9, 0) -- (-4.1,0) ; 
\draw [thick] (-3.9, 0) -- (-3.1,0) ; 
\draw [thick] (-2.05,0.07)--(-2.95,0.07);
\draw [thick] (-2.07,0.02)--(-2.93,0.02);
\draw [thick] (-2.07,-0.02)--(-2.93,-0.02);
\draw [thick] (-2.05,-0.07)--(-2.95,-0.07);
\draw [thick] (-2.6,0) -- (-2.4,-0.2);
\draw [thick] (-2.6,0) -- (-2.4,0.2);
\end{tikzpicture}\\ 
\hline 
SU(2)_{12}\times U(1) & (\frac{5}{2}, 4, 6)& 2& \begin{tikzpicture}
 \filldraw[fill= white] (0,0) circle [radius=0.1] node[below] {\scriptsize 1};
\filldraw[fill= white] (1,0) circle [radius=0.1] node[below] {\scriptsize 2};
\filldraw[fill= white] (2,0) circle [radius=0.1] node[below] {\scriptsize 1}; 
\node[] at (0.5,0.5) {};
\draw [thick] (0.05,0.07)--(0.95,0.07);
\draw [thick] (0.07,0.02)--(0.93,0.02);
\draw [thick] (0.07,-0.02)--(0.93,-0.02);
\draw [thick] (0.05,-0.07)--(0.95,-0.07);
\draw [thick] (1.05,0.07)--(1.95,0.07);
\draw [thick] (1.07,0.02)--(1.93,0.02);
\draw [thick] (1.07,-0.02)--(1.93,-0.02);
\draw [thick] (1.05,-0.07)--(1.95,-0.07);
\draw [thick] (0.4,0) -- (0.6,-0.2);
\draw [thick] (0.4,0) -- (0.6,0.2);
\end{tikzpicture}\\ 
\hline  
\end{array}$$ 
In principle our argument predicts the existence of a theory with CB spectrum $(2,4,6)$ which is a mass deformation of the $SU(4)_{16}$ SCFT in the above table. We are unsure about the interpretation of this theory and therefore we refrain from including it in our list.

\subsection{Theories with $\ell=5,6$} 

For $\ell=5,6$ we have only the S-fold theories studied in \cite{Giacomelli:2020gee} since other orbi-S-fold setups start contributing only at rank 4. We therefore only have two models to discuss which were called $\mathcal{T}_{\varnothing,5}^{(3)}$ and  $\mathcal{T}_{\varnothing,6}^{(3)}$ in \cite{Giacomelli:2020gee} for $\ell=5$ and $\ell=6$ respectively. The data of these theories were derived in  \cite{Giacomelli:2020gee} and we report them here for convenience, adding our proposal for the magnetic quiver. 
$$\begin{array} {|c|c|c|c|} 
\hline 
\text{Theory}& (\Delta_1, \Delta_2, \Delta_3)& 24(c-a)& \text{Magnetic Quiver}\\ 
\hline 
\mathcal{T}_{\varnothing,5}^{(3)}& (\frac{18}{5}, 6, 12)& \frac{18}{5}& \begin{tikzpicture}
 \filldraw[fill= white] (0,0) circle [radius=0.1] node[below] {\scriptsize 1};
\filldraw[fill= white] (1,0) circle [radius=0.1] node[below] {\scriptsize 3};
\draw [] (0.1,0)--(0.9,0);
\draw [] (0.07,0.03)--(0.93,0.03);
\draw [] (0.07,-0.03)--(0.93,-0.03);
\draw [] (0.04,0.07)--(0.96,0.07);
\draw [] (0.04,-0.07)--(0.96,-0.07);
\draw [thick] (0.4,0) -- (0.6,0.2);
\draw [thick] (0.4,0) -- (0.6,-0.2);
\draw [thick] (1.05, 0.05) to [out=45,in=315,looseness=15] (1.05,-0.05);
\end{tikzpicture}\\ 
\hline 
\mathcal{T}_{\varnothing,6}^{(3)}& (3, 6, 12)& 0& \begin{tikzpicture}
 \filldraw[fill= white] (0,0) circle [radius=0.1] node[below] {\scriptsize 1};
\filldraw[fill= white] (1,0) circle [radius=0.1] node[below] {\scriptsize 3};
\draw [] (0.08,0.015)--(0.92,0.015);
\draw [] (0.08,-0.015)--(0.92,-0.015);
\draw [] (0.06,0.045)--(0.94,0.045);
\draw [] (0.06,-0.045)--(0.94,-0.045);
\draw [] (0.04,0.075)--(0.96,0.075);
\draw [] (0.04,-0.075)--(0.96,-0.075);
\draw [thick] (0.4,0) -- (0.6,0.2);
\draw [thick] (0.4,0) -- (0.6,-0.2);
\draw [thick] (1.05, 0.05) to [out=45,in=315,looseness=15] (1.05,-0.05);
\end{tikzpicture}\\ 
\hline
\end{array}$$ 
Both theories are known to have a three-dimensional Higgs branch and $U(1)$ global symmetry. Furthermore, the $\ell=6$ model has $a=c$ reflecting its enhanced $\mathcal{N}=3$ supersymmetry \cite{Giacomelli:2020gee}. There are no descendants to be discussed since we can deform neither the 7-brane nor the tail.

\section{Discussion}\label{Sec:Disc}

In this paper we have generalized the S-fold construction of $\mathcal N=2$ 4d SCFTs initiated in \cite{Apruzzi:2020pmv, Giacomelli:2020jel, Giacomelli:2020gee, Bourget:2020mez} by including a orbifold in the Type IIB background. This leads to a much larger class of theories which, as we have seen, includes almost all known theories at low rank. All the theories we have constructed come in infinite families of increasing rank and, as we increase the orbifold order, we find more and more families, all connected by Higgs-branch RG flows for a given S-fold order $\ell$. By combining geometric and field-theoretic tools we can determine in detail the data specifying the 4d theory, including the description of the Higgs branch via magnetic quivers. Our expectation is that with this method we can reach all SCFTs whose Higgs branch is described by a unitary magnetic quiver. An important advantage of our approach is that we can rather easily analyze RG flows triggered by mass deformations which, unlike higgsings, cannot be studied using anomaly matching. In particular we have introduced and studied in detail two classes of mass deformations for which we can describe systematically the Coulomb-branch spectrum and 't Hooft anomalies of the infrared fixed point. For the theory of lowest rank within each family these two sequences of relevant deformations happen to coincide and therefore the tree of mass deformations simplifies. 

We recover all rank-2 theories with a unitary magnetic quiver (i.e.~roughly 90$\%$ of the known theories) and were able to derive new predictions at rank 2. Specifically, we claim the existence of a new RG flow providing the first stringy realization of the $USp(4)$ theory with a half-hypermultiplet in the {\bf 16}, and the existence of a new theory on the Coulomb branch of the $\mathcal N=2^*$ $G_2$ theory. At rank 3 we found 115 SCFTs, which include more than a half of the lagrangian theories. Along the way we found several examples of Argyres-Seiberg-like dualities.

As a general remark, we can observe that all the 4d SCFTs we can generate with our method have at least one Coulomb-branch operator with dimension $6$ or less. This property is shared, to our knowledge, by all Class-$\mathcal{S}$ theories and all models one can construct in the context of Type IIB compactifications on singular Calabi-Yau threefolds (a.k.a.~geometric engineering). At present the only candidate counterexample we are aware of is, modulo discrete gauging (see \cite{Amariti:2023qdq}), the $G_{31}$ $\mathcal N=3$ theory proposed in \cite{Kaidi:2022lyo}.

There are plenty of possible directions for future investigations. The most pressing question is to complete the analysis by studying systematically all possible mass deformations and derive the properties of the resulting theories. In particular, it would be important to understand what determines the Coulomb-branch spectra we have found. It would also be important to find general criteria to single out mass deformations leading to conformal theories in the infrared, and consequently to identify which FI deformations to perform at the level of the magnetic quiver. This question is especially relevant for the analysis of models associated to Type IIB backgrounds in which the 7-brane has been completely removed. We currently have only a partial understanding of these theories since often they do not obey the rules we have derived, and the determination of their data still requires a case-by-case analysis. It is also important to find a first-principle derivation from the stringy setup of the various empirical rules we have found in this work. In our scan of rank-3 models we have found several examples of Argyres-Seiberg-like dualities. By inspection, it seems that by gauging subgroups of the global symmetry of our models it is possible to find many other examples of such dualities, and it would be interesting to study them systematically. It would also be interesting to study in detail the full Higgs-branch Hasse diagram using the magnetic quivers we have proposed as a starting point, perhaps via the recent proposal of \cite{Bourget:2023dkj, Bourget:2024mgn}. Finally, it would be important to understand in detail higher-form symmetries for orbi-S-fold theories with trivial 7-brane, which we have initiated in Section \ref{Sec:1form}. 

Our results about rank-3 theories clearly show that with this method we can significantly improve our understanding of the landscape of higher-rank superconformal theories. In order to enlarge as much as possible the space of theories we can probe, one needs to further generalize the method we have presented in this paper. One obvious direction is to repeat the analysis we have carried out taking as a starting point orbi-instanton theories of type D or E \cite{DelZotto:2014hpa, Frey:2018vpw}. At least in the D case an analysis very similar to ours should be feasible since most of the required tools are available. In particular, the magnetic quivers for the parent 6d theories are known and are given by a family of orthosymplectic quivers \cite{Cabrera:2019dob}. 

Finally, there is at least another direction worth investigating. In \cite{Bhardwaj:2019fzv} twisted compactifications of 6d theories to 5d, where the twist involves outer automorphisms or permutations of the tensor multiplets of the 6d theory, were investigated. If we further reduce the 5d theory on a circle we can get new SCFTs in 4d which are indeed realized as twisted compactifications on $T^2$ of a 6d SCFT, but possibly not of Stiefel-Whitney type as in this work. We have an example of this at rank 2: The theory 47 in \cite{Martone:2021ixp} and its descendants can be realized as dimensional reduction of a 5d SCFT which UV completes SQCD with gauge group $G_2$ \cite{Martone:2021drm}. This 5d theory in turn can be obtained as the twisted compactification of the 6d SCFT UV completing $SU(3)$ SQCD with 12 flavors \cite{Bhardwaj:2019fzv}. The twist involves a $\bbZ_2$ outer-automorphism of the 6d theory. This construction is manifestly not a special case of what we have done in this paper: The 6d theory in question only admits a $\bbZ_3$ Stiefel-Whitney twist which leads to a rank-1 theory in 4d. The required $\bbZ_2$ twist which leads to theory 47 of  \cite{Martone:2021ixp} cannot be described within our framework. This clearly suggests that the twisted-compactification method we have used can be generalized to include the above-mentioned more general operation. We hope to come back to some of these open problems in the future.

\subsection*{Acknowledgements}

We would like to thank Philip Argyres, Antoine Bourget, Julius Grimminger, Mario Martone, and Noppadol Mekareeya for discussions. SG would like to thank the Physics Department of the University of Rome ``Tor Vergata'' for hospitality during the completion of this project. The work of SG is supported by the INFN grant ``Per attivit\`a di formazione per sostenere progetti di ricerca'' (GRANT 73/STRONGQFT).

\begin{appendix}

\section{List of 4d SCFTs from SW compactifications} \label{App:Mth}

In this appendix we list all the theories constructed for low values of $k$ together with their properties.

\subsection{$\ell =2$ quotients}

\subsubsection*{$\mathcal T^{(N)}_{E_6,2}(2,0,0,0,0)$}

\al{&\begin{tikzpicture}
\filldraw[fill= white] (0,0) circle [radius=0.1] node[below] {\scriptsize 1};
\filldraw[fill= white] (1,0) circle [radius=0.1] node[below] {\scriptsize 2};
\filldraw[fill= white] (2,0) circle [radius=0.1] node[below] {\scriptsize N};
\filldraw[fill= white] (3,0) circle [radius=0.1] node[below] {\scriptsize 2N};
\filldraw[fill= white] (4,0) circle [radius=0.1] node[below] {\scriptsize 3N};
\filldraw[fill= white] (5,0) circle [radius=0.1] node[below] {\scriptsize 4N};
\filldraw[fill= white] (6,0) circle [radius=0.1] node[below] {\scriptsize  2N};
%\filldraw[fill= white] (7,0) circle [radius=0.1] node[below] {\scriptsize  1};
\draw [thick] (0.1, 0) -- (0.9,0) ;
\draw [thick] (1.1, 0) -- (1.9,0) ;
\draw [thick] (2.1, 0) -- (2.9,0) ;
\draw [thick] (4.1, 0.05) -- (4.9,0.05) ;
\draw [thick] (4.1, -0.05) -- (4.9,-0.05) ;
\draw [thick] (3.1, 0) -- (3.9,0) ;
\draw [thick] (5.1, 0) -- (5.9,0) ;
%\draw [thick] (6.1, 0) -- (6.9,0) ;
\draw [thick] (4.4,0) -- (4.6,0.2);
\draw [thick] (4.4,0) -- (4.6,-0.2);
\end{tikzpicture} \\
&24(c-a) = 12 N + 2\,, \qquad 4(2a-c) = 12N^2 -16N+6\,,\\
&\Delta = \{\underbrace{(6k,6k+2)}_{k=1,\dots,N-2},6(N-1),3N \} \,,\qquad \mathfrak{g} = (\mathfrak{f}_4)_{6N}\oplus (\mathfrak{su}_2)_{16}\,.
}

\subsubsection*{$\mathcal T^{(N)}_{E_6,2}(1,0,0,0,1)$}

\al{&\begin{tikzpicture}
\filldraw[fill= white] (0,0) circle [radius=0.1] node[below] {\scriptsize 1};
\filldraw[fill= white] (1,0) circle [radius=0.1] node[below] {\scriptsize 2};
\filldraw[fill= white] (2,0) circle [radius=0.1] node[below] {\scriptsize N+1};
\filldraw[fill= white] (3,0) circle [radius=0.1] node[below] {\scriptsize 2N+1};
\filldraw[fill= white] (4,0) circle [radius=0.1] node[below] {\scriptsize 3N+1};
\filldraw[fill= white] (5,0) circle [radius=0.1] node[below] {\scriptsize 4N+1};
\filldraw[fill= white] (6,0) circle [radius=0.1] node[below] {\scriptsize  2N};
%\filldraw[fill= white] (7,0) circle [radius=0.1] node[below] {\scriptsize  1};
\draw [thick] (0.1, 0) -- (0.9,0) ;
\draw [thick] (1.1, 0) -- (1.9,0) ;
\draw [thick] (2.1, 0) -- (2.9,0) ;
\draw [thick] (4.1, 0.05) -- (4.9,0.05) ;
\draw [thick] (4.1, -0.05) -- (4.9,-0.05) ;
\draw [thick] (3.1, 0) -- (3.9,0) ;
\draw [thick] (5.1, 0) -- (5.9,0) ;
%\draw [thick] (6.1, 0) -- (6.9,0) ;
\draw [thick] (4.4,0) -- (4.6,0.2);
\draw [thick] (4.4,0) -- (4.6,-0.2);
\end{tikzpicture} \\
&24(c-a) = 12 N + 6\,, \qquad 4(2a-c) = 12N^2 +2N-3\,, \\
&\Delta = \{\underbrace{(6k,6k+2)}_{k=1,\dots,N-1},6N \}\,,\qquad \mathfrak{g} = (\mathfrak{sp}_3)_{6N+1}\oplus (\mathfrak{su}_2)_{16}\,.
}

\subsubsection*{$\mathcal T^{(N)}_{E_6,2}(0,1,0,0,0)$}

\al{&\begin{tikzpicture}
\filldraw[fill= white] (0,0) circle [radius=0.1] node[below] {\scriptsize 1};
\filldraw[fill= white] (1,0) circle [radius=0.1] node[below] {\scriptsize 2};
\filldraw[fill= white] (2,0) circle [radius=0.1] node[below] {\scriptsize N+1};
\filldraw[fill= white] (3,0) circle [radius=0.1] node[below] {\scriptsize 2N};
\filldraw[fill= white] (4,0) circle [radius=0.1] node[below] {\scriptsize 3N};
\filldraw[fill= white] (5,0) circle [radius=0.1] node[below] {\scriptsize 4N};
\filldraw[fill= white] (6,0) circle [radius=0.1] node[below] {\scriptsize  2N};
%\filldraw[fill= white] (7,0) circle [radius=0.1] node[below] {\scriptsize  1};
\draw [thick] (0.1, 0) -- (0.9,0) ;
\draw [thick] (1.1, 0) -- (1.9,0) ;
\draw [thick] (2.1, 0) -- (2.9,0) ;
\draw [thick] (4.1, 0.05) -- (4.9,0.05) ;
\draw [thick] (4.1, -0.05) -- (4.9,-0.05) ;
\draw [thick] (3.1, 0) -- (3.9,0) ;
\draw [thick] (5.1, 0) -- (5.9,0) ;
%\draw [thick] (6.1, 0) -- (6.9,0) ;
\draw [thick] (4.4,0) -- (4.6,0.2);
\draw [thick] (4.4,0) -- (4.6,-0.2);
\end{tikzpicture} \\
&24(c-a) = 12 N + 3\,, \qquad 4(2a-c) = 12N^2 -4N-3\,,  \\
&\Delta = \{\underbrace{(6k,6k+2)}_{k=1,\dots,N-1},3N \}\,, \qquad \mathfrak{g} = (\mathfrak{so}_7)_{6N}\oplus (\mathfrak{su}_2)_{12N-8}\oplus (\mathfrak{su}_2)_{16}\,.
}

\subsubsection*{$\mathcal T^{(N)}_{E_6,2}(0,0,0,0,2)$}

\al{&\begin{tikzpicture}
\filldraw[fill= white] (0,0) circle [radius=0.1] node[below] {\scriptsize 1};
\filldraw[fill= white] (1,0) circle [radius=0.1] node[below] {\scriptsize 2};
\filldraw[fill= white] (2,0) circle [radius=0.1] node[below] {\scriptsize N+2};
\filldraw[fill= white] (3,0) circle [radius=0.1] node[below] {\scriptsize 2N+2};
\filldraw[fill= white] (4,0) circle [radius=0.1] node[below] {\scriptsize 3N+2};
\filldraw[fill= white] (5,0) circle [radius=0.1] node[below] {\scriptsize 4N+2};
\filldraw[fill= white] (6,0) circle [radius=0.1] node[below] {\scriptsize  2N};
%\filldraw[fill= white] (7,0) circle [radius=0.1] node[below] {\scriptsize  1};
\draw [thick] (0.1, 0) -- (0.9,0) ;
\draw [thick] (1.1, 0) -- (1.9,0) ;
\draw [thick] (2.1, 0) -- (2.9,0) ;
\draw [thick] (4.1, 0.05) -- (4.9,0.05) ;
\draw [thick] (4.1, -0.05) -- (4.9,-0.05) ;
\draw [thick] (3.1, 0) -- (3.9,0) ;
\draw [thick] (5.1, 0) -- (5.9,0) ;
%\draw [thick] (6.1, 0) -- (6.9,0) ;
\draw [thick] (4.4,0) -- (4.6,0.2);
\draw [thick] (4.4,0) -- (4.6,-0.2);
\end{tikzpicture} \\
&24(c-a) = 12 N + 10\,, \qquad 4(2a-c) = 12N^2 +8N-2\,, \\
&\Delta = \{\underbrace{(6k,6k+2)}_{k=1,\dots,N-1},(6N,3N+1) \}\,, \qquad \mathfrak{g} = (\mathfrak{sp}_4)_{6N+2}\oplus (\mathfrak{su}_2)_{16}\,.
}

\subsubsection*{$\mathcal S^{(N)}_{E_6,2}(0,0,0,1,0)$}

\al{&\begin{tikzpicture}
\filldraw[fill= white] (0,0) circle [radius=0.1] node[below] {\scriptsize 1};
\filldraw[fill= white] (1,0) circle [radius=0.1] node[below] {\scriptsize 2};
\filldraw[fill= white] (2,0) circle [radius=0.1] node[below] {\scriptsize N+2};
\filldraw[fill= white] (3,0) circle [radius=0.1] node[below] {\scriptsize 2N+2};
\filldraw[fill= white] (4,0) circle [radius=0.1] node[below] {\scriptsize 3N+2};
\filldraw[fill= white] (5,0) circle [radius=0.1] node[below] {\scriptsize 4N+2};
\filldraw[fill= white] (6,0) circle [radius=0.1] node[below] {\scriptsize  2N+1};
%\filldraw[fill= white] (7,0) circle [radius=0.1] node[below] {\scriptsize  1};
\draw [thick] (0.1, 0) -- (0.9,0) ;
\draw [thick] (1.1, 0) -- (1.9,0) ;
\draw [thick] (2.1, 0) -- (2.9,0) ;
\draw [thick] (4.1, 0.05) -- (4.9,0.05) ;
\draw [thick] (4.1, -0.05) -- (4.9,-0.05) ;
\draw [thick] (3.1, 0) -- (3.9,0) ;
\draw [thick] (5.1, 0) -- (5.9,0) ;
%\draw [thick] (6.1, 0) -- (6.9,0) ;
\draw [thick] (4.4,0) -- (4.6,0.2);
\draw [thick] (4.4,0) -- (4.6,-0.2);
\end{tikzpicture} \\
&24(c-a) = 12 N + 11\,, \qquad 4(2a-c) = 12N^2 +14N \,,\\
&\Delta = \{\underbrace{(6k,6k+2)}_{k=1,\dots,N} \}\,, \qquad \mathfrak{g} = (\mathfrak{su}_2)_{6N+3}\oplus (\mathfrak{su}_4)_{12N+4}\oplus (\mathfrak{su}_2)_{16}\,.
}

\subsubsection*{$\mathcal T^{(N)}_{E_6,2}(3,0,0,0,0)$}

\al{&\begin{tikzpicture}
\filldraw[fill= white] (-1,0) circle [radius=0.1] node[below] {\scriptsize 1};
\filldraw[fill= white] (0,0) circle [radius=0.1] node[below] {\scriptsize 2};
\filldraw[fill= white] (1,0) circle [radius=0.1] node[below] {\scriptsize 3};
\filldraw[fill= white] (2,0) circle [radius=0.1] node[below] {\scriptsize N};
\filldraw[fill= white] (3,0) circle [radius=0.1] node[below] {\scriptsize 2N};
\filldraw[fill= white] (4,0) circle [radius=0.1] node[below] {\scriptsize 3N};
\filldraw[fill= white] (5,0) circle [radius=0.1] node[below] {\scriptsize 4N};
\filldraw[fill= white] (6,0) circle [radius=0.1] node[below] {\scriptsize  2N};
\draw [thick] (-0.9, 0) -- (-0.1,0) ;
\draw [thick] (0.1, 0) -- (0.9,0) ;
\draw [thick] (1.1, 0) -- (1.9,0) ;
\draw [thick] (2.1, 0) -- (2.9,0) ;
\draw [thick] (4.1, 0.05) -- (4.9,0.05) ;
\draw [thick] (4.1, -0.05) -- (4.9,-0.05) ;
\draw [thick] (3.1, 0) -- (3.9,0) ;
\draw [thick] (5.1, 0) -- (5.9,0) ;
%\draw [thick] (6.1, 0) -- (6.9,0) ;
\draw [thick] (4.4,0) -- (4.6,0.2);
\draw [thick] (4.4,0) -- (4.6,-0.2);
\end{tikzpicture} \\
&24(c-a) = 12 N + 5\,, \qquad 4(2a-c) = 18N^2 -41N+27\,,\\
&\Delta = \{\underbrace{(6k,6k+2,6k+3)}_{k=1,\dots,N-3},(6N-12,6N-10),6N-6,3N \}\,, \qquad \mathfrak{g} = (\mathfrak{f}_4)_{6N} \oplus (\mathfrak{su}_3)_{18}\,.
}

\subsubsection*{$\mathcal T^{(N)}_{E_6,2}(2,0,0,0,1)$}

\al{&\begin{tikzpicture}
\filldraw[fill= white] (-1,0) circle [radius=0.1] node[below] {\scriptsize 1};
\filldraw[fill= white] (0,0) circle [radius=0.1] node[below] {\scriptsize 2};
\filldraw[fill= white] (1,0) circle [radius=0.1] node[below] {\scriptsize 3};
\filldraw[fill= white] (2,0) circle [radius=0.1] node[below] {\scriptsize N+1};
\filldraw[fill= white] (3,0) circle [radius=0.1] node[below] {\scriptsize 2N+1};
\filldraw[fill= white] (4,0) circle [radius=0.1] node[below] {\scriptsize 3N+1};
\filldraw[fill= white] (5,0) circle [radius=0.1] node[below] {\scriptsize 4N+1};
\filldraw[fill= white] (6,0) circle [radius=0.1] node[below] {\scriptsize  2N};
\draw [thick] (-0.9, 0) -- (-0.1,0) ;
\draw [thick] (0.1, 0) -- (0.9,0) ;
\draw [thick] (1.1, 0) -- (1.9,0) ;
\draw [thick] (2.1, 0) -- (2.9,0) ;
\draw [thick] (4.1, 0.05) -- (4.9,0.05) ;
\draw [thick] (4.1, -0.05) -- (4.9,-0.05) ;
\draw [thick] (3.1, 0) -- (3.9,0) ;
\draw [thick] (5.1, 0) -- (5.9,0) ;
%\draw [thick] (6.1, 0) -- (6.9,0) ;
\draw [thick] (4.4,0) -- (4.6,0.2);
\draw [thick] (4.4,0) -- (4.6,-0.2);
\end{tikzpicture} \\
&24(c-a) = 12 N + 9\,, \qquad 4(2a-c) = 18N^2 -11N-1\,,\\
&\Delta = \{\underbrace{(6k,6k+2,6k+3)}_{k=1,\dots,N-2},(6N-6,6N-4),6N \} \,, \qquad \mathfrak{g} = (\mathfrak{sp}_3)_{6N+1} \oplus (\mathfrak{su}_3)_{18}\,.
}

\subsubsection*{$\mathcal T^{(N)}_{E_6,2}(1,0,0,0,2)$}

\al{&\begin{tikzpicture}
\filldraw[fill= white] (-1,0) circle [radius=0.1] node[below] {\scriptsize 1};
\filldraw[fill= white] (0,0) circle [radius=0.1] node[below] {\scriptsize 2};
\filldraw[fill= white] (1,0) circle [radius=0.1] node[below] {\scriptsize 3};
\filldraw[fill= white] (2,0) circle [radius=0.1] node[below] {\scriptsize N+2};
\filldraw[fill= white] (3,0) circle [radius=0.1] node[below] {\scriptsize 2N+2};
\filldraw[fill= white] (4,0) circle [radius=0.1] node[below] {\scriptsize 3N+2};
\filldraw[fill= white] (5,0) circle [radius=0.1] node[below] {\scriptsize 4N+2};
\filldraw[fill= white] (6,0) circle [radius=0.1] node[below] {\scriptsize  2N};
\draw [thick] (-0.9, 0) -- (-0.1,0) ;
\draw [thick] (0.1, 0) -- (0.9,0) ;
\draw [thick] (1.1, 0) -- (1.9,0) ;
\draw [thick] (2.1, 0) -- (2.9,0) ;
\draw [thick] (4.1, 0.05) -- (4.9,0.05) ;
\draw [thick] (4.1, -0.05) -- (4.9,-0.05) ;
\draw [thick] (3.1, 0) -- (3.9,0) ;
\draw [thick] (5.1, 0) -- (5.9,0) ;
%\draw [thick] (6.1, 0) -- (6.9,0) ;
\draw [thick] (4.4,0) -- (4.6,0.2);
\draw [thick] (4.4,0) -- (4.6,-0.2);
\end{tikzpicture} \\
&24(c-a) = 12 N + 13\,, \qquad 4(2a-c) = 18N^2 +7N-7\,,\\
&\Delta = \{\underbrace{(6k,6k+2,6k+3)}_{k=1,\dots,N-1},(6N,3N+1) \} \,, \qquad \mathfrak{g} = (\mathfrak{sp}_3)_{6N+2} \oplus (\mathfrak{su}_3)_{18}\,.
}

\subsubsection*{$\mathcal T^{(N)}_{E_6,2}(1,1,0,0,0)$}

\al{&\begin{tikzpicture}
\filldraw[fill= white] (-1,0) circle [radius=0.1] node[below] {\scriptsize 1};
\filldraw[fill= white] (0,0) circle [radius=0.1] node[below] {\scriptsize 2};
\filldraw[fill= white] (1,0) circle [radius=0.1] node[below] {\scriptsize 3};
\filldraw[fill= white] (2,0) circle [radius=0.1] node[below] {\scriptsize N+1};
\filldraw[fill= white] (3,0) circle [radius=0.1] node[below] {\scriptsize 2N};
\filldraw[fill= white] (4,0) circle [radius=0.1] node[below] {\scriptsize 3N};
\filldraw[fill= white] (5,0) circle [radius=0.1] node[below] {\scriptsize 4N};
\filldraw[fill= white] (6,0) circle [radius=0.1] node[below] {\scriptsize  2N};
\draw [thick] (-0.9, 0) -- (-0.1,0) ;
\draw [thick] (0.1, 0) -- (0.9,0) ;
\draw [thick] (1.1, 0) -- (1.9,0) ;
\draw [thick] (2.1, 0) -- (2.9,0) ;
\draw [thick] (4.1, 0.05) -- (4.9,0.05) ;
\draw [thick] (4.1, -0.05) -- (4.9,-0.05) ;
\draw [thick] (3.1, 0) -- (3.9,0) ;
\draw [thick] (5.1, 0) -- (5.9,0) ;
%\draw [thick] (6.1, 0) -- (6.9,0) ;
\draw [thick] (4.4,0) -- (4.6,0.2);
\draw [thick] (4.4,0) -- (4.6,-0.2);
\end{tikzpicture} \\
&24(c-a) = 12 N + 6\,, \qquad 4(2a-c) = 18N^2 -17N-1\,,\\
&\Delta = \{\underbrace{(6k,6k+2,6k+3)}_{k=1,\dots,N-2},(6N-6,6N-4),3N \} \,, \qquad \mathfrak{g} = (\mathfrak{so}_7)_{6N} \oplus (\mathfrak{su}_3)_{18}\,.
}

\subsubsection*{$\mathcal S^{(N)}_{E_6,2}(1,0,0,1,0)$}

\al{&\begin{tikzpicture}
\filldraw[fill= white] (-1,0) circle [radius=0.1] node[below] {\scriptsize 1};
\filldraw[fill= white] (0,0) circle [radius=0.1] node[below] {\scriptsize 2};
\filldraw[fill= white] (1,0) circle [radius=0.1] node[below] {\scriptsize 3};
\filldraw[fill= white] (2,0) circle [radius=0.1] node[below] {\scriptsize N+2};
\filldraw[fill= white] (3,0) circle [radius=0.1] node[below] {\scriptsize 2N+2};
\filldraw[fill= white] (4,0) circle [radius=0.1] node[below] {\scriptsize 3N+2};
\filldraw[fill= white] (5,0) circle [radius=0.1] node[below] {\scriptsize 4N+2};
\filldraw[fill= white] (6,0) circle [radius=0.1] node[below] {\scriptsize  2N+1};
\draw [thick] (-0.9, 0) -- (-0.1,0) ;
\draw [thick] (0.1, 0) -- (0.9,0) ;
\draw [thick] (1.1, 0) -- (1.9,0) ;
\draw [thick] (2.1, 0) -- (2.9,0) ;
\draw [thick] (4.1, 0.05) -- (4.9,0.05) ;
\draw [thick] (4.1, -0.05) -- (4.9,-0.05) ;
\draw [thick] (3.1, 0) -- (3.9,0) ;
\draw [thick] (5.1, 0) -- (5.9,0) ;
%\draw [thick] (6.1, 0) -- (6.9,0) ;
\draw [thick] (4.4,0) -- (4.6,0.2);
\draw [thick] (4.4,0) -- (4.6,-0.2);
\end{tikzpicture} \\
&24(c-a) = 12 N + 14\,, \qquad 4(2a-c) = 18N^2 +13N-5\,,\\
&\Delta = \{\underbrace{(6k,6k+2,6k+3)}_{k=1,\dots,N-1},(6N,6N+2) \} \,, \qquad \mathfrak{g} = (\mathfrak{su}_2)_{6N+3} \oplus (\mathfrak{su}_3)_{12N+4} \oplus (\mathfrak{su}_3)_{18}\,.
}

\subsubsection*{$\mathcal T^{(N)}_{E_6,2}(0,0,0,0,3)$}

\al{&\begin{tikzpicture}
\filldraw[fill= white] (-1,0) circle [radius=0.1] node[below] {\scriptsize 1};
\filldraw[fill= white] (0,0) circle [radius=0.1] node[below] {\scriptsize 2};
\filldraw[fill= white] (1,0) circle [radius=0.1] node[below] {\scriptsize 3};
\filldraw[fill= white] (2,0) circle [radius=0.1] node[below] {\scriptsize N+3};
\filldraw[fill= white] (3,0) circle [radius=0.1] node[below] {\scriptsize 2N+3};
\filldraw[fill= white] (4,0) circle [radius=0.1] node[below] {\scriptsize 3N+3};
\filldraw[fill= white] (5,0) circle [radius=0.1] node[below] {\scriptsize 4N+3};
\filldraw[fill= white] (6,0) circle [radius=0.1] node[below] {\scriptsize  2N};
\draw [thick] (-0.9, 0) -- (-0.1,0) ;
\draw [thick] (0.1, 0) -- (0.9,0) ;
\draw [thick] (1.1, 0) -- (1.9,0) ;
\draw [thick] (2.1, 0) -- (2.9,0) ;
\draw [thick] (4.1, 0.05) -- (4.9,0.05) ;
\draw [thick] (4.1, -0.05) -- (4.9,-0.05) ;
\draw [thick] (3.1, 0) -- (3.9,0) ;
\draw [thick] (5.1, 0) -- (5.9,0) ;
%\draw [thick] (6.1, 0) -- (6.9,0) ;
\draw [thick] (4.4,0) -- (4.6,0.2);
\draw [thick] (4.4,0) -- (4.6,-0.2);
\end{tikzpicture} \\
&24(c-a) = 12 N + 17\,, \qquad 4(2a-c) = 18N^2 +13N-5\,,\\
&\Delta = \{\underbrace{(6k,6k+2,6k+3)}_{k=1,\dots,N-1},(6N,6N+2) \} \,, \qquad \mathfrak{g} = (\mathfrak{sp}_4)_{6N+3} \oplus (\mathfrak{su}_3)_{18}\,.
}

\subsubsection*{$\mathcal T^{(N)}_{E_6,2}(0,1,0,0,1)$}

\al{&\begin{tikzpicture}
\filldraw[fill= white] (-1,0) circle [radius=0.1] node[below] {\scriptsize 1};
\filldraw[fill= white] (0,0) circle [radius=0.1] node[below] {\scriptsize 2};
\filldraw[fill= white] (1,0) circle [radius=0.1] node[below] {\scriptsize 3};
\filldraw[fill= white] (2,0) circle [radius=0.1] node[below] {\scriptsize N+2};
\filldraw[fill= white] (3,0) circle [radius=0.1] node[below] {\scriptsize 2N+1};
\filldraw[fill= white] (4,0) circle [radius=0.1] node[below] {\scriptsize 3N+1};
\filldraw[fill= white] (5,0) circle [radius=0.1] node[below] {\scriptsize 4N+1};
\filldraw[fill= white] (6,0) circle [radius=0.1] node[below] {\scriptsize  2N};
\draw [thick] (-0.9, 0) -- (-0.1,0) ;
\draw [thick] (0.1, 0) -- (0.9,0) ;
\draw [thick] (1.1, 0) -- (1.9,0) ;
\draw [thick] (2.1, 0) -- (2.9,0) ;
\draw [thick] (4.1, 0.05) -- (4.9,0.05) ;
\draw [thick] (4.1, -0.05) -- (4.9,-0.05) ;
\draw [thick] (3.1, 0) -- (3.9,0) ;
\draw [thick] (5.1, 0) -- (5.9,0) ;
%\draw [thick] (6.1, 0) -- (6.9,0) ;
\draw [thick] (4.4,0) -- (4.6,0.2);
\draw [thick] (4.4,0) -- (4.6,-0.2);
\end{tikzpicture} \\
&24(c-a) = 12 N + 10\,, \qquad 4(2a-c) = 18N^2 +N-8\,,\\
&\Delta = \{\underbrace{(6k,6k+2,6k+3)}_{k=1,\dots,N-1},6N \} \,, \qquad \mathfrak{g} = (\mathfrak{sp}_2)_{6N+1} \oplus (\mathfrak{su}_2)_{12N-6}\oplus (\mathfrak{su}_3)_{18}\,.
}

\subsubsection*{$\mathcal T^{(N)}_{E_6,2}(0,0,0,1,1)$}

\al{&\begin{tikzpicture}
\filldraw[fill= white] (-1,0) circle [radius=0.1] node[below] {\scriptsize 1};
\filldraw[fill= white] (0,0) circle [radius=0.1] node[below] {\scriptsize 2};
\filldraw[fill= white] (1,0) circle [radius=0.1] node[below] {\scriptsize 3};
\filldraw[fill= white] (2,0) circle [radius=0.1] node[below] {\scriptsize N+3};
\filldraw[fill= white] (3,0) circle [radius=0.1] node[below] {\scriptsize 2N+3};
\filldraw[fill= white] (4,0) circle [radius=0.1] node[below] {\scriptsize 3N+3};
\filldraw[fill= white] (5,0) circle [radius=0.1] node[below] {\scriptsize 4N+3};
\filldraw[fill= white] (6,0) circle [radius=0.1] node[below] {\scriptsize  2N+1};
\draw [thick] (-0.9, 0) -- (-0.1,0) ;
\draw [thick] (0.1, 0) -- (0.9,0) ;
\draw [thick] (1.1, 0) -- (1.9,0) ;
\draw [thick] (2.1, 0) -- (2.9,0) ;
\draw [thick] (4.1, 0.05) -- (4.9,0.05) ;
\draw [thick] (4.1, -0.05) -- (4.9,-0.05) ;
\draw [thick] (3.1, 0) -- (3.9,0) ;
\draw [thick] (5.1, 0) -- (5.9,0) ;
%\draw [thick] (6.1, 0) -- (6.9,0) ;
\draw [thick] (4.4,0) -- (4.6,0.2);
\draw [thick] (4.4,0) -- (4.6,-0.2);
\end{tikzpicture} \\
&24(c-a) = 12 N + 18\,, \qquad 4(2a-c) = 18N^2 +25N\,,\\
&\Delta = \{\underbrace{(6k,6k+2,6k+3)}_{k=1,\dots,N} \} \,, \qquad \mathfrak{g} = (\mathfrak{su}_4)_{12N+6} \oplus (\mathfrak{su}_3)_{18}\,.
}

\subsubsection*{$\mathcal T^{(N)}_{E_6,2}(0,0,1,0,0)$}

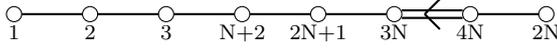
\al{&\begin{tikzpicture}
\filldraw[fill= white] (-1,0) circle [radius=0.1] node[below] {\scriptsize 1};
\filldraw[fill= white] (0,0) circle [radius=0.1] node[below] {\scriptsize 2};
\filldraw[fill= white] (1,0) circle [radius=0.1] node[below] {\scriptsize 3};
\filldraw[fill= white] (2,0) circle [radius=0.1] node[below] {\scriptsize N+2};
\filldraw[fill= white] (3,0) circle [radius=0.1] node[below] {\scriptsize 2N+1};
\filldraw[fill= white] (4,0) circle [radius=0.1] node[below] {\scriptsize 3N};
\filldraw[fill= white] (5,0) circle [radius=0.1] node[below] {\scriptsize 4N};
\filldraw[fill= white] (6,0) circle [radius=0.1] node[below] {\scriptsize  2N};
\draw [thick] (-0.9, 0) -- (-0.1,0) ;
\draw [thick] (0.1, 0) -- (0.9,0) ;
\draw [thick] (1.1, 0) -- (1.9,0) ;
\draw [thick] (2.1, 0) -- (2.9,0) ;
\draw [thick] (4.1, 0.05) -- (4.9,0.05) ;
\draw [thick] (4.1, -0.05) -- (4.9,-0.05) ;
\draw [thick] (3.1, 0) -- (3.9,0) ;
\draw [thick] (5.1, 0) -- (5.9,0) ;
%\draw [thick] (6.1, 0) -- (6.9,0) ;
\draw [thick] (4.4,0) -- (4.6,0.2);
\draw [thick] (4.4,0) -- (4.6,-0.2);
\end{tikzpicture} \\
&24(c-a) = 12 N + 8\,, \qquad 4(2a-c) = 18N^2 -5N-8\,,\\
&\Delta = \{\underbrace{(6k,6k+2,6k+3)}_{k=1,\dots,N-1},3N \} \,, \qquad \mathfrak{g} = (\mathfrak{su}_3)_{6N} \oplus (\mathfrak{su}_3)_{12N-6} \oplus (\mathfrak{su}_3)_{18}\,.
}

\subsection{$\ell =3$ quotients}

\subsubsection*{$\mathcal T^{(N)}_{D_4,3}(2,0,0)$}

\al{&\begin{tikzpicture}
\filldraw[fill= white] (0,0) circle [radius=0.1] node[below] {\scriptsize 1};
\filldraw[fill= white] (1,0) circle [radius=0.1] node[below] {\scriptsize 2};
\filldraw[fill= white] (2,0) circle [radius=0.1] node[below] {\scriptsize N};
\filldraw[fill= white] (3,0) circle [radius=0.1] node[below] {\scriptsize 2N};
\filldraw[fill= white] (4,0) circle [radius=0.1] node[below] {\scriptsize 3N};
\draw [thick] (0.1, 0) -- (0.9,0) ;
\draw [thick] (1.1, 0) -- (1.9,0) ;
\draw [thick] (2.1, 0) -- (2.9,0) ;
\draw [thick] (3.1, 0.07) -- (3.9,0.07) ;
\draw [thick] (3.1, -0.07) -- (3.9,-0.07) ;
\draw [thick] (3.1, 0) -- (3.9,0) ;
\draw [thick] (3.4,0) -- (3.6,0.2);
\draw [thick] (3.4,0) -- (3.6,-0.2);
\end{tikzpicture} \\
&24(c-a) = 6 N + 2\,, \qquad 4(2a-c) = 12N^2-18N+6\,,\\
&\Delta = \{\underbrace{(6k,6k+2)}_{k=1,\dots,N-2},6N-6,2N \}\,, \qquad \mathfrak g =  (\mathfrak{g}_2)_{4N}\oplus (\mathfrak{su}_2)_{16}\,.
}

\subsubsection*{$\mathcal S^{(N)}_{D_4,3}(1,0,1)$}

\al{&\begin{tikzpicture}
\filldraw[fill= white] (0,0) circle [radius=0.1] node[below] {\scriptsize 1};
\filldraw[fill= white] (1,0) circle [radius=0.1] node[below] {\scriptsize 2};
\filldraw[fill= white] (2,0) circle [radius=0.1] node[below] {\scriptsize N+1};
\filldraw[fill= white] (3,0) circle [radius=0.1] node[below] {\scriptsize 2N+1};
\filldraw[fill= white] (4,0) circle [radius=0.1] node[below] {\scriptsize 3N+1};
\draw [thick] (0.1, 0) -- (0.9,0) ;
\draw [thick] (1.1, 0) -- (1.9,0) ;
\draw [thick] (2.1, 0) -- (2.9,0) ;
\draw [thick] (3.1, 0.07) -- (3.9,0.07) ;
\draw [thick] (3.1, -0.07) -- (3.9,-0.07) ;
\draw [thick] (3.1, 0) -- (3.9,0) ;
\draw [thick] (3.4,0) -- (3.6,0.2);
\draw [thick] (3.4,0) -- (3.6,-0.2);
\end{tikzpicture} \\
&24(c-a) = 6 N + 5\,, \qquad 4(2a-c) = 12N^2 +2N-3\,,\\
&\Delta = \{\underbrace{(6k,6k+2)}_{k=1,\dots,N-1},6N\}\,, \qquad \mathfrak g =  (\mathfrak{su}_2)_{12N+2}\oplus (\mathfrak{su}_2)_{16}\,.
}

\subsubsection*{$\mathcal R^{(N)}_{D_4,3}(0,0,2)$}

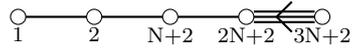
\al{&\begin{tikzpicture}
\filldraw[fill= white] (0,0) circle [radius=0.1] node[below] {\scriptsize 1};
\filldraw[fill= white] (1,0) circle [radius=0.1] node[below] {\scriptsize 2};
\filldraw[fill= white] (2,0) circle [radius=0.1] node[below] {\scriptsize N+2};
\filldraw[fill= white] (3,0) circle [radius=0.1] node[below] {\scriptsize 2N+2};
\filldraw[fill= white] (4,0) circle [radius=0.1] node[below] {\scriptsize 3N+2};
\draw [thick] (0.1, 0) -- (0.9,0) ;
\draw [thick] (1.1, 0) -- (1.9,0) ;
\draw [thick] (2.1, 0) -- (2.9,0) ;
\draw [thick] (3.1, 0.07) -- (3.9,0.07) ;
\draw [thick] (3.1, -0.07) -- (3.9,-0.07) ;
\draw [thick] (3.1, 0) -- (3.9,0) ;
\draw [thick] (3.4,0) -- (3.6,0.2);
\draw [thick] (3.4,0) -- (3.6,-0.2);
\end{tikzpicture} \\
&24(c-a) = 6 N + 8\,, \qquad 4(2a-c) = 12N^2 +14N\,,\\
&\Delta = \{\underbrace{(6k,6k+2)}_{k=1,\dots,N}\}\,, \qquad \mathfrak g =  (\mathfrak{su}_3)_{12N+4}\oplus (\mathfrak{su}_2)_{16}\,.
}

\subsubsection*{$\mathcal T^{(N)}_{D_4,3}(0,1,0)$}

\al{&\begin{tikzpicture}
\filldraw[fill= white] (0,0) circle [radius=0.1] node[below] {\scriptsize 1};
\filldraw[fill= white] (1,0) circle [radius=0.1] node[below] {\scriptsize 2};
\filldraw[fill= white] (2,0) circle [radius=0.1] node[below] {\scriptsize N+1};
\filldraw[fill= white] (3,0) circle [radius=0.1] node[below] {\scriptsize 2N};
\filldraw[fill= white] (4,0) circle [radius=0.1] node[below] {\scriptsize 3N};
\draw [thick] (0.1, 0) -- (0.9,0) ;
\draw [thick] (1.1, 0) -- (1.9,0) ;
\draw [thick] (2.1, 0) -- (2.9,0) ;
\draw [thick] (3.1, 0.07) -- (3.9,0.07) ;
\draw [thick] (3.1, -0.07) -- (3.9,-0.07) ;
\draw [thick] (3.1, 0) -- (3.9,0) ;
\draw [thick] (3.4,0) -- (3.6,0.2);
\draw [thick] (3.4,0) -- (3.6,-0.2);
\end{tikzpicture} \\
&24(c-a) = 6 N + 3\,, \qquad 4(2a-c) = 12N^2-6N-3\,,\\
&\Delta = \{\underbrace{(6k,6k+2)}_{k=1,\dots,N-1},2N\}\,, \qquad \mathfrak g =  (\mathfrak{su}_2)_{4N}\oplus (\mathfrak{su}_2)_{12N-8}\oplus  (\mathfrak{su}_2)_{16} \,.
}

\subsection{$\ell = 4$ quotients}

\subsubsection*{$\mathcal T^{(N)}_{\mathcal H_2,4} (2,0) $}

\al{&\begin{tikzpicture}
\filldraw[fill= white] (0,0) circle [radius=0.1] node[below] {\scriptsize 1};
\filldraw[fill= white] (1,0) circle [radius=0.1] node[below] {\scriptsize 2};
\filldraw[fill= white] (2,0) circle [radius=0.1] node[below] {\scriptsize N};
\filldraw[fill= white] (3,0) circle [radius=0.1] node[below] {\scriptsize 2N};
\draw [thick] (0.1, 0) -- (0.9,0) ;
\draw [thick] (1.1, 0) -- (1.9,0) ;
\draw [thick] (2.1, 0.06) -- (2.9,0.06) ;
\draw [thick] (2.1, 0.02) -- (2.9,0.02) ;
\draw [thick] (2.1,-0.02) -- (2.9,-0.02) ;
\draw [thick] (2.1, -0.06) -- (2.9,-0.06) ;
\draw [thick] (2.4,0) -- (2.6,0.2);
\draw [thick] (2.4,0) -- (2.6,-0.2);
\end{tikzpicture} \\
&24(c-a) = 3 N + 2\,, \qquad 4(2a-c) = 12N^2-19N+6\,,\\
&\Delta = \{\underbrace{(6k,6k+2)}_{k=1,\dots,N-2},6N-6,3N/2\}\,, \qquad \mathfrak g =  (\mathfrak{su}_2)_{3N}\oplus (\mathfrak{su}_2)_{16}\,.
}

\subsubsection*{$\mathcal T^{(N)}_{\mathcal H_2,4} (0,2) $}

\al{&\begin{tikzpicture}
\filldraw[fill= white] (0,0) circle [radius=0.1] node[below] {\scriptsize 1};
\filldraw[fill= white] (1,0) circle [radius=0.1] node[below] {\scriptsize 2};
\filldraw[fill= white] (2,0) circle [radius=0.1] node[below] {\scriptsize N+1};
\filldraw[fill= white] (3,0) circle [radius=0.1] node[below] {\scriptsize 2N};
\draw [thick] (0.1, 0) -- (0.9,0) ;
\draw [thick] (1.1, 0) -- (1.9,0) ;
\draw [thick] (2.1, 0.06) -- (2.9,0.06) ;
\draw [thick] (2.1, 0.02) -- (2.9,0.02) ;
\draw [thick] (2.1,-0.02) -- (2.9,-0.02) ;
\draw [thick] (2.1, -0.06) -- (2.9,-0.06) ;
\draw [thick] (2.4,0) -- (2.6,0.2);
\draw [thick] (2.4,0) -- (2.6,-0.2);
\end{tikzpicture} \\
&24(c-a) = 3 N + 3\,, \qquad 4(2a-c) = 12N^2-7N-3\,,\\
&\Delta = \{\underbrace{(6k,6k+2)}_{k=1,\dots,N-1},3N/2\}\,, \qquad \mathfrak g =  (\mathfrak{su}_2)_{12N-8}\oplus (\mathfrak{su}_2)_{16}\,.
}

\subsubsection*{$\mathcal S^{(N)}_{\mathcal H_2,4} (1,1) $}

\al{&\begin{tikzpicture}
\filldraw[fill= white] (0,0) circle [radius=0.1] node[below] {\scriptsize 1};
\filldraw[fill= white] (1,0) circle [radius=0.1] node[below] {\scriptsize 2};
\filldraw[fill= white] (2,0) circle [radius=0.1] node[below] {\scriptsize N+1};
\filldraw[fill= white] (3,0) circle [radius=0.1] node[below] {\scriptsize 2N+1};
\draw [thick] (0.1, 0) -- (0.9,0) ;
\draw [thick] (1.1, 0) -- (1.9,0) ;
\draw [thick] (2.1, 0.06) -- (2.9,0.06) ;
\draw [thick] (2.1, 0.02) -- (2.9,0.02) ;
\draw [thick] (2.1,-0.02) -- (2.9,-0.02) ;
\draw [thick] (2.1, -0.06) -- (2.9,-0.06) ;
\draw [thick] (2.4,0) -- (2.6,0.2);
\draw [thick] (2.4,0) -- (2.6,-0.2);
\end{tikzpicture} \\
&24(c-a) = 3 N + 4\,, \qquad 4(2a-c) = 12N^2+2N-3\,,\\
&\Delta = \{\underbrace{(6k,6k+2)}_{k=1,\dots,N-1},6N\}\,, \qquad \mathfrak g = (\mathfrak{su}_2)_{16}\,.
}

\section{List of 4d SCFTs from Type IIB orbi-S-folds} \label{App:IIBth}

In this appendix we list the theories we find after the deformation of the $T(SU(k))$ tail that breaks the $\mathfrak{su}_k$ symmetry. We will call these theories with the same notation as before, adding a small circle on top to distinguish them from their progenitors.

\subsection{$\ell =2$ quotients}

\subsubsection*{$\mathring{\mathcal T}^{(N)}_{E_6,2}(2,0,0,0,0)$}

\al{&\begin{tikzpicture}
\filldraw[fill= white] (2,0) circle [radius=0.1] node[below] {\scriptsize N-2};
\filldraw[fill= white] (3,0) circle [radius=0.1] node[below] {\scriptsize 2N-2};
\filldraw[fill= white] (4,0) circle [radius=0.1] node[below] {\scriptsize 3N-2};
\filldraw[fill= white] (5,0) circle [radius=0.1] node[below] {\scriptsize 4N-2};
\filldraw[fill= white] (6,0) circle [radius=0.1] node[below] {\scriptsize  2N};
\filldraw[fill= white] (7,0) circle [radius=0.1] node[below] {\scriptsize  1};
\draw [thick] (2.1, 0) -- (2.9,0) ;
\draw [thick] (4.1, 0.05) -- (4.9,0.05) ;
\draw [thick] (4.1, -0.05) -- (4.9,-0.05) ;
\draw [thick] (3.1, 0) -- (3.9,0) ;
\draw [thick] (5.1, 0) -- (5.9,0) ;
\draw [thick] (6.1, 0.05) -- (6.9,0.05) ;
\draw [thick] (6.1, -0.05) -- (6.9,-0.05) ;
\draw [thick] (4.4,0) -- (4.6,0.2);
\draw [thick] (4.4,0) -- (4.6,-0.2);
\draw [thick] (6.6,0) -- (6.4,0.2);
\draw [thick] (6.6,0) -- (6.4,-0.2);
\end{tikzpicture} \\
&24(c-a) = 12N-8\,, \qquad 4(2a-c) = 12N^2-24N+16\,,\\
&\Delta = \{\underbrace{(6k-2,6k)}_{k=1,\dots,N-2},6N-8,3N-1 \}\,,  \qquad \mathfrak{g} = (\mathfrak{f}_4)_{6N-2}\,.
}

\subsubsection*{$\mathring{\mathcal T}^{(N)}_{E_6,2}(1,0,0,0,1)$}

\al{&\begin{tikzpicture}
\filldraw[fill= white] (2,0) circle [radius=0.1] node[below] {\scriptsize N-1};
\filldraw[fill= white] (3,0) circle [radius=0.1] node[below] {\scriptsize 2N-1};
\filldraw[fill= white] (4,0) circle [radius=0.1] node[below] {\scriptsize 3N-1};
\filldraw[fill= white] (5,0) circle [radius=0.1] node[below] {\scriptsize 4N-1};
\filldraw[fill= white] (6,0) circle [radius=0.1] node[below] {\scriptsize  2N};
\filldraw[fill= white] (7,0) circle [radius=0.1] node[below] {\scriptsize  1};
\draw [thick] (2.1, 0) -- (2.9,0) ;
\draw [thick] (4.1, 0.05) -- (4.9,0.05) ;
\draw [thick] (4.1, -0.05) -- (4.9,-0.05) ;
\draw [thick] (3.1, 0) -- (3.9,0) ;
\draw [thick] (5.1, 0) -- (5.9,0) ;
\draw [thick] (6.1, 0.05) -- (6.9,0.05) ;
\draw [thick] (6.1, -0.05) -- (6.9,-0.05) ;
\draw [thick] (4.4,0) -- (4.6,0.2);	
\draw [thick] (4.4,0) -- (4.6,-0.2);
\draw [thick] (6.6,0) -- (6.4,0.2);
\draw [thick] (6.6,0) -- (6.4,-0.2);
\end{tikzpicture} \\
&24(c-a) = 12N-4\,, \qquad 4(2a-c) = 12N^2 -6N +1 \,,  \\
&\Delta = \{\underbrace{(6k-2,6k)}_{k=1,\dots,N-1},6N-2 \}\,,\qquad \mathfrak{g} = (\mathfrak{sp}_3)_{6N-1}\,.
}

\subsubsection*{$\mathring{\mathcal T}^{(N)}_{E_6,2}(0,1,0,0,0)$}

\al{&\begin{tikzpicture}
\filldraw[fill= white] (2,0) circle [radius=0.1] node[below] {\scriptsize N-1};
\filldraw[fill= white] (3,0) circle [radius=0.1] node[below] {\scriptsize 2N-2};
\filldraw[fill= white] (4,0) circle [radius=0.1] node[below] {\scriptsize 3N-2};
\filldraw[fill= white] (5,0) circle [radius=0.1] node[below] {\scriptsize 4N-2};
\filldraw[fill= white] (6,0) circle [radius=0.1] node[below] {\scriptsize  2N};
\filldraw[fill= white] (7,0) circle [radius=0.1] node[below] {\scriptsize  1};
\draw [thick] (2.1, 0) -- (2.9,0) ;
\draw [thick] (4.1, 0.05) -- (4.9,0.05) ;
\draw [thick] (4.1, -0.05) -- (4.9,-0.05) ;
\draw [thick] (3.1, 0) -- (3.9,0) ;
\draw [thick] (5.1, 0) -- (5.9,0) ;
\draw [thick] (6.1, 0.05) -- (6.9,0.05) ;
\draw [thick] (6.1, -0.05) -- (6.9,-0.05) ;
\draw [thick] (4.4,0) -- (4.6,0.2);
\draw [thick] (4.4,0) -- (4.6,-0.2);
\draw [thick] (6.6,0) -- (6.4,0.2);
\draw [thick] (6.6,0) -- (6.4,-0.2);
\end{tikzpicture} \\
&24(c-a) = 12N-7\,, \qquad 4(2a-c) = 12N^2 -12N+5\,,\\
&\Delta = \{\underbrace{(6k-2,6k)}_{k=1,\dots,N-1},3N-1 \}\,,  \qquad \mathfrak{g} = (\mathfrak{so}_7)_{6N-2}\oplus (\mathfrak{su}_2)_{12N-12}\,.
}

\subsubsection*{$\mathring{\mathcal T}^{(N)}_{E_6,2}(0,0,0,0,2)$}

\al{&\begin{tikzpicture}
\filldraw[fill= white] (2,0) circle [radius=0.1] node[below] {\scriptsize N};
\filldraw[fill= white] (3,0) circle [radius=0.1] node[below] {\scriptsize 2N};
\filldraw[fill= white] (4,0) circle [radius=0.1] node[below] {\scriptsize 3N};
\filldraw[fill= white] (5,0) circle [radius=0.1] node[below] {\scriptsize 4N};
\filldraw[fill= white] (6,0) circle [radius=0.1] node[below] {\scriptsize  2N};
\filldraw[fill= white] (7,0) circle [radius=0.1] node[below] {\scriptsize  1};
\draw [thick] (2.1, 0) -- (2.9,0) ;
\draw [thick] (4.1, 0.05) -- (4.9,0.05) ;
\draw [thick] (4.1, -0.05) -- (4.9,-0.05) ;
\draw [thick] (3.1, 0) -- (3.9,0) ;
\draw [thick] (5.1, 0) -- (5.9,0) ;
\draw [thick] (6.1, 0.05) -- (6.9,0.05) ;
\draw [thick] (6.1, -0.05) -- (6.9,-0.05) ;
\draw [thick] (4.4,0) -- (4.6,0.2);
\draw [thick] (4.4,0) -- (4.6,-0.2);
\draw [thick] (6.6,0) -- (6.4,0.2);
\draw [thick] (6.6,0) -- (6.4,-0.2);
\end{tikzpicture} \\
&24(c-a) = 12N\,, \qquad 4(2a-c) = 12N^2\,, \\
&\Delta = \{\underbrace{(6k-2,6k)}_{k=1,\dots,N-1},(6N-2,3N) \} \,,\qquad \mathfrak{g} = (\mathfrak{sp}_4)_{6N}\,.
}

\subsubsection*{$\mathring{\mathcal S}^{(N)}_{E_6,2}(0,0,0,1,0)$}

\al{&\begin{tikzpicture}
\filldraw[fill= white] (2,0) circle [radius=0.1] node[below] {\scriptsize N};
\filldraw[fill= white] (3,0) circle [radius=0.1] node[below] {\scriptsize 2N};
\filldraw[fill= white] (4,0) circle [radius=0.1] node[below] {\scriptsize 3N};
\filldraw[fill= white] (5,0) circle [radius=0.1] node[below] {\scriptsize 4N};
\filldraw[fill= white] (6,0) circle [radius=0.1] node[below] {\scriptsize  2N+1};
\filldraw[fill= white] (7,0) circle [radius=0.1] node[below] {\scriptsize  1};
\draw [thick] (2.1, 0) -- (2.9,0) ;
\draw [thick] (4.1, 0.05) -- (4.9,0.05) ;
\draw [thick] (4.1, -0.05) -- (4.9,-0.05) ;
\draw [thick] (3.1, 0) -- (3.9,0) ;
\draw [thick] (5.1, 0) -- (5.9,0) ;
\draw [thick] (6.1, 0.05) -- (6.9,0.05) ;
\draw [thick] (6.1, -0.05) -- (6.9,-0.05) ;
\draw [thick] (4.4,0) -- (4.6,0.2);
\draw [thick] (4.4,0) -- (4.6,-0.2);
\draw [thick] (6.6,0) -- (6.4,0.2);
\draw [thick] (6.6,0) -- (6.4,-0.2);
\end{tikzpicture} \\
&24(c-a) = 12N+1\,, \qquad 4(2a-c) = 12N^2+6\,, \\
&\Delta = \{\underbrace{(6k-2,6k)}_{k=1,\dots,N} \}\,,\qquad \mathfrak{g} = (\mathfrak{su}_2)_{6N+1}\oplus (\mathfrak{su}_4)_{12N}\,.
}

\subsubsection*{$\mathring{\mathcal T}^{(N)}_{E_6,2}(3,0,0,0,0)$}

\al{&\begin{tikzpicture}
\filldraw[fill= white] (2,0) circle [radius=0.1] node[below] {\scriptsize N-3};
\filldraw[fill= white] (3,0) circle [radius=0.1] node[below] {\scriptsize 2N-3};
\filldraw[fill= white] (4,0) circle [radius=0.1] node[below] {\scriptsize 3N-3};
\filldraw[fill= white] (5,0) circle [radius=0.1] node[below] {\scriptsize 4N-3};
\filldraw[fill= white] (6,0) circle [radius=0.1] node[below] {\scriptsize  2N-1};
\filldraw[fill= white] (5.5,1) circle [radius=0.1] node[above] {\scriptsize  1};
\draw [thick] (2.1, 0) -- (2.9,0) ;
\draw [thick] (4.1, 0.05) -- (4.9,0.05) ;
\draw [thick] (4.1, -0.05) -- (4.9,-0.05) ;
\draw [thick] (3.1, 0) -- (3.9,0) ;
\draw [thick] (5.1, 0) -- (5.9,0) ;
\draw [thick] (4.4,0) -- (4.6,0.2);
\draw [thick] (4.4,0) -- (4.6,-0.2);
\draw [thick] (5+0.05,0.1) -- (5.5,0.9);
\draw [thick] (6-0.05,0.1) -- (5.5,0.9);
\end{tikzpicture} \\
&24(c-a) = 12N-13\,, \qquad 4(2a-c) = 18N^2 -59N+60\,,\\
&\Delta = \{\underbrace{(6k-3,6k-1,6k)}_{k=1,\dots,N-3},(6N-15,6N-13),6N-9,3N-3/2 \}\,, \qquad \mathfrak{g} = (\mathfrak{f}_4)_{6N-3}\,.
}

\subsubsection*{$\mathring{\mathcal T}^{(N)}_{E_6,2}(2,0,0,0,1)$}

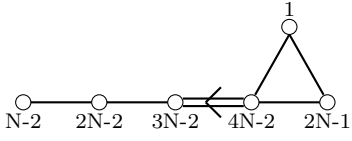
\al{&\begin{tikzpicture}
\filldraw[fill= white] (2,0) circle [radius=0.1] node[below] {\scriptsize N-2};
\filldraw[fill= white] (3,0) circle [radius=0.1] node[below] {\scriptsize 2N-2};
\filldraw[fill= white] (4,0) circle [radius=0.1] node[below] {\scriptsize 3N-2};
\filldraw[fill= white] (5,0) circle [radius=0.1] node[below] {\scriptsize 4N-2};
\filldraw[fill= white] (6,0) circle [radius=0.1] node[below] {\scriptsize  2N-1};
\filldraw[fill= white] (5.5,1) circle [radius=0.1] node[above] {\scriptsize  1};
\draw [thick] (2.1, 0) -- (2.9,0) ;
\draw [thick] (4.1, 0.05) -- (4.9,0.05) ;
\draw [thick] (4.1, -0.05) -- (4.9,-0.05) ;
\draw [thick] (3.1, 0) -- (3.9,0) ;
\draw [thick] (5.1, 0) -- (5.9,0) ;
\draw [thick] (4.4,0) -- (4.6,0.2);
\draw [thick] (4.4,0) -- (4.6,-0.2);
\draw [thick] (5+0.05,0.1) -- (5.5,0.9);
\draw [thick] (6-0.05,0.1) -- (5.5,0.9);
\end{tikzpicture} \\
&24(c-a) = 12N-9\,, \qquad 4(2a-c) = 18N^2-29N+17\,,\\
&\Delta = \{\underbrace{(6k-3,6k-1,6k)}_{k=1,\dots,N-2},(6N-9,6N-7),6N-3 \}\,, \qquad \mathfrak{g} = (\mathfrak{sp}_3)_{6N-2}\,.
}

\subsubsection*{$\mathring{\mathcal T}^{(N)}_{E_6,2}(1,0,0,0,2)$}

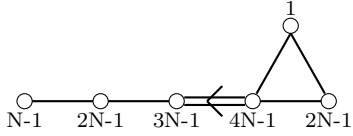
\al{&\begin{tikzpicture}
\filldraw[fill= white] (2,0) circle [radius=0.1] node[below] {\scriptsize N-1};
\filldraw[fill= white] (3,0) circle [radius=0.1] node[below] {\scriptsize 2N-1};
\filldraw[fill= white] (4,0) circle [radius=0.1] node[below] {\scriptsize 3N-1};
\filldraw[fill= white] (5,0) circle [radius=0.1] node[below] {\scriptsize 4N-1};
\filldraw[fill= white] (6,0) circle [radius=0.1] node[below] {\scriptsize  2N-1};
\filldraw[fill= white] (5.5,1) circle [radius=0.1] node[above] {\scriptsize  1};
\draw [thick] (2.1, 0) -- (2.9,0) ;
\draw [thick] (4.1, 0.05) -- (4.9,0.05) ;
\draw [thick] (4.1, -0.05) -- (4.9,-0.05) ;
\draw [thick] (3.1, 0) -- (3.9,0) ;
\draw [thick] (5.1, 0) -- (5.9,0) ;
\draw [thick] (4.4,0) -- (4.6,0.2);
\draw [thick] (4.4,0) -- (4.6,-0.2);
\draw [thick] (5+0.05,0.1) -- (5.5,0.9);
\draw [thick] (6-0.05,0.1) -- (5.5,0.9);
\end{tikzpicture} \\
&24(c-a) = 12N-5\,, \qquad 4(2a-c) = 18N^2-11N+2\,,\\
&\Delta = \{\underbrace{(6k-3,6k-1,6k)}_{k=1,\dots,N-1},(6N-3,3N-1/2) \}\,, \qquad \mathfrak{g} = (\mathfrak{sp}_3)_{6N-1}\,.
}

\subsubsection*{$\mathring{\mathcal T}^{(N)}_{E_6,2}(1,1,0,0,0)$}

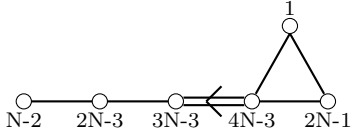
\al{&\begin{tikzpicture}
\filldraw[fill= white] (2,0) circle [radius=0.1] node[below] {\scriptsize N-2};
\filldraw[fill= white] (3,0) circle [radius=0.1] node[below] {\scriptsize 2N-3};
\filldraw[fill= white] (4,0) circle [radius=0.1] node[below] {\scriptsize 3N-3};
\filldraw[fill= white] (5,0) circle [radius=0.1] node[below] {\scriptsize 4N-3};
\filldraw[fill= white] (6,0) circle [radius=0.1] node[below] {\scriptsize  2N-1};
\filldraw[fill= white] (5.5,1) circle [radius=0.1] node[above] {\scriptsize  1};
\draw [thick] (2.1, 0) -- (2.9,0) ;
\draw [thick] (4.1, 0.05) -- (4.9,0.05) ;
\draw [thick] (4.1, -0.05) -- (4.9,-0.05) ;
\draw [thick] (3.1, 0) -- (3.9,0) ;
\draw [thick] (5.1, 0) -- (5.9,0) ;
\draw [thick] (4.4,0) -- (4.6,0.2);
\draw [thick] (4.4,0) -- (4.6,-0.2);
\draw [thick] (5+0.05,0.1) -- (5.5,0.9);
\draw [thick] (6-0.05,0.1) -- (5.5,0.9);
\end{tikzpicture} \\
&24(c-a) = 12N-12\,, \qquad 4(2a-c) = 18N^2-35N+20\,,\\
&\Delta = \{\underbrace{(6k-3,6k-1,6k)}_{k=1,\dots,N-2},(6N-9,6N-7),3N-3/2 \}\,, \qquad \mathfrak{g} = (\mathfrak{so}_7)_{6N-3}\,.
}

\subsubsection*{$\mathring{\mathcal S}^{(N)}_{E_6,2}(1,0,0,1,0)$}

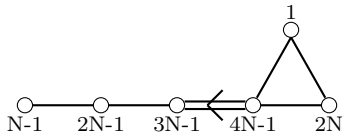
\al{&\begin{tikzpicture}
\filldraw[fill= white] (2,0) circle [radius=0.1] node[below] {\scriptsize N-1};
\filldraw[fill= white] (3,0) circle [radius=0.1] node[below] {\scriptsize 2N-1};
\filldraw[fill= white] (4,0) circle [radius=0.1] node[below] {\scriptsize 3N-1};
\filldraw[fill= white] (5,0) circle [radius=0.1] node[below] {\scriptsize 4N-1};
\filldraw[fill= white] (6,0) circle [radius=0.1] node[below] {\scriptsize  2N};
\filldraw[fill= white] (5.5,1) circle [radius=0.1] node[above] {\scriptsize  1};
\draw [thick] (2.1, 0) -- (2.9,0) ;
\draw [thick] (4.1, 0.05) -- (4.9,0.05) ;
\draw [thick] (4.1, -0.05) -- (4.9,-0.05) ;
\draw [thick] (3.1, 0) -- (3.9,0) ;
\draw [thick] (5.1, 0) -- (5.9,0) ;
\draw [thick] (4.4,0) -- (4.6,0.2);
\draw [thick] (4.4,0) -- (4.6,-0.2);
\draw [thick] (5+0.05,0.1) -- (5.5,0.9);
\draw [thick] (6-0.05,0.1) -- (5.5,0.9);
\end{tikzpicture} \\
&24(c-a) = 12N-4\,, \qquad 4(2a-c) = 18N^2 -5N+1\,,\\
&\Delta = \{\underbrace{(6k-3,6k-1,6k)}_{k=1,\dots,N-1},(6N-3,6N-1) \} \,, \qquad \mathfrak{g} = (\mathfrak{su}_2)_{6N} \oplus (\mathfrak{su}_3)_{12N-2}\,.
}

\subsubsection*{$\mathring{\mathcal T}^{(N)}_{E_6,2}(0,0,0,0,3)$}

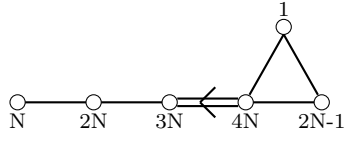
\al{&\begin{tikzpicture}
\filldraw[fill= white] (2,0) circle [radius=0.1] node[below] {\scriptsize N};
\filldraw[fill= white] (3,0) circle [radius=0.1] node[below] {\scriptsize 2N};
\filldraw[fill= white] (4,0) circle [radius=0.1] node[below] {\scriptsize 3N};
\filldraw[fill= white] (5,0) circle [radius=0.1] node[below] {\scriptsize 4N};
\filldraw[fill= white] (6,0) circle [radius=0.1] node[below] {\scriptsize  2N-1};
\filldraw[fill= white] (5.5,1) circle [radius=0.1] node[above] {\scriptsize  1};
\draw [thick] (2.1, 0) -- (2.9,0) ;
\draw [thick] (4.1, 0.05) -- (4.9,0.05) ;
\draw [thick] (4.1, -0.05) -- (4.9,-0.05) ;
\draw [thick] (3.1, 0) -- (3.9,0) ;
\draw [thick] (5.1, 0) -- (5.9,0) ;
\draw [thick] (4.4,0) -- (4.6,0.2);
\draw [thick] (4.4,0) -- (4.6,-0.2);
\draw [thick] (5+0.05,0.1) -- (5.5,0.9);
\draw [thick] (6-0.05,0.1) -- (5.5,0.9);
\end{tikzpicture} \\
&24(c-a) = 12N-1\,, \qquad 4(2a-c) = 18N^2 -5N+1\,,\\
&\Delta = \{\underbrace{(6k-3,6k-1,6k)}_{k=1,\dots,N-1},(6N-3,6N-1) \}\,, \qquad \mathfrak{g} = (\mathfrak{sp}_4)_{6N}
}

\subsubsection*{$\mathring{\mathcal T}^{(N)}_{E_6,2}(0,1,0,0,1)$}

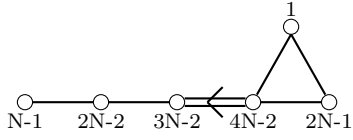
\al{&\begin{tikzpicture}
\filldraw[fill= white] (2,0) circle [radius=0.1] node[below] {\scriptsize N-1};
\filldraw[fill= white] (3,0) circle [radius=0.1] node[below] {\scriptsize 2N-2};
\filldraw[fill= white] (4,0) circle [radius=0.1] node[below] {\scriptsize 3N-2};
\filldraw[fill= white] (5,0) circle [radius=0.1] node[below] {\scriptsize 4N-2};
\filldraw[fill= white] (6,0) circle [radius=0.1] node[below] {\scriptsize  2N-1};
\filldraw[fill= white] (5.5,1) circle [radius=0.1] node[above] {\scriptsize  1};
\draw [thick] (2.1, 0) -- (2.9,0) ;
\draw [thick] (4.1, 0.05) -- (4.9,0.05) ;
\draw [thick] (4.1, -0.05) -- (4.9,-0.05) ;
\draw [thick] (3.1, 0) -- (3.9,0) ;
\draw [thick] (5.1, 0) -- (5.9,0) ;
\draw [thick] (4.4,0) -- (4.6,0.2);
\draw [thick] (4.4,0) -- (4.6,-0.2);
\draw [thick] (5+0.05,0.1) -- (5.5,0.9);
\draw [thick] (6-0.05,0.1) -- (5.5,0.9);
\end{tikzpicture} \\
&24(c-a) = 12N-8\,, \qquad 4(2a-c) = 18N^2-17N+4\,,\\
&\Delta = \{\underbrace{(6k-3,6k-1,6k)}_{k=1,\dots,N-1},6N-3 \}\qquad \mathfrak{g} = (\mathfrak{sp}_2)_{6N-2} \oplus (\mathfrak{su}_2)_{12N-12}\,.
}

\subsubsection*{$\mathring{\mathcal T}^{(N)}_{E_6,2}(0,0,0,1,1)$}

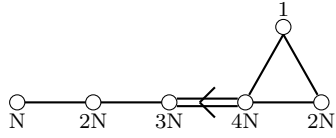
\al{&\begin{tikzpicture}
\filldraw[fill= white] (2,0) circle [radius=0.1] node[below] {\scriptsize N};
\filldraw[fill= white] (3,0) circle [radius=0.1] node[below] {\scriptsize 2N};
\filldraw[fill= white] (4,0) circle [radius=0.1] node[below] {\scriptsize 3N};
\filldraw[fill= white] (5,0) circle [radius=0.1] node[below] {\scriptsize 4N};
\filldraw[fill= white] (6,0) circle [radius=0.1] node[below] {\scriptsize  2N};
\filldraw[fill= white] (5.5,1) circle [radius=0.1] node[above] {\scriptsize  1};
\draw [thick] (2.1, 0) -- (2.9,0) ;
\draw [thick] (4.1, 0.05) -- (4.9,0.05) ;
\draw [thick] (4.1, -0.05) -- (4.9,-0.05) ;
\draw [thick] (3.1, 0) -- (3.9,0) ;
\draw [thick] (5.1, 0) -- (5.9,0) ;
\draw [thick] (4.4,0) -- (4.6,0.2);
\draw [thick] (4.4,0) -- (4.6,-0.2);
\draw [thick] (5+0.05,0.1) -- (5.5,0.9);
\draw [thick] (6-0.05,0.1) -- (5.5,0.9);
\end{tikzpicture} \\
&24(c-a) = 12N\,, \qquad 4(2a-c) = 18N^2+7N\,,\\
&\Delta = \{\underbrace{(6k-3,6k-1,6k)}_{k=1,\dots,N} \}\,, \qquad \mathfrak{g} = (\mathfrak{su}_4)_{12N}\,.
}

\subsubsection*{$\mathring{\mathcal T}^{(N)}_{E_6,2}(0,0,1,0,0)$}

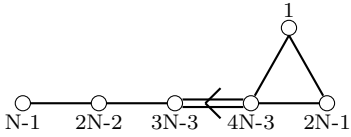
\al{&\begin{tikzpicture}
\filldraw[fill= white] (2,0) circle [radius=0.1] node[below] {\scriptsize N-1};
\filldraw[fill= white] (3,0) circle [radius=0.1] node[below] {\scriptsize 2N-2};
\filldraw[fill= white] (4,0) circle [radius=0.1] node[below] {\scriptsize 3N-3};
\filldraw[fill= white] (5,0) circle [radius=0.1] node[below] {\scriptsize 4N-3};
\filldraw[fill= white] (6,0) circle [radius=0.1] node[below] {\scriptsize  2N-1};
\filldraw[fill= white] (5.5,1) circle [radius=0.1] node[above] {\scriptsize  1};
\draw [thick] (2.1, 0) -- (2.9,0) ;
\draw [thick] (4.1, 0.05) -- (4.9,0.05) ;
\draw [thick] (4.1, -0.05) -- (4.9,-0.05) ;
\draw [thick] (3.1, 0) -- (3.9,0) ;
\draw [thick] (5.1, 0) -- (5.9,0) ;
\draw [thick] (4.4,0) -- (4.6,0.2);
\draw [thick] (4.4,0) -- (4.6,-0.2);
\draw [thick] (5+0.05,0.1) -- (5.5,0.9);
\draw [thick] (6-0.05,0.1) -- (5.5,0.9);
\end{tikzpicture} \\
&24(c-a) = 12N-10\,, \qquad 4(2a-c) = 18N^2 -23N+7\,,\\
&\Delta = \{\underbrace{(6k-3,6k-1,6k)}_{k=1,\dots,N-1},3N-3/2 \}\,, \qquad \mathfrak{g} = (\mathfrak{su}_3)_{6N-3} \oplus (\mathfrak{su}_3)_{12N-12}
}

\subsection{$\ell =3$ quotients}

\subsubsection*{$\mathring{\mathcal T}^{(N)}_{D_4,3}(2,0,0)$}

\al{&\begin{tikzpicture}
\filldraw[fill= white] (2,0) circle [radius=0.1] node[below] {\scriptsize N-2};
\filldraw[fill= white] (3,0) circle [radius=0.1] node[below] {\scriptsize 2N-2};
\filldraw[fill= white] (4,0) circle [radius=0.1] node[below] {\scriptsize 3N-2};
\filldraw[fill= white] (5,0) circle [radius=0.1] node[below] {\scriptsize 1};
\draw [thick] (2.1, 0) -- (2.9,0) ;
\draw [thick] (3.1, 0.07) -- (3.9,0.07) ;
\draw [thick] (3.1, -0.07) -- (3.9,-0.07) ;
\draw [thick] (3.1, 0) -- (3.9,0) ;
\draw [thick] (4.1, 0.05) -- (4.9,0.05);
\draw [thick] (4.1, -0.05) -- (4.9,-0.05);
\draw [thick] (3.4,0) -- (3.6,0.2);
\draw [thick] (3.4,0) -- (3.6,-0.2);
\end{tikzpicture} \\
&24(c-a) = 12N-6\,, \qquad 4(2a-c) = 12N^2 -26N + \frac{50}{3}\,,\\
&\Delta = \{\underbrace{(6k-2,6k)}_{k=1,\dots,N-2},6N-8,2N-2/3 \}\,, \qquad \mathfrak  g =  (\mathfrak{g}_2)_{4N-4/3}\,.
}

\subsubsection*{$\mathring{\mathcal S}^{(N)}_{D_4,3}(1,0,1)$}

\al{&\begin{tikzpicture}
\filldraw[fill= white] (2,0) circle [radius=0.1] node[below] {\scriptsize N-1};
\filldraw[fill= white] (3,0) circle [radius=0.1] node[below] {\scriptsize 2N-1};
\filldraw[fill= white] (4,0) circle [radius=0.1] node[below] {\scriptsize 3N-1};
\filldraw[fill= white] (5,0) circle [radius=0.1] node[below] {\scriptsize 1};
\draw [thick] (2.1, 0) -- (2.9,0) ;
\draw [thick] (3.1, 0.07) -- (3.9,0.07) ;
\draw [thick] (3.1, -0.07) -- (3.9,-0.07) ;
\draw [thick] (3.1, 0) -- (3.9,0) ;
\draw [thick] (4.1, 0.05) -- (4.9,0.05);
\draw [thick] (4.1, -0.05) -- (4.9,-0.05);
\draw [thick] (3.4,0) -- (3.6,0.2);
\draw [thick] (3.4,0) -- (3.6,-0.2);
\end{tikzpicture} \\
&24(c-a) = 12N-6\,, \qquad 4(2a-c) = 12N^2-6N+1\,,\\
&\Delta = \{\underbrace{(6k-2,6k)}_{k=1,\dots,N-1},6N-2\}\,, \qquad \mathfrak g =  (\mathfrak{su}_2)_{12N-2}\,.
}

\subsubsection*{$\mathring{\mathcal R}^{(N)}_{D_4,3}(0,0,2)$}

\al{&\begin{tikzpicture}
\filldraw[fill= white] (2,0) circle [radius=0.1] node[below] {\scriptsize N};
\filldraw[fill= white] (3,0) circle [radius=0.1] node[below] {\scriptsize 2N};
\filldraw[fill= white] (4,0) circle [radius=0.1] node[below] {\scriptsize 3N};
\filldraw[fill= white] (5,0) circle [radius=0.1] node[below] {\scriptsize 1};
\draw [thick] (2.1, 0) -- (2.9,0) ;
\draw [thick] (3.1, 0.07) -- (3.9,0.07) ;
\draw [thick] (3.1, -0.07) -- (3.9,-0.07) ;
\draw [thick] (3.1, 0) -- (3.9,0) ;
\draw [thick] (4.1, 0.05) -- (4.9,0.05);
\draw [thick] (4.1, -0.05) -- (4.9,-0.05);
\draw [thick] (3.4,0) -- (3.6,0.2);
\draw [thick] (3.4,0) -- (3.6,-0.2);
\end{tikzpicture} \\
&24(c-a) = 12N\,, \qquad 4(2a-c) = 12N^2+6N\,,\\
&\Delta = \{\underbrace{(6k-2,6k)}_{k=1,\dots,N}\}\,, \qquad \mathfrak g =  (\mathfrak{su}_3)_{12N}\,.
}

\subsubsection*{$\mathring{\mathcal T}^{(N)}_{D_4,3}(0,1,0)$}

\al{&\begin{tikzpicture}
\filldraw[fill= white] (2,0) circle [radius=0.1] node[below] {\scriptsize N-1};
\filldraw[fill= white] (3,0) circle [radius=0.1] node[below] {\scriptsize 2N-2};
\filldraw[fill= white] (4,0) circle [radius=0.1] node[below] {\scriptsize 3N-2};
\filldraw[fill= white] (5,0) circle [radius=0.1] node[below] {\scriptsize 1};
\draw [thick] (2.1, 0) -- (2.9,0) ;
\draw [thick] (3.1, 0.07) -- (3.9,0.07) ;
\draw [thick] (3.1, -0.07) -- (3.9,-0.07) ;
\draw [thick] (3.1, 0) -- (3.9,0) ;
\draw [thick] (4.1, 0.05) -- (4.9,0.05);
\draw [thick] (4.1, -0.05) -- (4.9,-0.05);
\draw [thick] (3.4,0) -- (3.6,0.2);
\draw [thick] (3.4,0) -- (3.6,-0.2);
\end{tikzpicture} \\
&24(c-a) = 12N-5\,, \qquad 4(2a-c) = 12N^2-14 N +\frac{11}{3}\,,\\
&\Delta = \{\underbrace{(6k-2,6k)}_{k=1,\dots,N-1},2N-2/3\}\,, \qquad \mathfrak g =  (\mathfrak{su}_2)_{4N-2/3}\oplus (\mathfrak{su}_2)_{12N-12}\,.
}

\subsection{$\ell = 4$ quotients}

\subsubsection*{$\mathring{\mathcal T}^{(N)}_{\mathcal H_2,4} (2,0) $}

\al{&\begin{tikzpicture}
\filldraw[fill= white] (2,0) circle [radius=0.1] node[below] {\scriptsize N-2};
\filldraw[fill= white] (3,0) circle [radius=0.1] node[below] {\scriptsize 2N-2};
\filldraw[fill= white] (4,0) circle [radius=0.1] node[below] {\scriptsize 1};
\draw [thick] (2.1, 0.06) -- (2.9,0.06) ;
\draw [thick] (2.1, 0.02) -- (2.9,0.02) ;
\draw [thick] (2.1,-0.02) -- (2.9,-0.02) ;
\draw [thick] (2.1, -0.06) -- (2.9,-0.06) ;
\draw [thick] (3.1, 0.06) -- (3.9,0.06) ;
\draw [thick] (3.1, 0.02) -- (3.9,0.02) ;
\draw [thick](3.1,-0.02) -- (3.9,-0.02) ;
\draw [thick] (3.1, -0.06) -- (3.9,-0.06) ;
\draw [thick] (2.4,0) -- (2.6,0.2);
\draw [thick] (2.4,0) -- (2.6,-0.2);
\end{tikzpicture} \\
&24(c-a) = 3N-5\,, \qquad 4(2a-c) = 12N^2 -27N +17\,,\\
&\Delta = \{\underbrace{(6k-2,6k)}_{k=1,\dots,N-2},6N-8,3/2 N -1/2 \}\,, \qquad \mathfrak g =  (\mathfrak{su}_2)_{3N-3}\,.
}

\subsubsection*{$\mathring{\mathcal T}^{(N)}_{\mathcal H_2,4} (0,2) $}

\al{&\begin{tikzpicture}
\filldraw[fill= white] (2,0) circle [radius=0.1] node[below] {\scriptsize N-1};
\filldraw[fill= white] (3,0) circle [radius=0.1] node[below] {\scriptsize 2N-2};
\filldraw[fill= white] (4,0) circle [radius=0.1] node[below] {\scriptsize 1};
\draw [thick] (2.1, 0.06) -- (2.9,0.06) ;
\draw [thick] (2.1, 0.02) -- (2.9,0.02) ;
\draw [thick] (2.1,-0.02) -- (2.9,-0.02) ;
\draw [thick] (2.1, -0.06) -- (2.9,-0.06) ;
\draw [thick] (3.1, 0.06) -- (3.9,0.06) ;
\draw [thick](3.1, 0.02) -- (3.9,0.02) ;
\draw [thick] (3.1,-0.02) -- (3.9,-0.02) ;
\draw [thick] (3.1, -0.06) -- (3.9,-0.06) ;
\draw [thick] (2.4,0) -- (2.6,0.2);
\draw [thick] (2.4,0) -- (2.6,-0.2);
\end{tikzpicture} \\
&24(c-a) = 3N-4\,, \qquad 4(2a-c) = 12N^2 -15 N +4 \,,\\
&\Delta = \{\underbrace{(6k-2,6k)}_{k=1,\dots,N-1},3/2N-1/2 \}\,, \qquad \mathfrak g =  (\mathfrak{su}_2)_{12N-12}\,.
}

\subsubsection*{$\mathring{\mathcal S}^{(N)}_{\mathcal H_2,4} (1,1) $}

\al{&\begin{tikzpicture}
\filldraw[fill= white] (2,0) circle [radius=0.1] node[below] {\scriptsize N-1};
\filldraw[fill= white] (3,0) circle [radius=0.1] node[below] {\scriptsize 2N-1};
\filldraw[fill= white] (4,0) circle [radius=0.1] node[below] {\scriptsize 1};
\draw [thick] (2.1, 0.06) -- (2.9,0.06) ;
\draw [thick] (2.1, 0.02) -- (2.9,0.02) ;
\draw [thick] (2.1,-0.02) -- (2.9,-0.02) ;
\draw [thick] (2.1, -0.06) -- (2.9,-0.06) ;
\draw [thick] (3.1, 0.06) -- (3.9,0.06) ;
\draw [thick](3.1, 0.02) -- (3.9,0.02) ;
\draw [thick] (3.1,-0.02) -- (3.9,-0.02) ;
\draw [thick] (3.1, -0.06) -- (3.9,-0.06) ;
\draw [thick] (2.4,0) -- (2.6,0.2);
\draw [thick] (2.4,0) -- (2.6,-0.2);
\end{tikzpicture} \\
&24(c-a) = 3N-3\,, \qquad 4(2a-c) = 12N^2 -6N+1\,,\\
&\Delta = \{\underbrace{(6k-2,6k)}_{k=1,\dots,N-1},6N-2 \}\,, \qquad \mathfrak g = \varnothing\,.
}

\end{appendix}

\bibliographystyle{JHEP}

\providecommand{\href}[2]{#2}\begingroup\raggedright\endgroup

\end{document}